\documentclass[12pt,a4paper]{report}
\usepackage{amssymb,graphicx,math,bezier}
\usepackage{epsfig}
\usepackage{color}
\UseRawInputEncoding
 \newskip\foobarskip
 \makeatletter
 \def\ps@headings{
 \def\@oddfoot{}%
 \def\@oddhead{\makebox[\textwidth][l]{\underline{\hbox to \textwidth{\sc
 \firstmark\hfill\thepage}}}}%
 \def\@evenfoot{}%
 \def\@evenhead{\makebox[\textwidth][r]{\underline{\hbox to \textwidth{\sc
 \thepage\hfill\@lhead}}}}%
 \def\chaptermark##1{\mark{}\def\@lhead{##1}}%
 \def\sectionmark##1{{\let\protect\noexpand\mark{\thesection
 \hskip 1em##1}}}
 }
\def \gv#1{\mbox{\boldmath $#1$}}
\renewcommand \vec \gv
\def\tder{\@ifnextchar[{\@itder}{\@itder[]}}
\begin{document}
 \baselineskip .5cm
\def\noi{\noindent}
\begin{titlepage}
  \begin{center}
      {\sc 
      {\LARGE
      Modelli idrodinamici per la verifica della\\
      dinamica di navi in avanzamento\\
      }}
      \vskip 3.0cm
      {\Large {\em Andrea Colagrossi}}
    \end{center}

  \vskip 2.0cm
\begin{figure}[htb]
      \epsfxsize=1.0\textwidth
      \epsfxsize=1.0\textwidth
      \makebox[1.0\textwidth]{\epsfbox{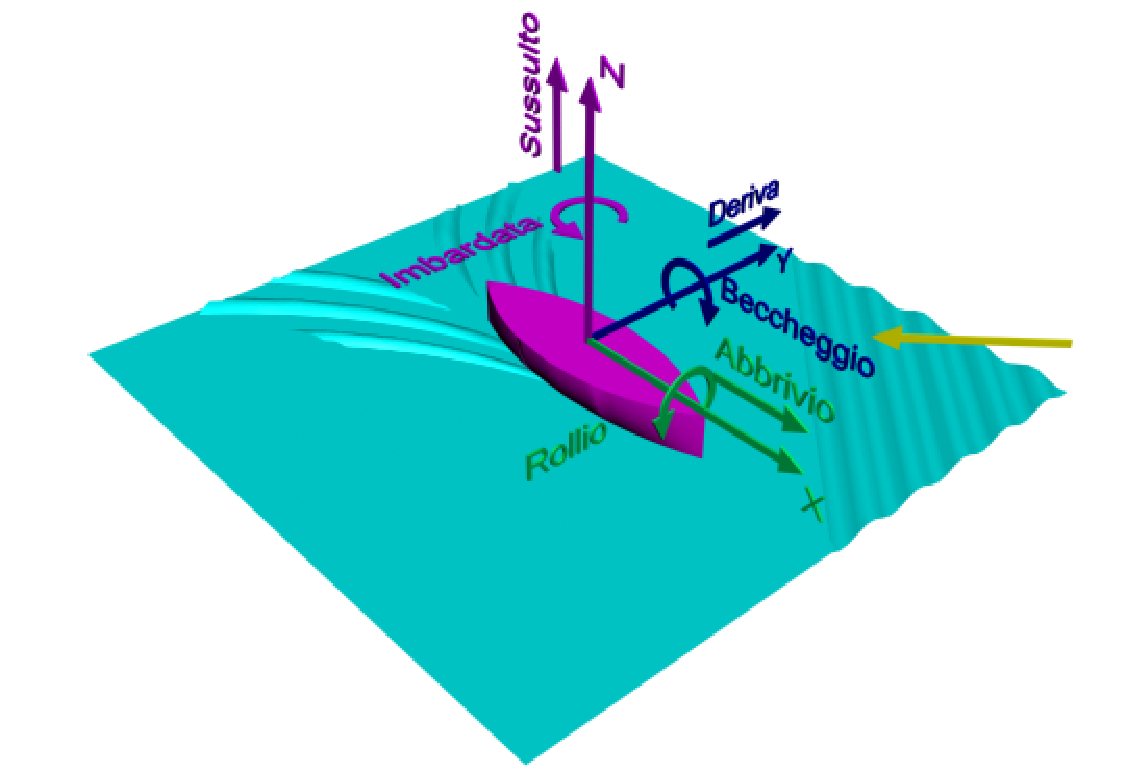}}
\end{figure}

  \vskip 2.0cm

\end{titlepage}
\def\a{\`a }
\def\e{\`e }
\def\i{\`i }
\def\o{\`o }
\def\u{\`u }
\def\h{{\'e} }
\def\be{\begin{equation}}
\def\ee{\end{equation}}
\def\dsty{\displaystyle}
\def\ssty{\scriptscriptstyle}
 \def\oneh{\frac{1}{2}}
 \def\imu{\mbox{i}\,}
 \def\K{\,{\cal K}(\hat z,z)\,}

 \def\J{{\cal J}}
 \def\Jc{{\cal J}_c(\theta)}
 \def\Jck{{\cal J}_{c\,_k}}
 \def\Kc{\,{\cal K}(\hat z,z(\theta))\,}
 \def\Ks{\,\frac{\imu \sigma(z)}{\tau_c(z)}\frac{1}{\hat{z}-z}\,}
 \def\Ksper{\, \frac{\imu \sigma(z)}{\tau_c(z)}
              \frac{\pi}{\Lambda}\cot(\frac{\pi}{\Lambda}(\hat{z}-z))\,}

 \def\Jsl{{\cal J}_{sl}(\xi)}
 \def\Ksl{\,{\cal K}(\hat z,z(\xi))\,}
 \def\Jsl{{\cal J}_{sl}(\xi)}
 \def\Kq{\,q(z)\frac{1}{\hat{z}-z}\,}
 \def\Kqper{\,q(z)\frac{\pi}{\Lambda}\cot(\frac{\pi}{\Lambda}(\hat{z}-z))\,}

 \def\SL{\partial \Omega_{sl}}
 \def\CR{\partial \Omega_{c}}
 \def\FO{\partial \Omega_{p}}
 \def\DO{\Omega}
 \def\SLCR{\partial \Omega_{sl} \cup \partial \Omega_{c}}
 \def\dxi{\partial_\xi}
 \def\dze{\partial_\zeta}
 \def\dl{d\ell}
 \newcommand{\D}[2]{\frac{D\,#1}{D\,#2}}
 \newcommand{\Dxi}[1]{\dsty \frac{d\,#1}{d\,\xi}}
 \newcommand{\Dze}[1]{\dsty \frac{d\,#1}{d\,\zeta}}
 \newcommand{\Dth}[1]{\dsty \frac{d\,#1}{d\,\theta}}

 \renewcommand \vec \gv
 \bibliographystyle{plain}

\newpage
\[
\begin{array}{lll}
\\\\ \\\\
\mbox{Si ringrazia l'Istituto Nazionale per Studi} \\
\mbox{ed Esperienze di Architettura Navale per }\\
\mbox{la disponibilit\a di materiale tecnologico e scientifico,} \\
\mbox{indispensabile per il buon svolgimento di questa Tesi.}
\\ \\
\mbox{Si ringraziano inoltre Claudio Lugni e Marilena Greco} \\  
\mbox{per la loro costante attenzione e per i numerosi suggerimenti forniti.} \\ 
\end{array}
\]

\vspace*{6.0cm}

\noindent
\textbf{ISBN  \hspace*{0.48cm}   : \hspace{0.650cm} 88-7617-021-9} \\ [0.25cm]
\noindent
\textbf{ISBN 13 : 978-88-7617-021-8} \\

\vskip 1.0cm
\begin{figure}[htb]
      \epsfxsize=0.33\textwidth
      \epsfxsize=0.33\textwidth
      \makebox[0.33\textwidth]{\epsfbox{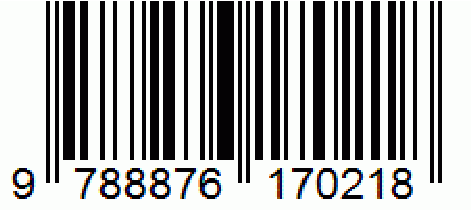}}
\end{figure} 

 \newpage
\begin{center} \textbf{SOMMARIO} \end{center}
\vspace*{-0.20cm}
\small{ 
In questa lavoro viene studiato il problema della previsione
dei carichi e dei moti indotti da sistemi ondosi su una nave in 
avanzamento {\em (tenuta al mare)}.

Assumendo che la carena sia indeformabile, il moto della nave 
\e descritto dalle equazioni della meccanica del corpo rigido.
Il fenomeno idrodinamico viene analizzato mediante lo schema di fluido 
non viscoso in moto irrotazionale e conduce ad un problema ai valori
iniziali per l'equazione di Laplace, accoppiato con quello del moto 
della nave, caratterizzato da una forte non linearit\a dovuta alla 
presenza di frontiere mobili (la superficie della carena e 
l'interfaccia aria--acqua).

Pertanto, il problema matematico \e stato ulteriormente semplificato supponendo
di piccola ampiezza il moto della nave e giungendo ad un modello lineare per 
il problema fluidodinamico.

In questo contesto vengono sviluppati due approcci: il primo nel
dominio della frequenza ed il secondo nel dominio del tempo.
Per entrambe le formulazioni sono stati implementati codici di calcolo 
ed \e stata ottenuta una ampia serie di risultati per carene di nave di 
complessit\a geometrica crescente. 

Nel caso del modello nel dominio della frequenza sono state studiate
sistematicamente alcune forme di carena ed il confronto con i dati 
sperimentali \e caratterizzato da un soddisfacente accordo.
Molto interessante \e anche il paragone con i risultati ottenibili mediante 
i modelli comunemente utilizzati nella progettazione basati
sull'ipotesi di locale bidimensionalit\a del flusso ({\em strip theory}).
In questo caso \e stato osservato un miglioramento delle previsioni ottenibili 
che evidenzia l'importanza quantitativa degli effetti tridimensionali nella 
tenuta al mare.
Globalmente i risultati ottenuti appaiono sufficientemente accurati da
suggerire l'applicazione del codice alla verifica in sede di progetto
della tenuta al mare di navi.

Pur se meno accurati, i modelli quasi tridimensionale sono  
decisamente meno onerosi da un punto di vista computazionale.
Per tale motivo \e stata impostata una formulazione del medesimo problema 
nel dominio del tempo. 
Infatti, rimanendo nell'ambito di un modello lineare, in linea di principio 
\e possibile dedurre le stesse informazioni valutate mediante i modelli in 
frequenza utilizzando un piccolo numero di 'prove numeriche' nel dominio
del tempo.
Il codice realizzato \e stato applicato al problema di riferimento della
partenza di una carena in assenza di onde ed il soddisfacente confronto con 
risultati sperimentali e numerici relativi a codici stazionari indica
le buone potenzialit\a del metodo. Tali propriet\a sono confermate dai i
primi risultati relativi alla simulazione del moto forzato di carene
in avanzamento.

Per quanto riguarda la prima formulazione, al crescere della frequenza \e 
stata osservata un aumento della difficolt\a nel riprodurre i risultati 
sperimentali, forse indicativa dell'insorgere di fenomeni non lineari 
non inclusi nel modello o, pi\u in generale, legati alla viscosit\a.
Per queste circostanze sembra opportuno un ulteriore approfondimento. \\
Il modello nel dominio del tempo non \e stato applicato in maniera
altrettanto estesa ma i primi risultati ne suggeriscono l'ulteriore
studio per verificarne o meno l'efficacia computazionale. \\
Infine, per entrambe le formulazioni, appare importante lo sviluppo
di tecniche di soluzione delle equazioni integrali che da un lato
ne accelerino la convergenza (elementi di ordine superiore) e
dall'altro riducano i tempi di calcolo (decomposizione in sottodomini
e sviluppo in multipoli con algoritmi di somma veloce).
Anche per questi aspetti \e doveroso suggerire un ulteriore sviluppo
dell'attivit\a per poter proporre i modelli tridimensionali come
efficaci metodi di verifica alla portata del progettista.
}
 \newpage
\begin{center} \textbf{ABSTRACT} \end{center}
\vspace*{-0.20cm}
\small{ 
This work studies the problem of predicting the loads and motions induced by wave systems on a ship in forward motion (seakeeping). Assuming that the hull is rigid, the motion of the ship is described by the equations of rigid body mechanics. The hydrodynamic phenomenon is analyzed using the inviscid fluid scheme in irrotational motion, leading to an initial value problem for Laplace¿s equation, coupled with the ship's motion, characterized by strong non-linearity due to the presence of moving boundaries (the hull surface and the air-water interface).

Therefore, the mathematical problem has been further simplified by assuming small amplitude ship motion, resulting in a linear model for the fluid-dynamic problem. In this context, two approaches are developed: one in the frequency domain and the other in the time domain. Computational codes have been implemented for both formulations, and a wide range of results has been obtained for ship hulls of increasing geometric complexity.

In the case of the frequency-domain model, several hull shapes were systematically studied, and the comparison with experimental data showed satisfactory agreement. Of particular interest is the comparison with results obtained from commonly used design models based on the assumption of local two-dimensional flow (strip theory). In this case, an improvement in the predictability was observed, highlighting the quantitative importance of three-dimensional effects in seakeeping. Overall, the results appear sufficiently accurate to suggest the application of the code for seakeeping verification in ship design.

Although less accurate, the quasi-three-dimensional models are significantly less computationally expensive. For this reason, a formulation of the same problem in the time domain was developed. Indeed, within the scope of a linear model, it is theoretically possible to obtain the same information as the frequency-domain models by performing a small number of ¿numerical trials¿ in the time domain. The developed code was applied to the reference problem of the departure of a hull in the absence of waves, and the satisfactory comparison with experimental and numerical results from stationary codes indicates the good potential of the method. These properties are confirmed by the first results related to the simulation of forced motion of hulls in forward motion.

Regarding the first formulation, as frequency increases, a greater difficulty in reproducing experimental results was observed, possibly indicating the onset of nonlinear phenomena not included in the model or, more generally, related to viscosity. Further investigation seems necessary in such cases.

The time-domain model has not been applied as extensively, but initial results suggest further study to verify its computational effectiveness. Finally, for both formulations, the development of solution techniques for integral equations appears important, both to accelerate convergence (higher-order elements) and to reduce computational times (domain decomposition and development in multipoles with fast summation algorithms). Further development of these aspects is also recommended to propose three-dimensional models as effective verification methods accessible to designers.
}
 \tableofcontents
\newpage
\pagenumbering{arabic}%
\chapter{Problemi di interesse in campo navale}
In questa lavoro si studia il problema della {\em tenuta al mare} delle navi, 
ossia la previsione dei moti e dei carichi indotti sulla carena di un veicolo 
marino che si muove in presenza di sistemi ondosi preesistenti.

Prima di addentrarci nei dettagli fisici e matematici che 
caratterizzano il problema idrodinamico e discutere gli aspetti numerici
della sua soluzione, sembra interessante descrivere per linee
generali il ruolo che strumenti teorico--numerici come  quelli
sviluppati ed applicati in questa tesi hanno o possono avere 
nei riguardi della progettazione di un veicolo marino.

\section{Attuali tendenze nella progettazione delle navi}
Storicamente, ci\o che maggiormente qualifica una nave agli occhi 
dell'armatore, ossia di colui che gestir\a il veicolo marino, sono le {\em
prestazioni}. Pi\u precisamente, questi \e interessato
al fatto che il {\em carico pagante} sia trasportato ad una data 
velocit\a di crociera e con un prestabilito consumo specifico.
Ulteriori requisiti sono spesso peculiari del particolare utilizzo che la 
nave dovr\a avere. Ad esempio il naviglio militare deve essere veloce e 
rimanere operativo in ogni condizione di mare.

Avendo tali requisiti {\em globali} da soddisfare e data la complessit\a dei 
fenomeni idrodinamici connessi con il moto di una nave, per poter ottimizzare 
il progetto si f\a spesso ricorso a prove su modelli in scala che, per ovvi 
motivi economici, dovranno essere in numero limitato.
Si pu\o facilmente immaginare come le prove sperimentali siano
tanto pi\u costose quanto pi\u si vuole approssimare il comportamento
in condizioni operative 'reali'. Ad esempio nelle prove di tenuta al
mare il modello si muove (spesso in autopropulsione ed in
assetto libero) attraversando le onde generate mediante un ondogeno 
posto all'estremit\a del bacino di prova.

Per ridurre i costi di progettazione si cerca allora di far affidamento 
su regole semi--empiriche basate sulle passate esperienze. 
Il risultato \e che i progetti evolvono lentamente e da un prototipo di nave 
al successivo raramente si hanno grosse novit\a nel disegno dello scafo.

In particolare, la carena \footnote{termine navale per indicare la parte 
immersa della nave, viene anche chiamata {\em opera viva}}
di una nave viene in prima istanza progettata basandosi su analisi statiche o 
quasi--statiche. Solo in una seconda fase, di verifica, si terr\a conto del
fatto che la struttura operer\a in condizioni ambientali che ben difficilmente
consentiranno alla nave il {\em moto rettilineo uniforme in acqua calma}.

A tale riguardo, in seguito all'irrigidimento delle normative internazionali 
sul naviglio convenzionale e alla crescente tendenza allo sviluppo di 
veicoli marini sempre pi\u veloci, un peso maggiore fra i parametri di 
progettazione \e stato assunto dalle caratteristiche di {\em manovrabilit\a} 
e di {\em tenuta al mare} del mezzo.

Per grandi linee nel {\em problema della manovrabilit\a} si
studia il comportamento intrinsecamente non stazionario di una nave, pensata 
come un corpo rigido, quando \e sottoposta ad azioni {\em di controllo}
come quelle causate dal timone, da pinne stabilizzatrici, o, ancora, 
dall'uso dei motori.
Quindi, in questo settore, rientrano sia il problema della stabilit\a della
rotta ({\em course--keeping}) sia quello della determinazione dell'abilit\a 
evolutiva ({\em turning ability}) della nave.
La manovra \e in generale caratterizzata da ampi moti di imbardata e di deriva. 
Da un punto di vista idrodinamico ci\o comporta una forte generazione
di vorticit\a da parte della carena che in prima approssimazione
pu\o essere immaginata come un'ala di piccolo allungamento in moto
non stazionario.
In circostanze molto frequenti, almeno per navi tradizionali, il moto di
manovra si svolge a basse velocit\a 
\footnote{
In ambito navale la velocit\a di avanzamento viene adimensionalizzata nel numero di Froude così definito: 
$Fr\,={\frac{U}{\sqrt{gL}}}$,  
$U$ =velocit\a di avanzamento $L$=lunghezza della carena, $g$ accelerazione gravitazionale
}  
,tali quindi da non provocare una significativa generazione di onde, 
almeno dal punto di vista della loro influenza sulla forza laterale e sul momento di imbardata necessari ad una virata.

Nei problemi di {\em tenuta al mare} ({\em seakeeping}) si \e interessati 
alla risposta della nave, pensata rigida, alle azioni esterne causate 
da sistemi di onde incidenti. 
Tale risposta pu\o essere studiata sia in campo deterministico che in campo 
stocastico.
In particolare la risposta in campo probabilistico \e richiesta come giudizio 
di merito quantitativo per un {\em sicuro} comportamento del veicolo marino.
L'obiettivo \e quello di determinare i carichi idrodinamici al fine di
valutare le massime ampiezze dei moti. 
Chiaramente, diversamente dal caso della manovrabilit\a di navi 
convenzionali, l'interazione della nave con la superficie libera
\e il fenomeno idrodinamico predominante.
Piuttosto,  in tali circostanze, gli effetti della viscosit\`a
si possono ritenere poco rilevanti e, nella descrizione matematica
del problema, viene solitamente trascurata la generazione della scia 
vorticosa che, a rigore, \e presente.

Una volta valutati i carichi agenti sulla carena, questa pu\o
essere studiata da un punto di vista strutturale. 
Tuttora, per navi sufficientemente lente e comunque in condizioni non estreme,
l'analisi strutturale adottata \e di natura statica o quasi statica.
In ogni caso, si assume che la sollecitazione idrodinamica non sia
influenzata dalla deformabilit\a della carena.

Con l'aumentare delle velocit\a e, comunque, in condizioni di 
esercizio gravose (moti di grande ampiezza) la deformabilit\a
dello scafo assume un ruolo quantitativamente importante 
nell'influenzare il campo idrodinamico. In tali circostanze diviene 
significativo lo studio del {\em problema idroelastico}. 
In questo la forma della struttura \e alterata dalle azioni
idrodinamiche e, a sua volta,  il campo idrodinamico risente
dei moti (rigidi e non) della carena. Pertanto
il problema fluidodinamico \e accoppiato non solo a quello
del moto 'rigido' della nave ma anche ad un problema di dinamica 
strutturale.

Addirittura, nella sua forma pi\u generale, l'idroelasticit\a 
comprender\a oltre alla tenuta al mare anche la manovrabilit\a, così  
come l'aeroelasticit\a comprende come caso speciale la meccanica del volo 
in cui l'aeromobile \e ipotizzato rigido.

\section{La tenuta al mare}
Le probabilit\a di cedimento strutturale o di comportamento
dinamico anomalo di una nave devono essere quanto pi\u possibile
contenute per numerosi, ovvi, motivi.
Basta immaginare i problemi di impatto ambientale conseguenti il
danneggiamento di nave adibite al trasporto di petrolio, 
agenti chimici o scorie di vario tipo o, ancora, la perdita di
vite umane.

Altri requisiti consistono nel contenimento delle ampiezze dei
moti indotti dal moto ondoso per motivi di benessere e operativit\a
degli occupanti. Un analogo requisito \e rilevante per navi posa--tubi
o per porta--aerei e porta--elicotteri.

Tali specifiche vengono solitamente espresse in termini di
probabilit\a che una data variabile 'ecceda' un limite prefissato
(massimo angolo di rollio, massime tensioni, ecc.). 

Al riguardo, i criteri di accettabilit\`a sono fissati dai governi, 
dagli istituti di classificazione e controllo o da altre autorit\a ufficiali 
internazionali per il tramite di 'regole' estremamente semplici e 
basate su decennale, quando non secolare, esperienza costruttiva.

In questo ambito si st\a sviluppando, in considerazione delle attuali
tendenze allo sviluppo di veicoli molto
veloci nei quali vengano impiegati materiali leggeri e quindi 
flessibili, il bisogno di criteri basati su metodi pi\u 'analitici'
e che non richiedano, almeno per grossa parte dell'attivit\a di sviluppo,
il ricorso a prove su modelli fisici.

Ci\o ha giustificato la definizione di modelli fluidodinamici per la 
descrizione dei fenomeni idrodinamici rilevanti per la tenuta al
mare e di metodi numerici per la soluzione dei relativi problemi
matematici.
Nel seguito, rinunciando a priori alla possibilit\a di sviluppare 
modelli generali basati sulle Equazioni di Navier--Stokes per la
descrizione del flusso non stazionario con frontiera libera,
si ricorrer\a ad una descrizione a potenziale del campo di flusso
sotto la ulteriore ipotesi di piccole ampiezze dei moti della nave
e piccola elevazione d'onda.

\subsection{Ipotesi di linearit\a e di flusso a 
            potenziale: limiti di validit\a}
Distinguiamo i moti rigidi della nave in moti longitudinali, ossia 
appartenenti al piano $y=0$, e trasversali.
I longitudinali sono indicati con: abbrivio, sussulto e beccheggio; 
quelli trasversali con: deriva, rollio, imbardata.

Quando una nave esegue una manovra i principali moti di interesse sono  
quelli nel piano orizzontale (deriva, imbardata e, meno rilevante, abbrivio)
cui, come accennato, \e associata la generazione di una scia vorticosa che 
rende necessari modelli in cui le caratteristiche rotazionali del campo 
fluidodinamico siano prese in opportuna considerazione.
In tali circostanze, almeno per navi convenzionali,
la velocit\a di manovra \e piccola rispetto a quella tipica di crociera
e gli effetti delle onde di gravit\a possono essere trascurati.

Al contrario, nei problemi di {\em tenuta al mare} l'effetto fluidodinamico 
predominante \e dato dall'azione prodotta dalle onde di gravit\a sullo scafo
e, pertanto, il moto ondoso \e la caratteristica fondamentale del fenomeno in 
esame.
Anzi, rispetto al caso di una nave in manovra, gli 'angoli di attacco'
che si realizzano sono relativamente piccoli e gli effetti legati alla
generazione di scie vorticose sono corrispondentemente minori.
In questo ambito si pu\o comprendere come una descrizione 
quantitativamente accurata del flusso attorno ad una carena in avanzamento 
in mare {\em mosso} possa essere ottenuta mediante il modello di
fluido non viscoso in moto irrotazionale.

\noindent
Pur con questa significativa semplificazione, il problema cui si giunge
\e di notevole difficolt\`a: la carena \e libera di muoversi sotto
l'azione delle forze idrodinamiche la cui entit\a dipende dal moto della
carena stessa. La non linearit\a insita nella 'deformabilit\a' del dominio
\e accentuata dalla presenza dell'interfaccia aria--acqua la cui 
configurazione \e a priori incognita e v\a determinata come parte 
della soluzione.

\noindent
In considerazione dell'eccezionale difficolt\a del problema sono
stati sviluppati modelli ulteriormente semplificati ipotizzando che
per piccole ampiezze delle onde incidenti sulla carena il conseguente
spostamento rispetto all'assetto `medio' sia piccolo: si giunge quindi a 
formulare il problema linearizzato della tenuta al mare di una
nave in avanzamento.

Vediamo ora, sulla base di considerazioni fenomenologiche, in quali 
circostanze ci si pu\o attendere che il modello linearizzato 
a potenziale sia quantitativamente affidabile. 

In generale in funzione della lunghezza d'onda, della 
provenienza e della ampiezza,
l'azione delle onde induce moti oscillatori della nave la
cui entit\a pu\o essere particolarmente rilevante.
Ad esempio, le navi dislocanti sono particolarmente
sensibili ad onde incidenti con lunghezza d'onda dell'ordine o maggiore 
della lunghezza della carena e rispondono soprattutto oscillando nel
piano verticale (sussulto e beccheggio).

Fra le possibili direzioni di interazione, il mare da prua o da 
poppa \e poi quello di maggior interesse perch\e, per evitare l'innesco di
moti con piccolo smorzamento (come il rollio), la nave viene governata
in maniera da mantenere quanto pi\u possibile questo angolo di interazione
con le onde.
I gradi di libert\a nel piano verticale sono infatti caratterizzati da 
rilevanti forze di richiamo di natura idrostatica che tendono a ripristinare
l'assetto indisturbato e sono accompagnati da una notevole generazione di 
onde che costituisce il principale meccanismo fisico di smorzamento
dell'oscillazione una volta innescata. 

Possiamo stimare facilmente l'ordine di grandezza delle pulsazioni di 
risonanza per una nave convenzionale identificandola con 
la pulsazione naturale dei modi rigidi sotto l'azione delle sole
forze di galleggiamento idrostatiche ({\em forze di richiamo}) 
e, quindi, in assenza di velocit\a di avanzamento.
In particolare, per il sussulto, considerando che la massa della nave 
\e proporzionale al volume della carena $\forall$ e che la forza di richiamo 
\e proporzionale all'area $S$ racchiusa dalla traccia della carena con 
il piano della superficie libera indisturbata, possiamo dire che la 
frequenza naturale \e stimata dalla
\be 
\dsty
 (\,\frac{gS}{\forall})^\oneh\,=\,{\cal O}(\,\frac{g}{T})^\oneh
\ee
che per navi convenzionali \e dell'ordine di 4--16 secondi \footnote{
Indicando con $L$, $B$ e $T$ le tre dimensioni della carena si ha
$\forall\simeq LBT$ e $S\simeq  LB$}.
Chiaramente il fenomeno \e {\em idrodinamico} e quindi  il 
comportamento inerziale della nave viene alterato e si osserva una 
tendenza alla riduzione della frequenza naturale pur rimanendo la stima
di cui sopra significativa.
Un regola analoga vale per il modo di beccheggio che,
in questo contesto, \e dinamicamente simile al sussulto.
L'esistenza di queste forze di richiamo permette di avere moti relativamente
piccoli in risposta all'azione di un'onda incidente eccetto che per pulsazioni
prossime alla risonanza, dove tali ampiezze sono elevate e dipendono
dall'entit\a delle forze di smorzamento.

Pertanto sembra possibile individuare nel modello
non viscoso uno strumento di analisi per la previsione 
del comportamento della carena almeno nei due principali 
modi di risposta.
Il problema fluidodinamico che ne risulta \e quello
di un flusso ideale il cui dominio ha come frontiera la carena, la 
superficie di discontinuit\a aria-acqua che \e libera di deformarsi
\footnote{l'azione della carena su questa superficie produce dell'onde che si 
propagano su di essa ({\em problema di radiazione});
mentre i sistemi ondosi preesistenti vengono deformati dalla presenza 
dello scafo ({\em problema della diffrazione}).}, e eventualmente 
il fondo marino.

Un altro moto di particolare rilievo nei riguardi della sicurezza 
\e quello di rollio.  In genere per navi convenzionali le 
{\em forze di richiamo} per tale grado di libert\a sono piccole rispetto 
a quelle che competono al sussulto e al beccheggio.
Inoltre il baricentro di solito viene posizionato in modo da ridurre
le forze di richiamo così da avere dei periodi di oscillazione naturali
pi\u lunghi e quindi delle minori accelerazioni angolari.
Anche l'effetto di dissipazione, attraverso la radiazione di sistemi 
d'onda, per il rollio \e molto limitato specialmente alle basse frequenze
naturali di risonanza. 
Il risultato \e che i moti che si ottengono per queste pulsazioni
hanno ampiezze relativamente grandi e il meccanismo di smorzamento 
predominante \e di natura viscosa.
Non esistono tutt'oggi metodi soddisfacenti per prevedere il moto di rollio
con adeguata accuratezza.
Le cose cambiano se consideriamo  dei catamarani in quanto per tali 
natanti le forze di richiamo nel modo di rollio non sono pi\u così
piccole. 

I rimanenti moti nel piano orizzontale  non sono contrastati dalle forze
di richiamo idrostatiche e quindi la risposta della nave \e di tipo
non risonante. Tuttavia i moti che conseguono possono essere ampi per 
basse frequenze e specie con onde da poppa. Particolarmente importante in 
questo contesto \e il {\em broaching} dovuto ad effetti di instabilit\a
dinamica sotto la prolungata azione di sistemi d'onda unidirezionali.
Così come nel caso del rollio, anche qui abbiamo che i modi
di abbrivio,  deriva e imbardata sono influenzati da effetti non lineari e 
viscosi significativi.
Pi\u precisamente, a differenza del moto di rollio in cui gli effetti
viscosi rappresentano l'elemento dominante nel determinare lo smorzamento,
per gli altri moti gli effetti non lineari sono i primi a dover essere
recuperati per ottenere una pi\u corretta previsione.

\subsection{Modelli quasi--tridimensionali:strip theory}
I modelli teorici per la previsione della {\em tenuta al mare} di navi si sono
evoluti in tre diverse fasi negli ultimi 50 anni.

Il pioniere del lavoro di ricerca in questo settore \e stato
Korvin--Kroukovsky che analizz\o il problema mediante un 
approccio semplificato ({\em Strip Theory}), stimolando così la ricerca 
nel settore.
Pi\u precisamente l'intuito fisico e la caratteristica 
{\em snellezza}\footnote{
Il pescaggio $T$ e la dimensione trasversale $B$ sono solitamente piccoli
rispetto alla lunghezza $L$ della carena.} delle carene convenzionali
suggerirono una analisi nella quale il campo fluidodinamico
veniva supposto localmente bidimensionale e tale da poter essere 
studiato indipendentemente dal flusso nelle sezioni adiacenti.
Le caratteristiche globali del fenomeno (per esempio i carichi)
risultavano quindi dalla semplice integrazione nella direzione
della lunghezza dei carichi agenti su ciascuna sezione.
La grande popolarit\a riscossa da questo tipo di approccio quasi-tridimensionale
\e giustificata
dai soddisfacenti risultati che si riescono ad ottenere per la previsione dei 
moti convenzionali delle navi e per la notevole semplicit\a computazionale.
Restavano per\o notevoli limitazioni sulle risposte in deriva, sui carichi
strutturali ed in genere sulle caratteristiche per la tenuta al mare per 
numeri di Froude elevati.

Negli anni '60 e '70 i numerosi studi analitici aeronautici sulla 
{\em Slender-Body Theory} hanno permesso di sviluppare una analoga teoria 
per navi con geometrie affusolate. 
La formulazione razionale del problema ha permesso di giustificare  
il perch\e la {\em Strip Theory} fosse un metodo valido per alte frequenze e 
moderati numeri di Froude. 
La restrizione delle alte frequenze per le prime {\em Slender Ship Theories}
fu poi rimossa da una teoria unificata presentata da Newman nel 1978 ed
estesa da Sclavounos (1980) per i problemi di diffrazione.
Nella met\a degli anni 80, le caratteristiche per la {\em tenuta al mare}
date dalla {\em Slender Body Theory} furono validate da misure sperimentali, 
ma divenne evidente che la {\em Slender-Body Theory} non modellava bene 
i casi ad alti numeri di Froude che richiedevano quindi la ricerca di una
soluzione completamente tridimensionale.  

Pregio essenziale dei modelli quasi--tridimensionali \e il loro ridotto
onere computazionale che li rende strumenti utilizzabili su computer
di modeste prestazioni.
Tuttavia, pur rimanendo nell'ambito di una teoria linearizzata, \e lecito
aspettarsi che gli effetti tridimensionali giuochino un
ruolo quantitativamente significativo per le carene commerciali
caratterizzate da un corpo cilindrico centrale e da brusche variazioni
di forma sia nella zona prodiera (che tipicamente termina con un bulbo)
sia nei quartieri poppieri ove trova alloggio l'apparato di propulsione o  
c'\e presenza di {\em transom}.
Queste ultime considerazioni suggeriscono lo sviluppo di modelli 
tridimensionali per lo studio del campo idrodinamico attorno ad
una nave.

\subsection{Modelli tridimensionali per la tenuta al mare}
L'approccio teorico alla base del modello sviluppato in questa tesi
\e, essenzialmente, il medesimo su cui si fondano i metodi strip theory.
Quindi il flusso viene considerato non viscoso ed irrotazionale, 
il moto della carena \e piccolo ed il disturbo della superficie libera
di piccola ampiezza rispetto alle lunghezze d'onda in gioco
(modello a potenziale linearizzato).
In questi termini le difficolt\a apparirebbero principalmente algoritmiche.
Inoltre, a differenza dei modelli strip theory in cui \e possibile
studiare i problemi bidimensionali utilizzando funzioni di Green
che soddisfano le condizioni di superficie libera linearizzate, la soluzione
del problema tridimensionale implica la discretizzazione della superficie
libera perch\'e la corrispondente funzione di Green 3D \e di particolare
onere computazionale.

A ci\o si unisce l'interessante risultato, evidenziato nell'ambito di studi
relativi al problema della resistenza d'onda,  per il quale
{\em differenti linearizzazioni portano a differenti risultati}

Queste due circostanze hanno suggerito (Nakos 1990) lo sviluppo di modelli
lineari in frequenza caratterizzati da particolari linearizzazioni.
In questa tesi verr\a impostata in maniera generale la linearizzazione
del problema (Newman 1978) e quindi verranno sviluppati due differenti 
tipi di linearizzazione.

Il problema formulato nel dominio della frequenza dipende parametricamente
da questa e richiede la soluzione di tanti problemi differenziali al
contorno quante sono le frequenze considerate.
Questa considerazione ha spinto allo studio di approcci alternativi.
In particolare, vista l'equivalenza fra analisi nel dominio del tempo 
ed analisi  nel dominio della frequenza, \e stato sviluppato un modello
linearizzato mirato allo studio di 'test numerici transitori' i quali
trasformati secondo Fourier forniscono risultati, in linea di principio,
su tutto lo spettro di frequenze.
Come controparte, l'analisi nel dominio del tempo comporta maggiori difficolt\a
legate al troncamento del dominio di calcolo.

\section{Struttura della tesi}
Prima di discutere i dettagli specifici del problema \e sembrato importante 
fornire nel capitolo 2 un quadro generale dell'analisi effettuabile mediante
i modelli che vengono sviluppati nel corso di questa tesi.
Di seguito il problema del moto arbitrario di una nave in presenza di onde
viene impostato nell'ambito della dinamica dei fluidi non viscosi.
Il problema che ne risulta, non stazionario e non lineare, viene
quindi linearizzato nel capitolo successivo ottenendone quindi una formulazione
nel dominio del tempo ed una nel dominio della frequenza.
Particolare attenzione \e stata posta nell'analisi teorica del legame 
esistente fra le due formulazioni.

Nel quinto capitolo viene introdotta la formulazione integrale per il
problema formulato nel dominio della frequenza.
Sulla base di questa formulazione discreta vengono studiati nel 
capitolo successivo numerose carene per le quali sono
disponibili risultati sperimentali di buona qualit\a e con i quali
\e stata possibile un'estesa  validazione del modello.

Infine, nel settimo capitolo, si discutono alcuni aspetti numerici 
relativi al problema formulato nel dominio del tempo e si mostrano
alcuni primi, promettenti, risultati sulla base dei quali, nell'ultimo
capitolo, vengono sviluppate alcune considerazioni conclusive e 
delineate le prospettive di sviluppo futuro dell'attivit\a di studio
numerica del problema della tenuta al mare.
\chapter[Modelli lineari per lo studio della dinamica di una nave]
          {Modelli lineari per lo studio del comportamento 
         dinamico di una nave} 

\section{La nave vista come un sistema dinamico.}
Vogliamo studiare il comportamento dinamico di una carena di nave di superficie
la cui opera viva \e immersa in un certo flusso idrodinamico;
tale flusso avr\a come parametri caratteristici
di nostro interesse l'insieme $Re\,, Fr\,, St$,
\footnote
{Numero di Reynolds  $\:Re\,=\,\frac{U_{\infty}\rho\,L}{\mu} $
 Numero di Froude    $\:Fr\,=\,\frac{U_{\infty}}{\sqrt{gL}} $\\
 Numero di Strouhal  $\:St\,=\,\frac{\omega}{2\pi}\,\frac{L}{U_{\infty}}$ 
 \\
 dove $\rho$ e $\mu$ sono rispettivamente la densit\a e la viscosit\a 
 del fluido e $\omega$ indica la pulsazione dominante nel campo fluidodinamico 
}
Parliamo di opera viva perch\h vogliamo trascurare
da subito le possibili interazioni del corpo in esame con l'aria;
tali interazioni, infatti, per i problemi che vogliamo trattare
sono di secondo ordine rispetto alle azioni idrodinamiche
\footnote{Ovviamente non \e così  per le navi la cui propulsione \e 
generata attraverso l'uso di vele}.
Per un sistema dinamico costituito da un aeromobile soggetto 
ad un certo campo aerodinamico
i numeri caratteristici saranno invece $U_{\infty}\,,Re\,, M\,, St$.
Sia il numero di Mach $M$ che il numero di Froude sono legati a fenomeni 
di propagazione ondosa, pur se
di origine diversa; la comprimibilit\a di un fluido implica la
genesi e la propagazione di onde nell'intero campo mentre per un fluido
incomprimibile che ha una superficie libera di deformarsi, 
l'origine della propagazione ondosa \e su questa frontiera del campo fluidodinamico.
Su tale osservazione si basa l'analogia Mach-Froude
secondo la quale  si possono studiare problemi di aerodinamica compressibile
bidimensionali attraverso analoghi problemi di superficie libera (con
l'ipotesi aggiuntiva di acqua bassa). Per tale analogia si ha che
\e possibile correlare l'altezza d'onda misurata con la
densit\a del problema compressibile;
la valutazione quantitativa che ne risulta non \e buona
ma si ha il vantaggio di poter osservare la propagazione ondosa
con una scala dei tempi dell'ordine dei secondi, infatti la
propagazione delle {\em onde di gravit\a} \e molto pi\u lenta
di quella dovuta alla compressione nell'aria 
\footnote{Parliamo di onde di gravit\a per sottolineare che 
stiamo trascurando la tensione superficiale dell'acqua,
gli effetti di questa si hanno infatti ad alte frequenze e sono di second'ordine
per problemi inerenti a navi o strutture con scala delle lunghezze
superiore al metro.}.
Abbiamo poi il numero di Strouhal che ci d\a l'ordine di grandezza del 
rapporto tra la frequenza dominante nel campo fluidodinamico e la 
frequenza caratteristica del trasporto dovuto alla corrente $U_{\infty}/L$.
\\
Sia per un aeromobile che per una nave siamo interessati
all'interazione di tre tipi di forze: le forze d'inerzia, le forze
aero-idrodinamiche e le forze dovute alla deformabilit\a del corpo.
Questo {\em "triangolo"} di interazione
fra le forze viene chiamato in ambito aeroelastico
{\em Triangolo di Collar}, e come concetto pu\o benissimo essere
applicato all'idroelasticit\a, basta infatti cambiare
forze aerodinamiche in forze idrodinamiche.\\
Se ci limitiamo a vedere soltanto due di queste tipi
di forze abbiamo dei sotto problemi importanti, per esempio,
se si ipotizza il corpo in esame come rigido, rimangono le azioni tra le forze
aerodinamiche e quelle inerziali: di questo tipo di problemi si
interessa la meccanica del volo in campo aeronautico.  
Se studiamo invece la sola interazione tra le forze d'inerzia e le forze
dovute alla deformabilit\a del corpo abbiamo dei problemi strutturali 
non forzati e tale studio \e di base per {\em l'analisi modale}
che permette di avere delle informazioni sulla dinamica della struttura libera;
tali informazioni saranno poi molto utili nel descrivere la dinamica
completa del sistema quando si introdurrano le azioni idro-aerodinamiche.\\

Ipotiziamo di poter descrivere il fenomeno che stiamo studiando,
con un certo insieme di grandezze che chiameremo {\em variabili di stato}
$\vec x$ del sistema; chiamiamo tale insieme {\em spazio delle fasi}.\\ 
Definiamo {\em Sistema Dinamico}:
{\em la famiglia delle trasformazioni dello spazio delle fasi in s\e stesso che
fanno passare da un certo stato "attuale" ad un altro "passato" o
"futuro", e che ha come indice un parametro reale "tempo".}\\
Si ipotizza inoltre che il fenomeno in questione sia sufficientemente
{\em regolare} da poter descrivere il sistema dinamico attraverso l'equazione
differenziale "Ordinaria":
\be    \label{sd}
 \left \{
  \begin{array}{cl}
  \dot{\vec x}\,=\,\vec v(t,\vec x)\\[.5cm]
  \vec x\,=\,\vec x(t_o) & \mbox{stato iniziale del sistema}
 \end{array}
 \right.
\ee
Ora rimane da trovare la metodologia e le ipotesi con le quali
poter arrivare a tale risultato. \\
Le variabili che descrivono lo stato della carena, vista come sistema dinamico,
in ogni istante sono date da una {\em opportuna} serie di parametri
lagrangiani, nel caso di ipotesi di corpo rigido tali
{\em variabili di stato} possono esserei, per esempio, la posizione di un punto
del corpo $\vec x_{P}\:$, l'assetto $\vec \Theta$,
la velocit\a di tale punto $\vec v_{P}$ e la velocit\a angolare $\vec W$; in
totale 2x6 variabili di stato per i 6 gradi di libert\a
\footnote{Un sistema dinamico {\em meccanico} pu\o essere determinato attraverso
la conoscenza delle posizioni dei suoi punti materiali e del
campo di velocit\a ad essi associato, la sola conoscenza delle posizioni non
renderebbe il processo deterministico. Infatti dalle equazioni di Lagrange
si vede che l'equazione al primo ordine (\ref{sd}) \e equivalente
ad una equazione nelle sole leggi orarie dei punti materiali ma del
secondo ordine nel tempo, quindi le condizioni iniziali per il problema
vanno date sia sulle posizioni iniziali dei punti che sulle relative
velocit\a }.
Indichiamo tali variabili con
\{$\vec q\,,\dot{\vec q}$\}, e quindi nel caso del corpo rigido avremo:
\be
 \left \{
 \begin{array}{c}
  \vec q\,=\,(\vec x_{P}\,,\vec \Theta)\\[.5cm]
  \dot{\vec q}\,=\,(\vec v_{P}\,,\vec W)  
  \end{array}
 \right.
\ee
\\
Per un corpo non rigido occorre allargare in modo opportuno lo
{\em spazio} delle variabili di stato per tener conto del maggior
numero di gradi di libert\a. \\ 
Passiamo quindi alla descrizione della dinamica della carena 
attraverso le equazioni di Lagrange.
Queste, come vedremo, ci permetteranno di descrivere {\em l'evoluzione}
del sistema attraverso un'equazione differenziale del tipo
(\ref{sd}). Inoltre ipotizzeremo la carena come un corpo deformabile
ed in particolare per fare ci\o useremo il modello del {\em continuo di Cauchy}.   
Anche se il lavoro di questa tesi \e incentrato sullo studio 
della tenuta al mare di una carena pensata rigida, vogliamo mostrare
nei prossimi paragrafi come l'approccio che useremo sia 
estendibile in maniera naturale anche a problemi idroelastici.
\subsection{Equazioni di Lagrange.}
Consideriamo un sistema meccanico e indichiamo con $\vec q$ le relative
variabile lagrangiane, la dimensione dello spazio delle \vec q rappresenta
il numero di gradi di libert\a del sistema.\\
Attraverso le $\vec q$ \e possibile descrivere il campo di spostamento
su ogni punto materiale del sistema, ad esempio con una legge del tipo:
\be
 \vec u(P,t)\,=\,\sum_{n=1}^{\infty}\, q_n(t)\vec \Phi_n(P)
\ee
ove $\vec \Phi_n(P)$ sono un campo vettoriale definito sui punti materiali e
sono dette funzioni di forma. Per un continuo di Cauchy
i gradi di libert\a sono infiniti. Pensiamo ora di discretizzare il
problema e quindi di scegliere N punti materiali per descrivere il moto
complessivo del continuo, ovviamente il numero N nella pratica dovr\a essere
molto elevato per poter descrivere con una certa accuratezza tale moto
$\cal O$($10^3$), $\cal O$($10^4$).
Ad ognuno di tali punti materiali conferiamo 6 gradi di libert\a,
il campo di spostamento sar\a quindi dato da:
\be \label{spos}
 \vec u(P,t)\,=\,\sum_{n=1}^{6N}\, q_n(t)\vec \Phi_n(P)
\ee
Le funzioni di forma $\vec \Phi_n(P)$ sono tali da {\em raccordare}
il moto tra i vari punti materiali e da rispettare gli eventuali vincoli
imposti su determinati punti materiali, per il resto rimane una certa
arbitrariet\a sulla scelta di tali funzioni.
\footnote{Quello che stiamo descrivendo in ambito strutturale va sotto il nome di 
{\em approccio agli elementi finiti}.}
Per i casi che vogliamo studiare, la struttura non sar\a vincolata
e quindi i gradi di libert\a relativi al moto rigido di insieme saranno
presenti.
In particolare per un corpo rigido N=1 le $q_n$ danno 
l'entit\a delle tre traslazioni e delle tre rotazioni, 
le 6 funzioni di forma sono costituite da
tre versori che danno le direzioni ed i versi delle tre traslazioni,
e da tre rotazioni con centro nel nodo scelto. \\
Con l'ipotesi che il campo di spostamento sia {\em piccolo} possiamo
usare le stesse funzioni di forma anche per il campo di velocit\a (cinematica
linearizzata):
\be \label{vel}
 \vec v(P,t)\,=\,\sum_{n=1}^{6N}\, \dot{q_n}(t)\vec \Phi_n(P)
\ee
Ora che abbiamo visto la parte cinematica del problema, passiamo
a quella dinamica; per semplicit\a noi consideriamo solidi iperelastici
adiabatici, ossia l'energia interna del continuo \e soltanto di natura
elastica, non esistono trasformazioni irreversibili all'interno del continuo. \\
Quindi per tale continuo la termodinamica \e banale ed \e sufficiente il 
bilancio dell'energia meccanica per descrivere la dinamica del sistema: 
\be
 \frac{d}{d t}\int_{V_M}(\rho e +\frac{1}{2}\rho  v^2) dV =\\
 \int_{V_M} \rho \vec f\cdot\vec v dV+ 
 \int_{S_M} \rho \vec t\cdot\vec v dS 
\ee 
\\
Dove $V_M$ \e il volume {\em  materiale} che racchiude il continuo,
     $e$ \e l'energia interna {\em elastica} del continuo, 
     $\vec f$ sono le forze di volume agenti sul continuo, 
     $\vec t$ sono le forze di superficie agenti sulla superficie del continuo
     e $\rho$ \e la densit\a di massa del continuo.  \\ 

Definiamo le  seguenti funzioni sul continuo:
\be
 \left \{
 \begin{array}{lc}
 \dsty
 T\,:=\,\frac{1}{2}\int_{V_M}\rho\, v^2dV
 & \mbox{Energia Cinetica del Sistema}\\ \\ 
 \dsty
 E\,:=\,\int_{V_M}\rho e dV
 & \mbox{Energia Potenziale elastica,}\\ \\
 \dsty
 e_n:=\int_{S}\vec t\cdot\vec \Phi_n dS
 &  \mbox{forze di superficie generalizzate} \\ \\
 \dsty
 f_n:=\int_{V_M}\vec f\cdot\vec \Phi_n dV
 &  \mbox{forze di volume generalizzate} \\ \\
 \end{array}
 \right.
\ee
\\
Attraverso delle leggi di spostamento, tipo la (\ref{spos}), l'energia
cinetica  \e una funzione delle variabili di stato $(\vec q,\dot{\vec q})$, 
e anche l'Energia potenziale elastica diviene una funzione delle $\vec q$.
Per quanto riguarda i campi $\vec f$,  $\vec t$, abbiamo che le forze
di volume per noi sono essenzialmente la forza di gravit\a, mentre 
le forze di superficie sono le forze idrodinamiche che agiscono sulla carena. \\
In particolare, se trascuriamo gli effetti della viscosit\a del fluido, possiamo
scrivere:
\be \label{fidrodyn0}   
 \left \{
 \begin{array}{c}
 \vec t(t,\vec x)\,=\,-\frac{1}{2}\rho U_{\infty}^2 
 c_p(t,\vec x) \vec n(t,\vec x) \\ \\
 \mbox{$\vec n$ normale della superficie, uscente dal corpo} 
 \end {array}
 \right.
\ee
\\
Ora sia $c_p(t,\vec x)$ sia $\vec n(t,\vec x)$ dipendono dallo stato del 
sistema dinamico, occorrer\a quindi, attraverso un modello per l'idrodinamica,
riuscire a trovare la dipendenza delle $\vec t$ dalle $(\vec q,\dot{\vec q})$;
per ora lasciamo indicata questa dipendenza  con la scrittura  
$e_n=e_n(\vec q,\dot{\vec q})$ per le forze di superficie generalizzate
\footnote{
Il termine {\em generalizzate} deriva dall'aver proiettato le forze sulle 
funzioni di forma, dato che queste descrivono sia moti di traslazione che 
di rotazione; nelle forze generalizzate abbiamo entrambe le azioni di forza 
e momento.}.\\
Va sottolineato il fatto che le forze idrodinamiche generalizzate dipendono
dalle variabili di stato del sistema dinamico e che d'altra parte il
campo idrodinamico si modifica al variare delle variabili di stato,   
si viene quindi a formare un ciclo chiuso di causa ed effetto. \\
Imponendo con un principio variazionale che l'energia totale del sistema
meccanico $T+E$ ammetta un minimo otteniamo le equazioni di Lagrange per il 
nostro sistema:
\be \label{lagrange}
 \begin{array}{lcr}
 \dsty
 \frac{d }{d t}\frac{\partial T}{\partial \dot{\vec q}} -
 \frac{\partial T}{\partial \vec q} +
 \frac{\partial E}{\partial \vec q} =
 e_n(\vec q,\dot{\vec q})\,+\,f_n(\vec q,\dot{\vec q}) \\ \\
 \end{array}
\ee
Con la (\ref{vel}) ci siamo messi nell'ipotesi di piccoli spostamenti,
ossia di piccole variazioni dei parametri lagrangiani, questo significa
che vogliamo studiare la dinamica di un sistema che si trova in una
posizione di equilibrio stabile, quindi lo stato del sistema 
indicato con $\{\vec q,\dot{\vec q}\}$ va inteso come lo scostamento
da uno stato di riferimento $\{\vec q_e,\dot{\vec q_e}\}$.
Con tale assunzione intendiamo dire che ad una variazione dello
stato corrisponde un'evoluzione del sistema dinamico in cui le
variabili di stato {\em orbitano} intorno al loro valore all'equilibrio.
Per questi motivi \e possibile effettuare uno sviluppo di Taylor
dell'energia potenziale elastica $ E(\vec q)$ ed \e possibile
scrivere
\footnote {L'energia potenziale elastica non agisce sulle variabili lagrangiane
relative al moto rigido di insieme ma su quelle legate ai
moti di deformazione della struttura, ci\o fa si che la matrice $\vec K$
delle costanti elastiche \e semi-definita positiva, qualora
i moti rigidi fossero impediti tale matrice risulterebbe definita positiva.}
\be
 E(\vec q)\,=\,\vec q\cdot\vec K\vec q\,+\,O(\vec q^2)
\ee
Ora sviluppiamo anche l'espressione dell'energia cinetica $T(\vec q,\dot{\vec q})$
nelle variabili lagrangiane ed otteniamo:
\be
 \begin{array}{c}
 \dsty
 T(\dot{\vec q})\,=\,\sum_{j=1}^{6N}\sum_{k=1}^{6N}\,\dot{q}_j\,
   \oneh \,\int_{V_M} (\vec \Phi_j\cdot\vec \Phi_k) dV\,\dot{q}_k \qquad 
 \mbox{ovvero:}\qquad
 T\,=\,\oneh \dot{\vec q}\cdot\,\vec M\dot{\vec q}
 \end{array}
\ee
La Matrice $\vec M$ prende il nome di Matrice di massa consistente.
\footnote{ Il fatto qui che l'energia cinetica dipenda soltanto dalle 
$\dot{\vec q}$ deriva  dalla scelta che \e stata fatta sui campi di
spostamento(\ref{spos}) e velocit\a (\ref{vel}).}   
Quindi le equazioni di Lagrange per il continuo diventano:
\\ 
\be
 \vec M\ddot{\vec q}\,+\,\vec K\vec q\,=\,
 \vec e(\vec q,\dot{\vec q})\,+\,\vec f(\vec q,\dot{\vec q}) 
\ee
Questa equazione differenziale ordinaria del secondo ordine, se 
portata al primo ordine \e proprio l'equazione differenziale (\ref{sd}) che 
cercavamo. \\ 
Rimangono due problemi da sviluppare: il primo \e trovare un'espressione 
per le forze idrodinamiche generalizzate, e questo punto prender\a  gran
parte di questa tesi a partire dal quarto paragrafo; il secondo 
punto \e che la dimensione dello spazio delle fasi che abbiamo 
scelto con l'approccio agli elementi finiti 
\footnote{che significa voler controllare i gradi di libert\a di un numero
elevato di punti materiali della struttura}
\e troppo elevato, e per questo si sceglie un altra via, che \e quella 
dell'analisi modale, descritta nel  prossimo paragrafo.
\subsection{L'analisi modale.}
Vogliamo trovare delle variabili lagrangiane che, pur non avendo 
un immediato senso fisico, siano pi\u {\em intrinseche} alla struttura, 
e che quindi con un numero limitato di  queste sia  possibile descrivere
la cinematica in maniera sufficientemente accurata, almeno quanto quella
descritta da un adeguato approccio agli elementi finiti.\\
Per far questo consideriamo la struttura priva  di forze esterne, sia 
di superficie che di volume, e per questo caso usiamo l'approccio
agli elementi finiti (con un numero {\em sufficientemente} elevato di nodi),
quindi l'equazione di Lagrange si scrive: 
\be \label{vuoto}
 \vec M\ddot{\vec q}\,+\,\vec K\vec q\,=\vec 0
\ee
Sappiamo che per le caratteristiche di cui godono $\vec M\,,\vec K$ 
l'equazione differenziale scritta \e diagonalizzabile in 6N equazioni 
scalari indipendenti  
\be 
  \begin{array}{lc}
  m_r\ddot{\eta_r}+k_r\eta_r=0
  & \mbox{con r=1..6N}
 \end{array}
\ee
la cui soluzione \e:
\be  
 \left \{
 \begin{array}{c}
 \dsty
 \eta_r(t)=\eta_{0r}\cos(\omega_rt)+
           \frac{\dot{\eta}_{0r}}{\omega_r}sin(\omega_rt) \\ \\
 \dsty
 \omega_r:=\sqrt{\frac{k_r}{m_r}} 
 \end{array}
 \right.
\ee
Tale soluzione pu\o essere scritta  in maniera compatta come: 
\be
 \vec \eta(t)=\vec C\vec \eta_0\,+\,
              \vec \Omega^{-1}\vec S\dot{\vec \eta_0}
\ee
Il passaggio dalle variabili $\vec \eta$ alle variabili $\vec q$ \e dato
da una relazione del tipo $\vec q\,=\,\vec Z\vec \eta$, con 
$\vec Z$ matrice reale non singolare. 
Quindi le variabili $\vec \eta$ sono delle legittime variabili lagrangiane, ed 
\e possibile esprimere il campo di spostamento anche attraverso loro:
\be \label{spos2}
 \dsty
 \vec u(P,t)\,=\,\sum_{n=1}^{6N}\, \eta_n(t)\tilde{\vec \Phi_n}(P)
\ee
Le $\tilde{\vec \Phi_n}$ sono delle nuove funzioni di forma, che possiamo
ricostruire dalle vecchie $\vec \Phi_n$ attraverso la matrice $\vec Z$ di
passaggio. Queste nuove funzioni di forma godono di una notevole caratteristica.
Infatti consideriamo il caso in cui le condizioni iniziali siano del tipo:
\be
 \left \{
 \begin{array}{c}
 \vec  \eta_0\,=\,\vec e_i \\ 
 \dot{\vec \eta_0}=\vec 0  
 \end{array}
 \right.
\ee
dove $\vec e_i$ \e un vettore che ha tutti 0 tranne che nella i-esima posizione 
dove c'\e un 1.  
In tale circostanza la soluzione in termini di $\vec \eta(t)$ \e :
\be
 \vec \eta(t)=\cos(\omega_it)\,\vec e_i
\ee
Questo significa che il campo di spostamento sar\a dato semplicemente da:
\be
 \left \{
 \begin{array}{c}
 \vec u(P,t)\,=\,\cos(\omega_it)\,\tilde{\vec \Phi_i}(P) \\ \\
 \vec u(P,0)\,=\,\tilde{\vec \Phi_i}(P)
 \end{array}
 \right.
\ee
Quindi se alla struttura viene dato un campo di spostamento iniziale proprio
uguale ad una di queste funzioni di forma, il campo di spostamento evolver\a
attraverso una sola variabile lagrangiana  $\eta$ relativa  a tale 
funzione di forma. 
Per $N\rightarrow\infty$ queste funzioni di forma hanno il nome di 
{\em Modi naturali di vibrazione della struttura (mnv)}, per N finito abbiamo
delle approssimazioni di tali modi ottenute con l'approccio agli elementi
finiti.\\
Le rispettive $\omega$ vengono invece chiamate {\em pulsazioni naturali della
struttura}.\\
I modi naturali di vibrazione sono intrinseci alla struttura, e le  
variabili lagrangiane $\vec \eta$ sono le variabili di stato che 
cercavamo, infatti sono sufficienti $\cal O$(10) di tali variabili 
per descrivere con adeguata accuratezza la cinematica del continuo, 
quelle  che vengono scartate sono le  $\eta$ relative alle pulsazioni
naturali {\em elevate}, che per i problemi trattati sono difficili da {\em eccitare}
e quindi hanno un peso minore.\\
In particolare le funzioni di forma relative ai moti rigidi sono
quei particolari {\em mnv} relativi alle pulsazioni naturali nulle, le quali
derivano dalla singolarit\a della matrice $\vec K$, come gi\a avevamo
sottolineato in una nota precedente.
Pertanto tali modi rimangono nella nostra trattazione. \\  
Per non appesantire la formulazione chiamiamo da ora con $\vec \Phi_n$ i 
{\em mnv} e con  $\vec q$ le relative variabili lagrangiane.\\ 
Noi siamo partiti con l'ipotesi di solidi iperelastici, e per 
tali solidi non esistono fenomeni dissipativi interni alla struttura, 
perch\h questi sono dei processi irreversibili. 
Spesso nelle analisi delle strutture 
si reintroducono tali fenomeni attraverso l' introduzione di una 
matrice di Smorzamento $\vec D$   
\footnote{Tale modello per i fenomeni dissipativi non descrive bene 
la realt\a  fisica, ma consente comunque di {\em riparare} in parte 
alla scelta {\em semplicistica} del solido iperelastico adiabatico}
\be
 \vec M\ddot{\vec q}\,+\,
 \vec D\dot{\vec q}\,+\,
 \vec K\vec q\,=\,
 \vec e(\vec q,\dot{\vec q})\,+\,\vec f(\vec q,\dot{\vec q}) 
\ee
\section{Espressione delle forze idrodinamiche nel dominio di Laplace.}
Possiamo sintetizzare quanto finora detto con il seguente schema:
\begin{figure}[htb] 
  \vskip 0.5cm
  \epsfxsize=0.7\textwidth
  \epsfysize=0.2\textwidth
  \makebox[\textwidth]{\epsfbox{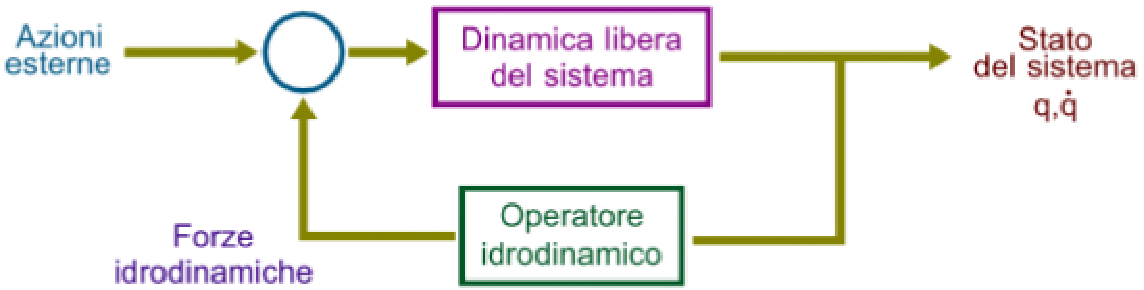}}
  \caption{Indicazione schematica del problema trattato.
           \label{schemaSD}}
\end{figure}
\\
Nella figura \ref{schemaSD} le azioni idrodinamiche sono state suddivise in 
forze idrodinamiche dovute al moto della carena e forze idrodinamiche 
dovute ad azioni esterne.   
Possiamo modellare le prime come il risultato di un certo operatore 
{\em idrodinamico} che agisce sul {\em moto} del sistema $\vec q(t)$:
\be
 \begin{array}{c}
 \vec F_{idrodyn}\,=\,\vec F_0\,-\,\vec E(t;Re,Fr)[\vec q(t)]
 \end{array}
\ee
Con $\vec F_0$ indichiamo la parte delle forze idrodinamiche che compete
alla configurazione di equilibrio.  
L'indagine che si vuole fare mira a linearizzare tale operatore
nelle variabili di stato, e quindi a costruire un operatore
integro-differenziale lineare
per le forze idrodinamiche che avr\a la forma:
\be \label{fidrodyn}
 \begin{array}{c}
 \vec F_{idrodyn}\,=\,\vec F_0\,-\,\vec L(t;Re,Fr)[\vec q(t)]
 \end{array}
\ee
Anche se la descrizione nel dominio del tempo risulta pi\u idonea per
la descrizione fisica del sistema dinamico, l'uso della trasformata
di Laplace presenta dei notevoli vantaggi nella formulazione del problema,
grazie alle sue propriet\a.
Un'ipotesi {\em sufficiente} per poter passare dal dominio $t$ al dominio $s$ 
\e che le funzioni considerate siano {\em generalmente continue}, e quindi
per i problemi che vogliamo considerare ci\o non pone grosse limitazioni. \\
Se esprimiamo la (\ref{fidrodyn}) nel dominio di Laplace abbiamo:
\be \label{fidrodyn2}
 \begin{array}{c}
 \vec F_{idrodyn}\,=\,\vec F_0\,-\,\vec L(s;Re,Fr)\vec q(s)
 \end{array}
\ee
L'operatore $\vec L[\cdot]$ \e diventato l'operatore $\vec L(s)\cdot$
di moltiplicazione della matrice \vec L(s) per un vettore.\\
Alla matrice $\vec L(s)$ diamo il nome di matrice delle forze idrodinamiche
generalizzate in analogia a quanto si fa in aeroelasticit\a.
Il {\em prodotto} di $\vec L(s)$ con $\vec q(s)$
\footnote{Per non appesantire la notazione indichiamo con $\vec q(s)$ la 
trasformata di Laplace del moto $\vec q(t)$}
genera dei prodotti tra funzioni nel dominio di Laplace che sono
dei prodotti di convoluzione nel dominio del tempo:
\be \label{fidrodyn3}
 \begin{array}{c}
 \dsty
 \vec F_{idrodyn}(t)\,=\,\vec F_0\,-\,\int_{0}^{t}\vec L(t-\tau)\vec q(\tau)\,d\tau
 \end{array}
\ee \\
Tale relazione pu\o essere scritta in modo migliore, infatti come vedremo
nei capitoli successivi sar\a possibile scrivere $\vec L(s)$ come
\be \label{GIF} 
 \vec L(s)\,=\,\vec A s^2 \,+\,\vec B s\,+\,\vec C \,+\,\vec H(s)
\ee
e quindi nel dominio del tempo avremo per le forze idrodinamiche,
\be \label{fidrodyn4}
 \begin{array}{c}
 \dsty
 \vec F_{idrodyn}(t)\,=\,\vec F_0\,-[\,
                         \vec A\ddot{\vec q}\,+\,
                         \vec B\dot{\vec q}\,+\,
                         \vec C\vec q\,+\, \\\\
 \dsty
 \qquad \qquad \int_{0}^{t}\vec H(t-\tau)\vec q(\tau)\,d\tau\,]
 \end{array}
\ee \\
La prima parte, dove compaiono $\vec A, \vec B e \vec C$ rappresenta le forze
idrodinamiche {\em istantanee} ossia quella parte delle forze idrodinamiche
complessive che risponde istantaneamente ad ogni variazione del campo
fluidodinamico.
Ci\o che vogliamo sottolineare qui \e il ruolo dell'integrale
di convoluzione che invece rappresenta quella parte delle forze idrodinamiche
che ad una variazione istantanea del campo fluidodinamico risponde
in un intervallo prolungato di tempo. La convoluzione nel tempo implica
che le forze idrodinamiche dipendono dall'{\em intera storia} del moto
$\vec q(t)$ attraverso il {\em nucleo} $\vec H(t)$.\\
Le cause fisiche per cui l'azione idrodinamica non
pu\o essere istantanea nelle $\vec q(t)$ sono riconducibili alla
presenza di una velocit\a di propagazione finita di qualche grandezza,
ad esempio:
\begin{itemize}
\item
In un campo compressibile il numero di Mach \e finito
e quindi si ha un certo ritardo nella propagazione dell'effetto di
un generico disturbo.
\item
La circolazione intorno ad un corpo portante 
$\Gamma\,=\,\oint\,\vec v\cdot d\vec l$
non pu\o cambiare istantaneamente sul corpo, infatti l'intero campo di
velocit\a $\vec v$ dipende da tutta la {\em scia} 
\footnote {Con il termine scia indichiamo la zona del campo fluidodinamico
a valle di un corpo dove si ha un'elevata vorticit\a $\vec \omega\,=\,rot(\vec v)$.
Nei corpi {\em portanti}, dove non si ha un distacco della {\em vena fluida},
tale zona risulta di piccolo spessore e soltanto molto a valle del corpo
si ha un ispessimento dovuto all'effetto diffusivo della viscosit\a su
$\vec \omega$}
a valle del corpo. Le scie presentano un effetto memoria sulla vorticit\a
che il corpo rilascia in ogni istante a valle. Infatti ad una variazione
istantanea del corpo corrisponde un disturbo di vorticit\a che viene convetto
nella scia dalla corrente e la velocit\a con cui questo disturbo viaggia
\e finita, ci\o spiega perch\h $\Gamma$ non pu\o variare istantaneamente.
\item
Se \e presente una superficie di discontinuit\a tra due fluidi, ad
esempio aria/acqua si ha che ad una variazione del corpo corrisponde una
generazione di un {\em sistema ondoso} su tale superficie di discontinuit\a,
anche qui la propagazione di queste {\em onde} \e finita, e ci\o fa si
che il loro effetto sul corpo risulta prolungato nel tempo.
\end {itemize}
In realt\a tutte e tre gli esempi rappresentano fenomeni di propagazione ondosa,
per la compressibilit\a abbiamo onde tridimensionali
nell'intero campo fluidodinamico, mentre nell'ultimo caso la
propagazione ondosa \e su una superficie, il secondo caso \e intermedio ma
se lo spessore della scia \e sufficientemente piccolo allora possiamo
vedere anche questa come una superficie di discontinuit\a tra due fluidi
che differiscono non per natura ma per la velocit\a tangenziale su di essa.
Noi consideremo sempre quest'ultimo caso, e vedremo nel capitolo 
del modello idrodinamico come la scia e
la superficie di discontinuit\a aria/acqua possano essere trattate
in maniera simile.\\
Vediamo invece quali sono le azioni fluidodinamiche {\em istantanee}.
Se consideriamo ad esempio una sfera in fluido non viscoso e attaccato
(e quindi non sono presenti scie a valle del corpo)
si ha nel caso stazionario il {\em paradosso di d'Alambert}
per cui le forze fluidodinamiche risultano
nulle, ma se invece consideriamo un caso non stazionario abbiamo che
l'effetto del campo fluidodinamico \e quello di far comparire una {\em massa
aggiunta} al corpo, infatti la forza fluidodinamica \e esprimibile come
\be
 \vec F(t)\,=\,\vec A\ddot{\vec q}(t)
\ee
\\
Un altro effetto istantaneo \e quello dovuto al termine:
\be
 \vec F(t)\,=\,\vec C\vec q(t)
\ee
Le matrici \vec A e \vec C che compaiono nella (\ref{GIF}) sono imparentate
rispettivamente a delle masse aggiunte e a delle forze idrostatiche, 
ma non possono essere viste come tali. 
Infatti per quanto riguarda le masse aggiunte,
nel caso di una nave la presenza della superficie libera cambia le cose. 
Alla matrice $\vec C$ possiamo dare il nome di matrice dell'idrodinamica
stazionaria. Infatti se consideriamo, ad esempio, una variazione 
a {\em gradino} di una qualche variabile lagrangiana, dopo un certo 
transitorio la carena si posizioner\a su un nuovo stato di 
equilibrio $\vec x(t)\,=\,\vec x_e$:
l'unico contributo delle forze idrodinamiche che rimane \e proprio 
quello dato dalla matrice $\vec C$ tale contributo va a sommarsi 
con le forze idrostatiche concorrendo così all'equilibrio
del nuovo stato insieme alle forze stazionarie $\vec F_0$. \\
Per quanto riguarda la matrice $\vec B$, questa compare come 
una matrice di smorzamento ma non ha un diretta
interpretazione fisica, ed \e comunque legata alla presenza di
superficie di discontinuit\a.
A tal punto possiamo riscrivere l'equazione di Lagrange:
\be \label{Lagdt}
 \begin{array}{c}
 (\vec M+\vec A)\ddot{\vec q}\,+\,
 (\vec D+\vec B)\dot{\vec q}\,+\,
 (\vec K+\vec C)\vec q\,+\, \\ \\
 \dsty
 \qquad \qquad \int_{0}^{t}\vec H(t-\tau)\vec q(\tau)\,d\tau\,=\,
 \vec F_{esterne}(t)+\,C.I.(\vec q_0\,,\dot{\vec q_0})
 \end {array}
\ee
E equivalentemente nel dominio di Laplace abbiamo: \\[.5cm]
\be \label{Lagdf}
 \begin{array}{c}
 [(\vec M+\vec A)s^2\,+\,
 (\vec D+\vec B)s  \,+\,
 (\vec K+\vec C)   \,+\,\vec H(s)]\vec q(s)\,=\,
 \vec F_{esterne}(s)\,+\,C.I.(\vec q_0\,,\dot{\vec q_0})
 \\[.5 cm]
 \end {array}
\ee
Abbiamo gi\a detto che stiamo studiando un sistema dinamico {\em nell'intorno}
di un suo equilibrio stabile; le forze stazionarie $\vec F_0$, insieme a 
quelle di gravit\a, partecipano alla realizzazione di tale stato 
$\vec x_e\,=\,\{\vec q_e\,,\dot{\vec q_e}\}$, e quindi non compaiono
nell'equazione \ref{Lagdf}.  
E' bene ricordare che tutte le matrici $\vec A\,,\vec B\,,\vec C\,e\,\vec H$
che derivano dall'operatore $\vec L[\cdot]$ hanno come parametri le grandezze
$Fr\,,Re\,$.\\
L'equazione così  scritta mette bene in evidenza 
una parte di dinamica tipica dei {\em classici} sistemi meccanici 
lineari del secondo ordine {\em massa-molla-smorzatore}; in 
tale parte intervengono anche le forze idrodinamiche in particolare
quelle istantanee e non legate agli effetti di memoria delle superfici
di discontinuit\a. Di questi ultimi ne tiene conto la matrice $\vec H$
che \e in genere una funzione molto complessa in $s$, e 
che racchiude in s\e la non stazionariet\a  del sistema dinamico.
La  presenza del termine  $\vec H(s)$ rende la 
dinamica del sistema lontana da quella dei sistemi 
{\em massa-molla-smorzatore}; basti pensare che i sistemi di questo tipo:
\be 
 \begin{array}{c}
 (\vec M\,+\,\vec A)\ddot{\vec q}\,+\,
 (\vec D\,+\,\vec B)\dot{\vec q}\,+\,
 (\vec K\,+\,\vec C)\vec q\,=\, 
 \vec F_{esterne}(t)+\,C.I.(\vec q_0\,,\dot{\vec q_0})
 \end {array}
\ee
se portati al primo ordine nel $tempo$ con la posizione 
$\vec x\,=\,\{\vec q,\dot{\vec q}\}$ diventano dei semplici 
sistemi di equazioni differenziali ordinarie lineari a coefficienti costanti.
\be \label{dxAx} 
 \begin{array}{c}
 \dot{\vec x}\,=\,
 \vec {\cal A}\vec x\,+\,
 \vec y(t)\,+\,C.I.(\vec x_0)
 \end {array}
\ee
A volte per non rinunciare ai  vantaggi di poter descrivere 
il sistema dinamico attraverso un'equazione del tipo \ref{dxAx}
si forza la \ref{Lagdf} ad essere scritta in questa forma, 
e ci\o viene fatto approssimando la $\vec H(s)$ con una 
funzione razionale fratta; tale procedimento va sotto il nome di
{\em approssimazione agli stati finiti}:
\be
 \left \{
 \begin{array}{c}
 \vec H(s)\,=\,(s\vec I-\vec P)^{-1}\,s\vec R \\ \\
 \vec I \qquad \mbox{matrice identit\a} \\ 
 \vec P \,,\vec R \qquad \mbox{Due opportune matrici}
 \end{array}
 \right.
\ee
Senza entrare nel dettaglio, vogliamo solo
sottolineare che con quest'ultima posizione nascono delle nuove
variabili lagrangiane legate alle forze idrodinamiche, che 
aumentano la dimensione dello spazio delle fasi e che 
non hanno una diretta interpretazione fisica 
$\vec x\,=\,\{\vec q\,,\dot{\vec q}\,,\vec r\}$;
non solo, la stessa variabile $t$ viene fatta entrare 
nello {\em spazio delle fasi}, infatti soltanto così si
pu\o passare da un sistema non stazionario  
\footnote{Non va confuso il termine non stazionario relativo al
sistema dinamico con quello relativo all'idrodinamica, infatti
per esempio il fenomeno della massa aggiunta \e legato alla non stazionariet\a
del campo fluidodinamico, ma la sua presenza non rende l'equazione
differenziale del sistema dinamico non stazionaria ossia del tipo 
$\dot{\vec x}\,=\,\vec v(t,\vec x)$, dove compare esplicitamente 
la funzione $t$.}
come il \ref{Lagdt} ad uno {\em autonomo} come in \ref{dxAx}.
Quindi se si riesce a descrivere il sistema dinamico con una
equazione del tipo 
\be
 \begin{array}{c}
 s\vec x(s)\,=\,\vec {\cal A}\vec x(s)\,+\,\vec y(s)\,+\,
              C.I.(\vec q_0\,,\dot{\vec q_0}\,,\vec r_0)
 \end{array}
\ee
la {\em matrice $\cal A$} ha come parametri non solo 
$Fr\,,Re\,$, ereditati dall'operatore idrodinamico ma anche $s$.\\
Nei prossimi paragrafi si esporranno i vantaggi di tale processo. \\ 
Per ora torniamo all'equazione \ref{Lagdf} che possiamo riscrivere
in forma compatta come:
\be 
 \vec G(s)^{-1}\vec q(s)\,=\,\vec F_{esterne}(s)\,+\,C.I.(\vec q_0\,,\dot{\vec q_0})
\ee
ovvero:   
\be \label{funz_trasf} 
 \begin{array}{c}
 \vec q(s)\,=\,\vec G(s)\vec F_{esterne}(s)\,+\,
               \vec G(s)[C.I.(\vec q_0\,,\dot{\vec q_0})]
 \\[.5 cm]
 \end{array}
\ee
La matrice $\vec G(s)$ \e {\em la Funzione di Trasferimento del Sistema}
ossia quella funzione che, dato un certo sistema di forze $\vec F_{esterne}(s)$ 
{\em indipendente dallo stato della carena} e
agente su questa, fornisce la legge del moto $\vec q(s)$ risultante. \\
E' importante sottolineare che le forze idrodinamiche dipendenti 
dallo stato della carena $\{\vec q\,,\dot{\vec q}\}$ non vengono
viste come delle forzanti esterne ma come parte integrante del sistema
dinamico; lo schema seguente aiuta a capire questo modo di procedere. \\ 
Nei prossimi paragrafi mostreremo quali {\em notevoli} vantaggi si hanno
nel poter descrivere la dinamica del sistema attraverso la \ref{funz_trasf}
ossia attraverso un modello lineare. 
\newpage
\section{La Stabilit\a e la risposta per un Sistema Dinamico.}

Per la maggior parte dei sistemi dinamici ci si interessa in particolar modo
a tre cose:
\begin{itemize}
\item
{\em La ricerca degli stati di equilibrio} 
\item
{\em Lo studio della stabilit\a, nell'intorno di uno stato
     di equilibrio del sistema dinamico}
\item
{\em Lo studio della risposta del sistema dinamico}
\end{itemize}
Uno stato $\vec x_e$ si dice di equilibrio per un sistema dinamico
descritto dalla $\dot{\vec x}=\vec v(t,\vec x)$ se soddisfa l'identit\a    
\be 
 \begin{array}{c}
 \vec v(t,\vec x_e)\,=\,\vec 0 \qquad \forall t 
 \end{array}
\ee
Un sistema dinamico \e stabile in una sua condizione di equilibrio e
per un dato insieme di condizioni iniziale 
se, perturbato con una di queste \{$\vec q_0\,,\dot{\vec q_0}$\},
si verifica che il moto conseguente $\{\vec q(t)\,,\dot{\vec q(t)}\}$
{\em orbita} intorno alla posizione di equilibrio 
\{$\vec q_e\,,\dot{\vec q_e}$\} 
\footnote{
Se si verifica che il moto $\vec x(t)$ non solo orbita intorno allo stato
di equilibrio ma ci si avvicina asintoticamente per $t\rightarrow\infty$ 
allora chiameremo tale stabilit\a come {\em stabilit\a asintotica}.
}.\\ 
Se invece il moto $\vec x(t)$ diverge continuamente dallo stato di equilibrio
allora diremo che il sistema \e instabile, per un dato insieme di condizioni
iniziali
\footnote{
E' importante sottolineare che a tale livello, stiamo considerando
l'assenza di forzanti esterne $\vec F_{esterne}$.
Quello che ci interessa \e la stabilit\a intrinseca del solo sistema dinamico
non forzato}.\\
Per un sistema dinamico non siamo interessati soltanto alla 
natura dello stato di equilibrio, stabile o instabile, ma 
anche alla sua {\em qualit\a}, per esempio quali sono i tempi
caratteristici con cui il moto $\vec x(t)$ orbita intorno allo stato
di equilibrio, se abbiamo una stabilit\a asintotica, quale \e il tempo
caratteristico per arrivare ad una data {\em distanza} dallo stato 
di equilibrio, etc.\\
Per i modelli lineari che vogliamo studiare la conoscenza di 
queste informazioni va a caratterizzare fortemente il tipo di risposta  
del sistema dinamico. 
Questo significa che la conoscenza della funzione $\vec G(s)$ 
caratterizza sia la dinamica libera, date certe condizioni iniziali,
e sia la risposta ad azioni esterne, come si vede dalla \ref{funz_trasf}.\\
Descriviamo brevemente alcuni problemi di studio della stabilit\a relativi 
al campo aeronautico e a quello navale. \\
Nella meccanica del volo si studia, per esempio, il caso in cui
la posizione di equilibrio dell'intero aeromobile \e il volo stazionario
livellato, e date certe condizioni iniziali sulle 2x6 variabili
di stato, si osserva come evolve il sistema. \\
La stessa cosa si pu\o fare per i natanti dove la condizione 
di equilibrio pu\o essere sia il semplice 
galleggiamento senza velocit\a di avanzamento, e sia il caso con $Fr\neq0$.\\
In questo tipo di studio il moto della nave genera dei sistemi d'onda
che vengono {\em irradiati} dal corpo e per questo tali problemi
vengono chiamati {\em problemi di Radiazione}.\\ 
Nel caso dell'aeroelasticit\a abbiamo il problema del {\em Flutter} di un'ala,
qui il fenomeno \e incentrato sui gradi di libert\a flessionali
che sotto certe condizioni portano l'ala a vibrare con ampiezze 
sempre crescenti fino alla rottura della struttura.
\footnote{
In tali problemi si valuta per quale valore del parametro aerodinamico
$U_{\infty}$ si realizza questa {\em instabilit\a.}}.\\
In ambito navale non si hanno in genere tali instabilit\a per\o l'influenza dei 
gradi di libert\a flessionali per certe strutture, quali ad esempio 
quelle degli S.W.A.T.H. (small water-plane area twin hull), va ad alterare
notevolmente i parametri relativi alla stabilit\a portando, inoltre,   
a delle notevoli sollecitazioni sulla struttura.\\ 
Altri problemi di stabilit\a in campo navale possono essere: 
il problema dello {\em sloshing} che riguarda navi che contengono
al loro interno delle masse liquide; qui l'accoppiamento con la dinamica 
di queste ultime pu\o alterare sensibilmente la stabilit\a del natante.\\
Problemi di {\em water shipment}, in cui a causa di ampi moti, la nave 
pu\o imbarcare dell'acqua con il risultato che il campo fluidodinamico 
viene alterato in maniera vistosa, e così anche la stabilit\a.

Questi ultimi problemi non sono ovviamente descrivibili
con un modello lineare come quello visto nei precedenti paragrafi,
in quanto la complessit\a del campo fluidodinamico
fa cadere molte delle ipotesi che sono alla base di tali modelli.\\ 
Vediamo ora come si traduce lo studio della stabilit\a  nell'ambito
dei modelli lineari.\\
Due importanti propriet\a dei sistemi lineari sono:
\begin{itemize}
\item
la stabilit\a non dipende dal dato iniziale, e quindi \e possibile
parlare di stabilit\a {\em globale del sistema}.
\item
la stabilit\a del sistema {\em omogeneo} 
\footnote{ossia non forzato}
garantisce la stabilit\a del sistema {\em non omogeneo}
\item
Se il sistema {\em omogeneo} \e asintoticamente stabile 
allora la risposta ${\vec q(t)\,,\dot{\vec q(t)}}$ ad un ingresso esterno
$\vec f(t)_{esterno}$ {\em limitato} sar\a anch'essa limitata.\\ 
Se \e garantit\a la sola stabilit\a allora ci possono essere 
fenomeni di {\em risonanza} per cui a certi ingressi possono corrispondere
risposte illimitate.  
\footnote{la {\em risonanza} non va vista come una instabilit\a del 
sistema dinamico}
\end{itemize}
Per sistemi dinamici lineari a coefficienti costanti i parametri
caratteristici della stabilit\a possono essere facilmente ricavati
dal calcolo degli autovalori della {\em matrice} $\vec {\cal A}$.
\be \label{stabxAx}  
 \left \{
 \begin{array}{c}
 \dot{\vec x}\,=\,
 \vec {\cal A}\vec x\\
 \vec x(t_0)=\vec x_0
 \end {array}
 \right.
\ee
Infatti la soluzione analitica del sistema omogeneo
\e costituita da funzioni del tipo $e^{(\alpha_i+j\omega_i)t}$
dove $(\alpha_i+j\omega_i)$ \e l' i-esimo autovalore; quindi
per la stabilit\a condizione sufficiente \e
che tutti gli autovalori abbiano la parte reale negativa,
anzi tale condizione garantisce anche la stabilit\a asintotica.
Questo stesso metodo si traduce, per i sistemi del tipo:
\be \label{stab lin2} 
 \begin{array}{c}
 \vec q(s)\,=\,\vec G(s)[C.I.(\vec q_0\,,\dot{\vec q_0})]
 \\[.5 cm]
 \end{array}
\ee
nel calcolo delle radici dell'equazione: \\ 
\be\label{radG} 
 \begin{array}{c}
  \det \vec G^{-1}(s)\,=\,0
 \end{array}
\ee
\footnote{
Si noti che l'equazione \ref{stabxAx} nel dominio di Laplace diventa  
\be
 \begin{array}{c}
 \vec x(s)\,=\,(s\vec I-\vec {\cal A})^{-1}C.I.(\vec x_0)
 \end{array}
\ee
quindi per tali sistemi si ha che $\vec G^{-1}(s)=(s\vec I-\vec {\cal A})$
il cui calcolo delle radici di  $\det \vec G^{-1}(s)\,=\,0$ coincide con la
ricerca degli autovalori per $\vec {\cal A}$.}
Dato che la $\vec G(s)$ \e una funzione complessa in $s$ il calcolo
delle radici della \ref{radG} risulta non facile  
e questo \e uno dei motivi per cui si usa {\em l'approssimazione
agli stati finiti} descritta nel precedente paragrafo:
\be
 \left \{
 \begin{array}{c}
 \vec H(s)\,=\,(s\vec I-\vec P)^{-1}\,s\vec R \\ \\ 
 \det \vec G^{-1}(s)\,=\,0\,\,\Leftarrow\,\,
 \det(s\vec I-\vec P)\,=\,0 
 \end{array}
 \right.
\ee
\\
Con tale condizione si ha che le caratteristiche della stabilit\a
sono governate dagli autovalori della matrice $\vec P$.\\ 
La conoscenza di queste radici $s_i\,=\,(\alpha_i+j\omega_i)$ 
ci d\a una visione chiara e quantitativa sulla stabilit\a del sistema:
la parte reale $\alpha$ ci dice quanto velocemente il sistema smorza 
o amplifica dei disturbi; la parte immaginaria $\omega$ ci dice con 
quale pulsazione ci\o accade. \\
Chiamiamo {\em poli del sistema} le radici di $\det \vec G^{-1}(s)\,=\,0$. 
La conoscenza dei poli non ci d\a soltanto informazioni sulla 
stabilit\a ma ci permette anche di valutare come il sistema risponde ad un 
dato {\em ingresso}, come vedremo nel prossimo paragrafo.  
Se si fanno variare i parametri 
$Fr\,,Re\,$ e si graficano i poli nel piano complesso
si ha il {\em luogo delle radici} in cui si pu\o valutare quale 
sia la sensibilit\a rispetto ai parametri. 
In corrispondenza dei poli, in modo analogo a quanto si \e fatto
nel terzo paragrafo per la struttura libera, possiamo chiamare 
i modi rispettivi con {\em modi di vibrazione in acqua}, 
e le pulsazioni relative con  {\em pulsazioni di vibrazione in acqua},
in modo analogo a quanto si fa in aeroelasticit\a.\\
\newpage
\section{La risposta ai comandi per una  nave in avanzamento }
Focalizziamo l'attenzione sui problemi di risposta; quelli che di solito sono
di interesse in ambito aeronautico e in ambito navale sono essenzialmente : 
\begin{itemize} 
\item
la {\em risposta ai comandi}, e 
\item
la {\em risposta alla raffica} per gli aeromobili e 
\newline
la {\em risposta ad un sistema di onde incidenti} per le navi.
\end{itemize}

Iniziamo dando una semplice panoramica sulla prima problematica.
In ambito aeronautico si studia ad esempio quale sia l'effetto 
di una deflessione {\em forzata} dei flap sullo stato del sistema 
per una certa $U_{\infty}$. 
In ambito navale pu\o  essere di interesse vedere quale sia la risposta del
sistema  ad una deflessione {\em forzata} del timone di 
direzione ad un dato $Fr$.\\ 
Ora tali azioni sono mal modellate se viste come azioni
indipendenti dallo stato del sistema, quindi ci\o che si fa \e 
includere nel sistema dinamico le {\em superfici di governo}.  
A tal punto occorre valutare la stabilit\a di questo nuovo sistema dinamico 
\footnote {Infatti, i parametri della stabilit\a possono variare notevolmente
in presenza di tali superfici.}.  
Va quindi opportunamente ampliato lo spazio delle variabili di stato per tener
conto dei gradi di libert\a relativi a tali superfici.
Entrambe queste discipline fanno uso della {\em "moderna" teoria dei controlli} 
e quindi diventa di base poter costruire dei modelli lineari del tipo:
\be
 \begin{array}{c}
 \dot{\vec x}\,=\,\vec {\sc A}\vec x\,+\,\vec {\sc B}\vec u\,+\vec f(\vec x)
 \end{array}
\ee
\newline
Dove lo stato $\vec x$ sar\a dato da 
$\vec x\,=\,\{\vec q\,,\dot{\vec q}\,,\vec r\,,\vec \delta\}$
in cui con $\vec \delta$ indichiamo i gradi di libert\a relativi 
alle superfici di controllo.
La $\vec {\sc A}$ \e la matrice di stato del sistema dinamico
completo, velivolo + superfici di controllo.
Con $\vec u$ indichiamo le variabili di controllo del sistema
(gli angoli di deflessione delle superfici libere etc..), la
matrice $\vec {\sc B}$ \e detta {\em matrice di distribuzione del controllo}.\\ 
Con $\vec f(\vec x)$ indichiamo i contributi non lineari
\footnote{ che possono essere di origine strutturale e/o di 
origine aero-idrodinamica}
, ottenuti con un opportuno metodo perturbativo; questa  \e una 
parte molto difficile da modellare ma la sua importanza pu\o essere cruciale
perch\h tali termini possono rendere stabile o instabile il sistema dinamico
per certe condizioni iniziali sulle variabili di stato
\footnote {per i sistemi non
lineari non si pu\o parlare di stabilit\a globale, ma solo locale quindi le
condizioni iniziali diventano parte integrante del problema}.
Ci\o avviene quando, per qualche valore dei parametri, la parte 
lineare del sistema presenta dei {\em poli con parte reale nulla}, ossia
a limite di stabilita'; in tali condizioni infatti i contributi non lineari
non sono trascurabili; come \e enunciato dal {\em teorema di Liapunov} sulla
stabilit\a lineare.
\newline
Un problema interessante \e quello di trovare se 
esistano delle leggi di controllo del tipo 
$\vec u\,=\,-\vec N\vec x$, da applicare a tali superfici,
che date alcune informazioni sullo stato del sistema siano tali da
{\em  governare} una intrinseca instabilit\a del sistema;
alla matrice $\vec N$ si da il nome di {\em matrice di controllo.}
Di tali problemi si occupa ad esempio la meccanica del volo; infatti
un aeromobile pu\o essere un sistema instabile ed \e solo
attraverso l'uso delle superfici di governo che \e possibile 
mantenere una data rotta     
\footnote{ molti aerei soprattutto militari sono fortemente instabili     
perch\'e, come si sa, pi\u \e forte la stabilit\a di un velivolo 
e pi\u \e scarsa la sua manovrabilit\a}.
\newline
Anche un'instabilit\a legata ai modi di vibrazione come il caso del Flutter
pu\o essere controllata attraverso un'opportuna legge di controllo sulle
superfici ({\em soppressione del Flutter}); di tali studi si occupa la
aero-servo-elasticit\a.\\
Anche se non entriamo nei dettagli di quest'ultima parte, quello che intendiamo
evidenziare \e l'importanza dello sviluppo di modelli lineari per poter
studiare un sistema dinamico con le tecniche proprie della teoria dei controlli.
\newline
Questo \e ci\o che si sta facendo nell'ambito aeronautico in questi ultimi
anni, e ora anche l'ingegneria navale si sta accingendo a sviluppare
modelli analoghi. 
In questa tesi non ci interesseremo del problema della risposta ai comandi,
e considereremo la nave priva di superfici di controllo, 
soggetta ad un sistema di forze esterne. 
\newpage
\section{La risposta a sistemi d'onde incidenti per una nave in avanzamento.}
Passiamo a parlare del secondo punto di interesse, quello della risposta
di sistemi stabili ad agenti esterni.
Il calcolo della risposta alle raffiche da parte di un aereo \e di fondamentale
importanza per la certificazione del velivolo. Quello che si vuole calcolare
\e a quale campo di accelerazione \e sottoposto il velivolo investito da
una raffica. Le accelerazioni, infatti, vanno valutate perch\h sollecitano
le strutture a fatica e perch\h possono causare problemi ai piloti,
ai passeggeri e agli oggetti trasportati.\\
Analogalmente si pu\o fare il discorso per una nave investita da un sistema
ondoso, anche qui siamo interessati al campo di accelerazione nonch\h 
all'ampiezza dei moti conseguenti.
\newline
Sia nell'ingegneria aeronautica che in quella
navale si utilizzano medesime tecniche per lo studio della risposta ad
azioni esterne.
\newline
Anche alcune ipotesi semplificative sono simili in questi due campi, infatti
per le raffiche spesso si fa l'ipotesi di {\em raffica congelata} in cui
si dice in buona sostanza che la presenza dell'aeromobile non modifica
il campo aerodinamico della raffica, nell'ambito navale tale assunzione, 
per il sistema d'onde incidente, prende il nome di
{\em ipotesi di Froude-Krylov}. C'\e da dire per\o che tale semplificazione
cade se il sistema d'onde ha delle lunghezze d'onda $\lambda$ paragonabili
alla larghezza massima della carena ({\em beam}), in tal caso si deve
tener conto di come la presenza del corpo modifichi {\em l'azione}
del sistema d'onda incidente; tale problema va sotto il nome
di {\em diffrazione}.\\
Per le raffiche quest'ultimo problema non si studia in quanto si ritiene
che la scala delle variazioni del fronte d'onda sia sempre grande
rispetto a quella dell'aeromobile. \\
L'effetto di queste due azioni esterne \e puntuale sulla superficie
del corpo che \e funzione dello stato, quindi in realt\a tali azioni
non sono indipendenti dalle variabili lagrangiane; per\o in un'analisi
lineare si pu\o trascurare tale effetto considerando le azioni agenti
su una configurazione media del corpo (per esempio quella che compete
allo stato di equilibrio).\\
Un'altra strada \e quella di applicare le azioni sulla posizione {\em attuale}
del corpo e riportarle, attraverso uno sviluppo di Taylor, sulla
configurazione {\em media}, in questo modo si modifica la funzione di
trasferimento $\vec G(s)$ del sistema che diviene del tipo 
$\vec G(s)\vec G_{guest}(s)$ in cui la seconda funzione \e il risultato 
del procedimento esposto. \\
Per i problemi di risposta a sistemi ondosi, noi considereremo sempre
l'esistenza della diffrazione ma ci metteremo nell'ipotesi di azione 
del sistema d'onde sulla configurazione media della carena, \footnote{infatti 
\e possibile dimostrare che i termini trascurati sono effettivamente di 
ordine superiore}; in particolare se $Fr=0$ allora quest'ultimo 
problema equivale allo studio della diffrazione di un corpo fisso. \\

Di solito come sistema ondoso in un problema di {\em tenuta al mare} 
si considera un sistema piano di onde di forma sinusoidale  
\footnote{ Come vedremo nel capitolo del modello idrodinamico, 
in una teoria potenziale linearizzata tale
forma soddisfa le condizioni al contorno della superficie libera.}: 
\be\label{onda}
  \eta(x,y,t)\,=\,A\cos(kx-\omega t) 
\ee
dove $\eta$ rappresenta l'elevazione d'onda rispetto al piano indisturbato
$z=0$, $A$ \e l'ampiezza d'onda, $k$ \e il numero d'onda.   
La lunghezza d'onda $\lambda$  \e data dal rapporto $\lambda\,=\,2\pi/k$.  
Tali onde viaggiano sulla superficie libera con una velocit\a che 
\e legata alla loro lunghezza d'onda dalla {\em relazione di dispersione}:
\be\label{disper}
 V_{fase}\,=\,\frac{\omega}{k}\,=\,\sqrt{\frac{g\lambda}{2\pi}} 
\ee
tale velocit\a \e chiamata {\em velocit\a di fase}.\\
Questo sistema d'onda in particolare \e monocromatico. Le forze  
indotte da questo {\em ecciteranno} il sistema con una pulsazione 
pari a quella che compete all'onda. In virt\u della sovrapposizione degli effetti,
garantita dalla linearit\a, \e possibile scomporre un generico sistema 
d'onda nelle sue armoniche e quindi possiamo valutare il suo effetto, 
sommando i vari contributi dati dalle diverse armoniche, 
tale procedimento prende il nome di {\em analisi armonica}
del sistema. 
In virt\u della relazione \ref{disper} si ha che in un sistema 
d'onde composto da pi\u armoniche, ognuna di queste avr\a una 
differente velocit\a di fase, e conseguentemente ci sar\a una 
continua {\em velocit\a di dispersione} fra le varie armoniche. \\
Un altro parametro per il sistema d'onde incidenti, oltre la lunghezza
d'onda $\lambda$, \e la direzione di propagazione sulla superficie libera, 
che indichiamo con l'angolo $\beta$. Al variare di $\beta$ si possono avere 
onde da prua, da poppa, laterali o combinazioni di queste.
La \ref{onda} si modificher\a in:
\be \label{ondabeta}
 \dsty
 \eta(x,y,t)\,=\,A\cos(kx\cos\beta+ky\sin\beta-\omega t)\\ \\
\ee
 o equivalentemente :\\ \\
\be
\dsty
 \eta(x,y,t)\,=\,\Re\{A\exp^{-jk(x\cos\beta+y\sin\beta)+j\omega t)}\}\\ \\
\ee
Alla funzione di trasferimento $\vec G(s)$, per un sistema stabile, 
\e possibile dare la seguente interpretazione fisica: \\

Se si fa la posizione $s\rightarrow\,j\omega$, le colonne della 
funzione complessa $\vec G(j\omega)$ rappresentano in modulo e fase la 
risposta a regime di un sistema stabile ad un ingresso sinusoidale 
unitario di pulsazione pari a $\omega$
\footnote{
La stabilit\a garantisce che la parte di risposta legata ad eventuali 
condizioni iniziali, tender\a ad annullarsi per $t\rightarrow\infty$, 
e quindi a regime avremo soltanto la parte di risposta che compete 
al sistema di forze esterne.}.\\

La {\em trasformata di Fourier} $\vec G(j\omega)$ altro non \e che 
una sezione della funzione $\vec G(s)$, che \e un concetto pi\u generale,
lungo l'asse immaginario del piano complesso
\footnote{Comunque sotto certe ipotesi di regolarit\a, la trasformata
di Fourier \e del tutto equivalente, in termini di quantit\a di informazioni,
a quella di Laplace}.\\
In particolare in corrispondenza delle pulsazioni dei poli di $\vec G(s)$
questa {\em sezione} presenter\a dei picchi sul modulo 
$\mid \vec G(j\omega)\mid$. Se la forzante esterna riesce ad {\em eccitare}
tali pulsazioni caratteristiche del sistema, si otterranno degli ampi moti
$\vec q(t)$. A queste pulsazioni diamo il nome di pulsazioni di {\em risonanza}
ed in particolare quelle che ci interessano in questa sede sono quelle 
relative ai {\em modi rigidi in acqua}.\\ 
Le pulsazioni di {\em risonanza} sono quelle pi\u {\em pericolose}  
per un sistema dinamico, e lo diventano tanto pi\u quanto pi\u i i
poli si avvicinano all'asse immaginario, ossia tanto pi\u il sistema 
\e prossimo all'instabilit\a.
D'altra parte se i poli hanno delle parti reali fortemente negative, la 
risposta del sistema ad azioni esterne sar\a molto lenta, e ci\o pu\o andare a
scapito della sua manovrabilit\a.\\
La risposta del sistema nel dominio della frequenza per un'onda 
monocromatica con pulsazione $\omega$, e ampiezza A, sar\a dunque data da:
\be
\left \{
\begin{array}{l}
\dsty 
 \vec q\,=\,\vec G(i\omega)\,A\vec X 
\\ \\ \dsty
 \vec q(t)\,=\,\Re\{\vec q\,e^{i \omega t}\} 
\\ \\ \dsty
 A\,\vec X(t)\,=\,A\,\Re\{\vec X\,e^{i \omega t}\} \qquad 
\mbox {{\em Azioni esterne causate dal sistema ondoso}}
\end{array}
\right.
\ee
La funzione $\vec Z(\omega)\,=\,\vec G(i\omega)\vec X\,=\,\frac{1}{A}\vec q$ 
in ambito navale prende il nome di $R.A.O.$ (Response Amplitude Operator).\\
Attraverso la sovrapposizione di pi\u onde del tipo \ref{onda}
\footnote{Nell'ambito lineare il sistema d'onde ottenuto continuer\a 
a soddisfare le equazioni e le condizioni al contorno e quindi sar\a 
anch'esso una soluzione del campo fluidodinamico} 
possiamo scrivere il generico 
sistema d'onde come: 
\be
\eta(x,y,t)\,=\,\sum_{n=1}^{N}\,\Re\,\left[A_n\,
                \exp[-ik_n\,x\,+\,i\omega_n\,t]\right] 
\ee
E quando il numero delle onde discreto $N$ tende a $\infty$ abbiamo 
\be
\eta(x,y,t)\,=\,\Re\,\int_{-\infty}^{+\infty}\,A(\omega)
                \exp[-i\,k(\omega)x\,+\,i\omega\,t]\,d\omega
\ee
Se inoltre consideriamo diversi sistemi d'onde che provengono da 
diverse direzioni $\beta$ abbiamo:
\be \label{ocean}
 \eta(x,y,t)\,=\Re\int_{0}^{\infty}d\omega\int_{0}^{2\pi}d\beta\,
                A(\beta,\omega)
             \exp[-jk(\omega)(x\cos\beta+y\sin\beta)+j\omega t]
\ee
E quindi per un generico sistema ondoso composto da pi\u armoniche,
ognuna delle quali ha inoltre una propria direzione di propagazione
e una propria ampiezza, la risposta del sistema sar\a data da:
\be
\dsty
 q_j\,=\,\Re\,\int\int \vec Z_j(\omega,\beta)\,e^{i\omega\,t}\,
                        A(\omega,\beta)d\beta\,d\omega
\ee
\\
Un'altra interpretazione fisica per la funzione di trasferimento $\vec G(s)$
e' quella di risposta impulsiva. Consideriamo il caso in cui la 
funzione $\vec f_{esterne}$ sia un impulso di {\em Dirac} $\delta(t)$ su 
una generica variabile lagrangiana, e consideriamo che le condizioni 
iniziali siano tutte nulle, dato che la trasformata di Laplace di 
un impulso \e data da $\delta(f)=1$ si ha che 
$\vec q(s)\,=\,\vec G(s)\vec e_i$ dove $\vec e_i$ \e un vettore che 
ha tutti gli elementi nulli tranne che l'i-esimo dove c'\e un uno  
\footnote{Si noti che la risposta impulsiva coincide con l'evoluzione 
del sistema a date condizioni iniziali.}.\\
Quindi le colonne della funzione di trasferimento sono le risposte
impulsive sulle diverse variabili di stato nel dominio di Laplace.\\
Da queste due interpretazioni fisiche della funzione di trasferimento
nascono due famiglie di metodi di indagine dei sistemi dinamici: 
i {\em metodi in frequenza} in cui si lavora nel dominio $\omega$, 
e i {\em metodi di risposta impulsiva} in cui si lavora 
nel dominio del tempo. \\ 
Tali metodi sono usati sia in ambito sperimentale che in ambito 
numerico, il loro scopo \e quello di riuscire a {\em catturare} il 
maggior numero di informazioni sulla funzione di trasferimento. \\
I metodi in frequenza sono da pi\u anni sperimentati e 
hanno il vantaggio di fornire una formulazione 
del problema relativamente semplice e compatta, inoltre la 
generazione pratica nelle prove sperimentali di onde monocromatiche 
da parte di un ondogeno risulta relativamente semplice. \\ 
Si ha lo svantaggio
che le prove sperimentali e quelle numeriche devono essere fatte per
ogni valore di $\omega$, e non sempre si riesce a prevedere 
in modo accurato un eventuale picco di risonanza della funzione 
di trasferimento. \\
Nel dominio del tempo si cerca, attraverso un'unica prova di trovare 
quelle informazioni che richiedono pi\u prove in frequenza.   
L'impossibilit\a pratica di realizzare ingressi ad {\em impulsi di Dirac}
porta alla ricerca di opportune funzioni in grado di avere un contenuto
energetico sufficientemente elevato nell'intervallo di frequenze caratteristico
del sistema
\footnote{ Essendo la trasformata dell'impulso di Dirac uguale a uno, questo
significa che un tale ingresso copre tutto lo spettro delle frequenze  
$\omega\,\in\,-\infty..+\infty$.}.\\
La realizzazione pratica di tali funzioni, 
in prove sperimentali da parte di un ondogeno, risulta per\o complessa.   
Un altro vantaggio delle prove nel dominio del tempo \e che queste 
richiedono dei tempi di acquisizione bassi, questo \e un evidente
beneficio dal punto di vista  sperimentale a causa dell'assenza 
di problemi legati alla riflessione di onde nelle vasche. 
\newpage
\section{La Tenuta al mare in campo stocastico.}
La risposta in campo stocastico \e di notevole importanza sia 
in campo aeronautico per la risposta alla raffica, sia in campo navale
per la {\em tenuta al mare}, infatti la natura delle raffiche e dei  
sistemi ondosi \e di tipo casuale; quindi la risposta in questo 
regime \e pi\u realistica rispetto a quella deterministica.
Il passaggio da campo deterministico
a campo stocastico pu\o essere effettuato con relativa semplicit\a qualora si
abbia un modello lineare che descriva la dinamica del sistema in un'intorno
di un suo stato di equilibrio. \\
E questo \e un altro notevole vantaggio di questi modelli. \\
In campo stocastico l'unica informazione che si pu\o dare sull'ingresso 
del sistema \e la densit\a spettrale di potenza 
del sistema d'onda che risulta una funzione della pulsazione, e pi\u in generale 
anche dell'angolo $\beta$ di direzione dell'onda incidente. Chiariamo meglio
questo concetto. \\
Nel precedente paragrafo abbiamo visto l'espressione di 
un generico sistema d'onde: 
\be \label{ocean2}
 \eta(x,y,t)\,=\Re\int_{0}^{\infty}d\omega\int_{0}^{2\pi}d\beta\,
               A(\beta,\omega)
               \exp[-jk(\omega)(x\cos\beta+y\sin\beta)+j\omega t]
\ee
In campo deterministico possiamo calcolare l'energia media associata a
tale sistema attraverso la:
\be
 \left \{
 \begin{array}{c}
 E_{media}=\rho g \overline{\eta^2}\\ \\
 \dsty
 \overline{\eta^2}=\frac{1}{2}\overline{\int\int dA(\beta,\omega)
                       \exp^{-jk(x\cos\beta+y\sin\beta)+j\omega t)}} 
		             \,\,\overline{\int\int dA^*(\beta',\omega')
                       \exp^{-jk(x\cos\beta'+y\sin\beta')+j\omega' t)}}=\\ \\
  \dsty
 \overline{\eta^2}=\frac{1}{2}\int\int dA(\omega,\beta)dA^*(\omega,\beta)\,=\,
                    \int_0^{\infty}\int_0^{2\pi}S(\omega,\beta)d\omega d\beta
  \\ \\
 \mbox{ dove $*$ indica il complesso coniugato }
  \end{array}
  \right.
\ee
Alla funzione $S(\omega,\beta)$ diamo il nome di {\em densit\a di energia 
spettrale direzionale}.
Questa funzione pu\o essere estesa nel caso di processi stocastici, infatti
per il teorema di {\em Weiner Khintchine} la funzione $S(\omega,\beta)$ \e
la trasformata di Fourier della funzione di correlazione per l'elevazione d'onda,
sotto l'ipotesi di processi stocastici {\em ergodici}.\\
La dipendenza direzionale $\beta$ \e molto difficile da misurare sperimentalmente,
in quanto richiederebbe un gran numero di punti adiacenti, per questo spesso
si fa l'ipotesi di unidirezionalit\a e si passa a studiare una {\em densit\a di
energia spettrale} $S(\omega)$, alle onde di tale spettro si d\a 
il nome di {\em long crest} giacch\h il moto \e bidimensionale e 
le creste delle onde sono parallele.
Ovviamente questa ipotesi \e lontana dal modellare i fenomeni ondosi che si hanno
per esempio in oceano aperto, dove l'aspetto direzionale \e evidente e per questo 
si usa parlare di onde {\em short crest}.\\
Per i problemi di raffica in ambito aeronautico esistono degli 
spettri di riferimento, ricavati con campagne di misura, come ad esempio 
lo {\em Spettro di von-Karman}, o da regole semi-empiriche come lo 
{\em Spettro di Dryden}, che con leggi {\em semplici} tentano di approssimare 
i primi. Anche per i problemi navali qui trattati esistono 
degli spettri di riferimento tra cui lo spettro di {\em Pierson-Moskowitz}.\\
La risposta deterministica ad un sistema d'onde come il \ref{ocean2} 
sar\a data attraverso la funzione di trasferimento $\vec G(i\omega)$ 
e l'espressione delle forze indotte sulla carena dal sistema ondoso 
$\vec X(\omega,\beta)$:
\be
\left \{
\begin{array}{l}
\dsty
\vec Z(\omega,\beta)\,:=\,\vec G(\omega)\,\vec X(\omega,\beta) 
\\ \\
\dsty
 q_j\,=\,\Re\,\int\int \vec Z_j(\omega,\beta)\,e^{i\omega\,t}\,dA(\omega,\beta)
\end{array}
\right.
\ee
Se il sistema d'onde \e un processo stocastico, anche le variabili $q_j$ lo
saranno. Una volta noto lo spettro per l'elevazione d'onda $S(\omega,\beta)$, 
\e possibile calcolare il valore quadratico medio delle variabili lagrangiane:
\be
 \bar{q_j^2}\,=\,\int_0^{\infty}\int_0^{2\pi}\,S(\omega,\beta)\,
		     \mid\,\vec Z_j(\omega,\beta)\,\mid^2 d\omega\,d\beta
\ee
E' molto importante in questa analisi la conoscenza di entrambi le funzioni
$S$ e $Z$ infatti pu\o accadere che in corrispondenza di una ampia risposta
di risonanza per la $Z(\omega,\beta)$ lo spettro di energia d'onda presenti
dei massimi, con ovvie conseguenze.\\
Nell'analisi in campo stocastico la fase della risposta $q_j$ \e una
variabile casuale di poca importanza. Ci\o che invece \e importante
sono le {\em fasi relative} tra la risposta $q_j(t)$ e l'elevazione d'onda
$\eta(t)$ o per esempio tra i vari modi come il moto relativo tra beccheggio e
sussulto. In particolar modo \e importante conoscere il
moto relativo tra il moto di sussulto del corpo e il moto 
della superficie libera $(q_3\,-\eta)$, questo
determina infatti la probabilit\a di immersione o emersione di parti dello
scafo, e quindi \e importante conoscere in quali condizioni queste 
due grandezze sono in fase o in opposizione. 
Attraverso l'argomento della funzione 
$\vec Z(\omega,\beta$) \e possibile correlare le varie fasi delle variabili
lagrangiane tra di loro e quindi fornire una risposta anche 
in quest'ultimo caso. \\

\chapter[Modello fisico e formulazione del problema lineare]
{Definizione del modello fisico e  
         formulazione del problema non lineare.}

 In questo capitolo il problema di un veicolo marino in moto
 di avanzamento arbitrario viene descritto nelle sue linee generali.
 Dopo alcune ipotesi semplificative sul fenomeno fisico in oggetto,
 verr\a introdotto un modello matematico idoneo allo studio
 del flusso non stazionario attorno ad una carena.

 \section{Descrizione generale del problema}
 Argomento di questa tesi \e lo studio del flusso non stazionario
 attorno ad un veicolo marino che avanza in modo arbitrario. 
 In questo contesto, ai fini del comportamento della nave, il ruolo 
 cruciale \e svolto 
 dalle forze idrodinamiche associate all'interazione del fluido con la 
 superficie {\em bagnata} della carena (la cosiddetta {\em opera viva}).

 \`E opportuno osservare come le forze idrodinamiche nel loro 
 insieme siano il risultato di differenti fenomeni che caratterizzano il 
 campo fluidodinamico. Ad esempio,
 l'attrito alla parete determina una significativa componente della
 resistenza all'avanzamento.
 La generazione ed il rilascio di vorticit\a accompagnano le rilevanti forze 
 laterali che consentono la manovra delle navi.
 L'interazione, infine, fra la superficie libera e la carena
 comporta la generazione di onde ({\em radiazione}) o
 l'alterazione di sistemi ondosi preesistenti ({\em diffrazione}):
 a ci\o sono associate azioni idrodinamiche rilevanti sia per la 
 resistenza al moto ({\em resistenza d'onda}), sia per il comportamento
 del veicolo marino in mare agitato ({\em tenuta al mare}).

 In questa tesi l'attenzione sar\a focalizzata sullo studio dell'interazione 
 tra la carena e la superficie libera circostante e sulle forze che 
 insorgono come conseguenza di tale interazione.

 Pertanto, nel seguito, si trascura il rilascio di vorticit\a che tipicamente
 \e associato ai moti nave non simmetrici di grande ampiezza (imbardata e 
 deriva), rinunciando quindi alla trattazione del problema della 
 manovrabilit\`a.  Anche le forze di attrito vengono trascurate.

 Ulteriori semplificazioni fisiche possono essere determinate
 considerando pi\u in dettaglio il fenomeno di nostro interesse.
 In primo luogo, si assumer\a l'incompressibilit\a dell'acqua e,
 considerando che la densit\a dell'aria \e molto inferiore 
 a quella del mezzo sottostante, si trascureranno completamente
 gli effetti {\em aerodinamici} nei riguardi della propagazione
 ondosa. Si introduce così il concetto di {\em superficie libera}, 
 intesa come superficie di separazione la cui evoluzione non ha vincoli
 dovuti all'aria. 

 Le lunghezze d'onda tecnologicamente rilevanti nell'ingegneria navale
 sono almeno dell'ordine dei metri. Pertanto \e possibile trascurare 
 completamente l'effetto della tensione
 superficiale\footnote{Le oscillazioni della superficie libera
 sono legate alle {\em forze di richiamo} che tendono
 a riportare l'interfaccia nella configurazione di equilibrio.
 In particolare, la tensione superficiale ha un ruolo dominante
 per lunghezze d'onda inferiori ai 2 cm.}, ponendo 
 quindi l'attenzione sulle {\em onde di gravit\`a}.

 Il ruolo principale della viscosit\a del fluido nei riguardi della 
 propagazione ondosa \e quello di causare la progressiva attenuazione
 dell'ampiezza delle onde.
 Ci\o si verifica su una scala temporale dipendente dal periodo
 dell'onda stessa. Nel caso delle onde di gravit\a i tempi caratteristici
 di attenuazione sono molto pi\u lunghi di quelli tipici
 dell'interazione onda--carena. Pertanto, almeno da questo punto di
 vista, gli effetti della viscosit\a del fluido sono trascurabili.

 Pi\u complesso \e il ruolo della viscosit\a nell'interazione
 fluido corpo. Infatti la formazione di strati limite a ridosso 
 di questo e la complessa dinamica della vorticit\a qui generata
 giuocano spesso un ruolo significativo nei confronti delle 
 forze agenti sulla carena. 
 Tuttavia, le forze idrodinamiche significative nell'ambito
 della tenuta al mare sono in buona approssimazione legate allo 
 scambio di quantit\a di moto ed energia fra onde e carena.
 Pertanto nel seguito si ipotizzer\a che il fluido sia non
 viscoso.
 Quest'ultima ipotesi, assieme a quella che il campo di velocit\a sia 
 inizialmente irrotazionale, consente di garantire l'irrotazionalit\a del 
 flusso anche negli istanti successivi.

 Si ipotizzer\a inoltre che la nave si comporti come un corpo rigido, che
 possieda cio\e sei differenti gradi di libert\a con altrettanti 
 possibili moti, tre traslazionali ({surge}, {sway}, {heave}) e tre 
 rotazionali ({pitch}, {roll}, {yaw}).
 In tal modo potr\a essere tralasciato il comportamento elastico della
 struttura, che avrebbe complicato la soluzione del problema associando 
 in generale alla carena un infinito numero di gradi di libert\a.

 Quanto detto delinea il problema fisico che verr\a affrontato
 e pone le basi della relativa formulazione matematica, 
 caratterizzando al contempo le fenomenologie descrivibili 
 e fissando i limiti intrinseci alla presente analisi.

  \section{Formulazione matematica}
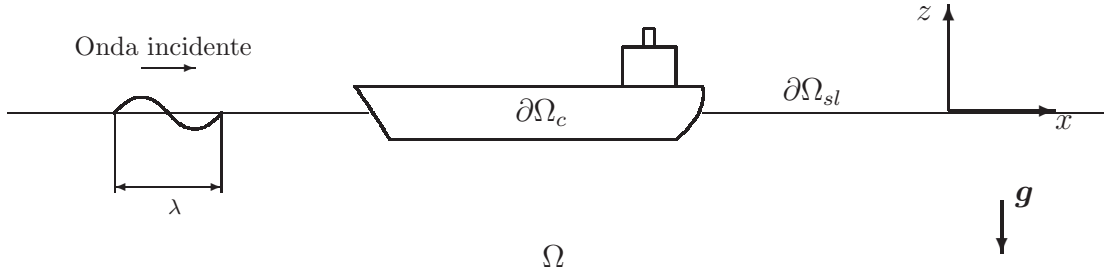
\begin{figure}[ht]
   \begin{center}
\begin{picture}(400,80)(0,00) 

    \put(290,63){$\SL$}
    \put(0,58){\line(1,0){135}}
    \put(260,58){\line(1,0){150}}

\thicklines 
     \bezier{50}(40,58)(50,70)(60,58)
     \bezier{50}(60,58)(70,46)(80,58)
\thinlines
     \put(40,58){\line(0,-1){30}}
     \put(80,58){\line(0,-1){30}}
     \put(60,30){\vector(-1,0){20}}
     \put(60,30){\vector(1,0){20}}
     \put(60,20){$\scriptstyle \lambda$}
\thinlines
     \put(50,75){\vector(1,0){20}}
     \put(25,80){\footnotesize Onda incidente}

\thicklines 
     \bezier{50}(130,68)(137,58)(143,48)
     \put(143,48){\line(1,0){107}}
     \bezier{50}(250,48)(262,58)(260,68)
     \put(260,68){\line(-1,0){130}}
     \put(230,68){\line(0,1){15}}
     \put(230,83){\line(1,0){20}}
     \put(250,83){\line(0,-1){15}}
     \put(238,83){\line(0,1){7}}
     \put(238,90){\line(1,0){4}}
     \put(242,90){\line(0,-1){7}}
     \put(190,55){$\CR$}
      
     \put(200,0){$\Omega$}

     \thicklines
     \put(352,59){\vector(0,1){40}}
     \put(352,59){\vector(1,0){40}}
     \put(392,52){$x$}
     \put(340,93){$z$}
     \put(372,25){\vector(0,-1){20}}
     \put(377,25){$\gv g$}
\end{picture}

   \end{center}
   \caption{Indicazione schematica del problema trattato
            e della simbologia adottata. \label{Pic_Prob}}
\end{figure}
 Il problema in esame \e quello di un corpo rigido $\cal B$
 in moto arbitrario in un fluido incompressibile, non viscoso e 
 inizialmente irrotazionale. 
 Il fluido occupa un dominio tridimensionale $\DO$ illimitato, 
 la cui frontiera \e costituita dalla porzione $\CR (t)$ della carena 
 bagnata dal fluido, dalla superficie libera $\SL$ e, eventualmente, da 
 un fondo.

 Per definire completamente il problema matematico, alle equazioni a derivate
 parziali che descrivono il campo fluidodinamico \S~\ref{Eq_Campo} occorre
 associare le opportune condizioni al contorno discusse nel \S~\ref{Con_Con}.
 Infine, trattandosi di un problema ai valori iniziali, verranno
 definite opportunamente le condizioni iniziali.

 \subsection{Equazioni di campo \label{Eq_Campo}}
 Nel problema in  esame, il campo fluidodinamico \e descritto in forma 
 differenziale da equazioni che esprimono puntualmente la 
 conservazione della massa 
 \be \label{Cont}
    \nabla\cdot\vec u = 0
 \ee
 e la conservazione della quantit\a di moto
 \be \label{Euler}
 \D{\vec u}{t} \,\,=\,-\,\frac{1}{\rho}\,\nabla p\,-\, g\,\nabla z\,\,.
 \ee
 In cui $\vec u$ \e la velocit\a del  fluido di densit\a $\rho$,
 $p$ \e la pressione e $\vec g$ \e l'accelerazione  di gravit\`a.
 
 Dalle ipotesi fatte il campo di velocit\`a, inizialmente irrotazionale,
 continuer\a ad essere tale anche negli istanti successivi. 
 Pertanto \e possibile esprimere la velocit\a come gradiente del 
 potenziale scalare $\Phi$ 
 \be \label{Def_Phi}
    \vec u = \nabla\Phi  \,\,.
 \ee
 Sostituendo la (\ref{Def_Phi}) nella equazione di continuit\a 
 (\ref{Cont}) si ottiene l'equazione ellittica a derivate parziali
 \be \label{Lapl}
     \nabla^2\,\Phi \,\,=\,\, 0\,\,,
 \ee
 in cui non compare esplicitamente il tempo $t$. Al riguardo
 si osservi come la soluzione dell'equazione di Laplace 
 (\ref{Lapl}) richieda dati sul contorno del dominio fluido $\DO$ e
 pertanto la dipendenza dal tempo del potenziale $\Phi$
 ha origine esclusivamente da tali condizioni e, come \e caratteristico
 dei problemi con superficie libera, dalla deformazione della frontiera 
 del dominio.

 Una volta noto il potenziale \e possibile ricavare la velocit\a  del fluido 
 per mezzo della (\ref{Def_Phi}).
 L'equazione di Bernoulli
 \be \label{Bern}
     \pder{\Phi}{t}+\oneh\mid\nabla\Phi\mid^2+g z+\frac{p}{\rho} =
       \mbox{C}(t)\,\,,
 \ee
 valida in ogni punto di $\DO$, permetter\a di ricavare la pressione, 
 descrivendo quindi completamente il campo fluidodinamico.

 In particolare, determinata la distribuzione di $p$ sulla superficie del 
 corpo, la forza ed il momento idrodinamici si possono ottenere con
 un'operazione di integrazione diretta.

 \subsection{Condizioni al contorno \label{Con_Con}}
 Per le situazioni che qui interessano, il dominio di integrazione presenta 
 una frontiera la cui configurazione dipende dall'istante temporale 
 considerato.

 In particolare, a causa del moto arbitrario della nave 
 e della deformazione continua della superficie libera circostante,
 la porzione di carena istantaneamente {\em bagnata} $\CR$ varia nel tempo.
 In ogni caso, il fluido non pu\o attraversare la superficie 
 impermeabile della carena e quindi, in un generico  istante,
 deve valere la condizione cinematica di impermeabilit\a
 \be \label{Cond_CR}
 (\vec u(P) - \vec V_P) \cdot \vec n(P) \,\,=\,\, 0 
 \qquad\qquad \forall P\in \CR
 \ee
 che sar\a anche l'unica a poter essere imposta su tale superficie.
 Nella (\ref{Cond_CR}) $\vec V_P$ indica la velocit\a di moto rigido
 del punto $\vec P$ della carena, il cui generico atto di moto \e del tipo
 \be \label{AdM}
  \vec V_P = \vec V_O  + \vec{OP}\times\omega\,\,.
 \ee

 Il moto della superficie libera $\SL$ \e a priori incognito  
 e va determinato come parte della soluzione del problema.
 In particolare la superficie libera, descritta in forma implicita dalla 
 ${\cal F}(\vec x,t)\,=\,0$, \e {\em definita} dalla propriet\a che il 
 fluido non la attraversi. Perci\o deve essere verificata la 
 {\em condizione cinematica}
 \be \label{CCon_Cin_SL}
    \begin{array}{lcr}
       w_n \,=\, \vec u \cdot \vec n_{sl} 
       &  \mbox{\hskip 1.cm} & 
       \forall t,\,\, \forall \vec x \in \SL\,\,,
    \end{array}
 \ee
 dove si \e indicato con 
 \[
 \vec n_{sl} \,=\, \frac{\nabla {\cal F}}{|\nabla {\cal F}|}
 \]
 il versore normale a $\SL$ in un suo punto e con 
 \[
 w_n \,=\, \frac{\dsty -\pder{{\cal F}}{t}}{|\nabla {\cal F}|}
 \]
 la {\em velocit\a di spostamento} nella medesima direzione.
 La (\ref{CCon_Cin_SL}) \e quindi equivalente alla
 \be 
       \pder{{\cal F}}{t} \,+\, \vec u \cdot \nabla {\cal F}
       \,=\,
       \D{{\cal F}}{t} \,=\, 0
 \ee
 e pertanto un elemento materiale che si trovi inizialmente su $\SL$ 
 non se ne distacca in tempi successivi.
 
 Trascurando le variazioni di pressione dovute al moto dell'aria, 
 si assume noto a priori il valore $p_a(\vec x,t)$ della pressione
 che agisce sull'interfaccia aria--acqua. Quindi, in assenza di 
 tensione superficiale, per l'equilibrio locale dalla superficie libera,
 si ha la {\em condizione dinamica}
 \be \label{CCon_Din_SL}
    \begin{array}{lcr}
       p \,=\, p_a (\vec x,t)
       &  \mbox{\hskip 1.cm} & 
       \forall t,\,\, \forall \vec x \in \SL\,\,.
    \end{array}
 \ee
 Le condizioni (\ref{CCon_Cin_SL}) e (\ref{CCon_Din_SL}) possono essere
 manipolate e combinate tra loro per ottenere condizioni equivalenti ma
 che, a seconda del caso trattato, possono risultare pi\u convenienti
 per le applicazioni. 
 Si noti che n\'e la condizione cinematica, n\'e 
 quella dinamica vincolano la componente tangenziale della velocit\a
 relativa fluido--superficie libera; quindi la velocit\a di un punto
 geometrico $P\in\SL$ pu\o essere definita arbitrariamente tramite la 
 \be \label{Def_Vel_SL}
    \vec w(P)\,=\, \vec w_t\,+\,u_n\vec n_{sl}\,\,,
 \ee
 dove $u_n=\vec u \cdot \vec n_{sl}$ \e la componente della velocit\a
 del fluido normale a $\SL$ e $\vec w_t$ \e il componente della 
 velocit\a (\ref{Def_Vel_SL}) contenuto nel piano tangente alla
 superficie libera.
 Quindi, sebbene la definizione (\ref{Def_Vel_SL}) sia 
 consistente con la condizione cinematica (\ref{CCon_Cin_SL}),
 la velocit\a $\vec w$ di un punto geometrico $P\in\SL$ coincide con la 
 velocit\a del fluido solo se $\vec w_t=\vec u_t=\vec u - u_n \vec n$.

 Nello studio dei fenomeni non stazionari  \e utile introdurre una 
 rappresentazione parametrica della superficie libera
 \be \label{Rap_Param}
    \SL\equiv
    \left\{
    \begin{array}{l}
       x \,=\, x(\xi_1,\xi_2,t)
       \\ \\
       y \,=\, y(\xi_1,\xi_2,t)
       \\ \\
       z \,=\, z(\xi_1,\xi_2,t)
    \end{array}
    \right.
 \ee
 in cui il significato dei parametri $\xi_1,\xi_2$ \e precisato dalla 
 condizione 
 \be \label{CCon_Cin_Lag_SL}
       \pder{\vec x(\xi_1,\xi_2,t)}{t} \,=\, \vec w (\xi_1,\xi_2,t)\,\,,
 \ee
 ossia la (\ref{CCon_Cin_Lag_SL}) individua la superficie libera tramite
 il moto dei suoi punti geometrici $P$, univocamente legati alla
 coppia di parametri $\xi_1,\xi_2$.
 Ovviamente, in accordo con l'arbitrariet\a nella definizione di 
 $\vec w$, la rappresentazione parametrica (\ref{Rap_Param}) non 
 \e unica.
 
 Anche la condizione dinamica (\ref{CCon_Din_SL}) pu\o essere riformulata
 ottenendo così un'equazione che descrive la dinamica 
 della superficie libera seguendo il moto dei suoi punti. 
 In particolare, sommando e sottraendo un termine {\em convettivo}
 del tipo $\vec w \cdot \nabla \Phi$, l'equazione di Bernoulli 
 scritta per i punti della superficie libera diviene
\be \label{CCon_Din_Lag_SL}
 \frac{D_{sl} \Phi}{D t} \,\,=\,\,
 \vec w \cdot \nabla \Phi \,-\,\oneh |\vec u|^2 \,-\,g \eta
 \,-\,\frac{1}{\rho} p_a
\ee
 nella quale compare l'operatore 
 $D_{sl} (\cdot) /D t:=\partial (\cdot)/\partial t+\vec w\cdot\nabla (\cdot)$
 di derivazione seguendo il moto dei punti di superficie libera.

 \subsection{Il problema non lineare nel dominio del tempo}
 Sulla base della breve descrizione riportata, il campo fluidodinamico
 attorno ad un veicolo marino \e
 descritto dalla equazione di Laplace, dalla condizione di
 impermeabilit\a sulla superficie bagnata della carena e 
 dalle condizioni di superficie libera:
\be \label{Prob_Completo}
     \begin{array}{lcl}
        \nabla^2\Phi \,\,=\,\,0 & \qquad & \forall \vec P \in \DO\\[0.5cm]
        \dsty
        \pder{\Phi}{n} = \vec V\cdot \vec n & \qquad &
        \forall \vec P \in \CR\\[0.5cm]
        \left\{
        \begin{array}{l}
           \dsty
          \pder{\vec x(\xi_1,\xi_2,t)}{t} \,\,=\,\, 
           \vec w (\xi_1,\xi_2,t)\\[0.5cm]
           \dsty
           \frac{D_{sl} \Phi}{D t} \,\,=\,\,
           \vec w \cdot \nabla \Phi \,-\,\oneh |\vec u|^2 \,-\,g \eta
           \,-\,\frac{1}{\rho} p_a
        \end{array}
        \right. & \qquad & \forall \vec P \in \SL
     \end{array}
\ee
 Ovviamente, il problema (\ref{Prob_Completo}) \e completamente
 definito solo specificando le condizioni iniziali ed il 
 comportamento asintotico della soluzione a grande distanza dalla
 carena.
 Al riguardo, nel seguito si considereranno problemi in cui si 
 pu\o assumere un ben preciso stato di riferimento per il
 campo fluidodinamico (per esempio fluido ovunque in quiete) e
 si supporr\a che, a distanza sufficientemente grande dalla carena,
 tale stato di riferimento rimanga indisturbato.

 \subsection{Moto del corpo}
 Se la legge secondo la quale il corpo si muove \e nota a priori
 le incognite del problema si esauriscono nelle sole variabili 
 fluidodinamiche.
 Diversamente accade nel caso in cui la cinematica del veicolo marino 
 non sia completamente prescritta. Infatti, in tali circostanze, il moto 
 del corpo \e strettamente  correlato a quello del fluido circostante per il
 tramite delle forze idrodinamiche che insorgono a causa del moto 
 relativo fluido--corpo. 
 La soluzione del problema completo richieder\a pertanto lo studio della 
 dinamica del corpo (supposto rigido)
 \be \label{Corpo_Rig}
    \left \{
    \begin{array} {lcl}
     m\dot{\vec V}\,\,=\,\,\vec F_{c}\,+\,\vec F_{i} \\[.5 cm]
    \dot{\vec K}\,\,=\,\,\vec M_{c}\,+\,\vec M_{i}        
    \end{array}
    \right.
 \ee
 in cui l'inerzia del veicolo ( traslazionale e rotazionale) \e
 funzione delle azioni agenti: nel presente caso le forze ed i momenti 
 idrodinamici ($\vec F_{i}$ e $\vec M_{i}$) sulla carena e le eventuali azioni
 di controllo ($\vec M_{c}$ e $\vec F_{c}$) dovute ad opportuni organi 
 che qui non vengono discusse in modo esplicito.
 Poich\'e le condizioni iniziali del corpo sono assegnate
 e assumendo note le azioni associate alla presenza degli organi di
 controllo, l'evoluzione del sistema corpo rigido--campo fluidodinamico 
 potr\a essere analizzata risolvendo i problemi accoppiati 
 (\ref{Prob_Completo}) e (\ref{Corpo_Rig}).

%
 
 Sembra importante sottolineare come  le forze idrodinamiche
 dipendano in generale dalla posizione del veicolo marino, 
 dall'atto di moto e dalla sua derivata temporale. Quest'ultima
 circostanza implica come l'inerzia {\em effettiva} del sistema
 possa variare nel tempo in funzione del campo fluidodinamico
 (massa aggiunta).  In generale ci\o aumenta le
 difficolt\a insite nell'integrazione numerica delle equazioni del moto
 accoppiate al problema fluidodinamico, dovendo trattare un sistema
 differenziale del secondo ordine con i secondi membri dipendenti dalla
 derivata di ordine pi\u alto.
 Nel seguito si illustra come esprimere le forze idrodinamiche in
 modo da fattorizzare la dipendenza dall'accelerazione generalizzata del
 veicolo marino.
 
  \section{Forze idrodinamiche}\label{Form_Forze}
 Le forze idrodinamiche 
\[
 \vec F = \int_{\CR} p \vec n dS
 \qquad \qquad 
 \vec M_O = \int_{\CR} p \vec r \times \vec n dS\,\,,
\]
 agenti sulla carena di una nave in movimento possono essere valutate per
 integrazione diretta della pressione 
 \[
  p = -\rho \left( \pder{\Phi}{t}+\oneh\mid\nabla\Phi\mid^2+g z\right)
 \]
 agente sulla superficie $\CR$ della carena.
 In particolare, il valore puntuale della pressione $p$ dipende, oltre che 
 dalla velocit\a e dal contributo idrostatico, anche dalla derivata 
 {\em euleriana} del potenziale.

 Il calcolo di quest'ultimo termine non \e 
 immediato quando si consideri una nave in  moto arbitrario.
 Come vedremo il problema viene radicalmente semplificato dalla
procedura di linearizzazione mentre, volendo affrontare il caso
non lineare, sono possibili alcune formulazioni che riducono il
problema.

\section{Natura non lineare del problema}
Nel caso dei flussi a potenziale per fluidi incomprimibili l'equazione
di campo che governa il fenomeno \e un'equazione lineare
che pertanto consente di costruire la soluzione del problema attraverso 
la {\em combinazione lineare} di soluzioni elementari, quali ad esempio quelle 
di sorgente, di doppietta e di vortice. Quanto detto pu\o essere visto 
in un ottica pi\u vasta, osservando che l'operatore $L\,:=\,\nabla^2(\cdot)$ 
\e un operatore lineare e quindi valgono tutta una serie di propriet\a che 
verranno esposte nel quinto capitolo.
Anche se l'equazione di campo \e lineare il problema complessivo trattato
\e non lineare. Pi\u precisamente
\begin{itemize} 
\item
La frontiera del dominio fluidodinamico, formata dalla superficie libera,
dalla superficie della carena e eventualmente da una scia vorticosa, \e una 
frontiera mobile.
\item
La configurazione della frontiera \e incognita. 
Questo vale anche per la carena, pensata rigida o meno, il suo stato infatti 
fa parte delle incognite del problema.  
\item
Il campo di spostamento della carena pensata rigida, \e non lineare: 
\be 
 \vec u(P)\,=\,\vec u(P_0)\,+\,\vec R\,(P-P_0)
\ee
l'operatore {\em di Rotazione} $\vec R$ contiene combinazioni di funzioni
trigonometriche; questo operatore compare nel bilancio del momento della 
quantit\a di moto 
\footnote{ A tal proposito ricordiamo che una delle prime ipotesi fatte 
per scrivere le equazioni di Lagrange fu proprio quella di considerare 
una cinematica linearizzata.}, e in $\vec n(t)$.                       
\item
L'operatore non lineare $\dsty D/Dt$ compare nell'equazione di 
Bernoulli e quindi anche nelle condizioni al contorno dinamiche per  
le superfici di discontinuit\a (scie e superficie libera). 
Anche nella condizione cinematica per la superficie libera compare 
questo operatore attraverso la $\eta(x,y,t)$.
\end{itemize}
Per ovviare alle difficolt\a collegate ai primi tre punti una
possibile tecnica consiste nello sviluppare in serie di Taylor le condizioni 
al contorno nell'intorno di una superficie nota e immobile.
In particolare si pu\o considerare la superficie libera indisturbata
coincidente con  il piano $x-y$, assumendo 
l'elevazione d'onda come funzione monodroma ad un sol valore,
mentre si considera la posizione che la carena assume nello
 {\em stato di equilibrio}.
Attraverso questi sviluppi si mettono bene in evidenza i contributi lineari da
quelli non lineari: nel paragrafo successivo 
se ne dar\a un esempio e, poi, tale approccio al problema verr\a applicato
sistematicamente nel capitolo seguente.

\section{Condizione unificata per la superficie libera}
Per quanto verr\a sviluppato nel capitolo seguente, 
risulta utile introdurre una condizione alternativa a  quelle
dedotte nei paragrafi precedenti.

In particolare,
indicata con $\eta(x,y,t)$ l'elevazione d'onda rispetto al piano $xy$,
la rappresentazione implicita della superficie libera ha la forma
$z\,-\,\eta(x,y,t)=0$ e la condizione cinematica diviene
\be 
\dsty
\pder{\Phi}{z}\,-\,\frac{D\eta}{D t}\,=\,0
\qquad \forall \vec x\in FS
\ee
che, sviluppata la derivata materiale, diviene
\be
\pder{\Phi}{z}\,-\,\dot{\eta}\,-\,\Phi_x\eta_x\,-\,\Phi_y\eta_y\,=\,0
\qquad \forall \vec x\in FS
\ee
Pertanto, introdotta la derivata sostanziale  
$\dsty \frac{D}{Dt}\,=\,\pder{(\cdot)}{t}\,+\,\vec v\cdot\nabla(\cdot)$, 
le due condizioni sulla superficie libera si possono scrivere
\be \label{FS} 
\left \{
\begin{array}{l}
 \dsty
 \frac{D\eta}{Dt}\,=\,\Phi_z
 \\ \\
 \dsty
 \dot{\Phi}\,+\,\frac{D\Phi}{Dt}\,+\,2\,g\,\eta\,=\,0
\end{array}
\right.
\ee
L'elevazione d'onda pu\o essere eliminata applicando la
derivata totale alla prima e sostituendo nella seconda 
\be
\dsty
\frac{D^2\Phi}{Dt^2}\,+\,\frac{D\dot{\Phi}}{Dt}\,+\,2g\Phi_z\,=\,0
\qquad \forall \vec x\in FS
\ee
Sviluppando allora l'operatore di derivata sostanziale
si ha la condizione unificata
\be \label{unificata}
\dsty
\ddot{\Phi}\,+\,g\,\Phi_z\,+\,2\nabla\dot{\Phi}\cdot\,\nabla\Phi\,+\,
  \oneh\nabla\Phi\cdot\nabla\,\left[\,\nabla\Phi\cdot\nabla\Phi\,\right]\,=\,0 
\qquad \forall \vec x\in FS
\ee
caratterizzata da termini non lineari quadratici in $\Phi$ e che
deve essere  soddisfatta sulla superficie libera incognita. 

Nel seguito, per ridurre le difficolt\a insite nel risolvere un
problema con condizioni non lineari su frontiera mobile, semplificheremo
le condizioni di superficie libera mediante uno sviluppo in serie di Taylor 
del potenziale nell'intorno del piano $xy$  
\be 
\dsty
\Phi(x,y,\eta,t)\,=\,\Phi(x,y,0,t)\,+\,\eta\,\left(\pder{\Phi}{z}\right)_{z=0}\,+\,
         \oneh\eta^2\left(\frac{\partial^2\Phi}{\partial z^2}\right)_{z=0}\,+...... 
\ee
ed introducendolo nella (\ref{unificata}), ottenendo 
\be \label{tayphi}
\left \{
\begin{array}{l}
 \dsty
 \ddot{\Phi}\,+\,g\,\Phi_z\,=\,0\,+{\cal O}(\Phi^2) \\\\
 \dsty
 \ddot{\Phi}\,+\,g\,\Phi_z\,+\,2\nabla\Phi\,\cdot\nabla\dot{\Phi}
       -\frac{1}{g}\dot{\Phi}\frac{\partial}{\partial z}
                             \left(\ddot{\Phi}\,+\,g\,\Phi_z\,\right)\,=\,0\,  
                            +\,{\cal O}(\Phi^3) \\\\
 \dsty
 \ddot{\Phi}\,+\,g\,\Phi_z\,+\,\oneh\,\nabla\Phi\cdot
                    \nabla\left(\,\nabla\Phi\cdot\nabla\Phi\,\right)\,=\,0\,
			    +\,{\cal O}(\Phi^4) \\\\
 \dsty
\qquad \forall \vec x\in (\,z\,=\,0\,)
\end{array}
\right.
\ee
L'ultima espressione \e stata  semplificata utilizzando l'equazione di Laplace 
e la prima espressione. 
Si osservi come la complessit\a delle condizioni ottenute aumenta 
notevolmente via via che l'ordine dello sviluppo cresce. 

In particolare uno sviluppo in serie di Taylor intorno al piano $z\,=\,0$ 
pu\o essere fatto per la sola condizione dinamica in modo da avere 
un'espressione per l'elevazione d'onda, in funzione di grandezze calcolate 
sul piano $xy$:
\be \label{tayeta}
\left \{
\begin{array}{l}
 \dsty
\eta\,=\,-\frac{1}{g}\left(\,\dot{\Phi}\,+\,
               \oneh\,\nabla\Phi\cdot\nabla\Phi\right)_{z\,=\,\eta}\,\\\\
\dsty
\qquad       \,=\,-\frac{1}{g}\left(\,\dot{\Phi}\,+\,
               \oneh\,\nabla\Phi\cdot\nabla\Phi\right)_{z\,=\,0}\,+\,\\\\
\dsty
\qquad \qquad
               \eta\frac{\partial}{\partial z}\left\{\,-\,\frac{1}{g}
                      \left(\,\dot{\Phi}\,+\,\oneh\,\nabla\Phi\cdot\nabla\Phi\right)
                      \right\}_{z\,=\,0}\,+\,..........\,=\, \\\\
\dsty
\qquad \qquad
      \,=\,-\,\frac{1}{g}\left(\,\dot{\Phi}\,+\,
                  \oneh\nabla\Phi\cdot\nabla\Phi\,-\,
                  \frac{1}{g}\dot{\Phi}\dot{\Phi}_z\right)\,+\,{\cal O}(\Phi^3)
\qquad \forall \vec x\in (\,z\,=\,0\,)
\end{array}
\right.
\ee

Cerchiamo ora una soluzione lineare valida in assenza della carena
per il problema di propagazione di onde piane.
Allora la \ref{tayphi} approssimata all'ordine ${\cal O}(\Phi)$ \e
\be
\dsty
 \ddot{\Phi}\,+\,g\,\Phi_z\,=\,0 \qquad su\,z\,=\,0
\ee
che adimensionalizzata diventa 
\footnote{
Dalla quale si vede che per $Fr\rightarrow\,0$ la condizione di 
superficie libera diventa la semplice condizione di impermeabilit\a
del piano $xy$. Questa \e in genere la condizione che viene soddisfatta
nell'ambito della {\em manovrabilit\a}.}:
\be
\dsty
 \ddot{\Phi}\,+\,\frac{1}{Fr^2}\,\Phi_z\,=\,0 \qquad su\,z\,=\,0
\ee
La soluzione cercata varia sinusoidalmente nel tempo con una pulsazione 
$\omega$ e si propaga con una velocit\a di fase $v_p$;
l'elevazione d'onda \e del tipo
\be \label{etalin}
 \dsty
 \eta(x,t)\,=\,A\cos(\,kx\,-\,\omega_0 t\,+\,\epsilon)
\ee
dove \e stato scelto l'asse x come asse di propagazione del sistema d'onde.
La velocit\a di propagazione \e data dal rapporto $v_p\,=\,\omega_0/k$.
$k$ prende il nome di numero d'onda ed \e legato alla lunghezza d'onda 
spaziale da $k\,=\,(2\pi)/ \lambda$. \\
La soluzione $\Phi(x,z,t)$ cercata, nel caso di profondit\a infinita, \e 
data da
\be \label{potlin}  
\dsty
 \Phi(x,z,t)\,=\,\frac{gA}{\omega_0}\,e^{kz}\,\sin(kx\,-\,\omega_0 t)
\ee 
Affinch\e l'espressione scritta sia soluzione del problema deve 
valere la {\em condizione di dispersione}
\be
\dsty
 k\,=\,\frac{\omega_0^2}{g}
\ee
da cui \e possibile calcolare la velocit\a di fase in funzione della pulsazione
$\omega$ o equivalentemente della lunghezza d'onda $\lambda$
\be
 \dsty
 v_p\,=\,\frac{\omega_0}{k}\,=\,\frac{g}{\omega_0}\,=\,
                                                   \sqrt{\frac{g\lambda}{2\pi}}
\ee
Adesso \e possibile dimostrare, attraverso le \ref{tayeta} al primo ordine,
che l'elevazione d'onda \e proprio del tipo indicato. 
E' possibile trovare anche una soluzione nel caso di presenza di un fondale,
in questo caso occorre soddisfare la condizione di impermeabilit\a per questa
superficie. L'elevazione d'onda sar\a sempre sinusoidale come nel caso 
precedente ma la velocit\a di fase delle onde dipende 
dalla profondit\a del fondale, e quando tale profondit\a \e 
dell'ordine dell'altezza d'onda, gli effetti non lineari diventano rilevanti, 
così come l'effetto della tensione superficiale.

Vogliamo ora vedere qual'\e l'effetto del passaggio da un riferimento fisso 
ad uno in moto traslatorio uniforme rispetto al primo con velocit\a
di avanzamento $\vec U_{\infty}$.
Nell'ambito di una teoria lineare questo significa che \e possibile 
sostituire l'operatore non lineare $D/Dt$ con l'operatore lineare
\be 
\dsty 
\frac{D}{Dt}\,=\,\pder{(\cdot)}{t}\,+\,\vec v\cdot\nabla(\cdot)  
\qquad \rightarrow \qquad
\frac{D}{Dt}\,=\,\pder{(\cdot)}{t}\,+\,\vec U_{\infty}\cdot\nabla(\cdot)
\ee
L'elevazione d'onda \ref{etalin}, per la soluzione con la condizione al contorno
linearizzata, in questo nuovo riferimento cambier\a la sua  pulsazione da 
$\omega_0$ a $\omega$, in quanto c'\e un {\em effetto Doppler}, infatti la
velocit\a di fase del sistema ondoso vista dalla nave \e diversa da quella 
vista da un osservatore fisso.
Per quanto detto la nuova pulsazione prende il nome di {\em pulsazione di incontro}
ed \e data dalla relazione:
\be
 \dsty
 \omega\,=\,\mid\,\omega_0\,-\,\frac{U_{\infty}\omega_0^2}{g}\cos\beta\,\mid
\ee
dove $\beta$ indica l'angolo formato dal vettore $\vec U_{\infty}$ di 
avanzamento della nave con la direzione di propagazione del sistema ondoso;
per $\beta\,=\,0$ abbiamo onde da poppa e quindi la pulsazione di incontro \e 
inferiore a quella propria del sistema d'onde, mentre per $\beta\,=\,\pi$ 
abbiamo onde da prua con effetto contrario sulla $\omega$.


\chapter{Linearizzazione del problema.}

In questo capitolo viene impostato il modello lineare per lo studio 
dell'interazione fra la nave in avanzamento e sistemi di onde originati a 
grande distanza dalla medesima.  
\`E abbastanza ovvio osservare che la linearizzazione del problema
\e di fatto imposta dalla difficolt\a nel risolvere il problema formulato
nella sua generalit\a nel precedente capitolo: non lineare ed intrinsecamente
non stazionario.
Tuttavia, come vedremo pi\u oltre, il problema linearizzato, 
oltre ad essere praticamente risolubile, \e suscettibile di una
pi\u completa analisi fisica e matematica che costituisce un indispensabile
supporto per ogni ulteriore sviluppo e approfondimento del problema in oggetto.

Gli sviluppi analitici riportati nei paragrafi seguenti sono, inevitabilmente,
di tediosa lettura. 
Sembra quindi importante delineare almeno per linee generali il percorso
logico che verr\a seguito.
Da un punto di vista fenomenologico possiamo osservare che
se la nave \e stabile e se l'ampiezza $A$ delle onde incidenti 
\e contenuta, la risposta del veicolo marino pu\o essere considerata
{\em piccola}.
Pi\u precisamente si supporr\a che le ampiezze delle oscillazioni
della nave nei suoi sei gradi di libert\a e l'entit\a della perturbazione
arrecata alla superficie libera circostante dalla presenza della 
nave (onde irradiate e diffrazione di quelle pre-esistenti) saranno 
proporzionali all'ampiezza d'onda $A$. 

Su questa base si opera una prima linearizzazione consistente nel
dire che il campo fluidodinamico generato dall'interazione onde--carena
differisce poco da quello {\em stazionario} che caratterizza il moto 
rettilineo ed uniforme della nave in {\em mare calmo}.
In questo modo si formula un problema per l'equazione di Laplace che 
dipende parametricamente, per il tramite delle condizioni al
contorno, dal campo fluidodinamico stazionario.

Purtroppo il {\em problema della resistenza d'onda}\footnote{La carena,
avanzando in moto rettilineo ed uniforme, genera con continuit\a un
sistema di onde di superficie. 
Alla energia spesa per generare il campo ondoso corrisponde una resistenza 
all'avanzamento che la nave deve vincere: la {\em resistenza d'onda}.
La sua determinazione \e tecnologicamente rilevante come si comprender\a
osservando che per carene commerciali pu\o variare, al crescere
della velocit\a, dal 5\% al 60\% della resistenza totale che il sistema
di propulsione deve vincere.} \e un problema non lineare di eccezionale 
difficolt\a pratica, oltre che teorica. 
Inoltre, il problema non stazionario sarebbe formulato in dominio con 
geometria dipendente dalla soluzione del problema stazionario e variazioni 
della velocit\a media di avanzamento richiederebbero la soluzione 
{\em ab initio} del problema di tenuta al mare.

Una ulteriore semplificazione viene allora introdotta linearizzando
anche il problema stazionario rispetto ad un flusso {\em base} giungendo
alla conclusione che i problemi stazionario e non stazionario si disaccoppiano
completamente.
Le soluzioni dei due problemi dipenderanno parametricamente dal flusso
base che verr\a scelto di tipo opportunamente semplice da non richiedere un 
significativo sforzo computazionale.

Il vantaggio, in termini pratici, \e significativo ma viene pagato al 
prezzo di una ulteriore semplificazione la cui portata \e difficilmente 
valutabile a priori e potr\a essere meglio analizzata mediante il confronto 
con dati sperimentali.

\section{Condizioni di superficie libera}
\subsection{Il problema stazionario}
Come accennato, si puo' supporre che la carena stia avanzando 
in una determinata direzione con velocit\a $\vec U$ opposta all'asse
delle $x$.
In assenza di onde incidenti e dopo un tempo sufficientemente
lungo, nel sistema di riferimento solidale alla carena,
il flusso appare stazionario e, nelle ipotesi gi\a discusse
in precedenza, possiamo descrivere il campo fluidodinamico in
termini di un potenziale stazionario
\be
	\Phi(\vec x)\,=-\,U\,x\,+\,\phi(\vec x)
\ee
ed indicheremo  con
\be
	\vec W\,=-\,U\,\vec e_x\,+\,\nabla\phi
\ee
il corrispondente campo di velocit\a.
Come noto, il potenziale di perturbazione $\phi(\vec x)$  soddisfa
l'equazione di Laplace  assieme alle condizioni al contorno 
imposte sulla superficie impermeabile della carena e sulla 
superficie libera.
In particolare, indicando con $\bar{S}$ la porzione bagnata di carena
nel suo assetto di regime,  la condizione di impermeabilit\a
richiede che
\be
 \begin{array}{c}
	\vec W\,\cdot\,\vec n\,=\,0\,
      \,\qquad su\,\,\bar{S}
 \end{array}
\ee \\
Analogamente, quando la superficie libera ha raggiunto la sua 
configurazione di equilibrio $\eta_s$,
il potenziale ivi soddisfa la condizione
\be\label{bcfsstaz}
	\oneh\,\vec W\,\cdot\,\nabla(W^2)\,+\,g\,\phi_z\,=\,0
      \qquad su\,z\,=\,\eta_s
\ee \\
La (\ref{bcfsstaz}) \e una condizione {\em unificata} per il potenziale
della velocit\a ottenibile dalle condizioni cinematica e dinamica 
eliminando l'elevazione d'onda che \e data implicitamente dalla relazione
\be
 \eta_s\,=\,-\,\frac{1}{2\,g}\,(W^2\,-\,U^2)_{z\,=\,\eta_s}
\ee

\subsection{Linearizzazione rispetto al problema stazionario}
In presenza di disturbi esterni, la nave inizier\a ad oscillare rispetto
alla condizione di riferimento sopra descritta.
In particolare, qui si assume che l'origine di tale disturbo sia
la presenza di onde incidenti.
Supponiamo ora che la soluzione del problema non stazionario sia
esprimibile nella forma
\be
 \left \{
 \begin{array}{c}
 \Phi(\vec x,t)\,=\,-\,U\,x\,+\,\phi(\vec x)\,+\,\varphi(\vec x,t) \\ \\
 \nabla\Phi(\vec x,t)\,=\,\vec W(\vec x)\,+\,\nabla\,\varphi(\vec x,t)
 \end{array}
 \right.
\ee \\
dove la dipendenza dal tempo \e interamente contenuta nel
potenziale $\varphi(\vec x,t)$ che si assume piccolo rispetto a quello 
stazionario.
Con questa ipotesi, la condizione al contorno sulla superficie libera 
pu\o essere semplificata nella forma
\be \label{bcfslin}
 \begin{array}{c}
 \dsty
 \oneh\,\vec W\,\cdot\,\nabla(W^2)\,+\,g\,\phi_z\,+\,
 \varphi_{tt}\,+\,2\vec W\,\cdot\,\nabla\varphi_t\,+\,
 \\ \\ \dsty
 \vec W\,\cdot\,\nabla(\vec W\,\cdot\,\nabla\varphi)\,+\,
 \oneh\,\nabla\varphi\,\cdot\,\nabla(W^2)\,+\,g\varphi_z\,+\,
 {\cal O}(\varphi^2)\,=\,0
 \qquad su\,\,z=\eta
 \end{array}
\ee
in cui sono stati trascurati termini di ordine superiore al primo
in $\varphi$.  
La corrispondente espressione per l'elevazione d'onda $\eta$,
trascurando termini di ordine superiore,
sar\a
\be \label{aux1}
 \begin{array}{c}
 \dsty
 \eta\,=\,-\frac{1}{g}\,\left [\,\frac{(W^2-U^2)}{2}\,+\,
 \dsty
        \varphi_t\,+\,\vec W\,\cdot\,\nabla\varphi\,\right]_{z\,=\,\eta}\,+\,
	  {\cal O}(\varphi^2)
 \end{array}
\ee 
Anche in questo caso l'elevazione \e data in forma implicita.
Tuttavia, consistentemente con l'ipotesi di piccole perturbazioni rispetto
al flusso stazionario, possiamo sviluppare la (\ref{aux1}) in serie di Taylor 
intorno alla $\eta_s$ ottenendo
\be
 \begin{array}{c}
 \dsty
    \eta\,\doteq\,\eta_s\,-\,\frac{1}{g}\,(\varphi_t
    \,+\,
    \vec W\,\cdot\,\nabla\varphi\,)_{z\,=\,\eta_s}
    \,-\,
    \frac{1}{g}\,(\eta\,-\,\eta_s)\,
    (\vec W\,\cdot\,\vec W_z)_{z\,=\,\eta_s}
 \end{array}
\ee
Che \e una equazione per la grandezza $(\,\eta\,-\,\eta_s)$
\be \label{etaty}
 \dsty
(\,\eta\,-\,\eta_s\,)\,=\,-\,\left [
 \frac{(\,\varphi_t\,+\,\vec W\,\cdot\,\nabla\varphi\,)}
      {(\,g\,+\,\vec W\,\cdot\,\vec W_z\,)}\right]_{z\,=\,\eta_s}
\ee
I primi due termini nella (\ref{bcfslin}),
\be
\left[\oneh\,\vec W\,\cdot\,\nabla(W^2)\,+\,g\,\phi_z\,\right]_{z\,=\,\eta}
\ee
costituiscono un contributo, non lineare, dovuto al campo stazionario $\vec W$
che viene supposto prolungabile analiticamente da $\eta_S$ ad $\eta$.
Allora, sviluppando in serie di Taylor ed utilizzando la (\ref{etaty}),
si ha
\be
\left[\oneh\,\vec W\,\cdot\,\nabla(W^2)\,+\,g\,\phi_z\,\right]
\,-\,
\frac{(\,\varphi_t\,+\,\vec W\,\cdot\,\nabla\varphi\,)}
     {(\,g\,+\,\vec W\,\cdot\,\vec W_z\,)}
\left(\,\oneh\,\frac{\partial}{\partial z}(\vec W\,\cdot\,\nabla W^2)\,+\,
             g\,\phi_{zz}\right)
\ee 
che va valutata per ${z\,=\,\eta_s}$.
Pertanto il primo addendo \e nullo in virt\u della (\ref{bcfsstaz}).

Quindi, al primo ordine, il potenziale di velocit\a non stazionario 
\e soluzione di un problema per l'equazione di Laplace con condizioni
al contorno da imporre sulla carena nel suo assetto medio e sulla 
superficie libera $\eta_S$ ove deve soddisfare la relazione 
\be \label{newlin}
 \begin{array}{lr}
 \dsty
   \varphi_{tt}\,+\,2\vec W\,\cdot\,\nabla\varphi_t
   \,-\,\frac{\varphi_t\,+\,\vec W\,\cdot\,\nabla\varphi}
             {g\,+\,\vec W\,\cdot\,\vec W_z}
   \left(\,\oneh\,\frac{\partial}{\partial z}(\vec W\,\cdot\,\nabla W^2)\,+\,
         g\,\phi_{zz}\right)
  \\
  & su\,\,z=\,\eta_s
  \\
 \dsty
 \,+\,\vec W\,\cdot\,\nabla(\vec W\,\cdot\,\nabla\varphi)\,+\,
 \oneh\,\nabla\varphi\,\cdot\,\nabla W^2 \,+\,g\varphi_z\,=\,0
 \end{array}
\ee 
\`E allora evidente come la soluzione del problema non stazionario 
dipenda parametricamente dal flusso 'base' stazionario, del quale
il potenziale $\varphi(\vec x,t)$ rappresenta un piccola perturbazione.

Pur se tecnicamente perseguibile, tale approccio \e reso oneroso
dalla necessit\a di valutare anche la soluzione del problema stazionario 
che \e, di per s\'e, di natura non lineare.
Ci\o giustifica una ulteriore semplificazione consistente nella
linearizzazione del problema stazionario che comporter\a alcuni
significativi vantaggi
\begin{itemize}
\item
 il potenziale stazionario potr\a essere risolto attraverso un 
 problema lineare.
\item
 i problemi stazionario e non stazionario risulteranno disaccoppiati
 e pertanto
\item
 le altezze d'onda relative alle due condizioni di flusso non rientrano 
 esplicitamente nelle equazioni di governo e possono essere calcolate una 
 volta risolti i due problemi.
\end{itemize}

\subsection{Disaccoppiamento dal problema stazionario}
Con lo scopo di semplificare la soluzione del problema stazionario
e, quindi, quella del problema non stazionario, introduciamo
una seconda linearizzazione consistente nel considerare il campo
fluidodinamico stazionario somma di un {\em flusso base}
e di un potenziale stazionario di perturbazione
\\
\be 
\left \{
\begin{array}{c}
\dsty
 \Phi(\vec x)\,=\,-\,U\,x\,+\,\phi(\vec x)\,=\,
                \Phi_B(\vec x)\,+\,\phi_{\epsilon}(\vec x)\\ \\
\dsty
 \vec W\,=\,-\,U\,\hat{i}\,+\,\nabla\phi\,=\,
 \nabla\Phi_B\,+\,\nabla\phi_{\epsilon}\,=\,\vec W_B\,+\,\nabla\phi_{\epsilon} 
 \\ \\
\dsty
 \nabla\Phi(\vec x,t)\,=\,\vec W_B(\vec x)\,+\,
                        \nabla\,(\phi_{\epsilon}(\vec x)\,+\,
                        \varphi(\vec x,t) )
 
\\ \\
\end{array}
\right.
\ee
\\ 
Il {\em Flusso Base}, in generale {\em grande} rispetto alla corrente
uniforme $-\vec U$, viene scelto attraverso delle considerazioni che
discuteremo pi\u avanti. 
Per il momento \e sufficiente richiedere che tale flusso soddisfi
la condizione
\be \label{base} 
 \oneh\,\vec W_B\,\cdot\,\nabla\,W_B^2\,+\,g\,
                              \frac{\partial}{\partial z}\Phi_B\,=\,0 
 \qquad su\,z\,=\,\eta_B
\ee
dove \e stata introdotta l'elevazione d'onda {\em fittizia}
\be \label{eta_B} 
 \eta_B(x,y)\,=\,-\,\oneh\,g\,(W_B^2\,-\,U^2)_{z\,=\,\eta_B} 
\ee
che compete al flusso base e che sommata al contributo associato alla 
perturbazione $\phi_{\epsilon}$ fornisce l'elevazione d'onda {\em fisica}
del problema stazionario.

La condizione al contorno (\ref{newlin}), relativa al flusso stazionario 
esatto, viene ora riscritta su $\eta_B$ ottenendo
\be \label{linb}
 \begin{array}{l}
 \dsty
   \varphi_{tt}\,+\,2\vec W_B\,\cdot\,\nabla\varphi_t
   \,-\,\frac{\varphi_t\,+\,
              \vec W_B\,\cdot\,\nabla\,(\,\varphi\,+\,\phi_{\epsilon}\,)}
             {g\,+\,\vec W_B\,\cdot\,\vec W_{Bz}}
   \left(\,\oneh\,\frac{\partial}{\partial z}(\vec W_B\,\cdot\,\nabla W_B^2)
         \,+\,g\,\Phi_{Bzz}\right)\\ \\
 \dsty
 \,+\,\vec W_B\,\cdot\,\nabla(\vec W_B\,\cdot\,\nabla(\,\varphi\,+\,\phi_{\epsilon})\,)
 \,+\, \\ \\
 \dsty
 \oneh\,\nabla(\,\varphi\,+\,\phi_{\epsilon})\,\cdot\,
        \nabla W_B^2\,+\,g\,(\varphi_z\,+\,\phi_{\epsilon\,z})\,+\,
	\oneh\,\vec W_B\,\cdot\,\nabla W_B^2 \,+\,g\Phi_{Bz}\,=\,0
 \qquad su\,\,z=\,\eta_B
 \end{array}
\ee 
in cui l'ultimo termine segue dalla (\ref{base}). 
\`E evidente il vantaggio numerico insito nel {\em trasferire}
la condizione (\ref{linb}) dalla superficie deformata secondo 
$\eta_B$ a quella indeformata $z=0$.
Pertanto, sviluppando secondo Taylor la relazione implicita che definisce 
l'elevazione di flussobase, si ha
\be 
 \begin{array}{l}
 \dsty
 \eta_B(x,y)\,=\,-\,\frac{1}{2\,g}\,(W_B^2\,-\,U^2)_{z\,=\,\eta_B}  \\ \\
 \dsty
 \qquad \qquad \doteq\,-\,\frac{1}{2\,g}\,(W_B^2\,-\,U^2)_{z\,=\,0}\,
 		      -\,\frac{1}{2\,g}\,\frac{\partial}{\partial\, z}
                       \,(W_B^2\,)_{z\,=\,0}\,(\,\eta_B\,-\,0\,)
 \end{array}
\ee
Da cui ricaviamo per $\eta_B$
\be
 \eta_B\,=\,-\,\left(\frac{W_B^2-U^2}{2g\,+\,W_{Bz}^2}\right)_{z\,=\,0}
\ee
E, in analogia a quanto fatto nella (\ref{newlin}), possiamo scrivere, 
osservando la (\ref{linb}), che
\be 
 \dsty
(\,\eta\,-\,\eta_B\,)\,=\,-\,\left [
 \frac{(\,\varphi_t\,+\,\vec W_B\,\cdot\,\nabla\,(\varphi\,+\,\phi_{\epsilon})}
      {(\,g\,+\,\vec W_B\,\cdot\,\vec W_{Bz}\,)}\right]_{z\,=\,\eta_B}
\ee
Lo sviluppo in serie di Taylor per tale grandezza porta infine a 
\be
(\,\eta\,-\,\eta_B\,)\,\doteq\,-\,\left [
 \frac{(\,\varphi_t\,+\,\vec W_B\,\cdot\,\nabla\,(\varphi\,+\,\phi_{\epsilon})}
      {(\,g\,+\,\vec W_B\,\cdot\,\vec W_{Bz}\,)}\right]_{z\,=\,0}\,-\,
 1\,(\,\eta_B\,-\,0\,)
\ee

I flussi base che verranno presi in considerazione nel seguito
soddisfano la condizione di impermeabilit\a in corrispondenza della
superficie libera indisturbata $(z\,=\,0)$, ossia
\\
\be \label{sempli}
 \frac{\partial}{\partial\,z}\Phi_B\,=\,0 \qquad su\,z\,=\,0\,\qquad
 \Rightarrow\, 
 \Phi_{Bxz}\,=\,\Phi_{Byz}\,=\,0 \qquad su\,z\,=\,0
\ee
Con tale scelta possiamo procedere alla linearizzazione della 
(\ref{linb}) intorno al piano $xy$
\be \label{linbz0}
 \begin{array}{l}
 \dsty
  \varphi_{tt}\,+\,2\vec W_B\,\cdot\,\nabla\varphi_t\,+\,
  \,-\,\left[\frac{\varphi_t\,+\,\vec W_B\,\cdot\,
                           \nabla\,(\,\varphi\,+\,\phi_{\epsilon}\,)}
   {\,g\,}\right]\,
   g\,\Phi_{Bzz}\,\,-\,g\,\Phi_{Bzz}\,\left(\frac{W_B^2-U^2}{2g\,}\right)
   \\ \\
 \dsty
 \,+\,\vec W_B\,\cdot\,\nabla(\vec W_B\,\cdot\,\nabla(\,\varphi\,+\,\phi_{\epsilon})\,)
 \,+\,\oneh\,\nabla(\,\varphi\,+\,\phi_{\epsilon})\,\cdot\,
        \nabla W_B^2 \,+\,g\varphi_z\,+\,g\phi_{\epsilon\,z}
   \\ \\
 \dsty
  \,+\,\oneh\,\vec W_B\,\cdot\,\nabla W_B^2 \,=\,0
 \qquad su\,\,z=\,0
 \end{array}
\ee

Non comparendo pi\u termini di ordine superiore al primo nei potenziali di 
perturbazione, \e ora possibile suddividere tale condizione in due espressioni
distinte relative, rispettivamente, al flusso stazionario 
\be \label{linstaz}
 \begin{array}{lr}
 \dsty
   \,-\,(\,\vec W_B\,\cdot\,\nabla\,\phi_{\epsilon}\,)\,
   \Phi_{Bzz}\,\,-\,\Phi_{Bzz}\,\left(\frac{W_B^2-U^2}{2\,}\right)
 \,+\,\vec W_B\,\cdot\,\nabla(\vec W_B\,\cdot\,\nabla\,\phi_{\epsilon}\,)
  \\ 
 & \mbox{su}\,\,z=\,0
  \\
 \dsty
 \,+\, \oneh\,\nabla\,\phi_{\epsilon}\,\cdot\,
        \nabla(W_B^2)\,+\,g\phi_{\epsilon\,z}\,+\,
        \oneh\,\vec W_B\,\cdot\,\nabla(W_B^2)\,=\,0
 \end{array}
\ee
 ed a quello non stazionario
\be \label{mother}
 \begin{array}{lr}
 \dsty
 \varphi_{tt}\,+\,2\vec W_B\,\cdot\,\nabla\varphi_t
 \,-\,(\,\varphi_t\,+\,\vec W_B\,\cdot\,\nabla\,\varphi\,)\,\Phi_{Bzz}
 \\
 & \mbox{su}\,\,z=\,0 
 \\
 \dsty
 \,+\,\vec W_B\,\cdot\,\nabla(\vec W_B\,\cdot\,\nabla\varphi\,)
 \,+\, \oneh\,\nabla\varphi\,\cdot\,\nabla(W_B^2)\,+\,g\varphi_z\,=\,0
 \end{array}
\ee

Abbiamo così ottenuto due problemi separati, uno per la parte stazionaria
ed un altro per quella non stazionaria; a questi due problemi 
rimangono associate le rispettive elevazioni d'onda. 
Seguendo un procedimento di sviluppo in serie di Taylor ed eliminando
termini di ordine superiore al primo nei due potenziali di perturbazione
si ha
\be
\left \{
\begin{array}{l}
\eta(x,y,t)\,\doteq \eta_{Staz}(x,y)\,+\, \eta_{NStaz}(x,y,t)
\\ \\
\dsty
\eta_{Staz}(x,y)\,=\,-\,\frac{1}{g}\,\left(\,\frac{W_B^2\,-\,U^2}{2}\,+\,
	                 \vec W_B\,\cdot\,\nabla\phi_{\epsilon}\,\right) 
\\  \\
\dsty 
\eta_{NStaz}(x,y,t)\,=\,-\,\frac{1}{g}\,\left(\dot{\varphi}\,+\,
                              \vec W_B\,\cdot\,\nabla\varphi\,\right) 
\end{array}
\right.
\ee

\subsection{Due scelte per il flusso base}
\paragraph{Corrente Uniforme}
La scelta pi\u semplice che possiamo fare per il flusso base
\e quella di considerarlo coincidente con la corrente e quindi supporre che 
il potenziale stazionario $\vec \phi_{\epsilon}$ sia una perturbazione 
di questa.

Da un punto di vista fisico questa scelta comporta delle restrizioni 
sulla geometria della carena.
Tale scelta vuole infatti dire che l'intera alterazione introdotta dalla
presenza della carena \e rappresentata da $\phi_\epsilon$ e perch\e ci\o
si realizzi la carena deve essere piuttosto sottile.
Nella letteratura scientifica del settore il problema così formulato \e
detto di {\em Neumann-Kelvin} e in tali condizioni si ha
$\phi\,\equiv\,\phi_{\epsilon}$, $\nabla \Phi_B\,\equiv\,-\,U\,\hat{i}$ e
l'elevazione d'onda relativa al flusso base risulta nulla.
In sintesi, per i due problemi le condizioni al contorno di superficie 
libera e le corrispondenti elevazioni d'onda diventano
\be
 \begin{array}{c}
  \fbox{Linearizzazione alla Neumann-Kelvin} \\ \\
   \begin{array}{ll}
   \mbox{Problema stazionario}
   &
   \left \{
    \begin{array}{c}
    \dsty
     U^2\,\phi_{\epsilon\,xx}\,+\,g\,\phi_{\epsilon\,z}\,=\,0
     \\ \\
     \dsty
     \eta(x,y)\,=\,-\,\frac{1}{g}\,\left(\,-\,U\,\hat{i}\,\cdot\,
                              \nabla\phi_{\epsilon}\,\right)
     \\ \\
    \end{array}
   \right.
   \end{array}
 \\ \\ \\
 \begin{array}{ll}
 \mbox{Problema non stazionario}
  &
 \left \{
  \begin{array}{c}
  \dsty
  \varphi_{tt}\,-\,2U\varphi_{xt}\,+\,
              U^2\varphi_{xx}\,+\,g\varphi_z\,=\,0
  \\ \\
  \dsty
  \eta_(x,y,t)\,=\,-\,\frac{1}{g}\,\left(\dot{\varphi}\,-\,
                                U\,\hat{i}\,\cdot\,\nabla\varphi\,\right)
  \\ \\
  \end{array}
  \right.
 \end{array}
\end{array}
\ee
Anche la \ref{newlin} d\a gli stessi risultati con la scelta 
$\vec W\,=\,-U\,\hat{i}$. 

\paragraph{Flusso di Doppio Modello}
Passiamo ora ad un'altra scelta per il flusso base, quella di {\em Doppio Modello}.
In questo caso il flusso base \e quel flusso generato da un corpo avente la
medesima velocit\a $\vec U$ della nave, che avanza in un fluido 
infinitamente esteso, senza pertanto la presenza di superfici di discontinuit\a.\\
Questo corpo \e costituito dalla carena, nella sua configurazione media,
e dalla sua immagine rispetto al  piano $(z\,=\,0)$, da cui il termine doppio 
modello. Sottolineiamo come per questo tipo di flusso valgono le ipotesi
semplificative \ref{sempli} usate per linearizzare il problema sul piano $xy$.\\
La scelta di un flusso base di questo tipo consente di fare delle ipotesi 
restrittive sulla  geometria della carena meno {\em forti} di quelle richieste
dalla linearizzazione alla {\em Neumann-Kelvin}. Per quest'ultima infatti
occorre considerare delle geometrie {\em sottili} e quindi 
avere le grandezze trasversali sufficientemente piccole rispetto 
alla lunghezza longitudinale $L$ della carena. \\ 
Mentre nel caso di geometrie non troppo sottili che avanzano con un $Fr$ non
elevato \e stato dimostrato che la scelta del doppio-modello risulta 
migliore di quella di {\em Neumann-Kelvin}. 
Il flusso di doppio modello recupera in parte le informazioni sul flusso
stazionario tridimensionale intorno alla carena.\\
Uno degli scopi di questa tesi \e proprio validare queste due linearizzazioni,
mostrando inoltre che per certe geometrie e per certi $Fr$ risulta migliore
la linearizzazione con il flusso di doppio-modello.\\   

\section{Condizione al contorno sulla Carena}
Indicata con
\be 
 \vec v(P)\,=\,\vec v(P_0)\,+\,\vec \Omega\,\times\,(\,P\,-\,P_0) 
\ee
la velocit\a di un punto $P$ della carena all'istante $t$,
la condizione di impermeabilit\a assume la forma
\be 
 \pder{\Phi(\vec x,t)}{\vec n}\,=\,
                           \vec v(P)\,\cdot\,\vec n(\vec x,t) \\ \\
\ee
Abbiamo gi\a discusso sulla non linearit\a di tale condizione.

Introducendo la relazione $\Phi\,=\,-U\,x\,+\,\phi\,+\,\varphi$
nella condizione al contorno e considerando
piccoli spostamenti\footnote{In particolare, per grandi
spostamenti, la derivata della rotazione $\Theta$ non fornirebbe
la velocit\a angolare}
\be
 \vec a(P)=\vec \xi\,+\,\vec \Theta\times(\,P\,-\,\vec \xi_0)
 \quad \Longrightarrow \quad
 \vec v(P)\simeq\dot{\vec \xi}\,+\,
               \dot{\vec \Theta}\,\times\,(\,P\,-\,\vec \xi_0) 
 \equiv \dot{\vec a}(P)
\ee
della carena rispetto alla configurazione media $\bar S$, la condizione
di impermeabilit\a assume la forma
\be \label{aux2}
 \pder{\varphi}{n}\,=\, \vec v(P)\,\cdot\,\vec n
                          -\,\vec W\,\cdot\,\vec n 
\ee
che v\a ancora imposta in corrispondenza della configurazione istantanea 
$S(t)$ della carena ed in cui $\vec W=\nabla(-U\,x\,+\,\phi)$.

Analogamente a quanto fatto per la condizione di superficie libera,
vogliamo semplificare il problema sviluppando la condizione (\ref{aux2})
intorno alla configurazione media $\bar S$ assunta nel moto stazionario.
In particolare abbiamo
\be \label{taycar} 
 \pder{\varphi}{n}\,\doteq\,
                      \left(\vec v(P)\,\cdot\,\vec n
                   -\,\vec W\,\cdot\,\vec n \right)_{\bar{S}}\,+\,
             \nabla \left(\vec v(P)\,\cdot\,\vec n
           -\,\vec W\,\cdot\,\vec n \right)_{\bar{S}}\,\cdot\,\vec a
\ee
Essendo la configurazione media, quella che compete al flusso stazionario 
si ha
\be 
 \vec W\,\cdot\,\vec n\,=\,0 \qquad \mbox{su}\,\bar{S}
\ee
e quindi possiamo riscrivere la (\ref{taycar})
\be
 \frac{\partial \varphi}{\partial \vec n}\,\doteq\,
                      \left(\dot{\vec a}\,\cdot\,\vec n \right)_{\bar{S}}\,+\,
             \nabla \left(\dot{\vec a}\,\cdot\,\vec n
           -\,\vec W\,\cdot\,\vec n \right)_{\bar{S}}\,\cdot\,\vec a
\ee
Il termine $\nabla(\dot{\vec a}\cdot\vec n\,)\,\cdot\,\vec a$ pu\o 
essere trascurato poich\'e di ordine superiore rispetto agli altri.
Rimane da sviluppare il termine 
\be
\nabla(\,\vec W\,\cdot\,\vec n)\,\cdot\,\vec a \,=\,
 (\nabla\vec W\,\cdot\,\vec n)\,\cdot\,\vec a\,+\,
 (\vec W\,\cdot\,\nabla\vec n)\,\cdot\,\vec a  
\ee
Per esplicitare $\nabla\vec n\,\cdot\,\vec a$ basta considerare che la 
grandezza $\vec n\,\cdot\,\vec a$ in un campo di spostamento rigido 
rimane costante, così come rimane costante il vettore $\vec \Theta$.
Quindi i gradienti spaziali di queste grandezze sono identicamente nulli 
e si ha
\be 
 (\nabla\vec n\,\cdot\,\vec a)\,+\,(\vec n\,\cdot\,\nabla\vec a)\,=\,
  \nabla(\,\vec n\,\cdot\,\vec a)\,\equiv\,0
\ee 
in cui, indicando con $\vec I$ il tensore identit\a,
\be 
  \nabla\vec a 
  \,=\,\nabla(\vec \xi+\vec \Theta\times(\vec P\,-\,\vec \xi_0)\,)
  \,=\, \nabla(\vec \Theta\times(P\vec \,-\,\vec \xi_0)\,)
  \,=\, -\vec \Theta\times\nabla(P\vec \,-\,\vec \xi_0)
  \,=\, -\vec \Theta\times\vec I
\ee 
In definitiva possiamo riscrivere la \ref{taycar} come
\be \label{aux3}
 \begin{array}{lcl}
 \dsty
 \pder{\varphi}{n}
 & \doteq &
 \dsty
                      \dot{\vec a}\cdot\vec n\,-\,
                      (\nabla\vec W\cdot\vec n)\cdot\vec a\,-\,
		      \vec W\cdot(\,\nabla\vec a\cdot\vec n\,)
\\ \\
& = & 
\dsty
\pder{\vec \xi}{t}\cdot\vec n
\,+\,   
\pder{\vec \Theta}{t}\cdot(\,\bar{\vec x}\times\,\vec n\,)
\\ \\
& - & 
( \nabla\vec W\cdot\vec n\cdot\vec \xi
  \,+\, 
  \nabla\vec W\cdot\vec n\cdot(\vec \Theta\times\,\bar{\vec x})
  \,+\,
  \vec W\cdot\vec \Theta\times\vec n\,)\,=\,
\\ \\
& - & 
\dsty
\pder{\vec \xi}{t}\cdot\vec n
\,+\,   
\pder{\vec \Theta}{t}\cdot(\,\bar{\vec x}\times\,\vec n\,)
\\ \\
& - & 
( \nabla\vec W\cdot\vec n\cdot\vec \xi
  \,+\, 
  \vec n\cdot\,(\bar{\vec x} \times \nabla\vec W)\cdot \vec \Theta
  \,+\,
  \vec n\times\,\vec W\cdot\, \vec \Theta\,) 

 \end{array}
\ee
in cui $\bar{\vec x}\,=\,(\vec P\,-\,\vec \xi_0)$ \e il vettore posizione 
nella configurazione media $\bar S$ e 
tutte le grandezze sono valutate relativamente a tale configurazione.

La(\ref{aux3}) \e una condizione di impermeabilit\a per la carena
riferita alla sua posizione media e {\em consistente} con le condizioni 
dedotte per la superficie libera.
Si osservi che nella (\ref{aux3}) compare esplicitamente la dipendenza 
{\em lineare} dai 6 gradi di libert\a, ossia dalle 12 variabili di stato 
$\{\vec q\,,\dot{\vec q}\}$. 

Definite le grandezze
 \be
  \left\{
  \begin{array}{lcl}  
  (q_1,q_2,q_3)\,:=\,\vec \xi \\[.5cm]
  \dsty
  (q_4,q_5,q_6)\,:=\,\vec \Theta  \\[.5cm] 
  \dsty
  (n_1,n_2,n_3)\,:=\,\vec n  \\[.5cm]   
  \dsty
  (n_4,n_5,n_6)\,:=\,\bar{\vec x}\,\times\,\vec n \\[.5cm]   
  \dsty
  (m_1,m_2,m_3)\,:=\,-\,(\vec n\,\cdot\,\nabla)\vec W \\[.5cm]   
  \dsty
  (m_4,m_5,m_6)\,:=\,-\,(\vec n\,\cdot\,\nabla)(\bar{\vec x}\times\vec W)    
  \end{array}  
   \right.
 \ee
la condizione di impermeabilit\a pu\o essere scritta nella
 forma pi\u compatta
 \be \label{carmj}
 \dsty
 \pder{\varphi}{n}\,=\,\sum_{j=1}^{6}\left[\pder{q_j}{t}\,n_j\,+\,
                     q_j\,m_j\right] \qquad \forall \vec x \in \bar{S}\,.
\ee
contenente i vettori {\em generalizzati} $\vec q\,,\vec n\,e\,\vec m$ 
di 6 componenti associati ai 6 gradi di libert\a della carena nel suo
moto {\em rigido}.

\subsubsection{I termini $m_j$}
Nella letteratura del settore, le componenti del vettore 
$\vec m$ sono comunemente denominati $m_j$--terms e furono introdotti 
in Ogilvie \& Tuck.
La peculiarit\a dei termini $m_j$ \e dovuta alla loro dipendenza dal gradiente 
$\nabla\vec W$ del campo di velocit\a stazionario che, pertanto, v\a risolto 
con un notevole grado di accuratezza sulla carena. 
Questo comporta dei problemi numerici di non facile trattazione, onde la loro 
{\em fama}. 

Cerchiamo di spiegare brevemente la natura di tali difficolt\a 
introducendo il flusso base, ossia sostituendo $\vec W$ con $\vec W_B$ 
nella condizione al contorno, e  facendo riferimento al suo calcolo 
mediante una formulazione integrale del problema.
Occorrer\a discretizzare la superficie del corpo in {\em pannelli}
per risolvere le opportune equazioni integrali.
L'accuratezza nel calcolo del potenziale sulla superficie del corpo
dipender\a dall'ordine $n$ del metodo usato nel descrivere il potenziale,
il suo gradiente normale e la geometria della carena (usualmente complessa
per navi commerciali).
Una volta ottenuto il potenziale sulla superficie occorrer\a valutarne
il gradiente secondo: ci\o che comporta la riduzione a $n-2$ dell'ordine di 
accuratezza.

Si pu\o allora facilmente comprendere come, utilizzando tecniche di semplice
implementazione come un metodo di ordine 0, il calcolo
dei termini $m_j$ possa essere estremamente poco accurato.
L'origine \e chiaramente legata sia alla 'perdita' dei gradienti 
tangenziali del potenziale sia alla semplificazione nella geometria
discreta che non consente di descrivere la curvatura della carena.
\`E  possibile mostrare che pur usando un elevato numero di pannelli
la convergenza a risultati analitici di riferimento \e estremamente
ridotta.

Non volendo rinunciare alla semplicit\a di implementazione di una 
tecnica di ordine zero, \e possibile sviluppare un procedimento
di estrapolazione.
Un approccio alternativo consiste nel considerare delle relazioni integrali 
sui termini $m_j$ che consentono di evitare il calcolo del gradiente 
di velocit\a sul corpo.

Data l'importanza pratica dell'argomento, un ampia trattazione
del problema del calcolo numerico dei termini $m_j$ \e riportata 
nell'Appendice, dove viene descritta la tecnica utilizzata 
in questo lavoro.

A conclusione di questa nota, si osservi che la grandezza $\nabla\vec W$ 
compare anche nella condizione al contorno per la superficie libera.
In questo caso, tuttavia, non si hanno problemi numerici rilevanti
poich\'e la condizione \e imposta su una superficie piana e la
'scarsa' convergenza associata alla non risoluzione della curvatura
e' di fatto eliminata.
Vedremo nel prossimo paragrafo che anche una parte delle forze idrodinamiche 
{\em istantanee} \e funzione del gradiente di velocit\a del flusso
stazionario o nel caso della linearizzazione di doppio modello 
del flusso base. 

\subsubsection{Dipendenza dal flusso base}
Come anticipato, scegliendo come flusso base il flusso di doppio modello,
si ottiene una condizione al contorno sulla carena formalmente 
uguale alla (\ref{newlin}) che non contiene $\phi_\epsilon$ ed in cui 
$\vec W\equiv\vec W_B$.

La linearizzazione alla $Neumann-Kelvin$ \e esente dal problema  
del calcolo dei termini $m_j$ in quanto il flusso base coincide con la
corrente uniforme e si ha, semplicemente,
\be 
 \left \{
 \begin{array}{l}
  \dsty
  (m_1,m_2,m_3)\,=\,0 \\ \\  
  \dsty
  (m_4,m_5,m_6)\,=\,\vec n\,\times\,U\,\hat{i} \\ 
 \end{array}
 \right.
\ee
in cui gli unici termini non nulli, $m_5$ ed $m_6$,
sono relativi al moto di beccheggio e a quello di imbardata.

Consistentemente con questa scelta per il flusso base,
la condizione al contorno per la carena diventa
\be 
 \begin{array}{lcl}
 \dsty
 \pder{\varphi}{n}
 & \doteq &
 \dot{\vec a}\cdot\vec n
\,+\,
 U\,\hat{i}\cdot(\,\nabla\vec a\cdot\vec n\,)
\\ \\
& = & 
\dsty
\pder{\vec \xi}{t}\cdot\vec n
\,+\,
\pder{\vec \Theta}{t}\cdot (\,\bar{\vec x}\times\,\vec n\,)
\,+\,
U\,\vec n\,\times\,\hat{i}\cdot\vec \Theta
 \end{array}
\ee
in cui l'ultimo addendo pu\o essere interpretato come il prodotto della 
velocit\a di avanzamento con l'angolo d'attacco dovuto al moto di beccheggio 
e al moto di imbardata.
\section{Espressione delle Forze idrodinamiche}
Le forze che agiscono sulla parte bagnata della carena ({\em opera viva})
possono essere ricavate, in base alla definizione, per integrazione della 
pressione agente su questa superficie. 

A tale scopo, la pressione \e ricavata  dall'equazione di Bernoulli scritta 
per i punti della superficie della carena,
\be \label{bernoulli}
\begin{array}{l}
\dsty
 p\,-\,p_a\,=\,-\rho\,\left[\dot{\Phi}\,+\,
                            \oneh\,\nabla\Phi\,\cdot\,\nabla\Phi\,-\,
                            \oneh\,U^2\,+\,g\,z\,\right] 
 \qquad \forall\,\vec x\,\in\,S \,\,,
\end{array}
\ee
e, per consistenza con quanto impostato in precedenza, deve essere trasferita
sulla configurazione media $\hat S$. Sviluppando in serie di Taylor 
si ha quindi
\be
\begin{array}{l}
\dsty
 p\,-\,p_a\,\doteq\,-\rho\,\left[\dot{\Phi}\,+\,
                            \oneh\,\nabla\Phi\,\cdot\,\nabla\Phi\,-\,
                            \oneh\,U^2\,+\,g\,z\,\right]_{\bar{S}}\,+\,
\\ \\
\dsty
           -\rho\,\nabla\,\left[\dot{\Phi}\,+\,
                            \oneh\,\nabla\Phi\,\cdot\,\nabla\Phi\,-\,
                            \oneh\,U^2\,+\,g\,z\,\right]_{\bar{S}}\,
                            \cdot\,\vec a
\end{array}
\ee
ed introducendo la decomposizione $\Phi\,=\,\phi\,+\,\varphi$ 
\be
\begin{array}{l}
\dsty
 p\,-\,p_a\,\doteq\,-\rho\,\left[\dot{\varphi}\,+\,
                            \oneh\,\vec W\,\cdot\,\vec W\,-\,
                            \oneh\,U^2\,+\, 
                            \vec W\cdot\,\nabla\varphi\,+\,
                             g\,z\,\right]_{\bar{S}}\,+\,
\\ \\
\dsty
           -\rho\,\nabla\,\left[\, \oneh\,\vec W\cdot\,\vec W\,-\,
                            \,g\,z\,\right]_{\bar{S}}\,
                            \cdot\,\vec a
\end{array}
\ee
Pertanto la parte di pressione indipendente dal tempo \e
\be
\dsty
    p\,-\,p_a\,=\,-\rho\left[\,\oneh\,\vec W\cdot\,\vec W\,-\,
                               \oneh\,U^2\,+\,g\,z\,\right]_{\bar{S}}
\ee
Si osservi che nella precedente, qualora venga introdotta la decomposizione
$\phi\,=\,\Phi_B\,+\,\phi_{\epsilon}$, la grandezza 
$\nabla\phi_{\epsilon} \cdot\,\nabla\phi_{\epsilon}$
non pu\o in generale essere trascurata ed il suo contributo all'integrale
delle forze pu\o essere rilevante.

Analogamente, 
per la parte della pressione dipendente dal tempo avremo
\be \label{aux6}
 \dsty
 p\,-\,p_a\,=\,-\rho\,\left[\,\dot{\varphi}\,+\,
                           \vec W\cdot\,\nabla\varphi\,\right]_{\bar{S}}\,-\,
	   \rho\,\vec a\,\cdot\,\nabla\left[\,\oneh\,\vec W\,\cdot\,\vec W\,+\,
                                           g\,z\,\right]_{\bar{S}}
\ee

Si osservi come il primo termine dipenda esplicitamente dal potenziale non 
stazionario. 
Come vedremo, questo primo termine \e responsabile dell'{\em effetto memoria}
delle forze idrodinamiche. 
Infatti, da  un lato la condizione di impermeabilit\a \ref{carmj} mostra 
come il valore istantaneo delle variabili $\vec q$ {\em forza} l'idrodinamica 
non stazionaria. Dall'altro
le forze idrodinamiche risentono della continua trasformazione del 
campo idrodinamico associata alla generazione e propagazione di onde di
superficie libera che hanno una velocit\a di propagazione finita a
dispetto della natura ellittica dell'equazione di Laplace.
Questa \e una sostanziale differenza rispetto al caso di flusso irrotazionale 
ovunque e in dominio illimitato per il quale le forze sono immediatamente note 
se \e noto l'atto di moto e il tensore di massa aggiunta che dipende solo dalla 
geometria del corpo.  \`E infine interessante osservare che 
nel caso di un modello non viscoso {\em rotazionale} la presenza di rilascio 
di vorticit\a e di scia introduce nuovamente un effetto memoria.

Nel secondo termine in (\ref{aux6}) notiamo anche un contributo di interazione 
tra potenziale non stazionario e potenziale stazionario 
$\vec W\cdot\nabla\varphi$: \e un contributo di forza dipendente dal tempo
associato al fatto che la carena oscilla nel campo di pressione
non uniforme associato al flusso base.
Analogamente, il termine $\vec a\cdot\nabla gz$ tiene conto dell'oscillazione 
della carena nel campo di pressione idrostatico.

Come detto la sollecitazione idrodinamica \e
\be
\left \{
\begin{array}{l}
\dsty
 F_{idrod\,j}\,=\,-\,\int_{\bar{S}}\,(\,p\,-\,p_a\,)\, n_j\,dS
\qquad con \,\, j\,=\,1\,..\,6
\\ \\
  \dsty
  (n_1,n_2,n_3)\,:=\,\vec n  \\[.5cm]   
  \dsty
  (n_4,n_5,n_6)\,:=\,\bar{\vec x}\,\times\,\vec n \\[.5cm]   
\end{array}
\right.
\ee
in cui $F_{idrod\,j}$ indica la $j$--esima componente delle forze 
{\em genaralizzate} agenti sulla carena ed $n_1,n_2,n_3$ sono le
componenti della normale uscente dalla carena. 
Esplicitando i vari contributi si ha
\be \label{fesplicita}
\left \{
\begin{array}{l}
\dsty
\vec F_{idrod}\,=\,\rho\,\int_{\bar{S}}\,\left[\,\dot{\varphi}\,+\, 
		    \vec W\cdot\,\nabla\varphi\,\right]\,\vec n dS\,+\,
\\ \\
\dsty
\qquad \qquad
              \vec \xi\,\cdot\,\rho\,\int_{\bar{S}}\,\left\{\,
              \nabla\left[\,\oneh\,\vec W\,\cdot\,\vec W\,+\,g\,z\,\right]
                          \,\right\}\vec n dS\,+\,
\\ \\
\dsty
\qquad \qquad
        \vec \Theta\,\cdot\,\rho\,\int_{\bar{S}}\,\left\{\,\bar{\vec x} 
           \times\nabla\left[\,\oneh\,\vec W\,\cdot\,\vec W\,+\,g\,z\,\right]
                          \,\right\}\vec n dS
\end{array}
\right.
\ee 

\section{Il problema linearizzato nel dominio della frequenza}
In questo paragrafo le condizioni al contorno linearizzate, sviluppate 
precedentemente, vengono utilizzate per impostare l'analisi armonica
del problema. 
In particolare, obiettivo di questo paragrafo \e esprimere
la trasformata di Fourier per le forze idrodinamiche in termini
dei potenziali di velocit\a stazionario $\phi(\vec x)$ e non 
stazionario, $\varphi(\vec x,t)$.
Risolto tale problema, tramite le equazioni di Lagrange potremo
scrivere la trasformata di Fourier della {\em Funzione di Trasferimento}
$\vec G(i\,\omega)$ e affrontare completamente 
il problema della {\em Tenuta al mare} nel dominio della frequenza.

Consideriamo la carena investita da un sistema regolare di onde 
monocromatiche di ampiezza $A$ e pulsazione $\omega_0$, la cui direzione di 
propagazione forma un angolo $\beta$ con quella di avanzamento della
carena\footnote{Ossia con il vettore $\vec U_{\infty}\,=\,-U\,\hat{i}$}
lungo l'asse $x$ del sistema di riferimento.
Il potenziale dell'onda incidente \e dato da
\be 
 \varphi_I\,=\,\frac{(i\,g\,A)}{\omega_0}\exp[k_0(z_0\,-\,i x_0 \cos\beta\,
					        -\,i y_0 \sin\beta\,)
						  +\,i\,\omega t]
\ee
dove $\omega$ rappresenta la frequenza apparente con la quale l'onda \e vista 
dalla nave per effetto della sua velocit\a di avanzamento, questa 
{\em frequenza di incontro} \e data da 
\be 
 \dsty
 \omega\,=\,\mid\,\omega_0\,-\,U_{\infty}\frac{\omega_0^2}{g}\cos\beta\,\mid
\ee
Assumiamo che sia trascorso un tempo sufficientemente lungo per 
il completo esaurimento del transitorio e che tutte le grandezze 
caratteristiche del problema abbiano raggiunto un ciclo limite di
regime. In particolare le variabili lagrangiane e le rispettive derivate 
varieranno armonicamente nel tempo con una pulsazione pari ad $\omega$ e 
quindi sar\a possibile tradurre la trasformata di Lorentz come
\be \label{Loreiw} 
 \begin{array}{c}
 \dsty
 \frac{\partial}{\partial t}\,-\,U\,\frac{\partial}{\partial x} \qquad 
 \rightarrow \qquad 
 \Re\left[\,i\,\omega\,-\,U\,\frac{\partial}{\partial x}\right] 
 \end{array}
\ee

Data la linearit\a del problema \e possibile introdurre una ulteriore 
decomposizione del potenziale $\varphi(\vec x,t)$ di cui 
discutiamo dapprima il significato fisico.

Si pu\o innanzitutto considerare il problema di una carena che, 
avanzando in mare calmo, sia forzata ad oscillare nei suoi sei gradi di 
libert\a con legge armonica e pulsazione $\omega$. 
\`E intuitivo che in tale circostanza la nave sar\a sorgente di
onde che si {\em irradieranno} lungo la superficie libera circostante 
sovrapponendosi al sistema ondoso stazionario. 
Indichiamo il potenziale soluzione del {\em problema di radiazione}
con $\varphi_R$.

Una differente situazione fisica si realizza quando la carena, 
avanzando di moto rettilineo uniforme e vincolata nel suo assetto
medio $\hat S$ interagisce con un sistema regolare di onde con 
potenziale $\varphi_I$. Dopo un tempo infinito osserveremo un 
campo ondoso modificato dall'interazione con la carena ({\em 
diffrazione}) la cui presenza \e rappresentata da un potenziale
di {\em scattering} $\varphi_S$, soluzione del {\em problema di diffrazione}

In definitiva la soluzione del problema non stazionario viene
decomposta nella forma
\be
\varphi(\vec x,t)\,=\, \varphi_R(\vec x,t) \,+\, \varphi_D(\vec x,t)
                 \,=\, \varphi_R(\vec x,t)
                 \,+\, \varphi_I(\vec x,t)\,+\, \varphi_S(\vec x,t)
\ee

\paragraph{Problema di Radiazione}
Nell'ambito di una analisi armonica, avremo sei problemi di radiazione, 
in ognuno dei quali la $i$--esima variabile lagrangiana $q_i$ 
varier\a nel tempo con legge sinusoidale, pulsazione pari a $\omega$ e 
ampiezza unitaria.
Le altre variabili lagrangiane saranno vincolate a zero.
Pertanto, in virt\u della sovrapposizione degli effetti, il 
potenziale non stazionario $\varphi_R(\vec x,t)$ \e dato dalla somma dei 
sei potenziali di radiazione ognuno {\em pesato} con la rispettiva variabile 
lagrangiana 
\be 
   \varphi_R(\vec x,t)\,=\,
          \Re\left[\,\,\sum_{j=1}^{6}\,q_j\,\varphi_j\,
           e^{i\omega\,t},\right]
\ee
in cui \e stata {\em separata} la dipendenza temporale dal tempo.
La parte spaziale $\varphi_j$ dei potenziali di radiazione 
deve soddisfare l'equazione di Laplace e le condizioni al contorno 
sulla superficie libera e sulla superficie di carena che, attraverso la 
\ref{Loreiw}, diventano
\be 
     \left\{
     \begin{array}{lcl}
        \dsty
        \pder{\varphi_j}{n} = i\,\omega\,n_j\,+\,m_j    
        \qquad \qquad \forall \vec P \in \bar{S}\\[0.3cm]
        \dsty
     \left.
     \begin{array}{lcl}
        \dsty
        g\,\pder{\varphi_j}{z}\,-\,\omega^2\,\varphi_j\,+\,2\,i\,\omega\,
        \vec W_B\,\cdot\,\nabla\varphi_j\,+\,\vec W_B\,\cdot\,
        \nabla(\vec W_B\,\cdot\,\nabla\varphi_j)\,+\,
        \\[0.25cm]
        \dsty
       +\,\oneh\,\nabla(W_B^2)\,\cdot\,\nabla\varphi_j\,
        -\,\frac{\partial^2\Phi_B}{\partial z^2}(i\,\omega\,\varphi_j\,+\,
        \vec W_B\,\cdot\,\nabla\varphi_j)\,=\,0
     \end{array}
      \right\}
              per\, z\,=\,0 
         \\ \\
        \dsty
        \mbox{con $j$ che va da $1$ a $6$} \\ \\
        \bullet\,condizione\, di\, radiazione \qquad 
     \end{array}
     \right.
\ee
e nelle quali la dipendenza dal tempo scompare lasciando una 
dipendenza parametrica dalla pulsazione $\omega$.
Nel problema continuo che stiamo considerando la superficie libera \e 
illimitata\footnote{Problemi di significativo interesse tecnologico e
scientifico sono anche quelli di {\em acque limitate}, in cui l'estensione
orizzontale \e limitata da ostacoli di varia natura, e di {\em fondale
finito}, in cui l'effetto del 'suolo' \e rilevante ai fini del fenomeno.
Un aspetto completamente differente \e quello di natura numerica
associato al troncamento del dominio di calcolo e di cui diremo pi\u avanti.} 
e per garantire l'unicit\a della soluzione occorre imporre il comportamento
asintotico per $\vec x\rightarrow\infty$ del potenziale di radiazione. 
In particolare, poich\e l'energia associata alle onde si distribuisce su 
superfici semisferiche di raggio crescente mentre il flusso di
energia irradiata si mantiene finito, la densit\a di energia deve andare a 
zero ed il potenziale di radiazione \e asintoticamente nullo.
\footnote{Kelvin deriv\o nella sua teoria sui sistemi ondosi che l'andamento
dei potenziali di radiazione doveva seguire una legge del tipo:
\be   
 \dsty
 \lim_{\parallel \vec x \parallel\rightarrow\infty} \varphi_j\,=\,
      c\,\parallel \vec x \parallel^{\oneh}\,
      e^{-i\,k\,\parallel \vec x \parallel} 
\ee
}.
\paragraph{Problema di Diffrazione}
Consideriamo ora il problema della diffrazione, in cui si vogliono
valutare le forze idrodinamiche indotte dalla presenza di un 
sistema regolare sulla carena nella sua posizione media.
Come detto il moto oscillatorio della carena \e nullo e pertanto
la forza di {\em eccitazione} indotta dall'onda incidente non dipende dalle 
variabili di stato del sistema\footnote{Questa osservazione e le sue 
conseguenze sono state gi\a trattate nel secondo capitolo, quando \e stato 
introdotto il problema della Tenuta al mare in ambito lineare.}.

Il potenziale di diffrazione $\varphi_D(\vec x,t)$ \e soluzione 
dell'equazione di Laplace e soddisfa le condizioni al contorno
\be 
     \left\{
     \begin{array}{lcl}
        \dsty
        \pder{\varphi_{D}}{n} =\,0\,     
        \qquad \qquad \forall \vec P \in \bar{S}\\[0.3cm]
        \dsty
     \left.
     \begin{array}{lcl}
        \dsty
        g\,\pder{\varphi_{D}}{z}\,-\,\omega^2\,\varphi_{D}\,+\,2\,i\,\omega\,
        \vec W_B\,\cdot\,\nabla\varphi_{D}\,+\,\vec W_B\,\cdot\,
        \nabla(\vec W_B\,\cdot\,\nabla\varphi_{D})\,+\,
        \\[0.25cm]
        \dsty
       +\,\oneh\,\nabla(W_B^2)\,\cdot\,\nabla\varphi_{D}\,
        -\,\frac{\partial^2\Phi_B}{\partial z^2}(i\,\omega\,\varphi_{D}\,+\,
        \vec W_B\,\cdot\,\nabla\varphi_{D})\,=\,0
     \end{array}
      \right\}
              per\, z\,=\,0 
         \\ \\
     \end{array}
     \right.
\ee
dipendenti da $\omega$. Infine
esplicitando il potenziale di scattering incognito 
\be 
     \left\{
     \begin{array}{lcl}
        \dsty
        \pder{\varphi_7}{n} =-\pder{\varphi_0}{n}   
        \qquad \qquad \forall \vec P \in \bar{S}\\[0.3cm]
        \dsty
     \left.
     \begin{array}{lcl}
        \dsty
        g\,\pder{\varphi_7}{z}\,-\,\omega^2\,\varphi_7\,+\,2\,i\,\omega\,
        \vec W_B\,\cdot\,\nabla\varphi_7\,+\,\vec W_B\,\cdot\,
        \nabla(\vec W_B\,\cdot\,\nabla\varphi_7)\,+\,
        \dsty
        \\[.25cm]
        \dsty
       +\,\oneh\,\nabla(W_B^2)\,\cdot\,\nabla\varphi_7\,
        -\,\frac{\partial^2\Phi_B}{\partial z^2}(i\,\omega\,\varphi_7\,+\,
        \vec W_B\,\cdot\,\nabla\varphi_7)\,=\, \\[.25cm]
        \dsty
        -g\,\pder{\varphi_0}{z}\,+\,\omega^2\,\varphi_0\,-\,2\,i\,\omega\,
        \vec W_B\,\cdot\,\nabla\varphi_0\,-\,\vec W_B\,\cdot\,
        \nabla(\vec W_B\,\cdot\,\nabla\varphi_0)\,+\,\\[.25cm]
        \dsty
       -\,\oneh\,\nabla(W_B^2)\,\cdot\,\nabla\varphi_0\,
        +\,\frac{\partial^2\Phi_B}{\partial z^2}(i\,\omega\,\varphi_0\,-\,
        \vec W_B\,\cdot\,\nabla\varphi_0)\,
     \end{array}
     \right\}
          per\, z\,=\,0 
         \\ \\
        \dsty
        \bullet\,condizione\, di\, radiazione 
     \end{array}
     \right.
\ee
in cui, da qui in poi, viene usata la notazione $\varphi_I=\varphi_0$, 
$\varphi_S=\varphi_7$.
Anche in questo caso occorre imporre una condizione di radiazione
sul potenziale di scattering $\varphi_7$ (il potenziale dell'onda incidente
\e ovviamente non nullo all'infinito perch\e corrispondente a fronti
d'onda piani con densit\a d'energia finita).

Ricapitolando il potenziale totale $\varphi(\vec x,t)$ \e
dato dalla sovrapposizione
\be
\varphi(\vec x,t)\,=\,\Re\left\{\,e^{i\omega\,t}\left[A\,
                              (\varphi_0\,+\,\varphi_7)\,+\,
                               \sum_{j\,=\,1}^{6}\,q_j\,\varphi_j\right]\,
					 \right\}
\ee
\subsection{Le forze idrodinamiche nel dominio della frequenza}
Una volta risolti i sei problemi di radiazione e quello di diffrazione,
\e possibile calcolare le forze idrodinamiche agenti sulla carena.
In particolare attraverso le $\varphi_j\,\,j\,=\,1,\ldots,6$ \e possibile 
esplicitare l'operatore idrodinamico
\be 
 \vec F_{idr}(t)\,=\,
                \vec F_0
                \,-\,[\vec A\ddot{\vec q}
                    \,+\,\vec B\dot{\vec q}
                    \,+\,\vec C\vec q
                    \,+\, 
                    \int_{0}^{t}\vec H(t-\tau)\vec q(\tau)\,d\tau
                   ]
\ee 
o equivalentemente nel dominio di Laplace, per la parte non stazionaria,
\be 
 \dsty
 \vec F_{idr}(s)\,=\,-\,\left[\,\vec A s^2 \,+\,\vec B s\,+\,\vec C \,+\,
                                \vec H(s)\,\right]\,\vec q(s)
\ee
che nel dominio della frequenza diventa
\be 
 \dsty
 \vec F_{idr}(i\omega)\,=\,-\,\left[\,
      -\,\vec A\omega^2\,+\,\vec B i\,\omega\,+\,\vec C \,\right]\,\vec q(i\omega)
                                \,-\,\vec H(i\omega)\,\vec q(i\omega)
\ee
Il primo termine dipende dall'atto di moto istantaneo mentre al
secondo, dove compare la matrice $\vec H(i\omega)$, \e associato
l'{\em effetto memoria}. 
Proprio per quest'ultima parte \e possibile
nel dominio della frequenza dare una particolare interpretazione fisica. 
Poich\`e
\be \label{gradino}
   \vec F_{mem}\,:=\,-\,\vec H(i\omega)\,\vec q(i\omega) \qquad \rightarrow \qquad
   \,-\,\int_{-\infty}^{t} \vec H(t\,-\,\tau) \vec q(\tau) d\tau 
\ee
allora la forza idrodinamica relativa ai processi di memoria dovuta
ad una variazione impulsiva delle variabili lagrangiane \e data dalle 
colonne della $\vec H(t)$.
Vogliamo ora vedere come questa grandezza sia legata ad una 
{\em variazione a gradino} delle $\vec q$, variazione che ha  
un'interpretazione fisica pi\u diretta 
\footnote{Ad esempio in ambito aeronautico 
si introduce la {\em funzione di Wagner} con la quale \e possibile valutare 
la variazione nel tempo delle forze aerodinamiche a seguito, ad esempio, 
di una variazione a gradino dell'angolo d'attacco.}. 
Integrando per parti l'integrale di convoluzione della (\ref{gradino}) si ha
\be
 \dsty
 \int_{-\infty}^{t} \vec H(t\,-\,\tau) \vec q(\tau) d\tau\,=\,
 \left[\,\vec K(t-\tau)\,\vec q(\tau)\,\right]_{-\infty}^{t}\,+\,
 \int_{-\infty}^{t} \vec K(t\,-\,\tau)\,\dot{\vec q}(\tau)\,d\tau
\ee
e, notato che 
$\left[\,\vec K(t-\tau)\,\vec q(\tau)\,\right]_{-\infty}^{t}\,=\,0$\footnote{
Infatti, come conseguenza del principio di determinismo per
un sistema dinamico reale, per $\tau\,=\,t$ si ha che $\vec K(t\,=\,0)\,=\,0$ 
mentre $\vec q(t\rightarrow\,-\infty)\,=\,0$.
Tale propriet\a vale per la risposta al gradino 
ma non per quella all'impulso.},
si deduce che la funzione $\vec -K(t)$, rappresenta la variazione nel tempo 
delle forze idrodinamiche $\vec F_{mem}$ a seguito di un impulso
in $\dot{\vec q}(t)$ ossia di un ingresso a gradino delle $\vec q(t)$. 
Quindi possiamo affermare che la conoscenza di $\vec K(t)$ \e equivalente a  
quella di $\vec H(t)$
\footnote{
Ovviamente vale anche il viceversa ossia:
\be
\begin{array}{l}
 \dsty
 \int_{-\infty}^{t} \vec K(t\,-\,\tau) \dot{\vec q}(\tau) d\tau\,=\,
 \left[\,\vec K(t-\tau)\,\vec q(\tau)\,\right]_{-\infty}^{t}\,+\,
 \int_{-\infty}^{t} \frac{\partial}{\partial t} \vec K(t\,-\,\tau)\,
                    \vec q(\tau)\,d\tau\,=\,
\\ \\
 \dsty
  \qquad \qquad \qquad \qquad \qquad \qquad \qquad 0\,\qquad \qquad 
  \,+\,\int_{-\infty}^{t} \vec H(t\,-\,\tau)\,\vec q(\tau)\,d\tau
 
\end{array}
\ee.}.
Possiamo quindi riscrivere la {\em trasformata di Fourier} per le forze 
idrodinamiche non stazionarie come
\be 
 \dsty
 \vec F_{idr}(i\omega)\,=\,-\,\left[\,
      -\,\vec A\omega^2\,+\,\vec B i\,\omega\,+\,\vec C \,\right]\,\vec q(i\omega)
                                \,-\,i\omega\,\vec K(i\omega)\,\vec q(i\omega)
\ee
Per le propriet\a di cui gode, \e possibile esprimere la
trasformata di Fourier $\vec K(i\omega)$ in termini delle trasformate
coseno e seno della medesima\footnote{La trasformata di Fourier di una generica 
funzione \e definita da 
$\dsty{\cal F}[f(t)]:=\int_{-\infty}^{+\infty}\exp(-i\omega\,t)\,f(t)\,dt$
se $f(t)\,=\,0$ per $t<0$ allora \e possibile scrivere
\be
	{\cal F}[f]\,=\,{\cal F}_c[f]\,-\,i\,{\cal F}_s[f].
\ee
dove sono state introdotte, rispettivamente, le trasformate coseno e seno
\be
   \begin{array}{lcr}
   \dsty
   {\cal F}_c[f]\,=\, \int_{0}^{\infty}f(t)\,\cos\omega\,t dt
   & \qquad &
   \dsty
   {\cal F}_s[f]\,=\, \int_{0}^{\infty}f(t)\,\sin\omega\,t dt
   \end{array}
\ee
che risultano essere funzioni {\em reali} della variabile $\omega$.
}
\be
 \vec K(i\omega)\,=\,\vec K_c(\omega)\,-\,i\,\vec K_s(\omega) 
\ee
e combinarle con le matrici $\vec A\,\,e\,\,\vec B$ dando
\be \label{ogilvie}
  \dsty
 \vec F_{idr}(i\omega)\,=\,\left[\,
      \,\omega^2\,(\,\vec A\,-\,\frac{1}{\omega}\vec K_s(\omega)\,)\,-\,
        i\,\omega\,(\,\vec B\,+\,\vec K_c(\omega)\,)\,-\,
                                 \vec C\,\right]\,\vec q(i\omega)
\ee
Possono  allora venir definite due nuove matrici
\be 
\left \{
\begin{array}{l}
\dsty
\vec A^*(\omega):=\vec A\,-\frac{1}{\omega}\,
                  \int_0^{\infty} \vec K(t)\sin\omega t\,dt
\\ \\
\dsty
\vec B^*(\omega):=\vec B\,+\,\int_0^{\infty} \vec K(t)\cos\omega t\,dt
\end{array}
\right.
\ee 
I coefficienti della $\vec A^*(\omega)$, sono chiamati, impropriamente,
{\em coefficienti di massa aggiunta} ed il loro {\em effetto fisico} 
\e quello di una massa aggiunta ma soltanto in un analisi armonica relativa
ad un solo grado di libert\`a.

I coefficienti della $\vec B^*(\omega)$ sono chiamati invece, 
{\em coefficienti di smorzamento}. A tal proposito consideriamo il caso in cui 
la j-esima variabile lagrangiana vari con la legge periodica  
\be
  q_j(t)\,=\,\Re\left\{q_j\,e^{i\omega\,t}\right\}\,=\,q_j\,\cos\omega\,t  
\ee
mentre le altre sono identicamente nulle.
La componente di forza data da $b^*_{kj}(\omega)\,i\omega\,q_je^{i\omega\,t}$ 
risulta in fase con la velocit\a del j-esimo modo, tale componente
pertanto compie lavoro, e tale lavoro \e correlato con il solo coefficiente
$b_{jj}$ che quindi \e propriamente un termine di trasferimento di
energia meccanica dal corpo verso il fluido.
Chiaramente, per l'ipotesi di fluido ideale, il sistema corpo--liquido
\e {\em conservativo}, tuttavia si pu\o avere {\em dissipazione} dell'energia
meccanica del corpo in favore di quella del liquido.

Le componenti delle forze idrodinamiche che invece risultano in fase
con l'accelerazione o con lo spostamento possono essere chiamate invece forze 
{\em reattive}, queste sono associate a disturbi locali sulla superficie libera
ma non sono correlati con il trasferimento medio di energia.

Se sono presenti due o pi\u modi armonici ad esempio $q_j$, $q_k$, allora 
esistono dei termini di accoppiamento tra questi modi dati da 
$c_{kj}$ e $a_{kj}$  con $j\neq\,k$, 
come risultato di un ingresso in accelerazione o in spostamento relativo al 
$k$---esimo modo, pu\o aver origine la $j$--esima componente delle forze 
in fase con la velocit\a $\dot{q_j}$.
Ossia lo smorzamento, nel caso di moti accoppiati, \e causato anche dai 
termini delle $\vec A^*$, $\vec C$.

Questo fa capire come anche il termine {\em coefficienti di smorzamento}
sia improprio, in quanto tutte le matrici $\vec A^*\,,\vec B^*\,,\vec C$ sono 
coinvolte nel fenomeno dello smorzamento, che \e insito nella radiazione 
di energia attraverso sistemi ondosi da parte della carena.
Non \e quindi da meravigliarsi se i termini fuori diagonale della matrice 
di massa aggiunta e della matrice di smorzamento possono essere negativi,
per certi valori della pulsazione $\omega$.
Le matrici  $\vec A$ e $\vec B$ sono, per costruzione, il limite  
delle matrici $\vec A^*$, $\vec B^*$ per frequenze infinite.
Infatti per il significato fisico della $\vec K(t)$ si ha che
\be
\lim_{\omega\rightarrow\infty}\,\int_{0}^{\infty} 
      \,\vec K(\tau)\,\sin\omega \tau\,d\tau
\,=\,
\lim_{\omega\rightarrow\infty}\,\int_{0}^{\infty} 
      \,\vec K(\tau)\,\cos\omega \tau\,d\tau 
\,=\,0
\ee
e quindi
\[
\left \{
\begin{array}{l}
\dsty
\lim_{\omega\rightarrow\infty}\, \vec A^*(\omega)\,=\,\vec A
\\ \\
\dsty
\lim_{\omega\rightarrow\infty}\, \vec B^*(\omega)\,=\,\vec B
\end{array}
\right.
\]

I coefficienti della $\vec C$ sono detti {\em coefficienti di richiamo}  
questi non dipendono dalla pulsazione $\omega$ e rappresentano un contributo 
alle forze idrodinamiche a pulsazione nulla, ossia a regime stazionario
\footnote{Nel dominio di Laplace abbiamo il teorema del valore finale, 
che aiuta a comprendere quanto detto
\[ 
\dsty
\lim_{s\rightarrow\,0} F(s)\,=\,\lim_{t\rightarrow\,+\infty} f(t)
\] 
}
\be
\dsty
\lim_{\omega\rightarrow\,0} F_{k}\,=\,\sum_{j\,=\,1}^{6}\, 
                    -\,c_{kj}\,q_j \qquad k\,=\,1,\ldots,6 
\ee
Se, ad esempio, consideriamo l'evoluzione del sistema a seguito di una
variazione a gradino nelle variabili $\vec q(t)$, dopo un certo transitorio
il sistema andr\a a regime verso un nuovo stato di equilibrio e il 
contributo dato dalle {\em forze di richiamo} andr\a a sommarsi
alle forze stazionarie $\vec F_0$ del precedente stato di equilibrio.

Rimane adesso da vedere come esprimere le forze idrodinamiche attraverso
i potenziali di velocit\a $\varphi_j$ dei problemi di radiazione.
L'espressione delle forze vista nel precedente paragrafo nel dominio della
frequenza diventa
\be
\left \{
\begin{array}{lr}
\begin{array}{lcl}
\dsty
\vec F_{idr} &=& \rho\,\int_{\bar{S}}\,\left[\,i\omega\,\varphi_j\,+\, 
		    \vec W\cdot\,\nabla\varphi_j\,\right]\,\vec n dS
\\ \\
 & + & 
\dsty
              \vec \xi\cdot\,\rho\,\int_{\bar{S}}\,\left\{\,
              \nabla\left[\,\oneh\,\vec W\,\cdot\,\vec W\,+\,g\,z\,\right]
                          \,\right\}\vec n dS
\\ \\
 & + & 
\dsty
        \vec \Theta\cdot\,\rho\,\int_{\bar{S}}\,\left\{\,\bar{\vec x} 
           \times\nabla\left[\,\oneh\,\vec W\,\cdot\,\vec W\,+\,g\,z\,\right]
                          \,\right\}\vec n dS
\end{array}
&
\mbox{con}\,\,j=1,\ldots,6
\end{array}
\right.
\ee 
In cui sono ben evidenziati i termini che non dipendono dalla frequenza
di incontro.
In particolare abbiamo dei {\em coefficienti di richiamo}: 
\be
\begin{array}{ll}
\dsty
 R_{kj}\,=\,
&
\left \{
\begin{array}{lcl}
\dsty
        \,-\,\rho\,\int_{\bar{S}}\,\left.
        \nabla\left[\,\oneh\,\vec W\,\cdot\,\vec W\,+\,g\,z\,\right]
        \,\right|_j\,n_k dS
        & &  j\,=\,1,\ldots,3 \,\, k\,=\,1,\ldots,6 
\\ \\
\dsty
        -\,\rho\,\int_{\bar{S}}\,\left. \bar{\vec x} 
        \times\nabla\left[\,\oneh\,\vec W\,\cdot\,\vec W\,+\,g\,z\,\right]
        \,\right|_{j-3}\,n_k dS
        & &  j\,=\,4,\ldots,6 \,\, k\,=\,1,\ldots,6 
\end{array}
\right.
\end{array}
\ee 
Le componenti del $R_{kj}$ date dall'accelerazione di gravit\a $\vec g$ 
sono dette {\em forze idrostatiche} e queste realizzano la parte pi\u 
consistente delle forze di richiamo. \\ \\ 
Passiamo all'espressione dei coefficienti di massa aggiunta e smorzamento 
\be 
\left \{
\begin{array}{l}
\dsty
\int_{\bar{S}}\,\left[\,i\omega\,\varphi_j\,+\,
                  \vec W\cdot\,\nabla\varphi_j\,\right]\,n_k dS\,=\,
\omega^2\,a^*_{kj}\,-c_{jk}\,+\,i\omega\,b^*_{kj}
\\ \\
\dsty
a^*_{kj}\,-\,\frac{c_{kj}}{\omega^2}\,=\,
                  \frac{\rho}{\omega^2}\Re\left\{\,
                  \int_{\bar{S}}\,\left[\,i\omega\,\varphi_j\,+\, 
		  \vec W\cdot\,\nabla\varphi_j\,\right]\,n_k dS\,
                  \right\}
\\ \\
\dsty
b^*_{kj}\,=\,-\,\frac{\rho}{\omega}\Im\left\{\,
                  \int_{\bar{S}}\,\left[\,i\omega\,\varphi_j\,+\, 
		  \vec W\cdot\,\nabla\varphi_j\,\right]\,n_k dS\,
                  \right\}
	          \qquad \qquad \mbox{con}\,\,j\,,k\,=\,1,\ldots,6
\end{array}
\right.
\ee 
I coefficienti di richiamo $c_{kj}$ sono ben pi\u modesti 
rispetto alle forze idrostatiche e la loro determinazione sperimentale
non \e semplice, per questi motivi spesso l'intero termine 
$(a_{kj}\,-\,c_{kj}/\omega^2)$ viene impropriamente chiamato   
{\em coefficiente di massa aggiunta}. \\
\newpage
Abbiamo così ricavato l'operatore idrodinamico ed \e possibile quindi
scrivere la funzione di trasferimento del sistema nel dominio della frequenza
\be
\dsty
 \vec G(j\omega)\,=\,\left[-\omega^2\,(\vec M\,+\,\vec A^*(\omega))\,+\,
                            i\omega\,\vec B^*(\omega)\,+\,\vec C 
                           \right]^{-1}
\ee

Una volta risolto il problema della diffrazione, attraverso 
il potenziale di velocit\a $\varphi_{D}\,=\,\varphi_0\,+\,\varphi_7$,
\e possibile 
calcolare con l'equazione di Bernoulli le forze di eccitazione $\vec X$
impresse sulla carena dalle onde incidenti
\be
\dsty
X_i\,:=\,\rho\,A\,\int_{\bar{S}} \left[\,i\omega\,(\varphi_0\,+\,\varphi_7)\,+\,
		   \vec W\,\cdot\,\nabla(\varphi_0\,+\,\varphi_7)\,\right]
                   \,n_i\, dS \qquad con\,i\,=\,1\,..\,6
\ee
Possiamo finalmente risolvere il problema della risposta 
ad un sistema ondoso monocromatico attraverso la
\be 
\left \{
\begin{array}{l}
\dsty
 \vec q\,=\,\vec G(i\,\omega)\,\vec X
 \\ \\
 \vec q(t)\,=\,\Re\left\{\,\vec q\,e^{i\omega t}\right\}
\qquad \qquad
 \vec X(t)\,=\,\Re\left\{\,\vec X\,e^{i\omega t}\right\}
\end{array}
\right.
\ee
\newpage
\section{Coefficienti di massa aggiunta e smorzamento}
In questo ultimo paragrafo evidenzieremo con maggior dettaglio alcune 
caratteristiche delle funzioni $\vec A^*(\omega)\,,\vec B^*(\omega)$ 
per la descrizione delle forze idrodinamiche nel dominio della frequenza:
\be
\dsty
\vec F_{idrod}(i\omega)\,=\,\left(\,\omega^2\,\vec A^*(\omega)\,-\,
                     i\omega\,\vec B^*(\omega)\,-\,\vec C\,\right)\vec q(i\,\omega) 
\ee
Per primo vogliamo vedere in che modo queste grandezze sono correlate
all'{\em energia media} irradiata dalla carena.  
Assumiamo che la carena sia forzata a muoversi secondo uno o pi\u modi 
in maniera armonica, attraverso un certo sistema di forze esterne, e che 
non siano presenti sistemi ondosi esterni. 
Dalla conoscenza  delle forze idrodinamiche e della velocit\a della carena, 
possiamo calcolare il lavoro medio che il sistema di forze esterno
esercita sulla nave, poich\h la carena non pu\o assorbire questo lavoro medio, 
questo verr\a ceduto al campo idrodinamico sotto forma di sistemi ondosi. 
Il sistema di forze esterne sar\a del tipo:
\be
F_j(t)\,=\,F_j\,\cos(\omega\,t\,+\,\delta_j) \qquad con\,j\,=\,1..6
\ee
e quindi possiamo scrivere l'equazione del moto come:
\be
\dsty
[\, \vec M\,+\,\vec A^*(\omega)\,]\,\ddot{\vec q}+\,
     \vec B^*(\omega)\,\dot{\vec q}\,+\,
     \vec C\,\vec q\,=\,\vec F\,\cos(\omega\,t\,+\,\vec \delta)
\ee
La potenza erogata \e data da:
\be
\begin{array}{l}
\dsty
W\,=\,\dot{\vec q}\,\cdot\,\vec F(t)\,=\, \\ \\
\dsty \qquad 
-\,\omega\,\sum_{j\,=\,1}^6\,\sum_{k\,=\,1}^6\,q_j\,q_k\,
                     \sin(\,\omega\,t\,+\,\epsilon_j\,) \\ \\
\dsty \qquad 
\left\{\,[\,-\omega^2\,(\,m_{jk}\,+\,a^*_{jk}\,)\,+c_{jk}\,]\,
           \cos(\,\omega t\,+\epsilon_k)\,-\,
\omega\,b^*_{jk}\sin(\omega\,t\,+\,\epsilon_k)\,\right\}
\end{array}
\ee
A noi interessa il valore medio su un intero ciclo di questa grandezza e quindi:
\be
\dsty
\bar{W}\,=\,\oneh\,\omega\,\sum_{j\,=\,1}^6\,\sum_{k\,=\,1}^6\,
                      q_j\,q_k\,\left\{\,\omega^2\,(\,a^*_{jk}\,-\,c_{jk}\,)
                      \sin(\epsilon_k\,-\,\epsilon_j)\,+\,\omega\,b^*_{jk}
                     \cos(\epsilon_k\,-\,\epsilon_j)\,\right\}
\ee
Il contributo dato dalla matrice simmetrica $m_{jk}$ si \e annullato 
con il fattore antisimmetrico $\sin(\,\epsilon_k\,-\epsilon_j\,)$,
d'altra parte nella $\bar{W}$ non possono comparire i contributi 
conservativi delle forze d'inerzia.
\\ 
Iniziamo ad analizzare questa relazione considerando che soltanto la j-esima
variabile lagrangiana sia diversa da zero. In questo caso si ha semplicemente:
\be 
\dsty
 \bar{W}\,=\,\oneh\,\omega^2\,q_j^2\,b^*{jj}
\ee
questa mostra come i termini sulla diagonale della matrice di smorzamento 
abbiano un significato fisico diretto.
Quindi nota questa potenza media $\bar{W}$ \e possibile calcolare questi 
coefficienti e viceversa. \\
Consideriamo ora il caso in cui soltanto due diverse variabili lagrangiane $q_j\,,q_k$
siano diverse da zero e che le loro fasi siano tali da assumere i due valori
$\,\,(\epsilon_j\,-\,\epsilon_k\,)\,=\,0\,oppure\,\pi/2\,\,$ in questi due casi abbiamo 
rispettivamente:
\be
 \begin{array}{l}
 \dsty
 \bar{W}\,=\,\frac{\omega}{2}\,q_j\,q_k\,\left[\,\omega^2(\,a^*_{jk}\,-\,a^*_{kj}\,)\,
 -\,(\,c_{jk}\,-\,c_{kj}\,)\,\right] \\ \\
 \bar{W}\,=\,\oneh\,\omega^2\,q_j\,q_k\,\left(\,b^*_{jk}\,+\,b^*_{kj}\,\right)
 \end{array}
\ee
Dove la prima delle relazioni scritte mette bene in evidenza come al fenomeno
dissipativo partecipino anche le matrici di massa aggiunta e delle forze di richiamo.
Attraverso la seconda relazione abbiamo che la conoscenza della potenza media, 
nel caso esposto, fornisce la somma dei termini di {\em Cross Coupling} per la 
matrice $\vec B^*(\omega)$. In realt\a attraverso la prima relazione possiamo 
conoscere anche la differenza tra questi coefficienti e quindi \e possibile 
arrivare a conoscere tutta la matrice dei coefficienti attraverso la misura della grandezza
$\bar{W}$ nei tre casi esposti. Per dimostrare tale assunto occorre ricordare che 
le matrici $\vec A^*\,,\vec B^*$ sono legate entrambe alla matrice $\vec K(t)$:
\be \label{matricek}
\left \{
\begin{array}{l}
\dsty
\vec A^*(\omega):=\vec A\,-\frac{1}{\omega}\,
                  \int_0^{\infty} \vec K(t)\sin\omega t\,dt
\\ \\
\dsty
\vec B^*(\omega):=\vec B\,+\,\int_0^{\infty} \vec K(t)\cos\omega t\,dt
\end{array}
\right.
\ee 
Invertendo ad esempio la trasformata per la matrice $\vec B^*$ abbiamo
\be
\dsty
\vec K(t)\,=\,\frac{2}{\pi}\,\int_0^{\infty}\,
     \left[\,\vec B^*(\omega)\,-\,\vec B\,\right]\,\cos (\omega\,t)\,d\omega
\ee
Inserendo questa nella prima delle \ref{matricek} abbiamo il legame 
tra le matrici di massa e quelle di smorzamento.
\be
\dsty
\vec A^*(\omega)\,=\,\vec A\,-\,\frac{2}{\pi\,\omega}\int_0^{\infty}\,\sin(\omega t)\,
                \int_0^{\infty}\,\left[\,\vec B^*(\bar{\omega}\,)\,-\,\vec B\,\right]
		    \cos(\bar{\omega}t)\,d\bar{\omega}\,dt
\ee
che pu\o essere scritta, dopo alcuni passaggi come:
\be \label{Hilbert}
\begin{array}{l}
\dsty
\vec A^*(\omega)\,-\,\vec A\,=\,\frac{2}{\pi}\int_0^{\infty}\,
                  \left[\,\vec B^*(\bar{\omega}\,)\,-\,\vec B\,\right]
		      \frac{d\,\bar{\omega}}{\bar{\omega}^2\,-\,\omega^2}
\\ \\ \mbox{Ed analogamente:} \qquad \qquad \qquad \\ \\ \dsty 
\vec B^*(\omega)\,-\,\vec B\,=\,-\,\frac{2}{\pi}\int_0^{\infty}\,
                  \left[\,\vec A^*(\bar{\omega}\,)\,-\,\vec A\,\right]
		      \frac{\bar{\omega^2}d\,\bar{\omega}}{\bar{\omega}^2\,-\,\omega^2}

\end{array}
\ee
Questa relazione \e nota in meccanica statistica come relazione di {\em Kramers-Kronig}
e pu\o essere interpretata come una trasformata di Hilbert. \\
Quindi torniamo alla misura della potenza media nel caso in cui soltanto due 
variabili lagrangiane siano diverse da zero e aventi tra loro uno sfasamento nullo,
qui abbiamo che $\bar{W}$ ci d\a:
\be 
\dsty
\bar{W}\,=\,\frac{\omega}{2}\,q_j\,q_k\,\left[\,\omega^2\,(\,a^*_{jk}\,-\,a^*_{kj}\,)\,-\,
(\,c_{jk}\,-\,c_{kj}\,)\,\right]
\ee
La conoscenza di $(\,c_{jk}\,-\,c_{kj}\,)$ ci permette di isolare la differenza
$(\,a^*_{jk}\,-\,a^*_{kj}\,)$ che ci permette attraverso la \ref{Hilbert} di trovare:
\be
\begin{array}{l}
\dsty
(\,b^*_{jk}\,-\,b^*_{kj}\,)(\omega)\,=\,-\,\frac{2}{\pi}\int_0^{\infty}\,
              \left[\,a^*_{jk}\,-\,a^*_{kj}\,\right](\bar{\omega})\,
              \frac{\bar{\omega^2}d\,\bar{\omega}}{\bar{\omega}^2\,-\,\omega^2}\,+\,
\\\\ \dsty
\qquad \qquad \qquad (\,a_{jk}\,-\,a_{kj}\,)\,-\,(\,b_{jk}\,-\,b_{kj}\,)
\end{array}
\ee
Quindi la misura della potenza media per i problemi radiativi visti, 
porta alla conoscenza  di tutti i coefficienti della matrice $\vec B^*(\omega)$. 
C'\e da sottolineare che la misura dell'energia media irradiata pu\o essere eseguita
attraverso delle superfici di controllo che possono essere poste ad una distanza arbitraria
dalla carena, e quindi c'\e in teoria la possibilit\a di ricavare alcune grandezze senza 
dover conoscere il complesso campo idrodinamico in prossimit\a della carena.
Inoltre non \e pi\u necessario tener conto che la carena si muove e quindi 
riportare tutte le grandezze sulla configurazione media dell'opera viva, 
attraverso sviluppi in serie di Taylor come \e stato fatto nei precedenti paragrafi. \\
In realt\a tale affermazione \e vera soltanto nel caso in cui la nave non abbia 
una velocit\a di avanzamento, infatti solo in questo caso la matrice $\vec C$ \e
data dalle semplici forze idrostatiche, in caso contrario le forze di richiamo 
dipendono anche loro dal campo idrodinamico. \\
C'\e da aggiungere poi che l'applicazione della relazione di Kramers-Kronig richiede
di conoscere in un ampio intervallo di frequenza la grandezza
$ \left[\,a^*_{jk}\,-\,a^*_{kj}\,\right](\bar{\omega})\, $per poter 
sviluppare l'integrale nel dominio della frequenza.\\ 
Quindi la trattazione fatta mostra i legami tra i coefficienti di massa aggiunta 
con i coefficienti di smorzamento nonch\h il legame di questi con la potenza 
media erogata dal sistema di forza, tuttavia l'utilizzo di queste relazioni per il calcolo 
di questi coefficienti non \e direttamente utilizzabile per quanto detto. \\
\indent
Vogliamo infine mostrare che l'espressione ricavata per le forze idrodinamiche 
pu\o essere ottenuta se si utilizza come soluzione generale per il potenziale di velocit\a 
la seguente espressione:
\be \label{Cummins} 
\begin{array}{l} 
\dsty
\Phi(\vec x,t)\,=\,-Ux\,+\,\phi(\vec x)\,+\,
\sum_{k=1}^6\,\dot{q}_k(t)\,\Psi_{1k}(\vec x)\,+
\sum_{k=1}^6\,q_k(t)\,\Psi_{2k}(\vec x)\,+\,
\\\\ \dsty
\sum_{k=1}^6\,\int_{-\infty}^t\,\chi_{1k}(\vec x,t-\tau)\,\dot{q}_k(t)\,d\tau\,+\,
\sum_{k=1}^6\,\int_{-\infty}^t\,\chi_{2k}(\vec x,t-\tau)\,q_k(t)\,d\tau
\end{array}
\ee
Il potenziale $\,-Ux\,+\,\phi(\vec x)$ \e il potenziale che compete al flusso stazionario 
ossia il moto di avanzamento con velocit\a costante; i potenziali 
$\Psi$ sono invece relativi alla risposta istantanea del flusso alle variazioni di stato 
del sistema, mentre i potenziali $\chi$ sono legati alle risposte prolungate nel tempo 
sempre a tali variazioni. Introducendo questa espressione nell'equazione di Bernoulli
per il calcolo della pressione e quindi delle forze, si trova che le matrici $\vec A$, 
$\vec B$ sono legate ai potenziali $\Psi$ e al potenziale stazionario
mentre la matrice $\vec K(t)$ \e legata all'insieme di tutti i potenziali ed in particolar
modo ai potenziali $\vec \chi$ che sono responsabili degli effetti di memoria del
fluido. 
La matrice $\vec C$ \e invece legata al valore asintotico per $t\rightarrow\infty$ 
dell'intero potenziale non stazionario $\varphi(\vec x,t)$. 
I potenziali $\Psi$ e $\varphi_{\infty}$ soddisfano le seguenti condizioni al contorno: \\\\
\be
\left \{
\begin{array}{l}
\dsty
\pder{\Psi_{1k}}{n}\,=\,n_k  \qquad \mbox{sulla superficie ${\bar{S}}$ } \\ \\
\dsty
\pder{\Psi_{2k}}{n}\,=\,m_k  \qquad \mbox{sulla superficie ${\bar{S}}$ } 
\qquad \mbox{ con $k\,=\,1\,..\,6$} \\ \\
\dsty
\pder{\varphi_{k\infty}}{n}\,=\,m_k  \qquad \mbox{sulla superficie ${\bar{S}}$ } 
\qquad \mbox{ con $k\,=\,1\,..\,6$} \\ \\
\dsty
\Psi_{1k}\,=\,\Psi_{2k}\,=\,0 \qquad \mbox{su  $z\,=\,0$} 
\end{array}
\right.
\ee
Senza sviluppare tutti i passaggi attraverso l'espressione delle forze 
\ref{fesplicita} e l'espressione del potenziale \ref{Cummins}, 
si ha per queste matrici:
\be
\left \{
\begin{array}{l}
\dsty
\vec A_{jk}\,=\,\rho\,\int_{\bar{S}}\,n_j\,\Psi_{1k}\, dS
\\\\ \dsty
\vec B_{jk}\,=\,\rho\,\int_{\bar{S}}\,n_j\,
\left[\,\Psi_{2k}\,+\,\vec W\,\cdot\,\nabla\Psi_{1k}\,\right]dS
\\\\ \dsty
\vec C_{jk}\,=\,\rho\,\int_{\bar{S}}\,n_j\,\vec W\,\cdot\,\nabla\varphi_{k\infty}\,dS
\end{array}
\right.
\ee
Nel caso in cui non esista un moto di avanzamento le matrici $\vec B$ 
e $\vec C$ sono identicamente nulle in quanto i potenziali 
$\Psi_2$ e $\varphi_{k\infty}$ dipendono dal vettore degli {\em m-terms}. \\ 
La matrice $\vec A$ rappresenta la massa aggiunta 
per il doppio modello  generato dall'unione delle carena con la sua 
riflessione rispetto al piano $(z\,=\,0)$; dove questo doppio modello \e immerso
completamente  nel fluido senza la presenza della superficie libera. 
Data la condizione al contorno $\varphi\,=\,0 \,\,$ su $z\,=\,0$, 
ci\o vale soltanto per i moti nel piano longitudinale ossia il sussulto, il beccheggio 
e il rollio, mentre per i moti nel piano antisimmetrico si ha sempre 
un problema di massa aggiunta ma per un doppio modello in cui le due parti 
eseguono movimenti opposti. \\
Occorre ricordare inoltre che queste due matrici sono il limite per 
$\omega\rightarrow\infty$ delle matrici $\vec A^*\,,\vec B^*$,
per le relazioni ricavate sulla potenza media erogata si ha che sia i coefficienti 
sulla diagonale della matrice $\vec B^*$ che la somma dei termini fuori diagonale 
$b^*_{kj}\,+\,b^*_{jk}$ vanno a zero con questo limite, in particolare quest'ultima 
propriet\a significa che la matrice $\vec B$ \e antisimmetrica. 
Mentre per le stesse propriet\a si ha che la matrice $\vec A$ deve essere simmetrica. \\
Se invece prendiamo in considerazione il limite opposto in cui $\omega\rightarrow\,0$,
allora la superficie libera \e una superficie rigida e quindi indeformabile; la matrice 
$\vec A^*(0)$ \e una matrice di massa aggiunta ancora una volta per un doppio modello
ma soltanto per i moti di abbrivio, deriva e imbardata,
in quanto la condizione al contorno sulla superificie libera diventa in questo 
caso $\partial{\varphi}/\partial{\vec n}\,=\,0$ su $z\,=\,0$ 
ossia una condizione di impermeabilit\a
per questo piano; per gli altri moti rigidi si ha 
un problema di un doppio modello che risulta non rigido e che oscilla dilatandosi.
Come abbiamo visto per la matrice $\vec A^*$ \e possibile calcolare i due limiti esposti 
per la pulsazione $\omega$, attraverso questi problemi in domini fluidodinamici 
illimitati e quindi senza la presenza della superficie libera.

\chapter{Formulazione integrale del problema 
         nel dominio della frequenza.}

Sia il problema non lineare (cfr. cap. 3) che quello linearizzato
e formulato nel capitolo precedente nel dominio della frequenza sono 
problemi per l'equazione di Laplace con condizioni al contorno di tipo Neumann 
sulla carena e, in generale, pi\u complesse sulla superficie 
libera\footnote{In una formulazione
non stazionaria per il problema non lineare, la condizione dinamica
di superficie libera costituisce una equazione di evoluzione per il
potenziale e quindi, per integrazione nel tempo, \e possibile disporre
di una condizione tipo Dirichlet sulla frontiera libera}.
Come visto, tali condizioni al contorno dipendono dal tempo o, nel
dominio della frequenza, dipendono parametricamente dalla pulsazione $\omega$.

In ogni caso, elemento essenziale nello studio numerico del campo 
fluidodinamico \e la soluzione di questo problema alle derivate parziali.
A tale scopo, nel seguito si far\a uso di una rappresentazione
integrale di semplice strato per il potenziale della velocit\a dalla quale si 
dedurranno equazioni agli integrali di contorno per la densit\a di
sorgente incognita e, quindi, tutte le grandezze fisicamente rilevanti
saranno espresse per il tramite di quest'ultima.

Prima di mostrare la formulazione discreta e di discutere alcuni dettagli
della formulazione numerica, si richiamano brevemente alcuni aspetti 
della formulazione integrale di problemi per l'equazione di Laplace

\section{La formulazione diretta e indiretta del problema}
Per brevit\a indichiamo $L\,:=\,\nabla^2(\cdot)$ l'operatore di Laplace
e indichiamo con $\cal D$ il dominio nel quale viene formulato un
generico problema per tale operatore.
Tale operatore \e {\em formalmente autoaggiunto}, ossia
\be
\dsty 
<\,L(G)\,,\Phi\,>\,-\,<\,G\,,L(\Phi)\,>\,:=\,
 \int_{\cal D}\,\left(\,L(G)\Phi\,-\,G\,L(\Phi)\,\right)\,dV\,=\,
 \int_{\cal D}\,L(G)\Phi\,dV 
\ee 
 dove la funzione $G$ soddisfa il problema {\em autoaggiunto}
\be
\left \{ 
\begin{array}{l}
\dsty 
 L[G(\vec x\,,\vec x^*)]\,=\,\delta(\vec x\,-\vec x^*) 
\\ \\ 
\dsty \lim_{\parallel\,\vec x\,\parallel\rightarrow\infty}\,
           G(\vec x\,,\vec x^*)\,=\,0 
\end{array}
\right.
\ee
La funzione $G$ \e la funzione di Green di spazio libero, regolare ovunque 
tranne che in $\vec x^\star$ dove \e data dalla {\em delta di Dirac},
senza alcuna  condizione al contorno se non la richiesta del comportamento 
asintotico a grande distanza da $\vec x^\star$.

La soluzione del problema autoaggiunto \e semplice in virt\u del fatto che
non occorre soddisfare le condizioni al contorno del problema differenziale 
reale, in particolare tale soluzione in uno spazio tridimensionale \e data 
semplicemente da
\be
\dsty
G(\vec x,\vec x^*)\,=\,-\frac{1}{4\pi\,\mid\,\vec x\,-\vec x^*\,\mid}
\ee
che pu\o essere interpretata fisicamente come un {\em pozzo}.
Date le sue propriet\a si ha
\be
 \int_{\cal D}\,L(G)\Phi\,dV 
 \,=\, 
 \int_{\cal D}\,\Phi(\vec x)\,\delta(\,\vec x-\,\vec x^*)dV
 \,=\,
 \Xi(\vec x^*)\Phi(\vec x^*) 
\ee
e applicando il teorema della divergenza all'integrale di volume 
\be
 \int_{\cal D}\,\nabla^2(G)\Phi\,dV
 \,=\,
 \,-\,\oint_{\partial {\cal D}}\,\left(\,\nabla G\,\Phi\,-\,G\nabla\Phi\,
                                \right)\cdot\vec n\, dS
\ee
e quindi il potenziale $\Phi$ pu\o essere rappresentato nella forma
\be \label{Green}
\Xi(\vec x^*)\Phi(\vec x^*)
\,=\, 
\oint_{\partial {\cal D}}\,
                  \left(
                  \,\pder{\Phi}{n}\,G\,-\,\pder{G}{n}\Phi\,
                  \right)\,dS
\ee
ossia in termini dei valori che assume sul contorno $\partial\cal D$
assieme alla sua derivata $\pder{\Phi}{n}$ in direzione normale.
La normale $\vec n$ \e stata scelta {\em entrante} nel dominio fluidodinamico 
e la funzione
\be 
 \Xi(\vec x^*)\,=\,\frac{\Omega(\vec x^*)}{4\pi}
\ee
tiene conto se il punto $\vec x^*$ si trova all'interno di ${\cal D}$, 
sulla sua frontiera $\partial{\cal D}$ o in $R^3\slash{\cal D}$ per il
tramite dell'angolo solido $\Omega(\vec x^*)$ con cui il punto $\vec x^*$ 
{\em vede} la frontiera del dominio\footnote{Se $\vec x^*$ \e nel dominio 
l'angolo solido visto \e $4 \pi$ steradianti, mentre \e nullo se il punto di 
osservazione \e esterno al dominio. Se 
$\vec x^\star \in \partial\cal D$ il valore dipende dalla regolarit\a
della frontiera: se \e sufficientemente {\em liscia} l'angolo solido vale 
$2 \pi$ altrimenti va valutato in funzione della {\em angolosit\a} della 
superficie nel punto in questione.}.

La relazione \ref{Green} \e alla base delle formulazioni
integrali per l'equazione di Laplace e seguendo un procedimento che 
generalizza quello sopra riportato \e possibile dare una formulazione 
integrale per problemi basati su operatori di natura pi\u complessa. 

In particolare, se $\vec x^\star\in \partial\cal D$,
la (\ref{Green}) \e una relazione di compatibilit\a agli integrali di contorno
per $\Phi|_{\partial {\cal D}}$, $\pder{\Phi}{n}|_{\partial {\cal D}}$
che, ove disponibile una condizione (puntuale) al contorno del tipo
\be \label{Robin}
\alpha \Phi + \beta \pder{\Phi}{n} = g 
\qquad\qquad \vec x^\star\in\partial\cal D
\ee
diventa una equazione differenziale per la parte incognita dei dati\footnote{
La condizione di Robin (\ref{Robin}) si riduce ad una condizione tipo Neumann o
Dirichlet per $\alpha=0$ e $\beta=0$ rispettivamente}.
\`E quindi evidente come il problema sia stato ridotto alla determinazione
dei dati incogniti su una variet\a bidimensionale malgrado il 
problema sia formulato in un dominio tridimensionale.

I metodi basati sulla soluzione delle
equazioni integrali che derivano dalla formula di Green (\ref{Green}) sono
usualmente detti metodi diretti.
\`E possibile dare formulazioni indirette in cui il potenziale \e
preliminarmente rappresentato in forma integrale per il tramite
di singolarit\a elementari incognite distribuite sul contorno del dominio
che, note dopo aver risolto le corrispondenti equazioni integrali, 
forniscono a posteriori l'incognita fisica del problema.

Rappresentazioni integrali in termini di singolarit\a di superficie
possono essere introdotte seguendo il procedimento illustrato da Lamb
in cui si considera un potenziale fittizio $\Psi$ in $R^3\slash\cal D$ e,
combinando linearmente la formula di Green scritta per $\Phi$ e $\Psi$ 
si ottiene la
\be
\left \{
\begin{array}{l}
\dsty
\Phi(\vec x^*) \,=\, 
           \int_{\partial{\cal D}} \left[\,\sigma(\vec x)G(\vec x,\vec x^*) 
           \,+\,
           \mu(\vec x)\pder{G}{\vec n}(\vec x,\vec x^*)\,\right]\, dS 
\\ \\
\dsty
\sigma\,:=\,\pder{\Phi}{\vec n}-\pder{\Psi}{\vec n} 
\qquad \qquad
\mu\,:=\,-(\Phi - \Psi)
\end{array}
\right.
\ee
Ossia il potenziale \e generato da una distribuzione superficiale
di sorgenti
\be  \label{aux7}
 \dsty 
 \Phi_S(\vec x^*)\,=\,\sigma\,G
\ee
e di doppiette
\be
 \dsty
 \Phi_D(\vec x^*)\,=\,\mu\,\pder{G}{n}
\ee
in cui l'orientamento del dipolo \e scelto coincidente con la 
normale alla frontiera.
L'incognita del problema non \e pi\u direttamente il potenziale di 
velocit\a ma l'intensit\a $\sigma$ e $\mu$ di tali distribuzioni.

Data la scelta arbitraria del potenziale $\Psi$, la rappresentazione
(\ref{aux7}) non \e unica ed in generale si possono avere rappresentazioni
di semplice strato (solo sorgenti) di doppio strato (solo doppiette)
o miste in cui una parte dei dati \e determinata mediante ulteriori 
condizioni. Pi\u in generale sono possibili rappresentazioni integrali
in termini di singolarit\a di ordine superiori (quadrupoli, ecc.).
Infine la scelta di una rappresentazione integrale piuttosto di un'altra
\e anche influenzata dal problema fisico in esame.

Nel seguito si \e scelto di rappresentare il potenziale mediante
sorgenti distribuite sulla superficie libera e sulla carena e quindi
faremo uso delle
\be
\left \{
\begin{array}{l}
\dsty
\Phi(\vec x^*)\,=\,\int_{\partial{\cal D}}\sigma G dS
\\ \\
\dsty
\pder{\Phi}{n^\star} 
\,=\, 
\pm \oneh\sigma(\vec x^*)
\,+\,\int_{\partial{\cal D}}\sigma(\vec x)\pder{G}{n^*} dS
\end{array}
\right.
\ee
dove, supponendo la frontiera sufficientemente regolare, \e stato
preso per l'angolo solido il valore di $2\pi$ steradianti ed il 
segno $+$ \e valido nel caso in cui il punto $\vec x^\star$
si accosti alla parte positiva della frontiera (la parte verso cui
punta la normale).
\section{Discretizzazione del problema.}
Nel seguito il problema linearizzato della tenuta al mare, formulato nel
dominio della frequenzai, viene risolto mediante il metodo delle
equazioni agli integrali di contorno.
Sembra opportuno precisare che, a differenza del problema del flusso attorno 
ad un corpo chiuso in un dominio illimitato, le equazioni integrali cui si
giunge sono di tipo completamente differente e non rientrano
in alcuna delle categorie per le quali esistono risultati rigorosi.
Infatti la frontiera del dominio \e infinitamente estesa, ossia 
l'operatore \e non compatto.
Inoltre, poich\'e sono coinvolte derivate seconde del potenziale 
tangenzialmente alla superficie libera, i nuclei di tali operatori non
sono n\'e di semplice n\'e di doppio strato.
Per tali problemi integrali non esiste una teoria che garantisca la
possibilit\a di una formulazione alle equazioni integrali di contorno.
Tuttavia, su un piano euristico, si potr\a constatare che le soluzioni
ottenibili sono ben confrontabili con dati sperimentali di riferimento
e, quindi, si pu\o assumere che l'approccio integrale seguito sia 
ammissibile.
Ulteriori discrepanze con i dati sperimentali andranno pertanto ascritte
a lacune fisiche del modello o insufficiente risoluzione numerica piuttosto
che ad una impossibilit\a teorica nell'utilizzare un operatore 
non di Fredholm per la rappresentazione della soluzione.

In particolare approssimiamo il potenziale e le sue derivate attraverso 
le 
\be \label{Hess}
 \left \{ 
 \begin{array}{lcccl}
 \dsty
 \varphi_j(\vec x^*) & = & \vec K_0(\vec x^*)\vec \sigma
                     & \simeq & \dsty \sum_{i=1}^{N}\sigma_i 
                               \int_{S_i} G(\vec x^*,\vec x) dS\,
\\ \\
 \dsty
 \nabla\varphi_j(\vec x^*)& = & \vec K_1(\vec x^*)\vec \sigma  
                          & \simeq & \dsty \sum_{i=1}^{N}\sigma_i 
                         \int_{S_i} \nabla G(\vec x^*,\vec x) dS\, 
 \\ \\ 
 \dsty
 \nabla \nabla\varphi_j(\vec x^*) & = & \vec K_2(\vec x^*)\vec \sigma 
                     & \simeq & \dsty \sum_{i=1}^{N}\sigma_i 
                     \int_{S_i} \nabla\nabla G(\vec x^*,\vec x) dS\,
 \end{array}
 \right.
\ee
in cui $N$ \e il numero di elementi $S_i$ utilizzato per la 
rappresentazione discreta della generica porzione della frontiera del 
dominio.  Si osserver\a che per ognuno degli elementi la
densit\a di sorgente $\sigma_i$ \e stata assunta costante e
quindi il metodo \e comunemente detto {\em di ordine zero}.
Il contributo $\dsty \int_{S_i} G(\vec x,\vec x^*) dS$
dell'i-esimo {\em pannello} in un generico punto del campo $\vec x^*$ 
pu\o essere calcolato analiticamente seguendo le formule derivate da
Hess \& Smith \cite{Hess} e, per derivazione analitica dei coefficienti di influenza 
rispetto alla variabile $\vec x^*$, verr\a ottenuta la velocit\a ed
il suo gradiente.
La disponibilit\a di tali formule ci ha fatto preferire l'operatore
di semplice strato ad una formulazione diretta.

Nel seguito indicheremo con i pedici $F$ e $B$ i termini relativi, 
rispettivamente, alla superficie libera e alla carena, ad esempio per il 
potenziale scriveremo
\be
 \varphi\,=\,\vec K_0\vec \sigma\,=\,\vec K_{0F}\vec \sigma_F +
			          \vec K_{0B}\vec \sigma_B 
\ee 
e analogamente per le sue derivate.
Le dimensioni delle {\em variabili} in $F$ sono date dal numero di pannelli
$N_F$ della superficie libera, mentre quelle con $B$ dal numero di pannelli
$N_B$ con cui \e discretizzata la superficie di carena.

Riscriviamo le condizioni al contorno per il problema di radiazione
\be 
     \left\{
     \begin{array}{lcl}
        \dsty
        \pder{\varphi_j}{n} = i\,\omega\,n_j\,+\,m_j    
        \qquad \qquad \forall \vec P \in \bar{S}\\[0.3cm]
        \dsty
     \left.
     \begin{array}{lcl}
        \dsty
        g\,\pder{\varphi_j}{z}\,-\,\omega^2\,\varphi_j\,+\,2\,i\,\omega\,
        \vec W_B\,\cdot\,\nabla\varphi_j\,+\,\vec W_B\,\cdot\,
        \nabla(\vec W_B\,\cdot\,\nabla\varphi_j)\,+\,
        \\[0.25cm]
        \dsty
       +\,\oneh\,\nabla(W_B^2)\,\cdot\,\nabla\varphi_j\,
        -\,\frac{\partial^2\Phi_B}{\partial z^2}(i\,\omega\,\varphi_j\,+\,
        \vec W_B\,\cdot\,\nabla\varphi_j)\,=\,0
     \end{array}
      \right\}
              per\, z\,=\,0 
         \\ \\
        \dsty
        \mbox{con $j$ che va da $1$ a $6$} \\ \\
        \bullet\,condizione\, di\, radiazione \qquad 
     \end{array}
     \right.
\ee
Utilizziamo le relazioni \ref{Hess} per approssimare i potenziali e le 
loro derivate e valutiamo le condizioni al contorno su tutti i punti 
di collocazione $\vec x^*$ dei pannelli. 
Per ognuno dei 6 problemi di radiazione avremo le seguenti  equazioni
discretizzate:
\be \label{raddis}
     \left\{
     \begin{array}{ll}
        \dsty 
              \vec K_{1B}\, \vec\sigma_B\, \cdot \vec n
              \,+\, 
              \vec K_{1F}\, \vec\sigma_F\, \cdot \vec n
              \,=\, i\,\omega\,n_j\,+\,m_j    
  & \forall \vec x^*\in \overline{\cal B}
  \\ \\
     \begin{array}{l}
        \dsty
        g\,\vec K_{1B}\vec \sigma_B\cdot\hat{k}
        \,-\,\omega^2 \vec K_{0B}\vec \sigma_B
        \,+\, 2i\omega\vec W_B\cdot\vec K_{1B}\vec \sigma_B
        \\
        \dsty
        \,+\, \vec W_B\cdot \nabla(\vec W_B\,\cdot\,\vec K_{1B}\vec \sigma_B)
        \,+\, \oneh\,\nabla(W_B^2)\,\cdot\,\vec K_{1B}\vec \sigma_B
        \\
        \dsty
        \,-\, \frac{\partial^2\Phi_B}{\partial z^2}(i\,\omega\,
                   \vec K_{0B}\vec \sigma_B
        \,+\, \vec W_B\,\cdot\,\vec K_{1B}\vec \sigma_B) 
 \\ + \\
        \dsty
        g\,\vec K_{1F}\vec \sigma_F\cdot\hat{k}
        \,-\,\omega^2 \vec K_{0F}\vec \sigma_F
        \,+\, 2i\omega\vec W_B\cdot\vec K_{1F}\vec \sigma_F
        \\
        \dsty
        \,+\, \vec W_B\cdot\nabla(\vec W_B\,\cdot\,\vec K_{1F}\vec \sigma_F)
        \,+\, \oneh\,\nabla(W_B^2)\,\cdot\,\vec K_{1F}\vec \sigma_F
        \\
        \dsty
        \,-\, \frac{\partial^2\Phi_B}{\partial z^2}(i\,\omega\,
                   \vec K_{0F}\vec \sigma_F
        \,+\, \vec W_B\,\cdot\,\vec K_{1F}\vec \sigma_F)
       \,=\, 0
     \end{array}
  & \forall \vec x^* \in\, (z\,=\,0) 
     \end{array}
     \right.
\ee
Le $\vec\sigma$ sono le incognite della precedente espressione e,
precisamente, il numero totale delle incognite \e $N=N_F+N_B$.
Pertanto  valutando le (\ref{raddis}) in altrettanti centroidi (uno per 
ogni pannello) si giunge alla formulazione di un problema algebrico della 
forma
\be
 \begin{array}{l}
  \vec A_{BB}\,\vec \sigma_{B}\,+\,\vec A_{BF}\,\vec \sigma_{F}\,=\,
  i\omega\,n_j\,+\,m_j
  \\ \\
  \vec A_{FB}\,\vec \sigma_{B}\,+\,\vec A_{FF}\,\vec \sigma_{F}\,=\,0
 \end{array}
\ee
Si osservi che la forma discreta del problema dipende dal 
grado di libert\a considerato per il tramite del vettore termine 
noto mentre le matrici di influenza possono essere calcolate una
volta per tutte.

Per il problema di diffrazione si avr\a
\be 
     \left\{
     \begin{array}{lcl}
        \dsty
        \pder{\varphi_7}{n} =-\pder{\varphi_0}{n}   
        & & \forall \vec P \in \bar{S}
        \\ \\
        \begin{array}{lcl}
           \dsty
           g\,\pder{\varphi_7}{z}\,-\,\omega^2\,\varphi_7\,+\,2\,i\,\omega\,
           \vec W_B\,\cdot\,\nabla\varphi_7\,+\,\vec W_B\,\cdot\,
           \nabla(\vec W_B\,\cdot\,\nabla\varphi_7)
           \\
           \dsty
          +\,\oneh\,\nabla(W_B^2)\,\cdot\,\nabla\varphi_7\,
           -\,\frac{\partial^2\Phi_B}{\partial z^2}(i\,\omega\,\varphi_7\,+\,
           \vec W_B\,\cdot\,\nabla\varphi_7) \,=\,
           \\ 
           \dsty
           -g\,\pder{\varphi_0}{z}\,+\,\omega^2\,\varphi_0\,-\,2\,i\,\omega\,
           \vec W_B\,\cdot\,\nabla\varphi_0\,-\,\vec W_B\,\cdot\,
           \nabla(\vec W_B\,\cdot\,\nabla\varphi_0)\,+\,\\[.25cm]
           \dsty
          -\,\oneh\,\nabla(W_B^2)\,\cdot\,\nabla\varphi_0\,
           +\,\frac{\partial^2\Phi_B}{\partial z^2}(i\,\omega\,\varphi_0\,-\,
           \vec W_B\,\cdot\,\nabla\varphi_0)\,
        \end{array}
        & & \mbox{per}\,z\,=\,0 
     \end{array}
     \right.
\ee
cui \e associata la forma discreta
\be
 \begin{array}{lcl}
 \dsty
  \vec A_{BB}\,\vec \sigma_{7B}\,+\,\vec A_{BF}\,\vec \sigma_{7F}
  & = & \dsty -\pder{\varphi_0}{n}
  \\ \\
  \dsty
  \vec A_{FB}\,\vec \sigma_{7B}\,+\,\,\vec A_{FF}\vec \sigma_{7F}
  & = & 
  \dsty
   -g\,\pder{\varphi_0}{z}\,+\,\omega^2\,\varphi_0\,-\,2\,i\,\omega\,
   \vec W_B\,\cdot\,\nabla\varphi_0\,-\,\vec W_B\,\cdot\,
   \nabla(\vec W_B\,\cdot\,\nabla\varphi_0)
  \\ & & 
  \dsty
   -\,\oneh\,\nabla(W_B^2)\,\cdot\,\nabla\varphi_0\,
   +\,\frac{\partial^2\Phi_B}{\partial z^2}(i\,\omega\,\varphi_0\,-\,
    \vec W_B\,\cdot\,\nabla\varphi_0)\,
 \end{array}
\ee
Si osserver\a che la matrice di influenza del problema di
diffrazione \e la stessa del precedente problema di radiazione e,
anche in questo caso, la differenza \e contenuta a livello dei
termini noti: nei primi compaiono i vettori $\vec n$ e $\vec m$,
legati al moto della carena, mentre nel problema di diffrazione compare 
l'effetto {\em forzante} dovuto al potenziale dell'onda incidente.

Finora non \e stato discusso il problema del troncamento del dominio
di calcolo che, in generale, se non si utilizzano apportuni
accorgimenti, impedisce l'ottenimento di soluzioni fisicamente 
sensate.

\section{La modellazione numerica della condizione di radiazione}
Nei problemi considerati, la superficie libera \e, nel problema continuo,
considerata di estensione infinita. Nel problema discreto sar\a
ovviamente necessario considerare una porzione limitata della 
frontiera libera.
Inoltre, per ovvi motivi di contenimento dell'onere computazionale,
tale porzione deve essere quanto  pi\u possibile ristretta nell'intorno della 
carena.

Per contro, da un punto di vista fenomenologico ci si pu\o attendere che
le onde generate dal moto della carena e dalla diffrazione dell'onda 
incidente si propaghino verso il confine numerico della superficie libera 
dove, se non opportunamente  trattate, daranno luogo a riflessioni
fisicamente prive di significato.

Da un punto di vista fisico, \e gi\a stato discusso come
la perturbazione ondosa deve essere asintoticamente nulla.
Inoltre, in generale, non si possono porre restrizioni sulla direzione
di propagazione di tale perturbazione che, a rigore, pu\o anche precedere
la carena che l'ha generata.

In questa tesi, 
nell'ambito dei problemi formulati nel dominio della frequenza,
sono considerati flussi caratterizzati da una frequenza ridotta
\be
 \tau\,:=\,\frac{\omega \,U_{\infty}}{g} 
\ee
superiore al valore critico $\tau_{cr}\,=\,\frac{1}{4}$ al di sotto
del quale \e possibile mostrare l'esistenza di un sistema di onde
che precede la carena.

\begin{figure}[htb]
      \vskip -2.cm
      \epsfxsize=0.9\textwidth
      \makebox[\textwidth]{\epsfbox{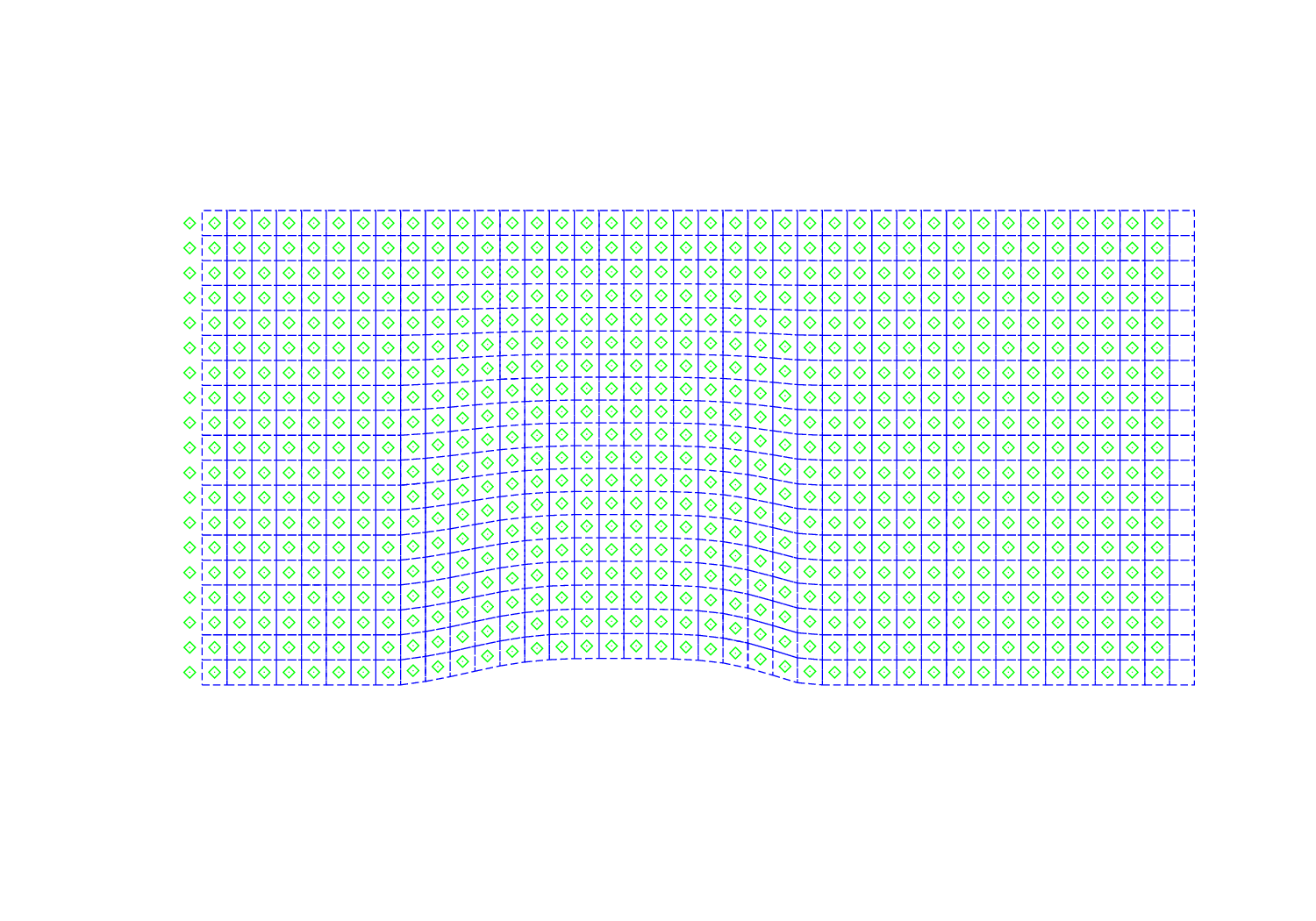}}
    \vskip -2.5cm
    \caption{Traslazione dei punti di collocazione rispetto ai pannelli di
             {\em superficie libera} \label{Griglia}
             }
\end{figure}

La tecnica utilizzata nel problema in frequenza, valida 
sotto la restrizione di cui sopra,  \e quella di
introdurre uno smorzamento di natura numerica mediante lo
spostamento dei punti di collocazione sulla superficie libera
di un certo valore nella direzione di avanzamento della carena
(cfr. fig. \ref{Griglia})

Si consideri per semplicit\a la linearizzazione alla Neumann--Kelvin
della condizione di superficie libera.
 Decomponendo il generico potenziale $\varphi_T$ nella somma dei
 contributi $\varphi$, $\varphi_{\cal B}$ dovuti alle singolarit\a di
 {\em superficie libera} e di carena, rispettivamente, la condizione su
 $\SL$ assume la forma  
\be \label{P_CRAD}
    [(i\,\omega\,-\,U\,\frac{\partial}{\partial x})^2 \,+\,
    g\,\frac{\partial}{\partial z}]\,\varphi\,=\,-
    [(i\,\omega\,-\,U\,\frac{\partial}{\partial x})^2 \,+\,
    g\,\frac{\partial}{\partial z}]\,\varphi_{\cal B}\,=\,\cal F_{\cal B}\,
\ee
 e, nel dominio dei numeri d'onda ($k_x , k_y$),
\be \label{DOM_K}
  \dsty
  \underbrace{[(-\,\omega^2\,+2\,U\,\omega\,k_x\,-\,U^2\,k_x^2\,+\,
    g\,|k|]}_{\dsty \tilde W }\,\tilde{\varphi}\,=\,\tilde{\cal F_{\cal B}}\,.
\ee
 La condizione $\tilde{W}=0$ fornisce, infine, la relazione di 
 dispersione mediante la quale si ricava una frequenza
 $\omega\in {\cal R}$ avente espressione
\be
  \begin{array}{lcl}
   \omega\,=\,U\,k_x\,\pm\,\sqrt{g\,|k|}\,.
  \end{array}
 \ee
 Supponendo ora di voler scrivere la condizione su $\SL$ in corrispondenza del 
 punto $\vec P$ e di applicarla invece in $\vec P_0=\vec P+ \delta \vec e_1$
 (con $\delta$ 'piccolo' e positivo nel verso di avanzamento della carena)
 la condizione effettivamente imposta diviene 
\be \label{P_CRAD1}
         [(i\,\omega\,-\,U\,\frac{\partial}{\partial x})^2 \,+\,
         g\,\frac{\partial}{\partial z}]\,\varphi_0\,=\,\cal F_{\cal B}\,
\ee
essendo 
$\varphi_{P_0}\simeq \varphi_P+\frac{\partial \varphi}{\partial x}|_P \delta$ 
ed il termine $U^2\,\frac{\partial^3 \varphi}{\partial x^3}\delta$ 
trascurabile.
Nel piano dei numeri d'onda si avr\a pertanto
\be \label{DOM_K2}
  \hat{W}\,\tilde{\varphi}\,=\,\tilde{\cal F_{\cal B}}
\ee
 e la relazione di dispersione $(\hat{W}=0)$ presenter\a la forma  
\be \label{CON_RA}
 \,\omega^2\,(-\,1-\,i\,k_x\,\delta)-2\,U\,\omega\,k_x\,
 (-\,1-\,i\,k_x\,\delta)\,-\,U^2\,k_x^2\,-
 \,g\,|k|(-\,1-\,i\,k_x\,\delta)\,=\,0\,.
\ee
 Diversamente da prima, la frequenza  
 \be
  \begin{array}{lcl}
   \omega\,=\,U\,k_x\,+\sqrt{\underbrace{U^2\,k_x^2\,+\,g\,|k|\,-\,
   \frac{U^2\,k_x^2}{1\,+\,k_x^2\,\delta^2}}_{a>0}+\underbrace
   {\frac{U^2\,k_x^3\,\delta}{1\,+\,k_x^2\,\delta^2}}_{b>0}\,i}
  \end{array}
 \ee
 risulta complessa ed il radicando $a\,+\,i\,b$
 ha fase $0 \leq \alpha \leq \frac{\pi}{2}$; pertanto, le sue
 radici hanno $0 \leq \alpha_1 \leq \frac{\pi}{2}$ e,
 rispettivamente, $\pi \leq \alpha_2 \leq \frac{3}{2}\pi$. In particolare, 
 per $\sqrt{a\,+\,i\,b}=\sqrt{\rho} e^{i\alpha_2}$ la
 parte immaginaria di $\omega$ \e negativa e determina pertanto una
 amplificazione delle onde di superficie. 
 Essendo inoltre negativa la sua parte reale le onde si muovono nel verso 
 della carena, cosa che si verifica in condizioni subcritiche della frequenza 
 ridotta ($\tau<\frac{1}{4}$).
 Le condizioni supercritiche corrispondono invece alla 
 situazione $\sqrt{a\,+\,i\,b}=\sqrt{\rho} e^{i\alpha_1}$, la traslazione 
 introduce, in questo caso, un termine smorzante 
 ($i\omega t =i(\omega_R+i\omega_I)t$). 
 Se si considera, pertanto, la regione
 supercritica di $\tau$ lo spostamento descritto dei punti di {\em superficie  
 libera} realizza una condizione di non propagazione a monte.

\section{Calcolo numerico delle forze idrodinamiche}
Risolti i problemi algebrici per le $\vec \sigma_j$ di radiazione,
\e possibile 
calcolare le matrici delle {\em masse aggiunte}, degli {\em smorzamenti}
e costruire, per una  $\omega$ assegnata la matrice dell'operatore 
idrodinamico, una volta calcolate numericamente le {\em forze di richiamo}
\be
  R_{ij}\,:=\,
  \left \{
  \begin{array}{c}
  \dsty
       \rho\,\int_{\overline{\cal B}}\,\nabla_j
      (\oneh\,W_B^2\,+\,g\,z)\,
       n_i d S_{\overline{\cal B}} \,\,\,\qquad j=1,..,3 \qquad i=1,..,6 \\ \\
       \dsty
       \rho\,\int_{\overline{\cal B}}\,\vec x \times \nabla_j
      (\oneh\,W_B^2\,+\,g\,z)\,
       n_i d S_{\overline{\cal B}} \,\,\, j=4,..,6 \qquad i=1,..,6 
 \end{array}
 \right.
\ee
Nota numericamente la $\vec \sigma_7$ invece possiamo calcolare la
forza di eccitazione dovuta all'onda incidente avente una 
{\em frequenza di incontro} pari ad $\omega$

In definitiva le
\be
  \left \{
  \begin{array}{l}
   \dsty
   a_{ij}\,:=\,-\,\frac{\rho}{\omega^2}\,\Re\{\int_{\overline{\cal B}}
   [i\,\omega\,\varphi_j\,+\,\vec W_B\,\cdot\,\nabla \varphi_j]\,
   n_i d S_{\overline{\cal B}}\} \\ \\ 
   \dsty
   b_{ij}\,:=\,\frac{\rho}{\omega}\,\Im\{\int_{\overline{\cal B}}
   [i\,\omega\,\varphi_j\,+\,\vec W_B\,\cdot\,\nabla \varphi_j]\,
   n_i d S_{\overline{\cal B}}\} 
     \qquad \qquad \qquad \qquad i,j=1,..,6 \\ \\
   \dsty
   X_i\,:=\,-\,\rho\,A\,\int_{\overline{\cal B}}
   [i\,\omega\,(\varphi_0\,+\,\varphi_7)\,+\,\vec W_B\,\cdot\,
   \nabla(\varphi_0+\,\varphi_7)]\,n_i d S_{\overline{\cal B}}
  \end{array}
   \right. 
\ee
diventano
\be
  \left\{
  \begin{array}{l}
   \dsty
   a_{ij}\,:=\,-\,\frac{\rho}{\omega^2}\,\Re\{\int_{\overline{\cal B}}
   [i\,\omega\,\vec K_0\,+\,\vec W_B\,\cdot\,\vec K_1]\,
   \vec \sigma_j\,n_i d S_{\overline{\cal B}}\} \\ \\
   \dsty
   b_{ij}\,:=\,\frac{\rho}{\omega}\,\Im\{\int_{\overline{\cal B}}
   [i\,\omega\,\vec K_0\,+\,\vec W_B\,\cdot\,\vec K_1]\,
   \vec \sigma_j\,n_i d S_{\overline{\cal B}}\} 
     \qquad \qquad \qquad \qquad i,j=1,..,6 \\ \\
   \dsty
   X_i\,:=\,-\,\rho\,A\,\int_{\overline{\cal B}}
   [(\,i\,\omega\,\varphi_0\,+\,\vec W_B\,\cdot\,\nabla\varphi_0\,)\,+\,
    (\,i\,\omega\,\vec K_0\,+\,\vec W_B\,\cdot\,\vec K_1\,)\,]\,
   \vec \sigma_7 n_i d S_{\overline{\cal B}}
  \end{array}
   \right. 
\ee
e, calcolate le azioni agenti sulla carena, \e possibile infine risolvere
il problema relativo al moto della nave mediante il sistema 
\be
 \dsty
 \sum_{j=1}^{6}\,[-\,\omega^2\,(m_{ij}\,+\,a_{ij})\,+\,i\,\omega\,
 b_{ij}\,+\,c_{ij}]\,\xi_j\,=\,X_i \qquad i=1,..,6\,
\ee

Le elevazioni d'onda relative ai sette problemi risolti sono date
dalla
\be
 \eta(x,y,\omega)\,=\,-\frac{1}{g}\,\Re[\,i\,\omega\,\vec K_0\,+\,
                                         \vec W_B\,\cdot\,
  						       \vec K_1]\,\vec \sigma_j
 \qquad con\,j\,=\,1..,7\,
\ee

\chapter{Risultati nel dominio della Frequenza.}

\section{Introduzione al fenomeno fisico}
In questo capitolo verr\a analizzato estesamente il campo idrodinamico
generato da una carena in moto di avanzamento in mare formato e
libera di rispondere alla sollecitazione idrodinamica che ne consegue.

Dapprima si analizzeranno gli aspetti numerici del problema. In
particolare \e importante determinare il ruolo dei vari parametri
che definiscono il problema {\em discreto}.
Successivamente si discuteranno problemi relativi a forme di carena
di complessit\a geometrica crescente e l'enfasi sar\a posta principalmente
sul confronto con i dati sperimentali.

Prima di ci\o sembra opportuno descrivere per grandi linee gli aspetti
fenomenologici che caratterizzano il problema della tenuta  al mare
delle navi.
Consideriamo il caso molto semplice di un galleggiante sferico 
che \e in condizione di equilibrio su una superficie libera indisturbata.
Se sollecitato al sussulto con legge armonica, per questioni di simmetria,
verr\a generato un sistema di onde sferiche concentriche che si
allontaneranno espandendosi in direzione radiale.

Consideriamo ora il caso in cui il galleggiante trasla con velocit\a
$U$ contenuta nel piano orizzontale, senza oscillare verticalmente.
Anche in questo caso \e intuitivo pensare al sistema di onde 
che, in condizioni di regime, inviluppano una 'V' con apice in 
corrispondenza del corpo e aperta a valle del medesimo.

Si pensi, infine, al caso in cui al galleggiante in sussulto
venga impartito il moto di traslazione.
Si avr\a allora una dissimmetrizzazione del sistema di onde: 
in particolare i fronti d'onda saranno addensati nella direzione di 
avanzamento. Se la velocit\a di avanzamento \e inferiore alla velocit\a
di fase delle onde l'inviluppo dei fronti d'onda rimarr\a aperto sia
a monte sia a valle del galleggiante. 
Nell'ambito di un'analisi lineare, si pu\o mostrare che ci\o accade 
fintanto che la frequenza di oscillazione $\omega$ \e inferiore alla 
frequenza $\frac{1}{4}g/U$ delle onde che hanno una velocit\a di fase 
$c=\sqrt{g/k}\equiv U$: ossia delle onde a regime {\em solidali} con il 
disturbo
\footnote{Questo \e vero per i fronti d'onda piani. 
In realt\a, nel caso tridimensionale che stiamo considerando, anche i fronti 
d'onda obliqui rispetto alla direzione di avanzamento possono rimanere 
solidali al disturbo purch\'e la proiezione della velocit\a di fase lungo 
la direzione di avanzamento eguagli $U$. 
\`E questa l'origine fisica del sistema ondoso 
discusso per la prima volta da Kelvin.}
\\
Nelle applicazioni che considereremo si assume sempre una frequenza ridotta 
$\tau=\omega U/g$ maggiore di $1/4$.
Pertanto, oltre alla parte stazionaria dovuta al solo avanzamento,
anche i fronti circolari sono inviluppati in una 'V' la cui ampiezza dipende
dalla velocit\a di avanzamento stessa.

\begin{figure}[htb]

\vspace*{0.5cm}

\begin{tabular}{lr}
      \hspace*{-4.25cm}
      \epsfxsize=.5\textwidth
      \epsfxsize=.5\textwidth
      \makebox[.9\textwidth]{\epsfbox{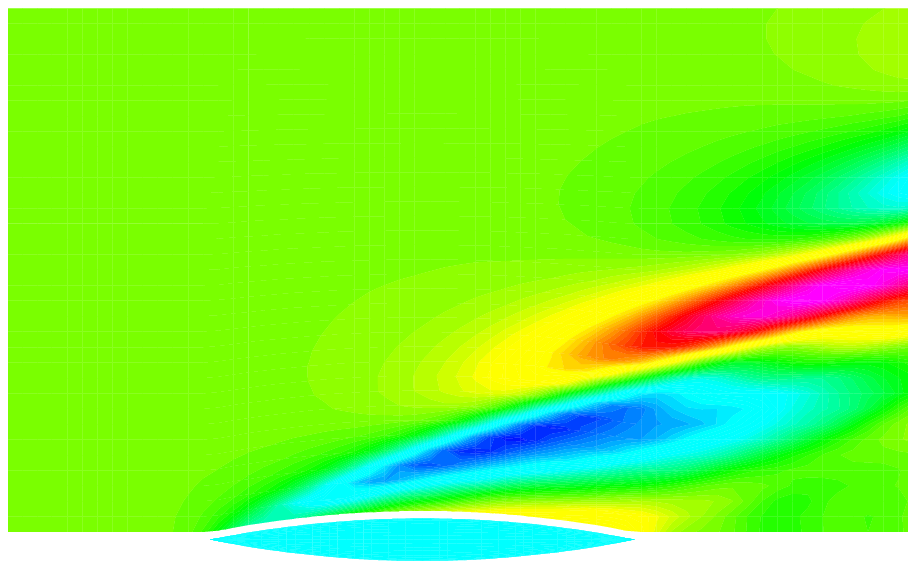}}
&
      \hspace*{-6.5cm}
      \epsfxsize=.5\textwidth
      \epsfxsize=.5\textwidth
      \makebox[.9\textwidth]{\epsfbox{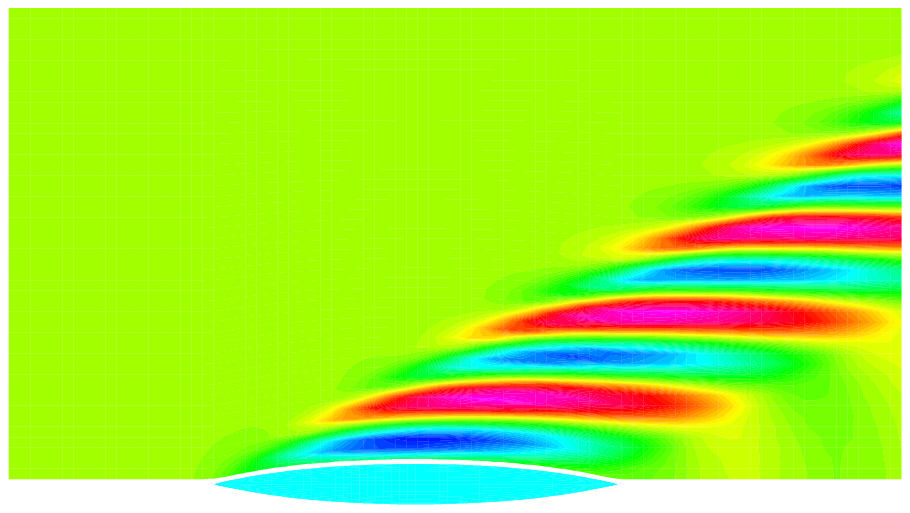}}
\end{tabular}

\vspace*{-1.0cm}

      \caption{Carena in avanzamento da destra verso sinistra con $Fr=0.3$
               in sussulto.  Solo il campo di radiazione \e rappresentato.
               Sulla sinistra $\omega=3$, sulla destra $\omega=5$.
               \label{Eff_Omega}
              }
\end{figure}

\begin{figure}[htb]
\begin{tabular}{lr}
      \hspace*{-4.25cm}
      \epsfxsize=.5\textwidth
      \epsfxsize=.5\textwidth
      \makebox[.9\textwidth]{\epsfbox{./TESIFIG/wighvG.ps}}
&
      \hspace*{-6.5cm}
      \epsfxsize=.5\textwidth
      \epsfxsize=.5\textwidth
      \makebox[.9\textwidth]{\epsfbox{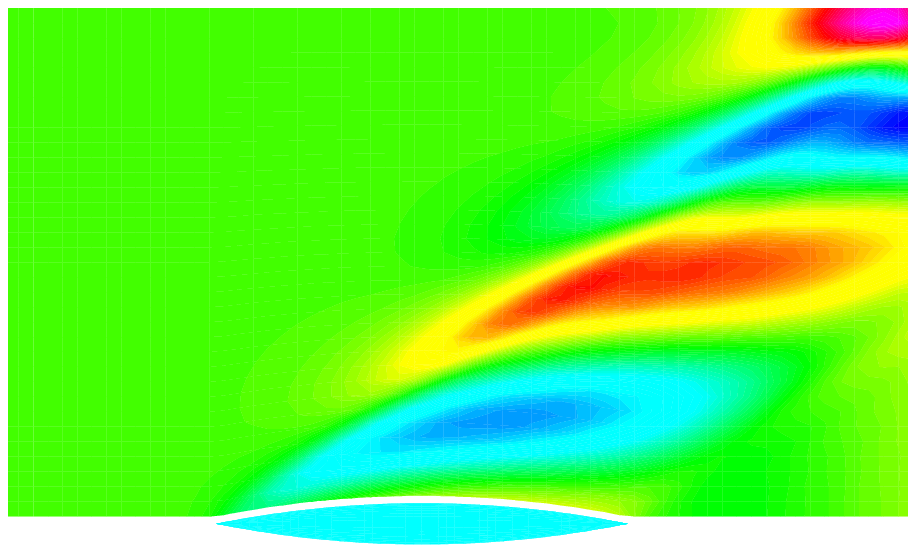}}
\end{tabular}
      \caption{Carena in avanzamento da destra verso sinistra
               in sussulto con $\omega=3$. Solo il campo di
               radiazione \e rappresentato. A sinistra $Fr=0.3$,
               a destra $Fr=0.2$.
               \label{Eff_Fr}
              }
\end{figure}

Questa descrizione pu\o essere 'ritrovata' osservando il sistema ondoso 
ottenuto numericamente per il caso di una carena in moto di avanzamento.
In particolare, anticipando il settimo capitolo,
nella figura \ref{staz3} \e riportato 
il sistema di onde generate da una carena che si muove con velocit\a 
$U=Fr\sqrt{gL}$ da destra verso sinistra e bloccata nel suo assetto medio.
La figura \ref{Eff_Omega} mostra, per il numero di Froude pari a $0.3$, 
le onde generate dal sussulto verticale della carena con frequenza angolare
$\omega=3$: queste si propagano trasversalmente con una lunghezza d'onda 
$\lambda\simeq 2\pi g/\omega^2$ tanto pi\u piccola quanto maggiore \e 
la frequenza (cfr. con il diagramma a destra relativo a $\omega=5$).
Si osservi che solo le onde irradiate dovute al sussulto sono rappresentate:
il sistema totale pu\o essere immaginato come somma di queste e di quello
stazionario sopra discusso.
La loro ampiezza sar\a funzione (lineare in una teoria lineare)
dell'ampiezza del moto imposto e dipendente non linearmente della geometria 
della carena oltre che della frequenza considerata: ossia il 
{\em potere radiativo} della carena dipende dalla frequenza di oscillazione.
Per la medesima pulsazione, ma per una maggiore velocit\a di avanzamento, si 
osserva quanto anticipato dall'analisi
intuitiva: maggiore \e il numero di Froude minore \e l'angolo dell'inviluppo
(cfr. fig. \ref{Eff_Fr}, $Fr=0.3$ a sinistra e $Fr=0.2$ a destra).

\begin{figure}[htb]
\begin{tabular}{lr}
      \hspace*{-4.25cm}
      \epsfxsize=.5\textwidth
      \epsfxsize=.5\textwidth
      \makebox[.9\textwidth]{\epsfbox{./TESIFIG/wighvG.ps}}
&
      \hspace*{-6.5cm}
      \epsfxsize=.5\textwidth
      \epsfxsize=.5\textwidth
      \makebox[.9\textwidth]{\epsfbox{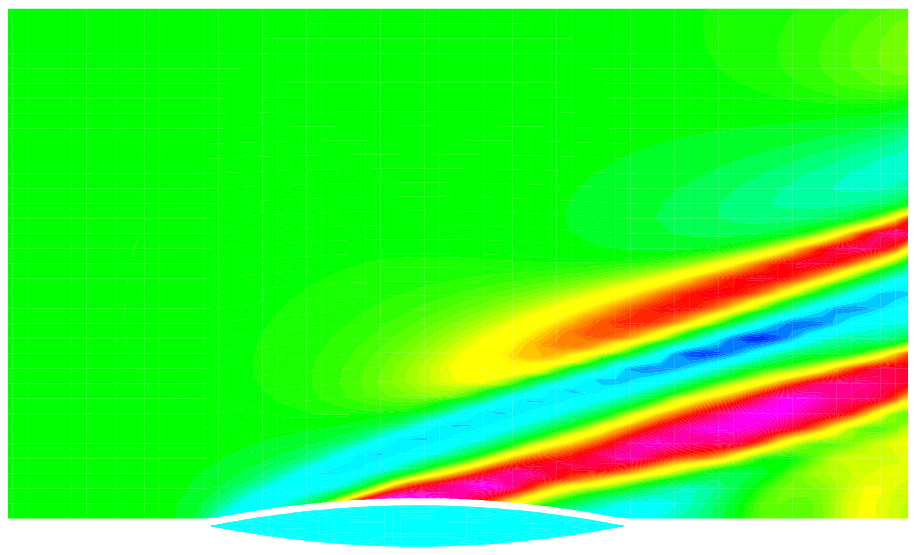}}
\end{tabular}
      \caption{Carena in avanzamento da destra verso sinistra con $Fr=0.3$.
               A sinistra in sussulto con $\omega=3$ a destra 
               in beccheggio con la medisima pulsazione.
               \label{Hv_vs_Pc}
              }
\end{figure}

Concludiamo l'analisi degli aspetti 'macroscopici' del fenomeno precisando
che, nel caso di beccheggio, le principali caratteristiche qualitative
rimangono inalterate. In aggiunta, in questo caso va precisato che, se l'asse 
di beccheggio \e in mezzeria, il moto \e antisimmetrico e i fronti d'onda
irradiati lateralmente risultano caratterizzati da una modulazione
delle sopraelevazioni e delle depressioni della superficie libera,
ossia seguendo una cresta si hanno due massimi e seguendo un cavo
si hanno due minimi, (cfr. fig. \ref{Hv_vs_Pc}).

Le caratteristiche qualitative dei sistemi ondosi potranno facilmente
essere riscontrate in tutti i risultati mostrati nel seguito.

\clearpage

\section{Analisi di sensibilit\a dei dati.}
\begin{figure}[hp]
\vskip  2.5cm
 \vspace*{-2cm}
      \epsfxsize=0.75\textwidth
      \epsfysize=0.45\textwidth
      \makebox[\textwidth]{\epsfbox{./TESIFIG/vsmall.ps}}
      \epsfxsize=0.75\textwidth
      \epsfysize=0.45\textwidth
      \makebox[\textwidth]{\epsfbox{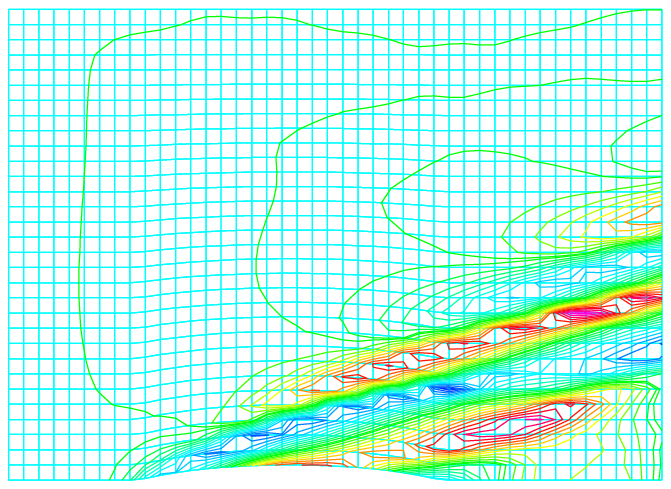}}
      \epsfxsize=0.75\textwidth
      \epsfysize=0.45\textwidth
      \makebox[\textwidth]{\epsfbox{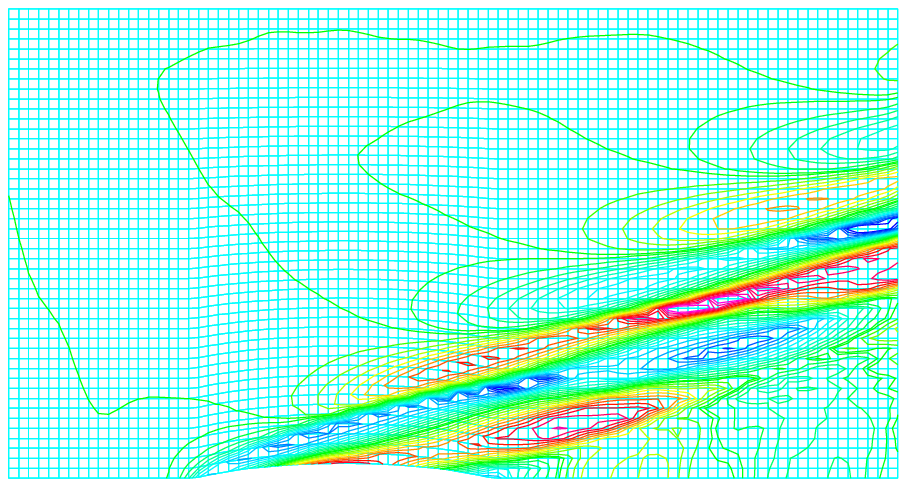}}
    \vskip -1.0cm
    \caption{Alcuni reticoli usati per le prove di convergenza dello schema numerico 
             nel dominio della frequenza. In particolare qui sono riportate le elevazioni 
             d'onda relative al problema radiativo per il moto di beccheggio con una 
             pulsazione $\omega\,=\,3.5$ con un numero di $Froude$ pari a 0.3 .
             \label {reticoli}}
\end{figure}

Prima di fare un'analisi su carene che presentano delle geometrie via via pi\u 
{\em complesse}, su un ampio intervallo di pulsazioni $\omega$ e per diversi
numeri di $Froude$, vogliamo fare delle prove del codice in frequenza, 
per valutare la sensibilit\a ad alcuni parametri del problema numerico.
In particolare i parametri che vogliamo far variare sono : la discretizzazione 
della griglia di superficie libera e la sua estensione. Quindi fisseremo il 
numero di $Froude$ e la pulsazione di incontro $\omega$, nonch\h la geometria
della carena. La carena {\em Wigley} utilizzata sar\a poi la prima di 
quelle analizzate nel prossimo paragrafo al variare della pulsazione $\omega$
e del numero di $Froude$.
Il tipo di analisi preliminare che vogliamo fare in questo paragrafo \e 
importante perch\h prima di passare ad una completa analisi armonica, occorre 
valutare le dimensioni della griglia di superficie libera, e il numero di pannelli
su questa. Al variare di questi due parametri variano sensibilmente la quantit\a di 
memoria che occorre allocare per il codice e il tempo di calcolo; quanto detto
stabilisce un limite massimo per i due parametri del reticolo di superficie libera
che vogliamo variare. 
L'influenza sul corpo dei pannelli della superficie libera 
va via via diminuendo all'aumentare della distanza 
ed \e quindi lecito trascurare i pannelli pi\u lontani, tuttavia occorre 
avere un'estensione sufficientemente ampia per permettere lo smorzamento 
delle onde uscenti, in modo da evitare delle riflessioni  
indesiderate all'interno del dominio di calcolo.
Al diminuire del numero di Froude aumenta l'angolo di divergenza 
del sistema ondoso
che la nave si lascia dietro, quindi in questi casi, per evitare riflessioni, occorrer\a
aumentare la dimensione trasversale del reticolo di superficie libera. 
L'angolo di divergenza dipende inoltre dalla geometria della carena ed in particolare 
dalla forma pi\u o meno ampia della prua; nel caso di carene {\em sottili} quest'angolo
sar\a contenuto facilitando il problema dell'estensione laterale della superficie libera. 
Consideriamo ora la grandezza dei singoli pannelli, questa  deve permettere la buona 
risoluzione delle lunghezze d'onda pi\u piccole generate dalla carena, per radiazione o 
diffrazione. 
Consideriamo per esempio un problema di radiazione per il
moto di sussulto, le onde generate dal moto della carena avranno delle lunghezze d'onda 
trasversali sempre pi\u piccole al crescere della $\omega$, e quindi al crescere della 
pulsazione occorrer\a utilizzare dei pannelli sempre pi\u piccoli per non 
avere effetti di {\em aliasing} dovuti alla discretizzazione. Vedremo che il caso pi\u 
critico sar\a dato dal problema radiativo per il moto di beccheggio, infatti in tal caso
si generano onde con lunghezza d'onda trasversale pi\u piccola rispetto a quella 
delle onde generate
nel moto di sussulto.  
Le prove di sensibilit\a ai parametri geometrici del reticolo di superficie libera sono 
state fatte utilizzando tre diversi tipi di estensione che riportiamo nella 
seguente tabella dove le lunghezze sono tutte adimensionalizzate con la 
lunghezza della carena:
\newpage
\begin{center}
\begin{table}[htb]
 \hskip 25mm \begin{tabular}{||l||l|l|l||} \hline \hline
                     & {\sc Prima }  & {\sc Seconda } & {\sc Terza} \\ \hline \hline\hline 
{\em Lungh.Monte} & $0.40$ & $0.40$ &$ 0.80$ \\ \hline
{\em Lungh.Valle} & $0.80$ & $0.80$ &$ 1.40$\\ \hline
{\em Trasversale} & $1.00$ & $1.60$ &$ 1.60$\\ \hline
  \end{tabular}  
\end{table} 
\end{center}
Lunghezza monte e valle sono rispettivamente la distanza tra la prua e l'inizio del 
reticolo e la distanza tra la poppa e la fine del reticolo; mentre la lunghezza trasversale
\e proprio la lunghezza trasversale del reticolo di superficie libera.  
Anche la posizione longitudinale della carena dovrebbe essere vista come un parametro 
del problema numerico. Comunque \e ovvio che la parte pi\u importante \e quella a valle
in quanto siamo in un regime supercritico e quindi non esistono onde generate dalla carena 
verso monte; per questo motivo la lunghezza dietro la carena \e stata presa
pi\u grande di quella davanti per tutte e tre le estensioni dei reticoli. 
Per la grandezza dei pannelli abbiamo considerato
varie ampiezze limitandoci per\o al caso di pannelli quadrati. La dicretizzazione 
pi\u rada \e di soli dieci pannelli per lunghezza di carena mentre la pi\u fitta arriva
fino a 35 pannelli per lunghezza di carena.\\
Abbiamo considerato un $Froude$ pari a 0.3 e una $\omega$ pari a $3.5$ (anche questa 
adimensionalizzata) tale scelta \e stata fatta perch\h per questi due valori siamo 
prossimi ad una condizione di risonanza del sistema dinamico.  
Per la carena abbiamo preso un reticolo di $50$x$10$, \footnote{reticolo di met\a carena per 
le propriet\a di simmetria del problema che vogliamo trattare.} che riteniamo pi\u che 
sufficiente per descrivere questa geometria. Tutti i casi di queste prove di convergenza 
sono stati fatti con l'utilizzo della linearizzazione di Doppio Modello, e per il 
calcolo dei termini $m_j$ \e stata utilizzata una tecnica di estrapolazione a seguito
di una attenta analisi riportata nell'appendice di questa Tesi.\\ 
Sono stati quindi risolti, per ognuno di questi reticoli, il problema di diffrazione 
con un'onda avente pulsazione di incontro pari a $3.5$ e tre problemi di radiazione per 
i tre modi nel piano longitudinale ossia l'abbrivio, il sussulto e il beccheggio.
I risultati che riportiamo sono soltanto per sussulto e beccheggio, poich\'e come gi\a 
sottolineato, \e per questi modi che si realizza la risonanza;
quindi i parametri principali per la tenuta al mare sono i coefficienti di massa aggiunta
e smorzamento relativi a tali modi rigidi. 
Il sistema d'onde incidente considerato proviene dalla prua della nave ed ha una 
lunghezza d'onda pari a $\lambda\,=\,1.380$ come si 
pu\o ricavare dalla relazione che fornisce la $\omega$ di incontro. Risolti i problemi
di radiazione abbiamo calcolato le matrici di massa aggiunta e smorzamento e quindi   
abbiamo risolto l'equazione della dinamica ricavando l'ampiezza dei moti e la loro 
relativa fase rispetto all'onda incidente.
Nella figura \ref{reticoli} sono riportate le tre superfici libere considerate discretizzate
rispettivamente con $15\,,20\,,30$ pannelli per lunghezza di carena. \\ 
Nella tabella \ref{contab1} sono riportati i coefficienti di massa aggiunta e smorzamento
$a_{33}\,,a_{55}\,,b_{33}\,,b_{55}$ e nelle ultime due righe abbiamo dei dati sperimentali 
per due valori delle pulsazioni {\em prossimi} a quella che stiamo analizzando.
La prima cosa che notiamo \e che i risultati ottenuti sono estremamente buoni 
per questa carena e per il $Fr$ e la $\omega$ scelti,
quindi possiamo ritenere soddisfacente il modello lineare sviluppato, almeno per 
questi valori dei parametri. 
Per tutte e quattro le grandezze si ha una giusta convergenza verso il valore sperimentale,
e in aggiunta vediamo che anche con la discretizzazione pi\u bassa otteniamo dei 
valori sufficientemente prossimi a quelli forniti dalle altre discretizzazioni. 
Il codice sviluppato fornisce quindi dei buoni risultati anche con una {\em densit\a} bassa
del reticolo. Tuttavia si nota che i coefficienti di smorzamento, legati all'energia 
media irradiata, crescono con il crescere di questa densit\a, e quindi si vede che 
i reticoli meno densi filtrano parte dell'energia irradiata dalla carena. Quanto detto
\e ben evidenziato per il moto di $beccheggio$ e una motivazione qualitativa pu\o essere 
tratta dall'osservazione del sistema ondoso mostrato nelle figure \ref{reticoli} dove si
vede come il reticolo meno denso deforma l'elevazione d'onda rispetto a quello pi\u denso. 
Tuttavia anche con pochi pannelli si riescono a vedere i {\em cavi} e i {\em picchi} 
principali.
All'aumentare della pulsazione, come vedremo nel prossimo paragrafo, l'energia persa
da un reticolo poco denso diviene sempre pi\u evidente. 
L'effetto dell'estensione della superficie libera \e 
veramente minimo e quindi i pannelli trascurati dal reticolo pi\u {\em piccolo} 
hanno effettivamente un'influenza trascurabile sulla carena, inoltre, sempre per 
questo reticolo, non si nota la presenza di onde riflesse e quindi anche la 
condizione di radiazione possiamo ritenere che sia stata sufficientemente soddisfatta. 
In particolare paragonando le griglie con densit\a pari a 30 pannelli per lunghezza di
carena al diminuire dell'estensione del reticolo si ottiene una massa aggiunta 
$a_{33}$ sovrastimata rispetto al dato sperimentale, mentre lo smorzamento 
$b_{55}$ viene sottostimato, mostrando che la sensibilit\a maggiore all'estensione 
del reticolo si ha ancora per quest'ultimo coefficiente. 
Per quanto riguarda i termini {\em incrociati} ({\em Cross-coupling}) i risultati 
ottenuti sono riportati nella successiva tabella \ref{contab2}, qui ci limitiamo
ad osservare un buon risultato complessivo, sia perch\h non ci sono andamenti irregolari
al variare del numero dei pannelli e dell'estensione del reticolo, sia perch\h 
queste grandezze non hanno un'interpretazione fisica marcata come le precedenti. 
Non abbiamo riportato invece i risultati ottenuti in diffrazione, che come vedremo 
nel prossimo paragrafo sono quelli che danno minori problemi per il confronto con
i dati sperimentali. 
Come risultato finale riportiamo il $R.A.O$ ossia l'ampiezza e la fase dei moti
di sussulto e beccheggio adimensionalizzati con l'ampiezza dell'onda incidente. 
Qui si nota un andamento convergente verso i valori sperimentali per le fasi mentre 
manca lo stesso andamento per le ampiezze dei moti ed inoltre per i reticoli pi\u 
fitti si hanno dei valori pi\u bassi dei dati sperimentali. Come vedremo, i risultati 
al variare della pulsazione sono migliori di quelli che vengono presentati qui, il motivo
\e che essendo vicini alla zona di risonanza si hanno delle forti variazioni delle 
ampiezze dei moduli al variare della $\lambda$ e quindi \e facile avere degli scarti 
come quelli riportati nella tabella pur avendo dei risultati globalmente buoni.  

\newpage
  \clearpage
{\large
\vskip -1.0cm 
\begin{table}[htb]
 \hskip 15mm \begin{tabular}{||l||l|l|l|l||} \hline \hline
                     & {\sc a33}  & {\sc b33} & {\sc a55} & {\sc b55}\\ \hline \hline\hline 

{\sc $ 10$}  &$0.44810$ &$1.5602$ &$0.20768$E$-01$ &$0.40912$E$-01$ \\ \hline
{\sc $ 15$}  &$0.47554$ &$1.7324$ &$0.21692$E$-01$ &$0.46461$E$-01$ \\ \hline
{\sc $ 20$}  &$0.48629$ &$1.7867$ &$0.22105$E$-01$ &$0.47599$E$-01$ \\ \hline
{\sc $ 25$}  &$0.48949$ &$1.8137$ &$0.23421$E$-01$ &$0.53380$E$-01$ \\ \hline
{\sc $ 30$}  &$0.49292$ &$1.8182$ &$0.22350$E$-01$ &$0.47542$E$-01$ \\ \hline
{\sc $ 35$}  &$0.49177$ &$1.8301$ &$0.23563$E$-01$ &$0.53166$E$-01$ \\ \hline
{} & & & & \\ \hline
{\sc $ 10$}  &$0.45102$ &$1.5405$ &$0.21794$E$-01$ &$0.44133$E$-01$ \\ \hline
{\sc $ 15$}  &$0.47607$ &$1.7330$ &$0.21713$E$-01$ &$0.46543$E$-01$ \\ \hline
{\sc $ 20$}  &$0.48544$ &$1.7892$ &$0.23231$E$-01$ &$0.53004$E$-01$ \\ \hline
{\sc $ 25$}  &$0.48972$ &$1.8136$ &$0.23433$E$-01$ &$0.53408$E$-01$ \\ \hline
{\sc $ 30$}  &$0.49133$ &$1.8247$ &$0.23527$E$-01$ &$0.53367$E$-01$ \\ \hline
{\sc $ 35$}  &$0.49190$ &$1.8299$ &$0.23572$E$-01$ &$0.53177$E$-01$ \\ \hline
{} & & & & \\ \hline
{\sc $ 10$}  & $0.45023$ & $1.5380$ & $0.21750$E$-01$ & $0.44151$E$-01$  \\ \hline
{\sc $ 15$}  & $0.47479$ & $1.7319$ & $0.21687$E$-01$ & $0.46711$E$-01$  \\ \hline
{\sc $ 20$}  & $0.48402$ & $1.7882$ & $0.23180$E$-01$ & $0.53178$E$-01$  \\ \hline  
{\sc $ 25$}  & $0.48780$ & $1.8103$ & $0.23367$E$-01$ & $0.53600$E$-01$  \\ \hline
{\sc $ 30$}  & $0.48974$ & $1.8245$ & $0.23483$E$-01$ & $0.53590$E$-01$  \\ \hline
{} & & & & \\ \hline
{\sc $3.35$}  & $0.485$   &  $1.843$ & $0.252$E$-01$   & $0.56$E$-01$ \\ \hline
{\sc $3.91$}  & $0.442$   &  $1.659$ & $0.237$E$-01$   & $0.55$E$-01$ \\ \hline 
  \end{tabular}  
    \vskip 0.5cm
       \caption{Prove di convergenza per lo schema numerico nel dominio della frequenza. 
                In questa tabella sono riportati i principali coefficienti di massa 
                aggiunta e smorzamento {\em adimensionalizzati}. 
                Nella prima colonna c'\e il numero di pannelli per lunghezza di carena.
                Le prove sono svolte con un numero di $Froude$ di 0.3 e con una 
                pulsazione $\omega\,=\,3.5$ quindi la lunghezza d'onda dell'onda 
                incidente \e pari a $\lambda\,=\,1.380$.
                Le ultime due righe sono i dati sperimentali per due diverse $\omega$.
       \label{contab1}}
\end{table}
}
\newpage
  \clearpage
{\large
\vskip -1.0cm 
\begin{table}[htb]
 \hskip 15mm \begin{tabular}{||l||l|l|l|l||} \hline \hline
                     & {\sc a35}  & {\sc b35} & {\sc a53} & {\sc b53}\\ \hline \hline\hline 

{\sc $ 10$}  &$-0.33735$E$-02$ &$0.85555$E$-01$  &$0.85746$E$-02$ &$-0.88692$E$-01$\\ \hline
{\sc $ 15$}  &$-0.29071$E$-01$ &$0.11504$E$-00$ &$-0.23821$E$-02$ &$-0.40504$E$-01$\\ \hline
{\sc $ 20$}  &$-0.95849$E$-02$ &$0.79405$E$-01$  &$0.11961$E$-01$ &$-0.91124$E$-01$\\ \hline
{\sc $ 25$}  &$-0.10153$E$-01$ &$0.78626$E$-01$  &$0.12066$E$-01$ &$-0.90967$E$-01$\\ \hline
{\sc $ 30$}  &$-0.10290$E$-01$ &$0.77543$E$-01$  &$0.11584$E$-01$ &$-0.90192$E$-01$\\ \hline
{} & & & & \\ \hline

{\sc $ 10$}  &$-0.34169$E$-02$ &$0.85320$E$-01$  &$0.83090$E$-02$ &$-0.87887$E$-01$\\ \hline
{\sc $ 15$}  &$-0.29198$E$-01$ &$0.11515$E$-00$ &$-0.25458$E$-02$ &$-0.38976$E$-01$\\ \hline
{\sc $ 20$}  &$-0.97356$E$-02$ &$0.79208$E$-01$  &$0.11790$E$-01$ &$-0.89611$E$-01$\\ \hline
{\sc $ 25$}  &$-0.10337$E$-01$ &$0.78260$E$-01$  &$0.11675$E$-01$ &$-0.89201$E$-01$\\ \hline
{\sc $ 30$}  &$-0.10478$E$-01$ &$0.77444$E$-01$  &$0.11487$E$-01$ &$-0.88521$E$-01$\\ \hline
{\sc $ 35$}  &$-0.10481$E$-01$ &$0.76823$E$-01$  &$0.11281$E$-01$ &$-0.87821$E$-01$\\ \hline
{} & & & & \\ \hline
{\sc $ 10$}  &$-0.21811$E$-01$ &$0.11515 $ &$-0.44331$E$-02$ &$-0.40551$E$-01$\\ \hline
{\sc $ 15$}  &$-0.29122$E$-01$ &$0.11496 $ &$-0.25402$E$-02$ &$-0.38723$E$-01$\\ \hline
{\sc $ 20$}  &$-0.31449$E$-01$ &$0.11506 $ &$-0.25192$E$-02$ &$-0.37586$E$-01$\\ \hline
{\sc $ 25$}  &$-0.10326$E$-01$ &$0.078188$ &$ 0.11692$E$-01$ &$-0.89123$E$-01$\\ \hline
{\sc $ 30$}  &$-0.32684$E$-01$ &$0.11445 $ &$-0.29479$E$-02$ &$-0.35534$E$-01$\\ \hline
{\sc $ 35$}  &$-0.10480$E$-01$ &$0.076794$ &$ 0.11301$E$-01$ &$-0.87813$E$-01$\\ \hline
{} & & & & \\ \hline
{\sc $3.35$}  &$-0.18$E$-01$    &$ 0.064$  & $0.11$E$-01$  & $-0.69$E$-01$\\ \hline
{\sc $3.90$}  &$-0.06$E$-01$    &$ 0.080$  & $0.00$E$-01$  & $-0.89$E$-01$\\ \hline

  \end{tabular}  
    \vskip 0.5cm
       \caption{Prove di convergenza per lo schema numerico nel dominio della frequenza. 
                In questa tabella sono riportati i coefficienti di massa 
                aggiunta e smorzamento {\em Cross-Coupling adimensionalizzati}. 
                Nella prima colonna c'\e il numero di pannelli per lunghezza di carena.
                Le prove sono svolte con un numero di $Froude$ di 0.3 e con una 
                pulsazione $\omega\,=\,3.5$ quindi la lunghezza d'onda dell'onda 
                incidente \e pari a $\lambda\,=\,1.380$.
                Le ultime due righe sono i dati sperimentali per due diverse $\omega$.
       \label{contab2}}
\end{table}
}
{\large
\vskip -1.0cm 
\begin{table}[htb]
 \hskip 15mm \begin{tabular}{||l||l|l|l|l||} \hline \hline
   & {\sc $\xi_3$}  & {\sc $arg(\xi_3)$} & {\sc $\xi_5$} & {\sc arg($\xi_5$)}\\ 
     \hline \hline\hline 

{\sc $ 10$}   &$1.1300$  &$ 6.4492$ &$9.2037$ &$-115.33$ \\ \hline
{\sc $ 15$}   &$1.2387$  &$ 4.3591$ &$9.0539$ &$-118.61$ \\ \hline
{\sc $ 20$}   &$1.2770$  &$ 3.7802$ &$9.0529$ &$-119.56$ \\ \hline
{\sc $ 25$}   &$1.0448$  &$-6.4600$ &$8.5832$ &$-118.34$ \\ \hline
{\sc $ 30$}   &$1.3017$  &$ 3.6120$ &$9.1118$ &$-119.92$ \\ \hline
{\sc $ 35$}   &$1.0497$  &$-6.4980$ &$8.6118$ &$-118.53$ \\ \hline
{} & & & & \\ \hline

{\sc $ 10$}   &$0.95371$ &$-4.5438$ &$8.9054$ &$-113.45$ \\ \hline
{\sc $ 15$}   &$1.2400$  &$ 4.3492$ &$9.0514$ &$-118.67$ \\ \hline
{\sc $ 20$}   &$1.0358$  &$-6.3268$ &$8.5863$ &$-118.00$ \\ \hline
{\sc $ 25$}   &$1.0451$  &$-6.4575$ &$8.5827$ &$-118.37$ \\ \hline
{\sc $ 30$}   &$1.0486$  &$-6.4969$ &$8.5967$ &$-118.51$ \\ \hline
{\sc $ 35$}   &$1.0499$  &$-6.4978$ &$8.6122$ &$-118.55$ \\ \hline
{} & & & & \\ \hline

{\sc $ 10$}   &$0.95282$  &$-4.5555$   &$8.8936$ &$-113.42$ \\ \hline
{\sc $ 15$}   &$1.2378$   &$ 4.2474$   &$9.0272$ &$-118.73$ \\ \hline
{\sc $ 20$}   &$1.0339$   &$-6.3608$   &$8.5652$ &$-117.99$ \\ \hline
{\sc $ 25$}   &$1.0428$   &$-6.4743$   &$8.5588$ &$-118.35$ \\ \hline
{\sc $ 30$}   &$1.0465$   &$-6.5525$   &$8.5734$ &$-118.53$ \\ \hline
{} & & & & \\ \hline

{\sc $1.250$}  &$ 1.38$      &$  -19$   &$  11.0$    &$ -169$ \\ \hline
{\sc $1.384$}  &$ 1.11$      &$  -6 $   &$  9.99$    &$ -129$ \\ \hline

  \end{tabular}  
    \vskip 0.5cm
       \caption{Prove di convergenza per lo schema numerico nel dominio della frequenza. 
                In questa tabella sono riportati i moti di sussulto e beccheggio 
                in modulo (adimensionale) e fase (in gradi).  
                Nella prima colonna c'\e il numero di pannelli per lunghezza di carena.
                Le prove sono svolte con un numero di $Froude$ di 0.3 e con una 
                pulsazione $\omega\,=\,3.5$ quindi la lunghezza d'onda dell'onda 
                incidente \e pari a $\lambda\,=\,1.380$.
                Le ultime due righe sono i dati sperimentali per due diverse $\lambda$.
       \label{contab3}}
\end{table}
}

\clearpage
\newpage
\section{Risultati in Frequenza \\ per sei carene di superficie.}
 In questa sezione viene studiata la risposta di alcune carene che avanzano con 
 certi numeri di $Froude$ e che vengono investite da vari sistemi d'onda. 
 Relativamente a questi considereremo treni d'onda monocromatici 
 di ampiezza $A$, lunghezza $\lambda$ e frequenza $\omega_0$ che raggiungono la 
 nave da prua. Al modificarsi di questi parametri, vengono quindi ricavate 
 la formazione ondosa, le forze idrodinamiche e i moti nave che ne risultano.\\
 Quest'analisi viene realizzata, in particolare, per sei carene diverse che presentano
 una geometria sempre pi\u {\em "complessa"}. 
 Identificheremo queste carene con alcuni parametri geometrici, in particolare:
 \begin{itemize} 
 \item 
 il rapporto $L/B$ tra la lunghezza e la massima larghezza ({\em beam}), 
 \item
 il rapporto $L/T$  dove $T$ indica il massimo {\em pescaggio} ({\em draft}) 
 \item 
 ed infine il rapporto $Cb\,:=\,Volume\,della\,carena/(B L T)$  
 detto in ambito navale {\em coefficiente di finezza totale (block coefficient)}.
 \end{itemize}
 \indent 
 Inizieremo con una carena {\em Wigley} avente $Cb\simeq 0.46\,,L/B\,=\,10\,,L/T\,=\,16$,
 da questi parametri si vede che questa carena \e sufficientemente snella ed inoltre
 vedremo che presenta una superficie con variazioni di curvatura molto contenute. Quindi 
 ci aspettiamo dei buoni risultati per entrambe le linearizzazioni viste sia quella di 
 Neumann-Kelvin che quella di doppio modello.  \\ \indent 
 La seconda carena presenta gli stessi parametri ma ha un coefficiente di blocco  
 $Cb\,\simeq\,0.56$ che indica una superficie con una forte variazione di curvatura lungo
 le sezioni 
 \footnote{All'aumentare del coefficiente di blocco le sezioni tendono a riempire 
il rettangolo che ha per lati il pescaggio e la larghezza della sezione, si ha quindi 
la presenza di un {\em ginocchio} ossia una zona con piccolo raggio di curvatura che 
raccorda due zone di bassa curvatura.}, 
 quindi questa \e ancora una carena {\em snella} ma il campo fluidodinamico
 che ne risulta sar\a pi\u complesso che nella precedente. 
 \\ \indent    
 La terza carena \e generata con le stesse sezioni della prima e quindi 
 presenta una superficie {\em semplice} ma ha un rapporto $L/B\,=\,5$ e quindi doppio
 rispetto alle altre. Per questa carena non possiamo considerare l'ipotesi di 
 snellezza, e ci aspettiamo che la differenza tra i due modelli lineari presenti 
 degli scostamenti maggiori rispetto ai casi precedenti, in particolare ci aspettiamo
 dei risultati migliori per la linearizzazione di doppio modello. 
 \\ \indent 
 La quarta carena riprende le sezioni della seconda ma, in analogia al precedente caso,
 presenta un rapporto $L/B\,=\,5$ che corrisponde quindi ad un corpo non snello ed in
 pi\u con una superficie che comporta un campo fluidodinamico di una certa difficolt\a. 
 \\ \indent
 La quinta carena \e ancora una carena {\em Wigley} che presenta un coefficiente di
 blocco $Cb\,=\,0.63$, il pi\u alto delle carene studiate. I rapporti 
 $L/B\,,L/T$ sono uguali a quelli delle prime due carene, e quindi anche questa 
 \e una carena snella, ma la geometria \e qui ancora pi\u complessa 
 con una variazione della curvatura molto accentuata nelle sezioni centrali della carena.
 \\ \indent
 L'ultima carena fa parte della famiglia della Serie 60, ha un $Cb\,=\,0.6$ 
 ed i rapporti $L/B\,=\,10\,,L/T\,\simeq\,18$; la superficie di questa carena 
 ha una complessit\a ancora maggiore ed in particolare \e l'unica carena 
 tra quelle considerate a non essere simmetrica rispetto al piano $yz$, 
 e a non avere le normali di superficie nella prima linea d'acqua 
 \footnote {con linee d'acqua indichiamo le sezioni di carena con piani paralleli
 al piano $xy$, in particolare la prima \e proprio quella ottenuta con l'intersezione 
 di quest'ultimo piano} contenute nel piano $z\,=\,0$.  
 \\ \indent
 Per tali problemi viene riportato il confronto con dati sperimentali 
 disponibili in letteratura; per le prime carene Wigley abbiamo anche il confronto
 con la Strip-Theory, mentre per la quinta Wigley abbiamo un confronto
 con un codice che utilizza pannelli di ordine superiore.
\newpage
 \clearpage
\section{Casi esaminati: Prima Carena Wigley.} 
Iniziamo l'analisi armonica con questa prima carena di cui ricordiamo 
i tre parametri geometrici definiti nel precedente paragrafo:
$L/B\,=\,10\,,L/T\,=\,16\,,Cb\,\simeq\,0.46$. \\ 
Nella figura sottostante \e riportato il reticolo utilizzato per i
descrivere questa geometria: 
\begin{figure}[htb]
    \vskip 0.5cm
      \epsfxsize=.5\textwidth
      \epsfxsize=.5\textwidth
      \makebox[.9\textwidth]{\epsfbox{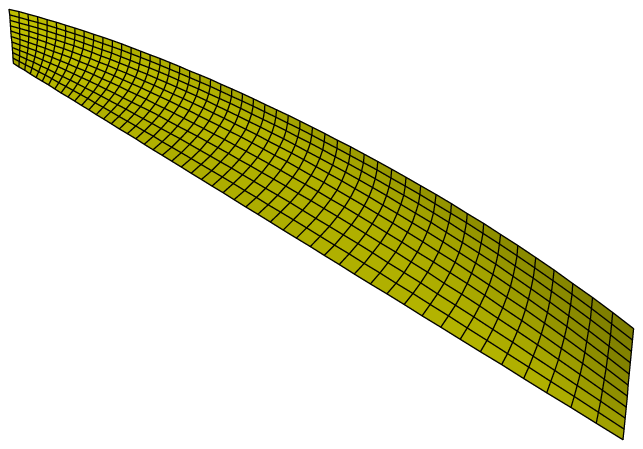}}
    \vskip -0.5cm
    \caption{Esempio di un tipico reticolo di calcolo per la prima 
             carena Wigley }
\end{figure}
 \vskip .5cm

\begin{figure}[htb]
      \epsfxsize=\textwidth
      \makebox[\textwidth]{\epsfbox{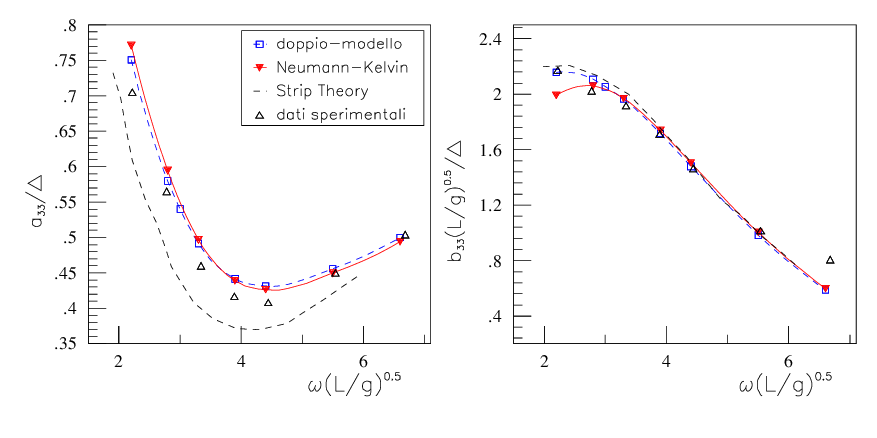}}
      \epsfxsize=\textwidth
      \makebox[\textwidth]{\epsfbox{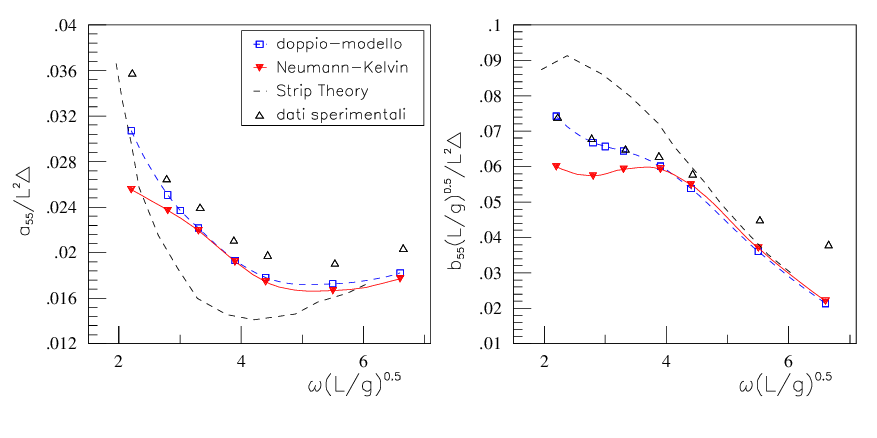}}
    \vskip -0.5cm
    \caption{Prima carena Wigley: confronto numerico--sperimentale 
             per {\em massa aggiunta} e {\em smorzamento} (Fr =0.2).
             Nel termine di adimensionalizzazione $\Delta := \rho\nabla\,$,
             con $\nabla\,=\,Volume\,carena/L^3$
             \label{wjou353f2}
             }
\end{figure}

\begin{figure}[htb]
      \epsfxsize=\textwidth
      \makebox[\textwidth]{\epsfbox{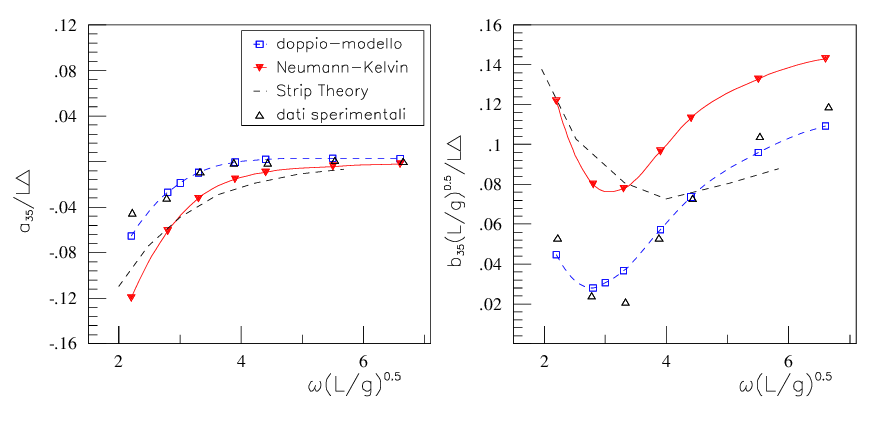}}
      \epsfxsize=\textwidth
      \makebox[\textwidth]{\epsfbox{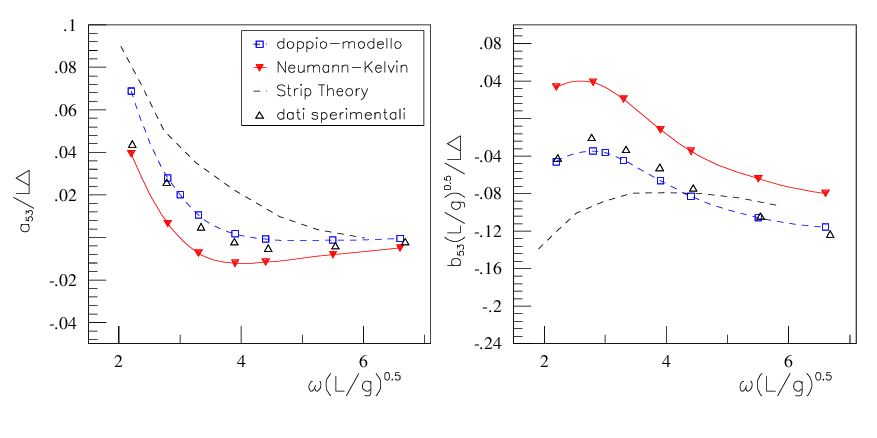}}
    \vskip -0.5cm
    \caption{Prima carena Wigley: confronto numerico--sperimentale 
             per i termini di {\em cross--coupling} (Fr =0.2).
             Nel termine di adimensionalizzazione $\Delta := \rho\nabla\,$.
             \label{wjou533f2}
             }
\end{figure}

\begin{figure}[htb]
      \epsfxsize=\textwidth
      \makebox[\textwidth]{\epsfbox{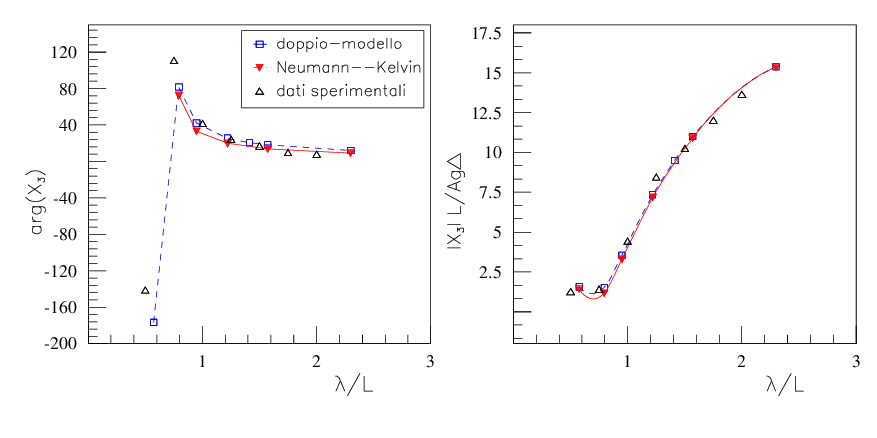}}
      \epsfxsize=\textwidth
      \makebox[\textwidth]{\epsfbox{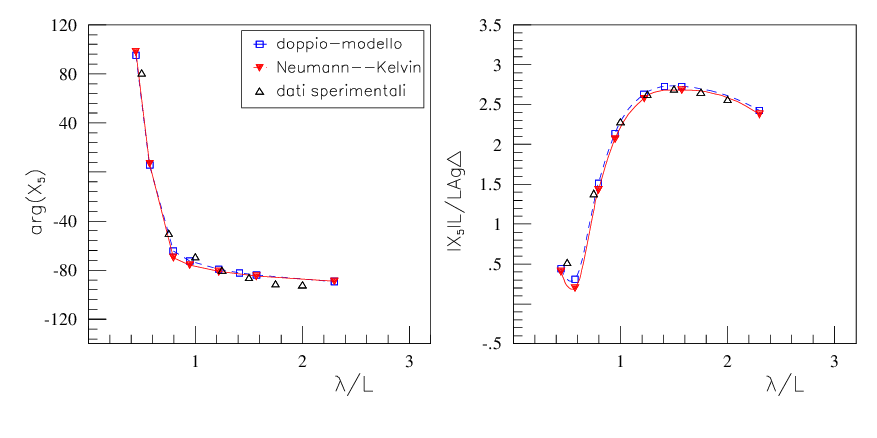}}
    \vskip -0.5cm
    \caption{Prima carena Wigley: confronto numerico--sperimentale 
             per le forze di eccitazione in {\em heave} e {\em pitch}  (Fr =0.2).
             Nel termine di adimensionalizzazione $\Delta := \rho\nabla\,$.
             \label{jouex3f2}
             }
\end{figure}

\begin{figure}[htb]
      \epsfxsize=\textwidth
      \makebox[\textwidth]{\epsfbox{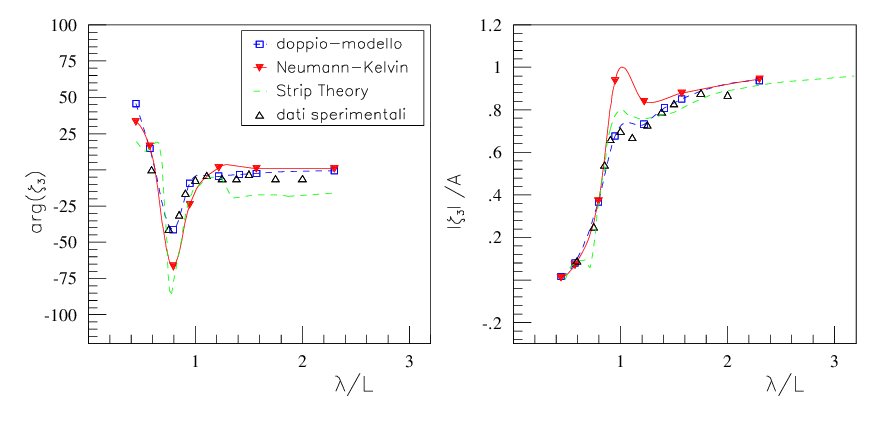}}
      \epsfxsize=\textwidth
      \makebox[\textwidth]{\epsfbox{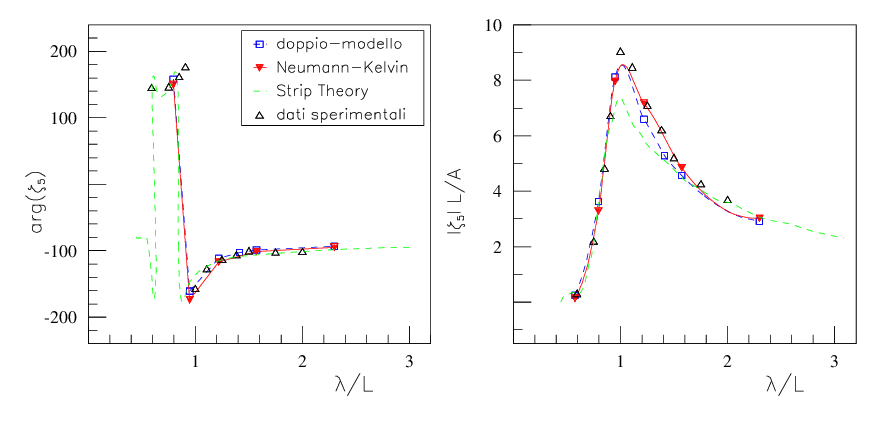}}
    \vskip -0.5cm
    \caption{Prima carena Wigley: confronto numerico--sperimentale 
             per l'ampiezza dei moti nave in {\em heave} e {\em pitch} (Fr =0.2).
             \label{jourao3f2}
             }
\end{figure}

\begin{figure}[htb]
      \epsfxsize=\textwidth
      \makebox[\textwidth]{\epsfbox{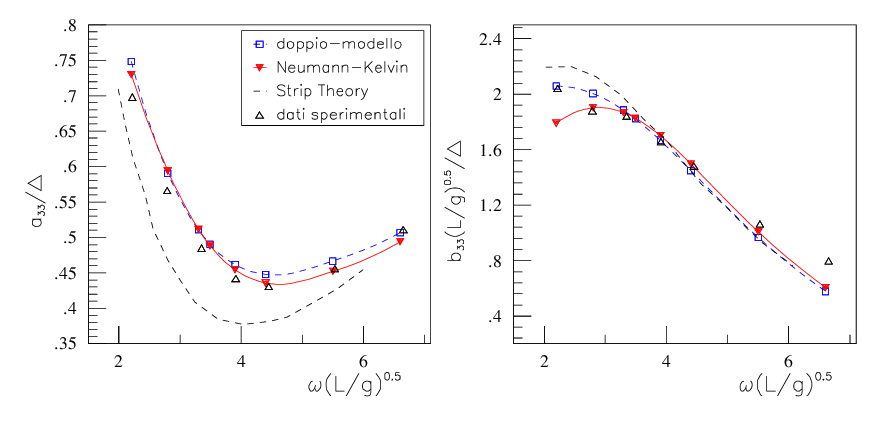}}
      \epsfxsize=\textwidth
      \makebox[\textwidth]{\epsfbox{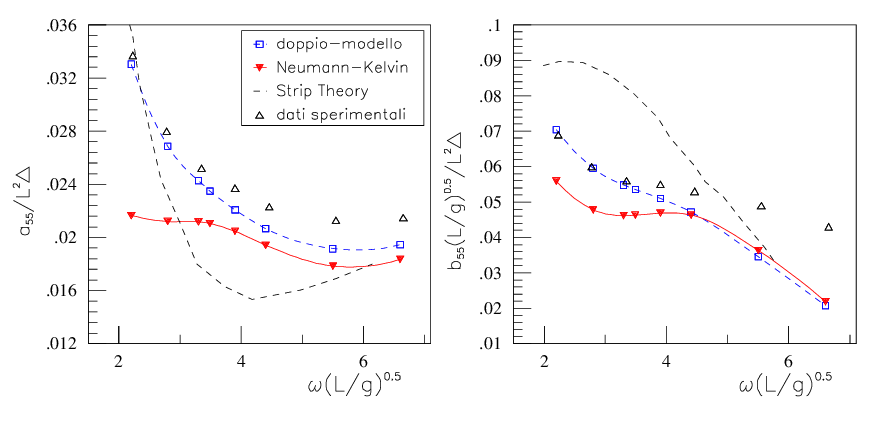}}
    \vskip -0.5cm
    \caption{Prima carena Wigley: confronto numerico--sperimentale 
             per {\em massa aggiunta} e {\em smorzamento} (Fr =0.3).
             Nel termine di adimensionalizzazione $\Delta := \rho\nabla\,$.
             \label{wjou353f3}
             }
\end{figure}

\begin{figure}[htb]
      \epsfxsize=\textwidth
      \makebox[\textwidth]{\epsfbox{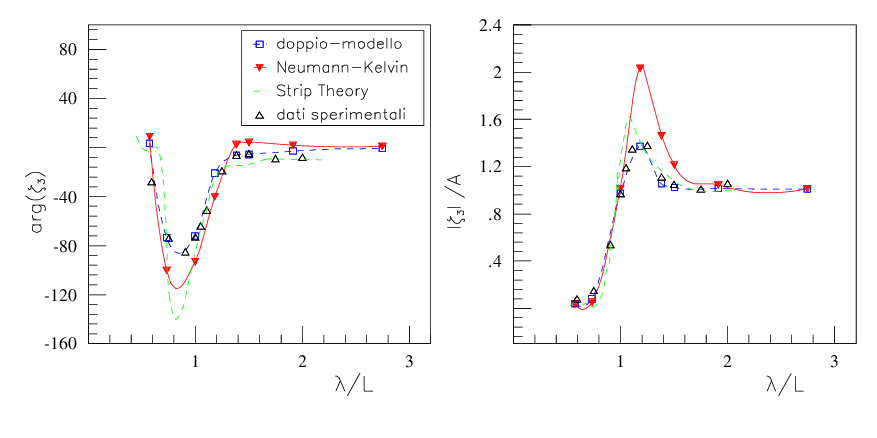}}
      \epsfxsize=\textwidth
      \makebox[\textwidth]{\epsfbox{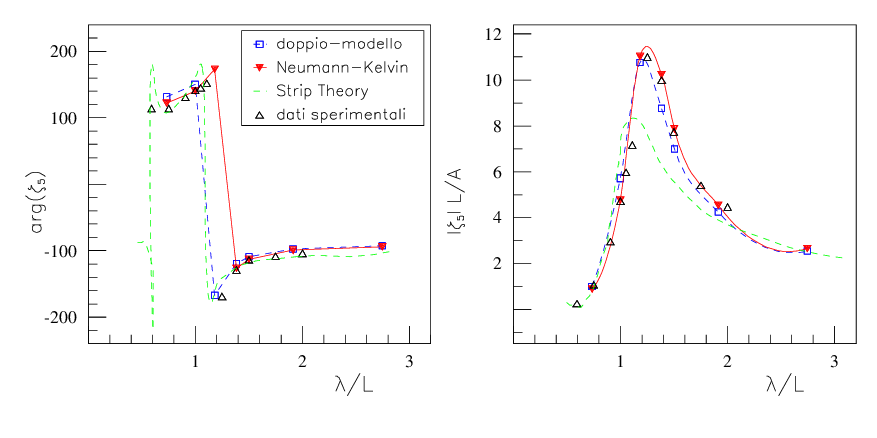}}
    \vskip -0.5cm
    \caption{Prima carena Wigley: confronto numerico--sperimentale 
             per l'ampiezza dei moti nave in {\em heave} e {\em pitch} (Fr =0.3).
             \label{jourao3f3}
             }
\end{figure}

 Abbiamo iniziato con una velocit\a di avanzamento corrispondente a 
 $Fr\,=\,.2$ e in tale condizione sono stati risolti i 
 i problemi di radiazione e di diffrazione relativi a una certa gamma di frequenze. 
 Nelle figure \ref{wjou353f2}--\ref{wjou533f2} sono mostrati i coefficienti 
 $a_{ij}$ e $b_{ij}$ 
 \footnote{
 Per utilizzare simbologie conformi alla letteratura indichiamo questi coefficienti
 senza $*$ 
 \be
 \vec A^*(\omega)\,,\vec B^*(\omega)\rightarrow a_{ij}\,,b_{ij} 
 \ee }
 derivanti da oscillazioni della carena in heave e pitch,in un intervallo di 
 pulsazioni $\omega\sqrt{L/g}\in(1.5;7.1)\,$.
 Per tale intervallo la frequenza ridotta $\tau=\omega U_0/g$ risulta maggiore 
 del valore critico $0.25\,$. \\
 Come si pu\`o notare, il coefficiente $a_{33}$ sperimentale \cite{Journ}
 si presenta descrescente fino ad un certo valore di $\omega$ oltre il quale riscresce, 
 e tale andamento \e ben approssimato dai risultati numerici.
 Diversamente accade per la massa aggiunta per il moto di beccheggio 
 che si mantiene invece decrescente. Qui per basse frequenze il risultato 
 di doppio modello \e notevolmente pi\u accurato di quello ottenuto con Neumann Kelvin. \\ 
 Per quanto riguarda i coefficienti di smorzamento, si ottengono degli andamenti via 
 via decrescenti con la pulsazione, come gi\a discusso nell'ultimo
 paragrafo del quarto capitolo; all'aumentare della frequenza l'energia rilasciata
 in sistemi ondosi \e sempre pi\u bassa. Nel moto di sussulto abbiamo 
 una perfetta sovrapposizione tra i due modelli e i dati sperimentali per il
 coefficiente $b_{33}$, tranne che i primi punti a bassa frequenza dove, ancora una volta 
 il doppio modello presenta un migliore risultato. 
 Si nota inoltre che anche il risultato ottenuto dalla {\em Strip Theory} \e ottimo, 
 ci\o si pu\o giustificare considerando che in questo tipo di moto non intervengono 
 in maniera incisiva gli effetti tridimensionali. \\ 
 Tutt'altro accade nel moto di beccheggio dove tali effetti sono presenti
 in misura maggiore. 
 Infatti in maniera analoga a quanto visto per il termine $a_{55}$, per basse frequenze 
 soltanto la linearizzazione di doppio modello fornisce valori prossimi ai dati sperimentali. 
 Per alte frequenze, così come per il sussulto, i due modelli e quello di Strip Theory
 convergono su un unica curva che si discosta sempre pi\u dai dati sperimentali.
 Il notevole errore commesso nella previsione del coefficiente $b_{55}$ nel moto di 
 beccheggio \e dovuto in parte alla perdita di energia dovuta alla discretizzazione 
 della superficie libera ed in parte al probabile intervento di fenomeni dissipativi
 di altra natura. Questo stesso effetto si presenta anche per il sussulto per\o qui 
 avviene per pulsazioni pi\u alte come si vede dai grafici. \\ 
 Passiamo quindi ai termini di {\em Cross-coupling}. Data la simmetria della carena 
 deve valere la relazione di reciprocit\a, dimostrata in \cite{TimNew}, 
 tra $a_{35}$ e $a_{53}$ e, rispettivamente, $b_{35}$ e $b_{53}\,$. 
 Tale relazione \e sufficientemente soddisfatta sia dai risultati sperimentali che da 
 quelli di doppio--modello. Diversamente accade per l'approssimazione Neumann--Kelvin, 
 la cui valutazione dei termini $m_j\,$ filtra alcune informazioni sulle caratteristiche 
 della carena che risultano importanti per questo tipo di problemi. 
 Per quanto detto nell'ultimo paragrafo del capitolo quattro, abbiamo che per 
 il limite $\omega\rightarrow\infty$ la matrice di massa aggiunta deve diventare 
 simmetrica e dato che vale anche la relazione di reciprocit\a si ha 
 che i termini di cross-coupling $a_{53}\,,a_{35}$ per pulsazioni crescenti 
 devono tendere a valori sempre pi\u prossimi allo zero. 
 Mentre, sempre per tale limite, la matrice di smorzamento deve diventare antisimmetrica
 ma per la relazione di reciprocit\a si ha che per ogni pulsazione vale la 
 $b_{53}\,+\,b_{53}\,=\,0$. 
 Anche la strip-theory fornisce dei risultati che si discostano molto da quelli 
 sperimentali. Possiamo concludere dicendo che per i termini di cross-coupling gli effetti 
 della tridimensionalit\a del problema e la presenza di un flusso base che non sia la 
 semplice corrente uniforme, non sono trascurabili. \\ \indent     
 Le successive figure \ref{jouex3f2} forniscono gli andamenti delle
 forze di eccitazione, sempre per il sussulto e il beccheggio.
 Ricordiamo che questi grafici hanno come ascissa la $\lambda$ del sistema d'onde 
 incidente e quindi la pulsazione $\omega$ qui cresce da destra verso sinistra. 
 Come si vede entrambi i modelli lineari utilizzati danno un'ottima approssimazione 
 dei dati sperimentali, sia in fase che in ampiezza (le fasi sono riferite al sistema 
 d'onde incidente).
 Il motivo per cui le forze di eccitazione descrescono in modulo all'aumentare della 
 pulsazione, \e che man mano che la lunghezza d'onda diventa pi\u piccola della lunghezza 
 della carena sono presenti simultaneamente diverse onde lungo il corpo che producono 
 quindi una serie di cancellazioni. Inoltre si deve tenere presente che l'effetto
 del potenziale dell'onda incidente \e rilevante solo in una regione di fluido 
 prossima alla superficie libera, infatti dalla relazione che fornisce questo potenziale
 per acque profonde
 \footnote{ Abbiamo visto questa relazione nel terzo capitolo:
 \be
 \dsty
  \Phi(x,z,t)\,=\,\frac{gA}{\omega_0}\,e^{kz}\,\sin(kx\,-\,\omega_0 t)
 \ee},
 compare un termine esponenziale che fa descrescere
 notevolmente gli effetti nelle zone di fluido sottostanti la superficie libera. 
 Un'altra motivazione fisica viene data dalla relazione di {\em Haskind} che 
 dimostra l'esistenza di un legame tra le ampiezze delle forze di eccitazione 
 ed i coefficienti di smorzamento; dato che questi tendono a zero per pulsazioni 
 crescenti anche tali forze dovranno avere questo medesimo andamento pur se con 
 legge diversa. \\ \indent 
 Passiamo quindi alla risoluzione del problema dinamico. Per calcolare i moti 
 conseguenti all'onda incidente occorre conoscere la matrice di massa, e le forze 
 di richiamo per la data velocit\a di avanzamento. 
 Riguardo la matrice di massa, nei casi che stiamo trattando risulta diagonale
 e quindi non ci sono accoppiamenti {\em inerziali} fra i vari modi rigidi. 
 Anche qui le fasi $arg(\zeta_i)$ sono riferite al sistema d'onda incidente che 
 investe la carena da prua.  
 Nell'intervallo di frequenze considerato, si determina un fenomeno di risonanza; 
 date le caratteristiche di massa ($m_{ij} + a_{ij}$), di smorzamento ($b_{ij}$) e  
 di richiamo ($c_{ij}$) associate al sistema fluido--carena, la risonanza 
 per i due modi si realizza all'incirca alla stessa frequenza.
 L'ordine di grandezza delle relative frequenze naturali
 deriva infatti sostanzialmente dai termini $c_{ij}$ che assicurano uno stesso
 ordine per i due modi come abbiamo gi\a detto nei primi capitoli(cfr. \cite{MarHy1}).   
 Per questa geometria, la linearizzazione di doppio--modello fornisce risultati 
 decisamente in buon accordo con i dati sperimentali. Sufficientemente vicini
 risultano anche i valori ottenuti tramite la formulazione di Neumann--Kelvin
 come si vede nei grafici \ref{jourao3f2}. 
 Tuttavia per quest'ultima si riscontra un'ampiezza di risonanza pi\u elevata di
 quella sperimentale nel caso del moto di sussulto  
 \footnote{ Si noti che i grafici relativi ai moti nave hanno come ascissa
 la lunghezza d'onda del sistema d'onda incidente, quindi per tali 
 grafici si ha che le pulsazioni crescono da destra verso sinistra.}.
 Anche la strip theory, malgrado i maggiori errori visti nella previsione 
 dei vari coefficienti, presenta una stima dei moti con una buona approssimazione 
 la quale, come vedremo, andr\a peggiorando all'aumentare del numero di $Froude$. 
 \\ 
 Passando ad una velocit\a di avanzamento pi\u elevata $Fr\,=\,0.3$ 
 non si hanno dei cambiamenti qualitativi per i coefficienti di massa aggiunta 
 e smorzamento, mentre si ha per i moti un'intervallo delle pulsazioni di risonanza 
 pi\u ristretto (figure \ref{wjou353f3},\ref{jourao3f3}). 
 La sovrastima dell'ampiezza di risonanza data dalla linearizzazione 
 di Neumann Kelvin nel moto di sussulto \e aumentata con il numero di $Froude$. 
 Quest'errore si spiega osservando i dati della tabella \ref{cij} 
 la quale mostra i termini $c_{ij}$ forniti da tale semplificazione e la 
 correzione aggiunta dal doppio--modello; quest'ultima \e soprattutto rilevante 
 proprio per $c_{33}\,$.
 Sempre per i termini di richiamo, sia questa carena che  
 la terza (appartengono alla stessa famiglia) presentano un baricentro al 
 di sotto del piano $(z\,=\,0)$, questo comporta un termine aggiuntivo per le 
 forze di richiamo ed in particolare l'effetto conseguente \e quello di aumentare 
 il coefficiente $c_{55}$ legato al moto di beccheggio con un aumento della 
 pulsazione naturale per questo modo rigido. Comunque tale effetto \e in questo caso 
 molto ridotto in quanto il baricentro \e abbassato soltanto di una distanza 
 pari al $10\%$ del pescaggio.
 Per $Fr\,=\,0.3$ osserviamo inoltre un peggioramento dei risultati della 
 strip theory, soprattutto per quanto riguarda il beccheggio;
 nella figura \ref{jourao3f2} si pu\o vedere come per entrambi i moti 
 la lunghezza d'onda corrispondente alla pulsazione di risonanza 
 \e valutata in difetto.
\newpage
\clearpage
{\large
\begin{center}
\begin{table}[htb]
 \hskip 15mm \begin{tabular}{||l||r|r|r||} \hline 
            & { $c_{ij}$ N-K}  & { $\Delta c_{ij}$ del D.M.} & { $c_{ij}$ tot.} \\ \hline \hline
  $c_{11}$ {\em $(N/m)$}  & $0.000     $ & $64.110 $  & $64.110 $   \\ \hline
  $c_{13}$ {\em $(N/m)$}  & $0.000     $ & $0.000     $  & $0.000     $   \\ \hline
  $c_{15}$ {\em $(N)$  }  & $762.800 $ & $-12.830$  & $750.000 $   \\ \hline
  $c_{31}$ {\em $(N/m)$}  & $0.000     $ & $0.000     $  & $0.000     $   \\ \hline
  $c_{33}$ {\em $(N/m)$}  & $6119.000  $ & $129.700 $  & $6249.000  $   \\ \hline
  $c_{35}$ {\em $(N)$  }  & $-12.450$ & $0.107 $  & $-12.340$   \\ \hline
  $c_{51}$ {\em $(N)$  }  & $0.000     $ & $39.520 $  & $39.520 $   \\ \hline
  $c_{53}$ {\em $(N)$  }  & $0.000     $ & $0.000     $  & $0.000     $   \\ \hline
  $c_{55}$ {\em $(N m)$ }  & $2821.000  $ & $72.770 $  & $2894.000  $   \\ \hline
  
  \end{tabular}  
  \vskip 0.5cm
      \caption{Prima carena Wigley: Confronto dei 
               coefficienti di richiamo ottenuti con Neumann--Kelvin ($c_{ij}$ 
               N-K.) e con il doppio--modello ($c_{ij}$ tot.),
               la seconda colonna da' lo scarto tra i due.
               \label{cij}}
\end{table}
\end{center}
}
\newpage
\clearpage
\newpage
 \clearpage
\newpage
 \clearpage
\newpage
 \clearpage
\newpage
 \clearpage
\newpage
 \clearpage
\begin{figure}[htb]
      \epsfxsize=\textwidth
      \makebox[\textwidth]{\epsfbox{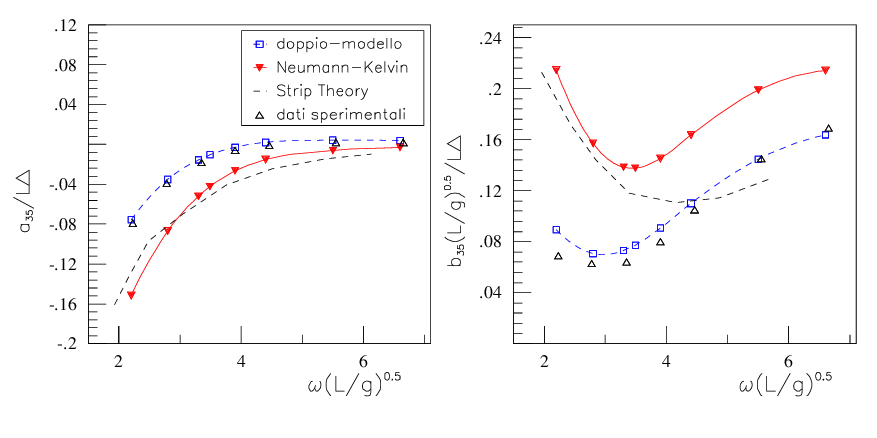}}
      \epsfxsize=\textwidth
      \makebox[\textwidth]{\epsfbox{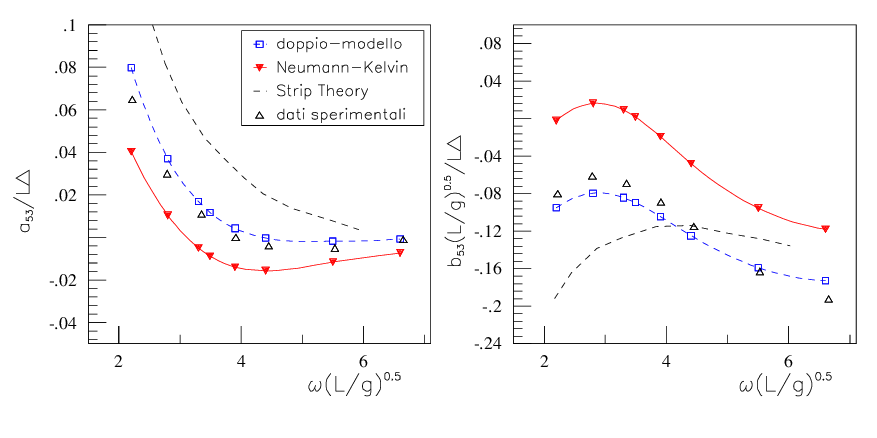}}
    \vskip -0.5cm
    \caption{Prima carena Wigley: confronto numerico--sperimentale 
             per i termini di {\em cross--coupling} (Fr =0.3).
             Nel termine di adimensionalizzazione $\Delta := \rho\nabla\,$.
             \label{wjou533f3}
             }
\end{figure}
\newpage
 \clearpage
\begin{figure}[htb]
      \epsfxsize=\textwidth
      \makebox[\textwidth]{\epsfbox{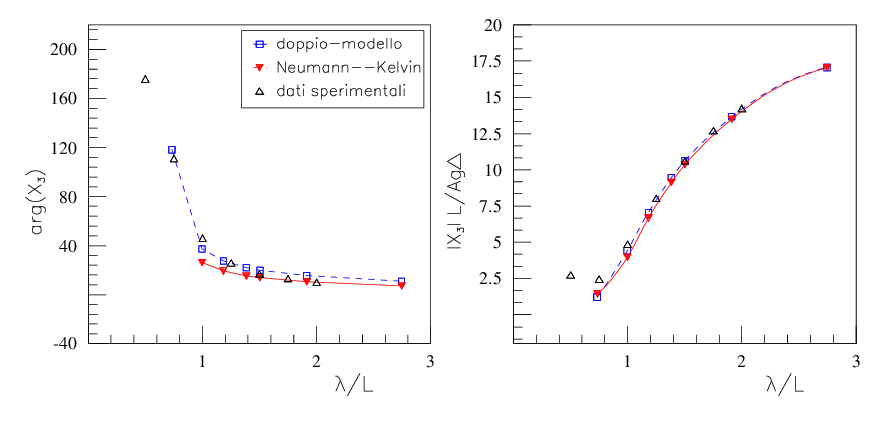}}
      \epsfxsize=\textwidth
      \makebox[\textwidth]{\epsfbox{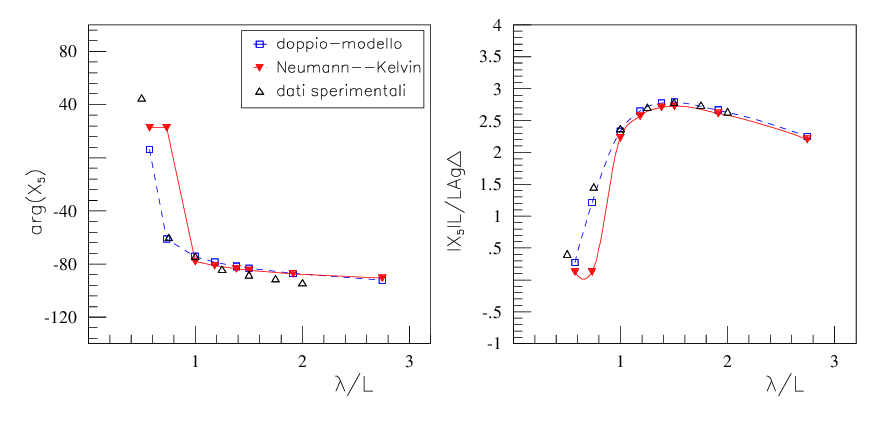}}
    \vskip -0.5cm
    \caption{Prima carena Wigley: confronto numerico--sperimentale 
             per le forze di eccitazione in {\em heave} e {\em pitch}  (Fr =0.3).
             Nel termine di adimensionalizzazione $\Delta := \rho\nabla\,$.
             \label{jouex3f3}
             }
\end{figure}
\newpage
 \clearpage
\newpage
\clearpage
\section{Casi esaminati: seconda Carena Wigley.}
\begin{figure}[htb]
      \epsfxsize=\textwidth
      \makebox[\textwidth]{\epsfbox{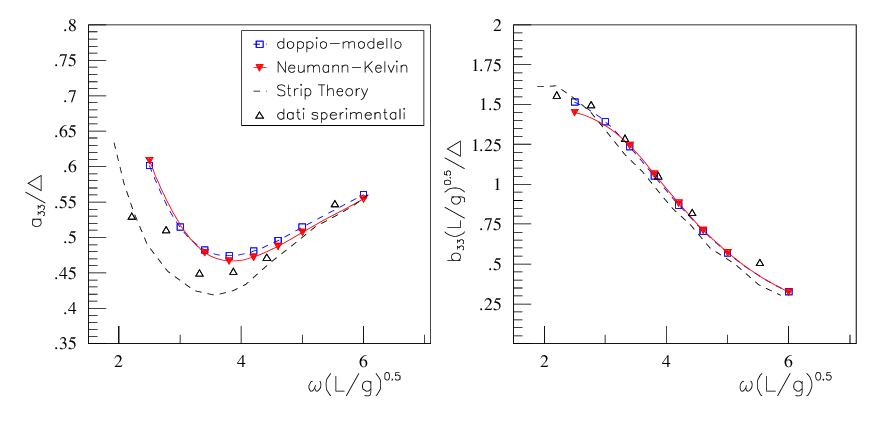}}
      \epsfxsize=\textwidth
      \makebox[\textwidth]{\epsfbox{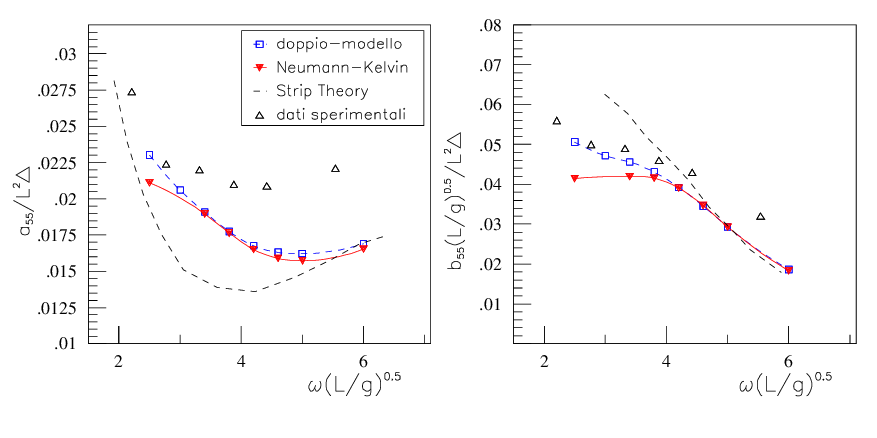}}
    \vskip -0.5cm
    \caption{Seconda carena Wigley: confronto numerico--sperimentale 
             per {\em massa aggiunta} e {\em smorzamento} (Fr =0.2).
             Nel termine di adimensionalizzazione $\Delta := \rho\nabla\,$,
             con $\nabla\,=\,Volume\,carena/L^3$.
             \label{wjou351f2}
             }
\end{figure}

\begin{figure}[htb]
      \epsfxsize=\textwidth
      \makebox[\textwidth]{\epsfbox{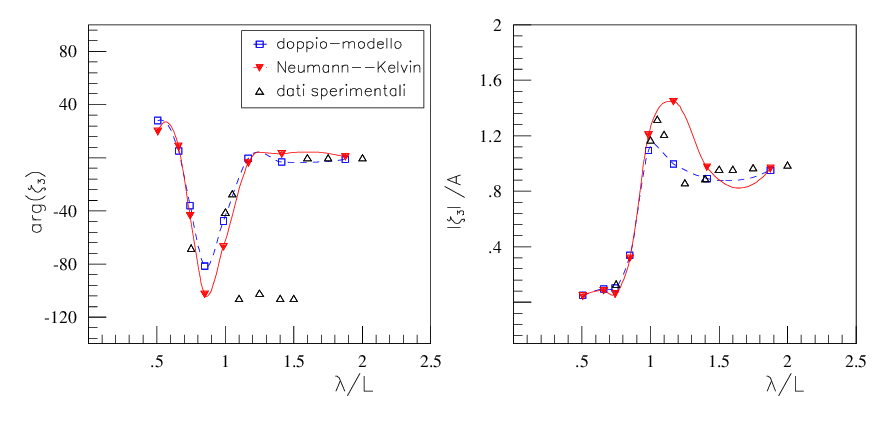}}
      \epsfxsize=\textwidth
      \makebox[\textwidth]{\epsfbox{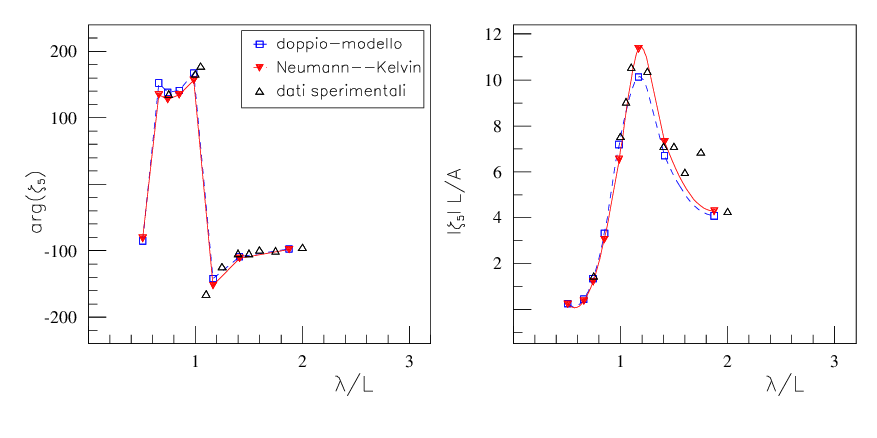}}
    \vskip -0.5cm
    \caption{Seconda carena Wigley: confronto numerico--sperimentale 
            per l'ampiezza dei moti nave in {\em heave} e {\em pitch} (Fr =0.2).
            \label{jourao1f2}
            }
\end{figure}

\begin{figure}[htb]
      \epsfxsize=\textwidth
      \makebox[\textwidth]{\epsfbox{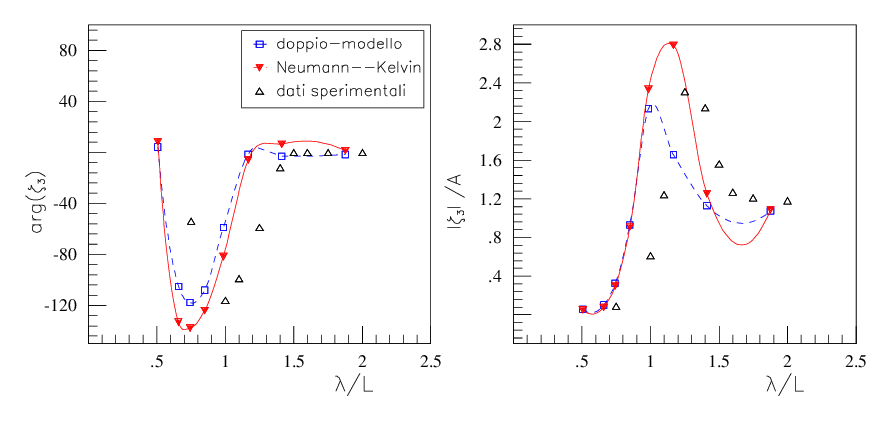}}
      \epsfxsize=\textwidth
      \makebox[\textwidth]{\epsfbox{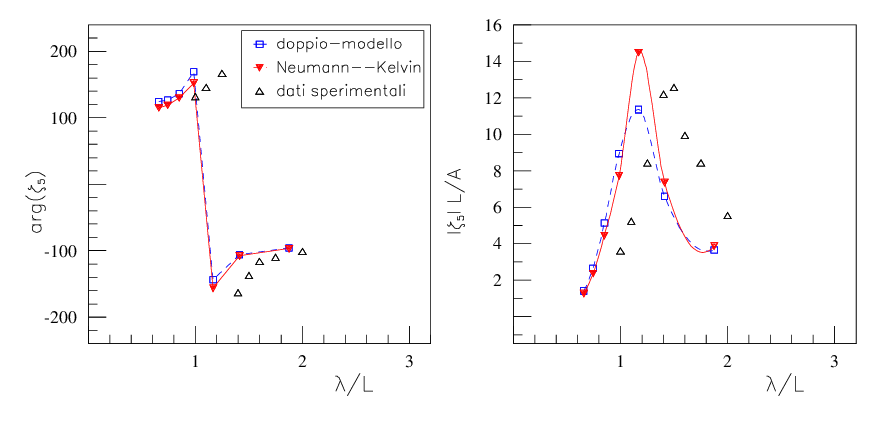}}
    \vskip -0.5cm
    \caption{Seconda carena Wigley: confronto numerico--sperimentale 
             per l'ampiezza dei moti nave in {\em heave} e {\em pitch} (Fr =0.3).
             \label{jourao1f3}
             }
\end{figure}

La seconda carena che abbiamo esaminato \e una carena Wigley con 
\be 
\dsty
\frac{L}{B}\,=\,10\qquad \frac{L}{T}\,=\,16\qquad Cb\,\simeq\,0.56 
\ee
Nella figura sottostante \e riportato il reticolo con cui \e stata 
discretizzata questa geometria. Si pu\o osservare come le sezioni 
presentino delle variazioni di curvatura, nella parte centrale, 
pi\u accentuate rispetto alla carena precedente:
\begin{figure}[htb]
    \vskip  0.5cm
      \epsfxsize=.5\textwidth
      \epsfxsize=.5\textwidth
      \makebox[.9\textwidth]{\epsfbox{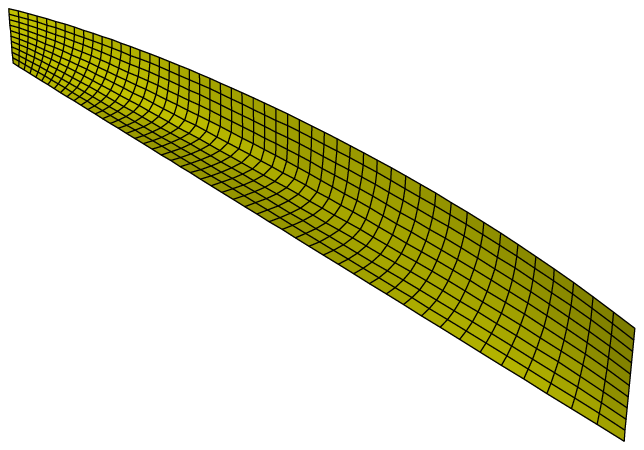}}
    \vskip -0.5cm
    \caption{Esempio di un tipico reticolo di calcolo per la seconda Wigley 
             }
\end{figure}
\\
Iniziamo l'analisi sempre con un valore del numero di $Froude$ pari a $0.2$.  
Nella figura \ref{wjou351f2} si osserva che per questa carena le previsioni 
fatte per i coefficienti $a_{33}\,,a_{55}$ sono peggiorate rispetto alla 
precedente carena. \`E evidente una sottostima per la massa 
aggiunta relativa al beccheggio ed ancora una volta per basse frequenze si ha
un miglioramento dei risultati per quanto riguarda la linearizzazione di doppio
modello. Mentre per il sussulto l'errore della strip theory \e dell'ordine di 
quello commesso dai codici tridimensionali, per il modo di beccheggio si ha 
un errore pi\u marcato, gli effetti tridimensionali per questa seconda carena
sono quindi pi\u importanti per questo modo rigido. \\
Invece per gli smorzamenti si hanno dei risultati migliori, ancora una volta
entrambi i tre metodi analizzati danno dei risultati coincidenti per il
moto di sussulto mentre per il beccheggio \e evidente la maggior accuratezza
del doppio modello. Sottolineamo come in questo caso anche i dati sperimentali  
mostrano un rateo di descrescita per il coefficiente $b_{55}$ pi\u elevato
rispetto alla precedente carena. E' quindi ragionevole pensare che per questa
seconda $Wigley$ gli effetti di dissipazione che non sono tenuti in conto in 
questo modello, non intervengono in maniera accentuata come per la prima carena. \\  
Per i coefficienti di {\em Cross Coupling} ancora una volta soltanto la 
linearizzazione di doppio modello segue con ottima approssimazione i dati sperimentali.
Anche per questa carena risulta sufficientemente soddisfatta la relazione di reciprocit\a  
per questi coefficienti. \\
Per le forze di eccitazione abbiamo una perfetta sovrapposizione delle stime 
fatte da entrambi i modelli, i dati sperimentali sono riprodotti ottimamente 
tranne che la fase della forza di eccitazione per il moto di beccheggio. \\   
Nella figura \ref{jourao1f2} abbiamo le ampiezze dei moti nave e le relative fasi;
i risultati ottenuti non sono ottimi come per la precedente carena 
ma dimostrano comunque un buon accordo con i dati sperimentali nel predire 
la pulsazione di risonanza e l'ampiezza dei moti per questa.
Passando a $Fr\,=\,0.3$ si ha un peggioramento dei risultati in diffrazione 
che comporta un peggioramento dei risultati per il $R.A.O.$. Nella figura
\ref{jourao1f3} si vede come le lunghezza d'onda corrispondenti alle 
pulsazioni di risonanza vengono sottostimate da entrambe le linearizzazioni. 
\newpage
 \clearpage
\newpage
\clearpage
\begin{figure}[htb]
      \epsfxsize=\textwidth
      \makebox[\textwidth]{\epsfbox{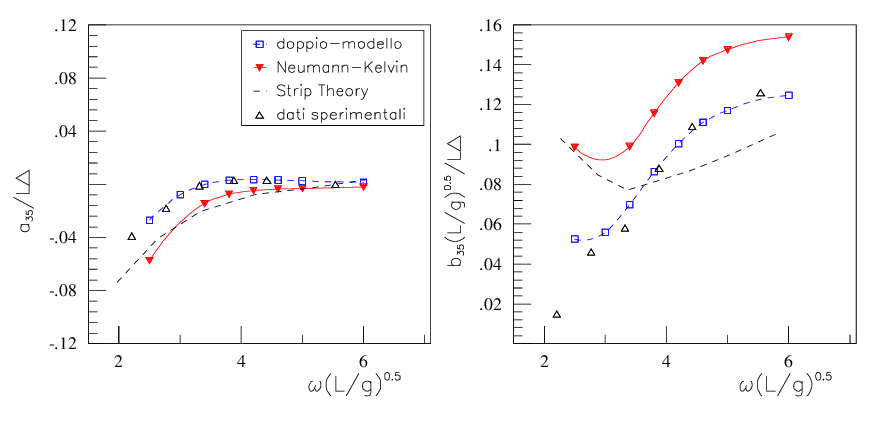}}
      \epsfxsize=\textwidth
      \makebox[\textwidth]{\epsfbox{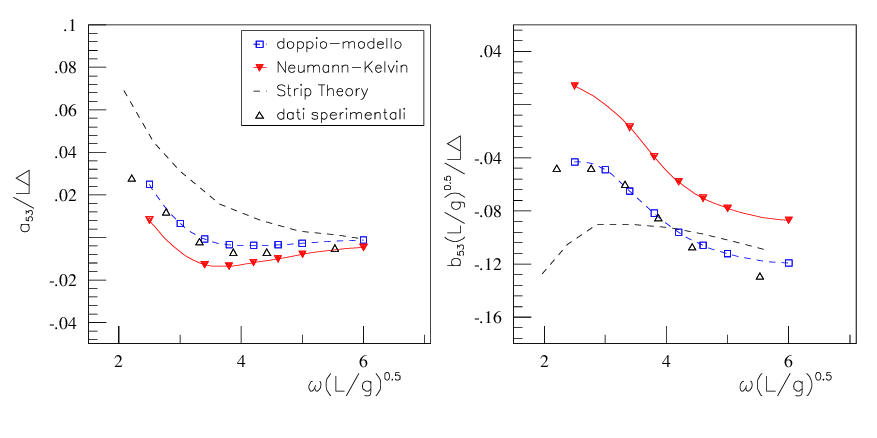}}
    \vskip -0.5cm
    \caption{Seconda carena Wigley: confronto numerico--sperimentale 
             per i termini di {\em cross--coupling} (Fr =0.2).
             Nel termine di adimensionalizzazione $\Delta := \rho\nabla\,$.
             \label{wjou531f2}
             }
\end{figure}
\newpage
 \clearpage
\begin{figure}[htb]
      \epsfxsize=\textwidth
      \makebox[\textwidth]{\epsfbox{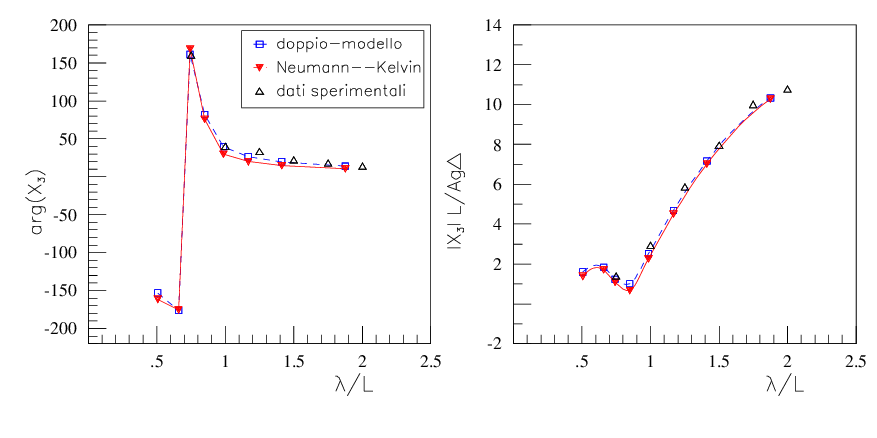}}
      \epsfxsize=\textwidth
      \makebox[\textwidth]{\epsfbox{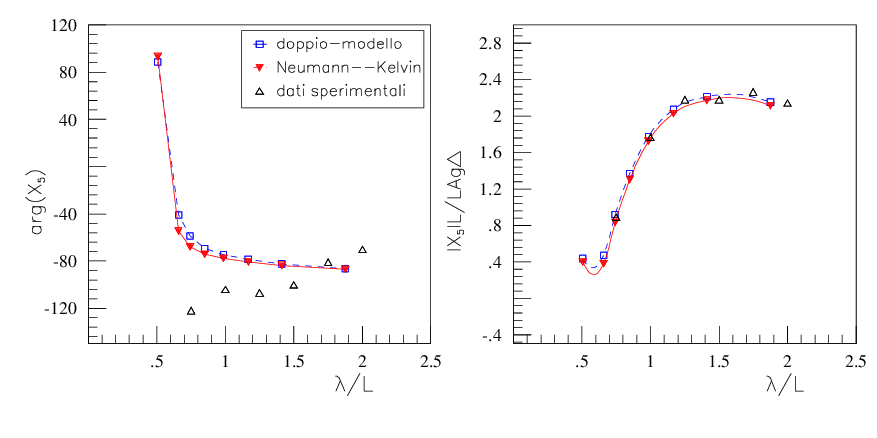}}
    \vskip -0.5cm
    \caption{Seconda carena Wigley: confronto numerico--sperimentale 
            per le forze di eccitazione in {\em heave} e {\em pitch}  (Fr =0.2).
            Nel termine di adimensionalizzazione $\Delta := \rho\nabla\,$.
            \label{jouex1f2}
            }
\end{figure}
\newpage
 \clearpage
\newpage
 \clearpage
\begin{figure}[htb]
      \epsfxsize=\textwidth
      \makebox[\textwidth]{\epsfbox{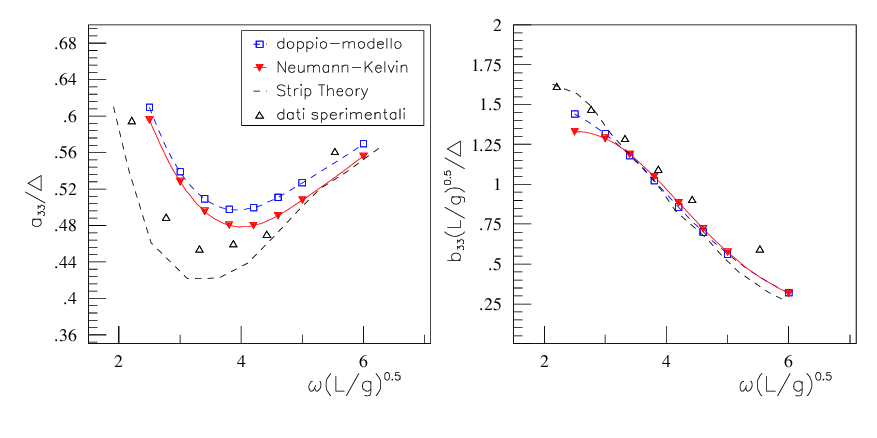}}
      \epsfxsize=\textwidth
      \makebox[\textwidth]{\epsfbox{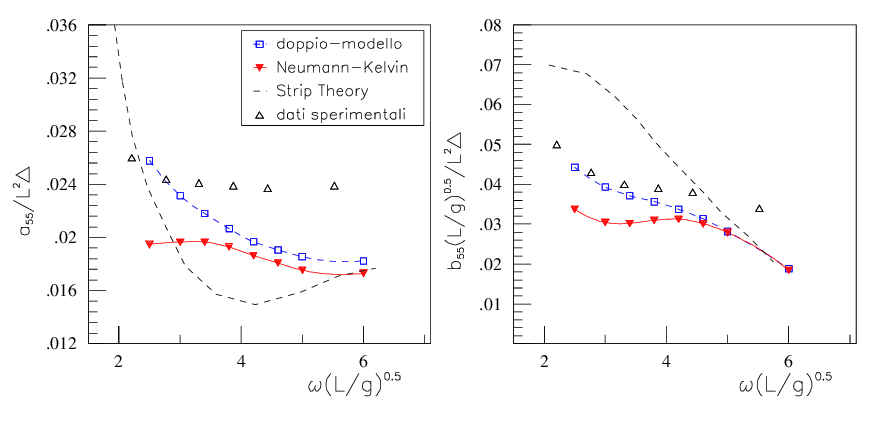}}
    \vskip -0.5cm
    \caption{Seconda carena Wigley: confronto numerico--sperimentale 
             per {\em massa aggiunta} e {\em smorzamento} (Fr =0.3).
             Nel termine di adimensionalizzazione $\Delta := \rho\nabla\,$.
             \label{wjou351f3}
             }
\end{figure}
\newpage
 \clearpage
\begin{figure}[htb]
      \epsfxsize=\textwidth
      \makebox[\textwidth]{\epsfbox{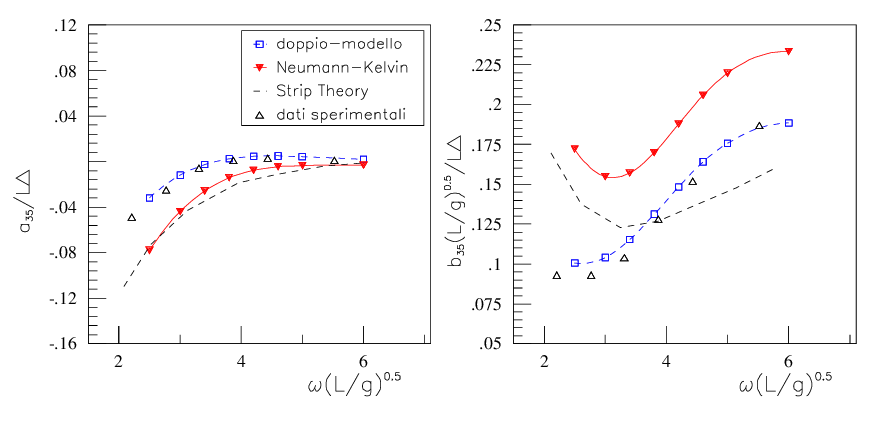}}
      \epsfxsize=\textwidth
      \makebox[\textwidth]{\epsfbox{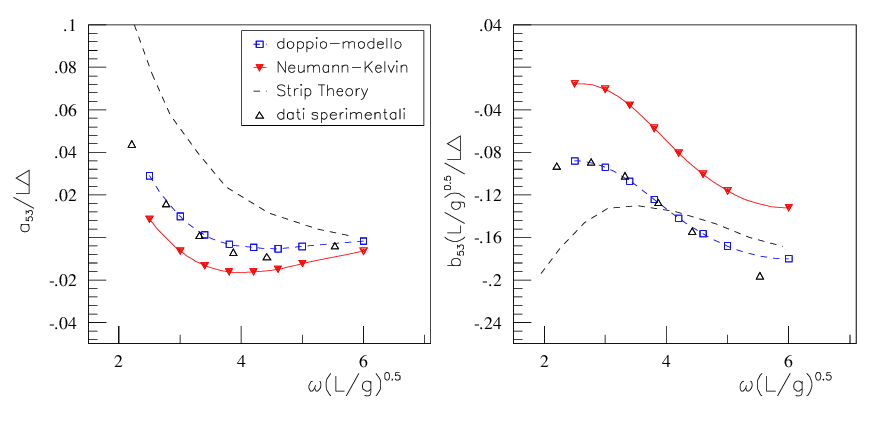}}
    \vskip -0.5cm
    \caption{Seconda carena Wigley: confronto numerico--sperimentale 
             per i termini di {\em cross--coupling} (Fr =0.3).
             Nel termine di adimensionalizzazione $\Delta := \rho\nabla\,$.
             \label{wjou531f3}
             }
\end{figure}
\newpage
 \clearpage
\begin{figure}[htb]
      \epsfxsize=\textwidth
      \makebox[\textwidth]{\epsfbox{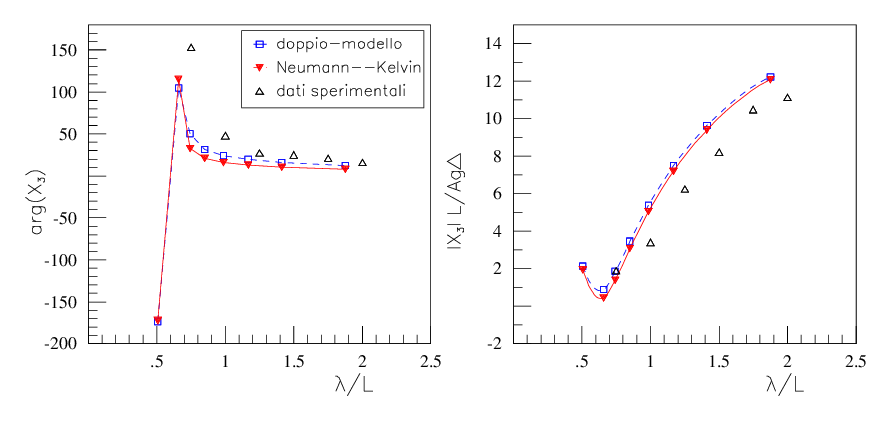}}
      \epsfxsize=\textwidth
      \makebox[\textwidth]{\epsfbox{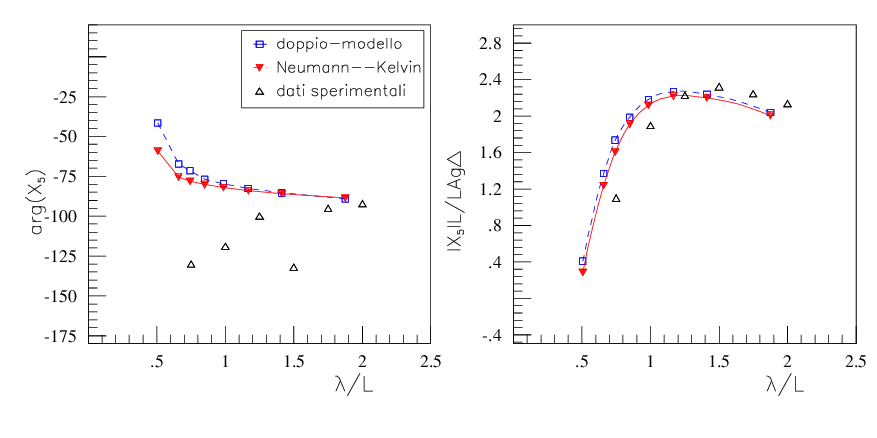}}
    \vskip -0.5cm
    \caption{Seconda carena Wigley: confronto numerico--sperimentale 
             per le forze di eccitazione in {\em heave} e {\em pitch}  (Fr =0.3).
             Nel termine di adimensionalizzazione $\Delta := \rho\nabla\,$.
             \label{jouex1f3}
             }
\end{figure}
\newpage
 \clearpage
\newpage
\clearpage
\section{Casi esaminati: terza Carena Wigley.}
\begin{figure}[htb]
      \epsfxsize=\textwidth
      \makebox[\textwidth]{\epsfbox{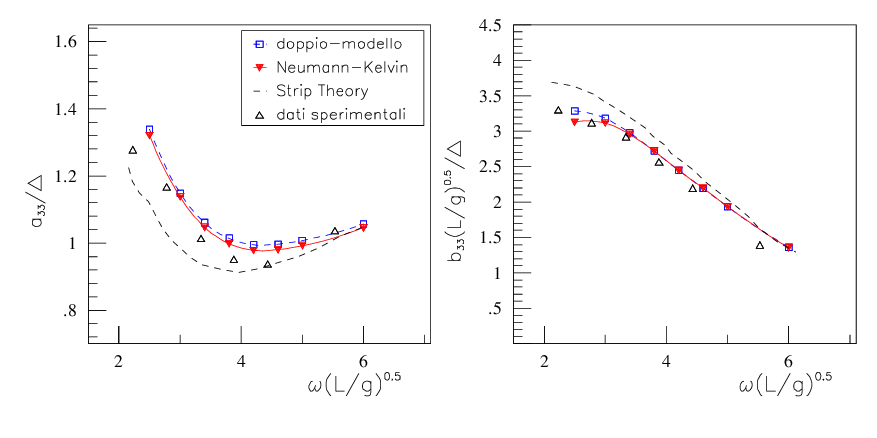}}
      \epsfxsize=\textwidth
      \makebox[\textwidth]{\epsfbox{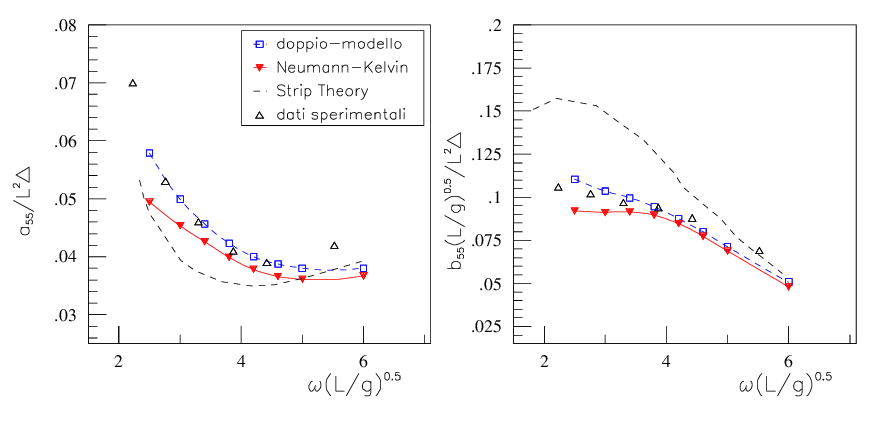}}
    \vskip -0.5cm
    \caption{Terza carena Wigley: confronto numerico--sperimentale 
             per {\em massa aggiunta} e {\em smorzamento} (Fr =0.2).
             Nel termine di adimensionalizzazione $\Delta := \rho\nabla\,$,
             con $\nabla\,=\,Volume\,carena/L^3$.
             \label{wjou354f2}
             }
\end{figure}

\begin{figure}[htb]
      \epsfxsize=\textwidth
      \makebox[\textwidth]{\epsfbox{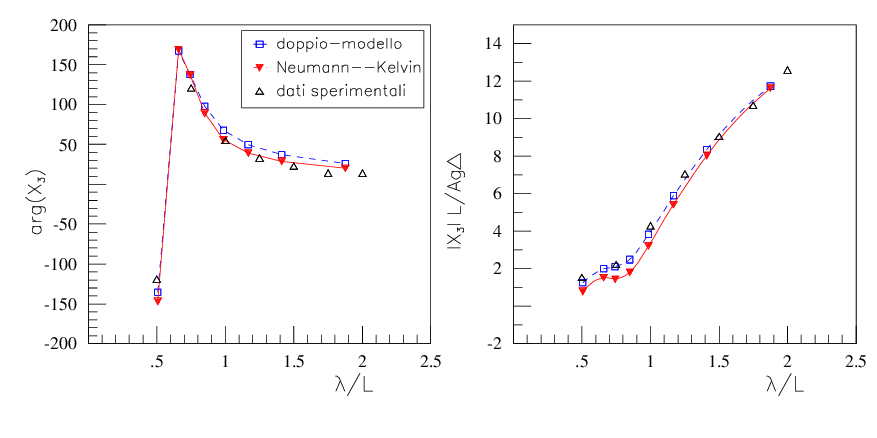}}
      \epsfxsize=\textwidth
      \makebox[\textwidth]{\epsfbox{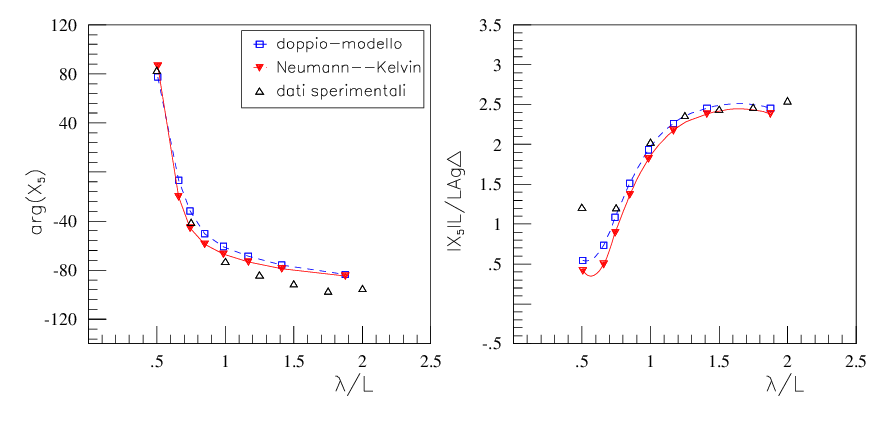}}
    \vskip -0.5cm
    \caption{Terza Carena Wigley: confronto numerico--sperimentale 
             per le forze di eccitazione in {\em heave} e {\em pitch}  (Fr =0.2).
             Nel termine di adimensionalizzazione $\Delta := \rho\nabla\,$.
             \label{jouex4f2}
             }
\end{figure}

\begin{figure}[htb]
      \epsfxsize=\textwidth
      \makebox[\textwidth]{\epsfbox{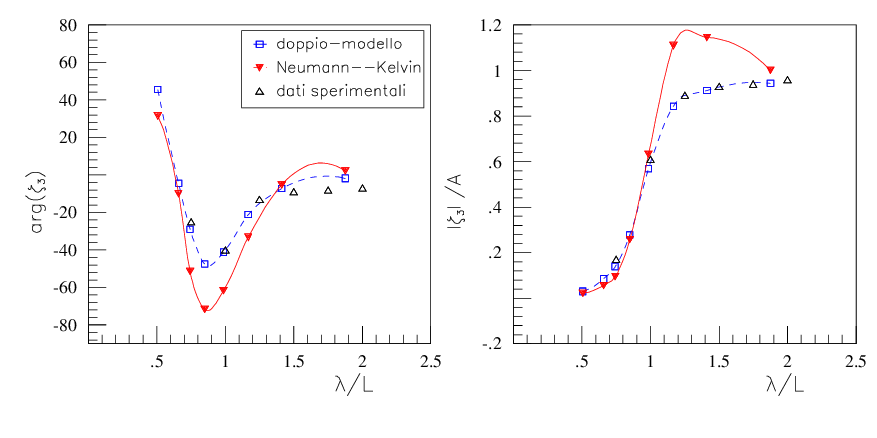}}
      \epsfxsize=\textwidth
      \makebox[\textwidth]{\epsfbox{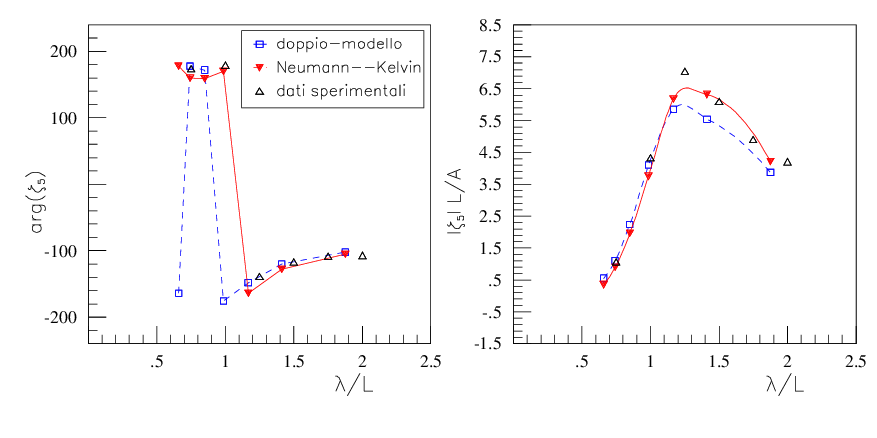}}
    \vskip -0.5cm
    \caption{Terza carena Wigley: confronto numerico--sperimentale 
             per l'ampiezza dei moti nave in {\em heave} e {\em pitch} (Fr =0.2).
             \label{jourao4f2}
             }
\end{figure}

\begin{figure}[htb]
      \epsfxsize=\textwidth
      \makebox[\textwidth]{\epsfbox{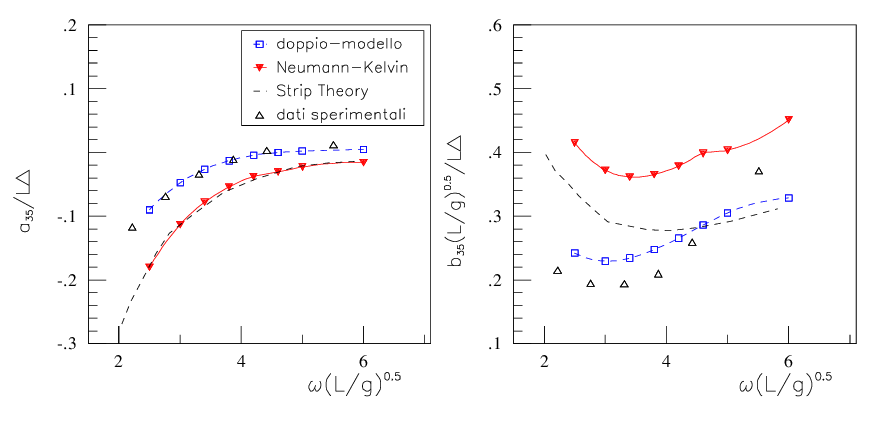}}
      \epsfxsize=\textwidth
      \makebox[\textwidth]{\epsfbox{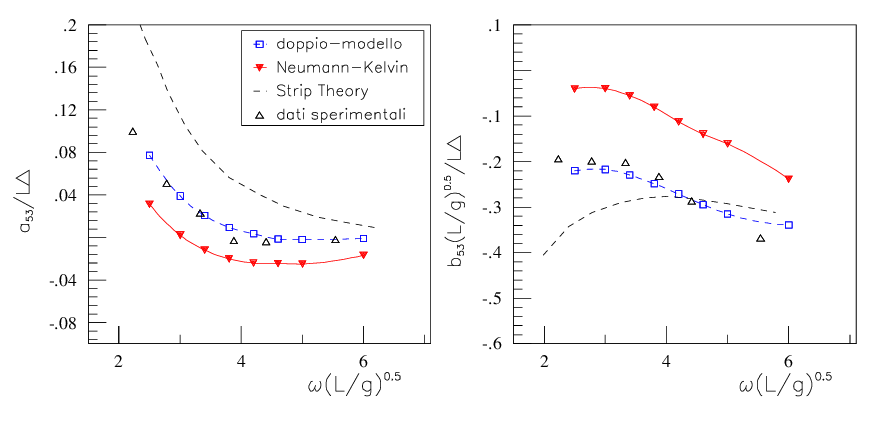}}
    \vskip -0.5cm
    \caption{Terza carena Wigley: confronto numerico--sperimentale 
             per i termini di {\em cross--coupling} (Fr =0.3).
             Nel termine di adimensionalizzazione $\Delta := \rho\nabla\,$.
             \label{wjou534f3}
             }
\end{figure}

\begin{figure}[htb]
      \epsfxsize=\textwidth
      \makebox[\textwidth]{\epsfbox{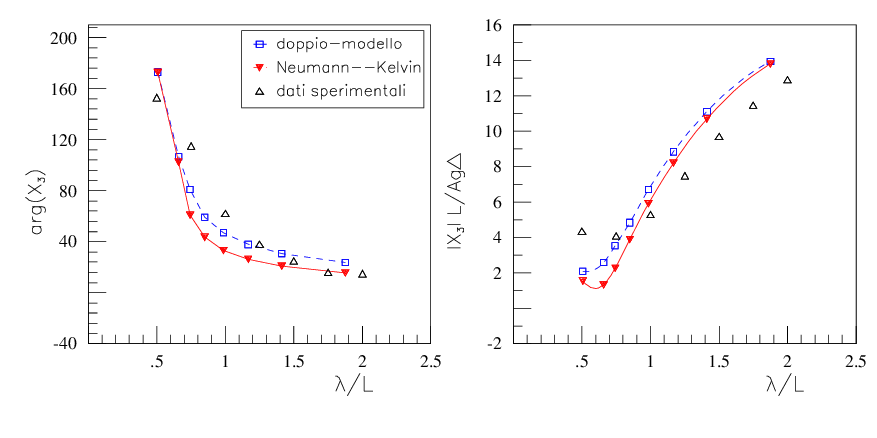}}
      \epsfxsize=\textwidth
      \makebox[\textwidth]{\epsfbox{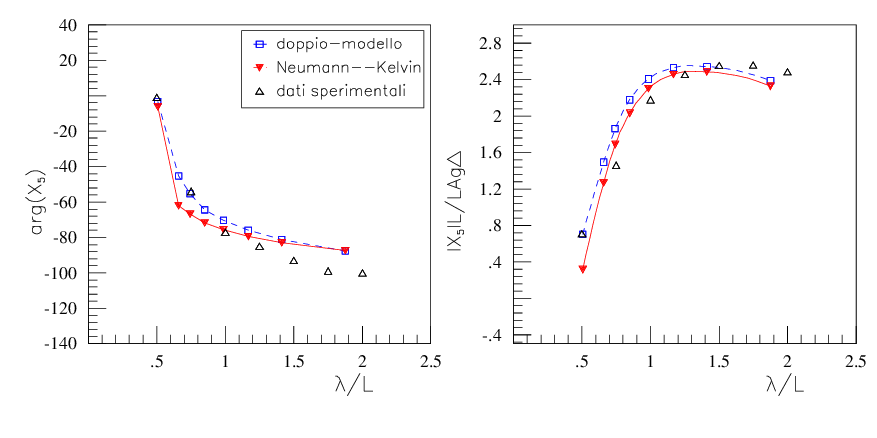}}
    \vskip -0.5cm
    \caption{Terza Carena Wigley: confronto numerico--sperimentale 
             per le forze di eccitazione in {\em heave} e {\em pitch}  (Fr =0.3).
             Nel termine di adimensionalizzazione $\Delta := \rho\nabla\,$.
             \label{jouex4f3}
             }
\end{figure}

\begin{figure}[htb]
      \epsfxsize=\textwidth
      \makebox[\textwidth]{\epsfbox{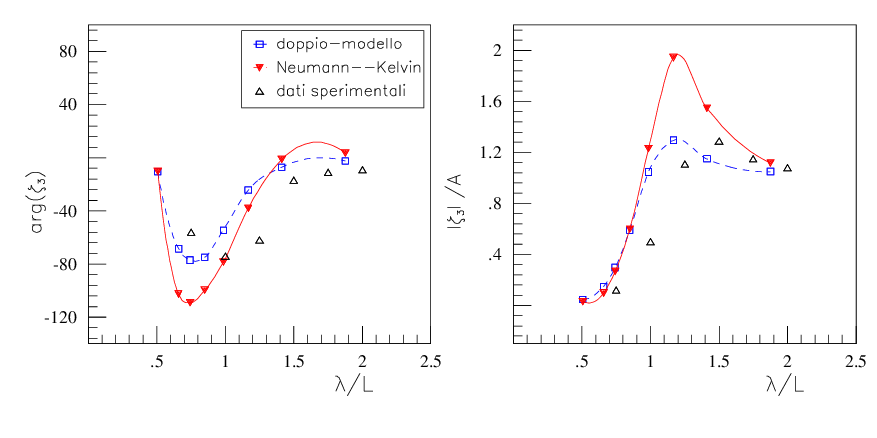}}
      \epsfxsize=\textwidth
      \makebox[\textwidth]{\epsfbox{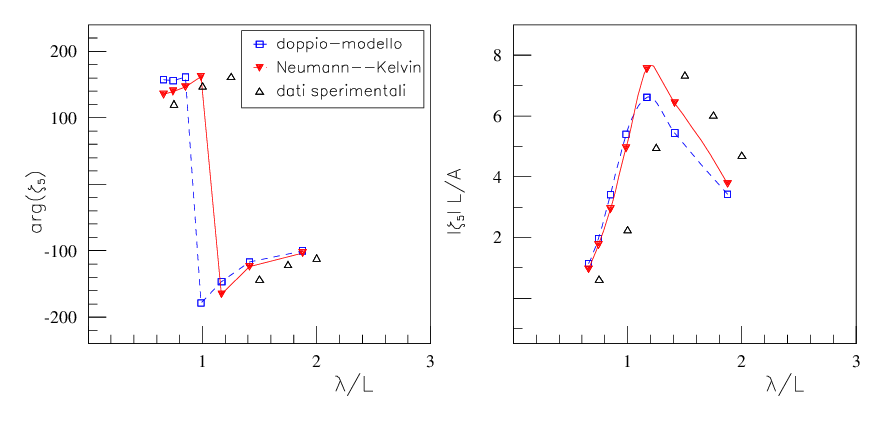}}
    \vskip -0.5cm
    \caption{Terza carena Wigley: confronto numerico--sperimentale 
             per l'ampiezza dei moti nave in {\em heave} e {\em pitch} (Fr =0.3).
             \label{jourao4f3}
             }
\end{figure}

La Terza carena che abbiamo esaminato \e una carena Wigley con 
\be 
\dsty
\frac{L}{B}\,=\,5\qquad \frac{L}{T}\,=\,16\qquad Cb\,\simeq\,0.46 
\ee
Nella figura sottostante \e riportato il reticolo con cui \e stata 
discretizzata questa geometria e dove si pu\o notare la maggiore 
larghezza rispetto alle prime due carene.  
\begin{figure}[htb]
    \vskip  0.5cm
      \epsfxsize=.5\textwidth
      \epsfxsize=.5\textwidth
      \makebox[.9\textwidth]{\epsfbox{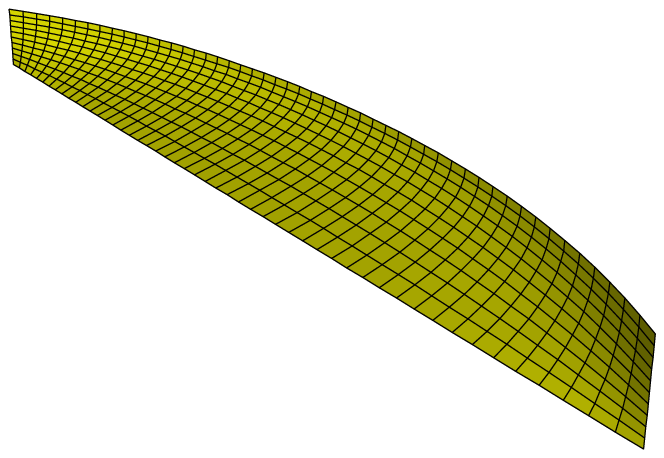}}
    \vskip -0.5cm
    \caption{Esempio di un tipico reticolo di calcolo per la terza Wigley 
             }
\end{figure}
\\
Partiamo dalla figura \ref{wjou354f2} in cui notiamo che 
si hanno degli ottimi risultati nella previsione delle masse aggiunte 
$a_{33}\,,a_{55}$ e degli smorzamenti $b_{33}\,,b_{55}$. In particolare 
per $a_{33}$ le due linearizzazioni danno la medesima previsione e ancora una 
volta anche la strip theory fornisce dei buoni risultati. Per $a_{55}$ \e 
la linearizzazione di doppio modello a fornire i migliori risultati e per questa
carena ci\o non accade solo a basse frequenze. Tale risultato  
era stato previsto in quanto il minor rapporto $L/B$ implica un maggior peso 
del flusso di doppio modello. Anche per questa carena, come per la precedente, 
i fenomeni dissipativi che non sono tenuti in conto da questo modello,
non intervengono in maniera "pesante" per alte frequenze, come per la prima carena,
e quindi le previsioni per i coefficienti di smorzamento sono buone per tutto
l'intervallo di pulsazioni di interesse. \\ 
Per i termini di cross-coupling vale ancora quanto detto per le precedenti carene.\\
La stima per le forze di eccitazione \e buona tranne per pulsazioni alte 
(piccole $\lambda$) dove viene sottostimato il modulo di $X_5$ (figura \ref{jouex4f2}). \\ 
Tale errore si ripercuote sulla previsione dei moti nave, infatti l'ampiezza del
moto di beccheggio risulta sottostimata, come si pu\o vedere nella figura \ref{jourao4f2}.
Comunque i risultati per il $R.A.O$ risultano in questo caso molto buoni, soprattutto 
per la linearizzazione di doppio modello. \\
Passando a $Fr\,=\,0.3$ si ha un peggioramento dei risultati numerici. In
particolare il valore sperimentale del coefficiente $a_{33}$ all'aumentare della pulsazione 
viene previsto meglio dalla linearizzazione di Neumann Kelvin e dalla strip theory
mentre soltanto per basse pulsazioni la linearizzazione di doppio modello rimane pi\u 
vicina alla realt\a. Invece per $a_{55}$ \e ancora quest'ultima linearizzazione 
a dare la previsione pi\u accurata. 
Il coefficiente $b_{33}$ \e ben previsto da tutte e tre le tecniche illustrate, 
e per basse frequenze questa volta il dato sperimentale \e pi\u vicino alla 
linearizzazione di Neumann Kelvin che non a quella di doppio modello.
Per lo smorzamento $b_{55}$ nel moto di beccheggio la previsione fatta 
utilizzando il flusso di doppio modello \e nettamente superiore alle altre due 
valutazioni numeriche. Notiamo per\o che per alte pulsazioni i fenomeni dissipativi 
non previsti dai nostri modelli iniziano a non essere pi\u trascurabili e quindi 
oltre $\omega\,=\,5$ anche il risultato di doppio modello non risulta pi\u soddisfacente. \\
Per la previsione dei termini di cross coupling abbiamo un peggioramento 
per il coefficiente di smorzamento $b_{35}$. Qui il risultato del codice in 
frequenza non rispetta la reciprocit\a di questo termine con $b_{53}$, soprattutto 
per pulsazioni elevate (figura \ref{wjou534f3}). \\ 
Dalla figura \ref{jouex4f3} si pu\o vedere che anche i risultati per le 
forze di eccitazione non sono molto accurati ed il tipo di errore commesso 
\e simile a quello visto, sempre per $Fr\,=\,0.3$, 
per la precedente carena, con il risultato che per i moti nave si ha che le 
lunghezza d'onda del sistema d'onde incidente,  
corrispondeti alle pulsazioni di risonanza, 
risultano sottostimate rispetto ai valori sperimentali 
(figura \ref{jourao4f3}).  
\newpage
 \clearpage
\newpage
 \clearpage
\begin{figure}[htb]
      \epsfxsize=\textwidth
      \makebox[\textwidth]{\epsfbox{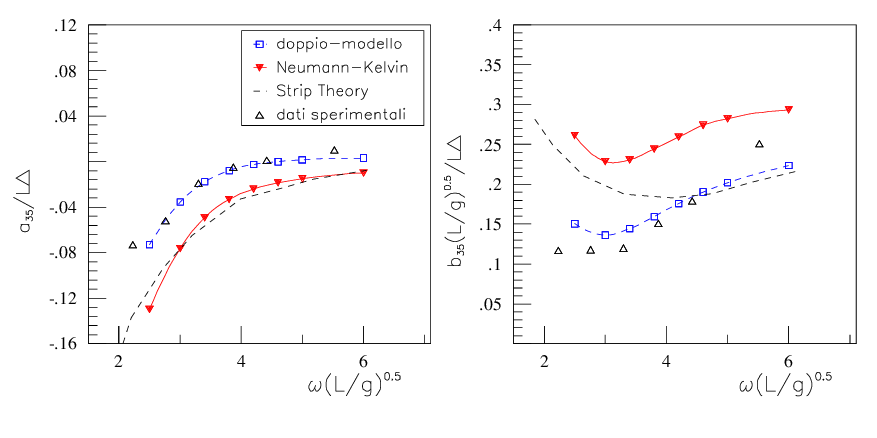}}
      \epsfxsize=\textwidth
      \makebox[\textwidth]{\epsfbox{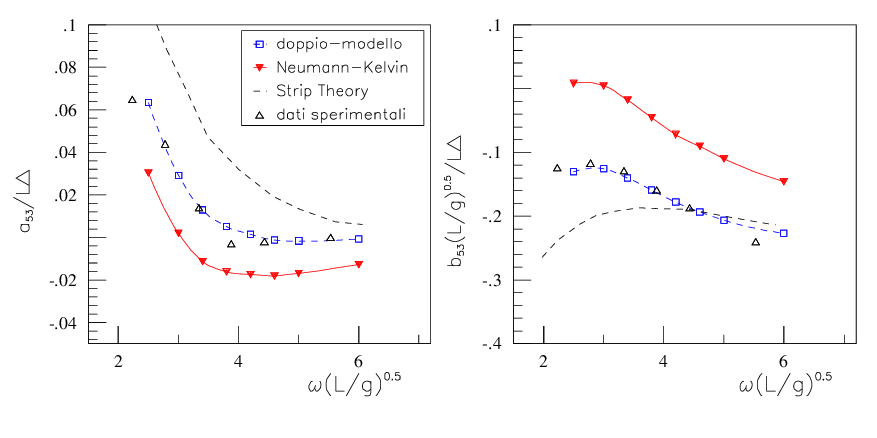}}
    \vskip -0.5cm
    \caption{Terza carena Wigley: confronto numerico--sperimentale 
             per i termini di {\em cross--coupling} (Fr =0.2).
             Nel termine di adimensionalizzazione $\Delta := \rho\nabla\,$.
             \label{wjou534f2}
             }
\end{figure}
\newpage
 \clearpage
\newpage
 \clearpage
\newpage
 \clearpage
\begin{figure}[htb]
      \epsfxsize=\textwidth
      \makebox[\textwidth]{\epsfbox{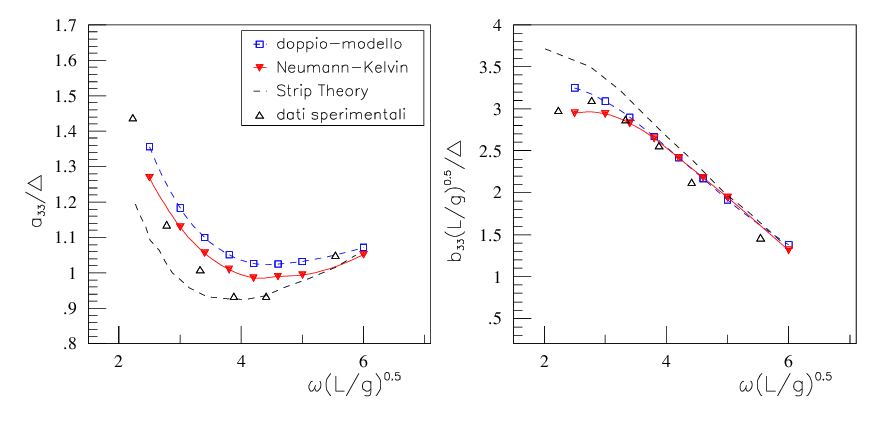}}
      \epsfxsize=\textwidth
      \makebox[\textwidth]{\epsfbox{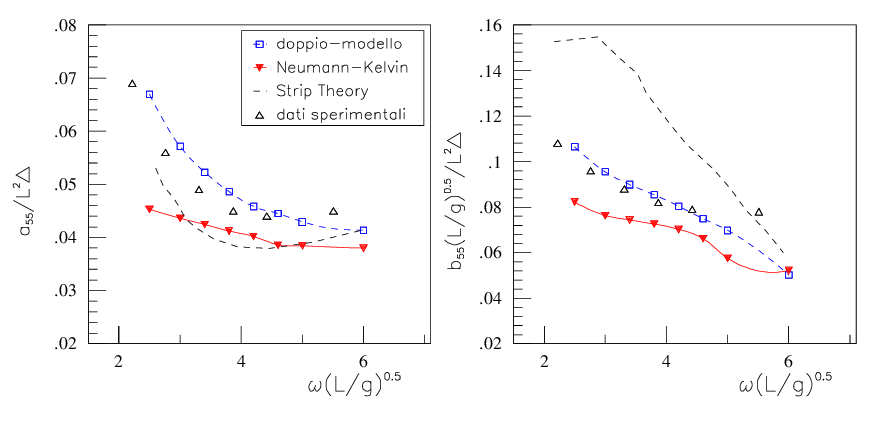}}
    \vskip -0.5cm
    \caption{Terza carena Wigley: confronto numerico--sperimentale 
             per {\em massa aggiunta} e {\em smorzamento} (Fr =0.3).
             Nel termine di adimensionalizzazione $\Delta := \rho\nabla\,$.
             \label{wjou354f3}
             }
\end{figure}
\newpage
 \clearpage
\newpage
 \clearpage
\newpage
 \clearpage
\newpage
\clearpage
\section{Casi esaminati: Quarta Carena Wigley.}
\begin{figure}[htb]
      \epsfxsize=\textwidth
      \makebox[\textwidth]{\epsfbox{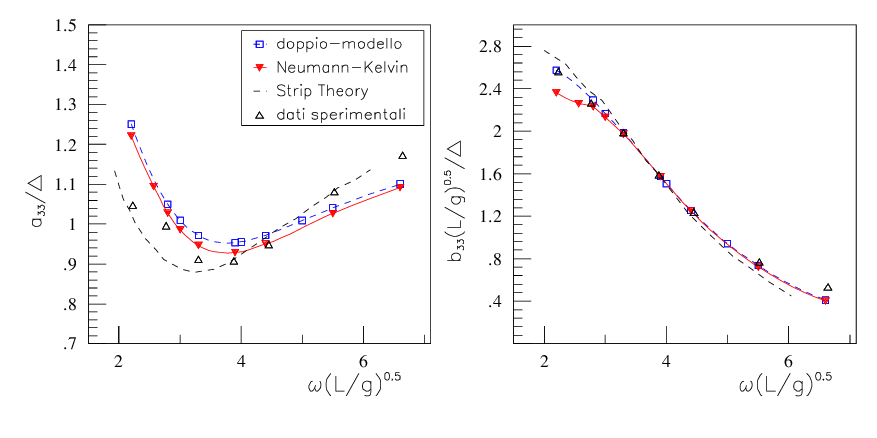}}
      \epsfxsize=\textwidth
      \makebox[\textwidth]{\epsfbox{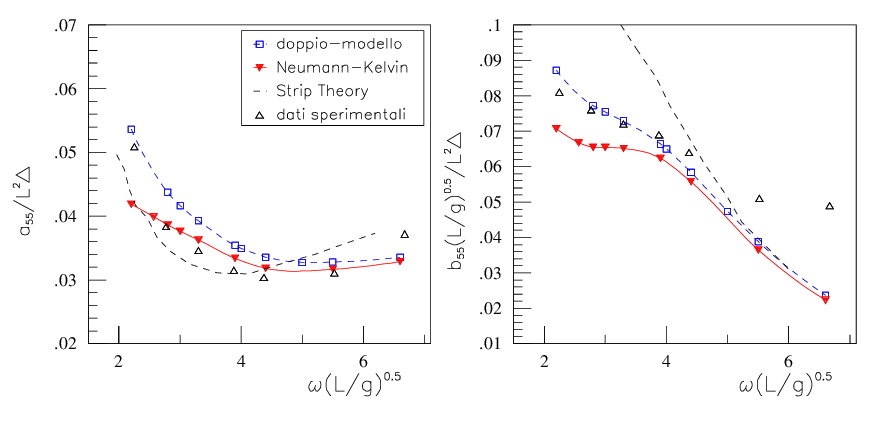}}
    \vskip -0.5cm
    \caption{Quarta carena Wigley:
             confronto numerico--sperimentale 
             per {\em massa aggiunta} e {\em smorzamento} (Fr =0.2).
             Nel termine di adimensionalizzazione $\Delta := \rho\nabla\,$,
             con $\nabla\,=\,Volume\,carena/L^3$.
             \label{wjou235f2}
             }
\end{figure}

\begin{figure}[htb]
      \epsfxsize=\textwidth
      \makebox[\textwidth]{\epsfbox{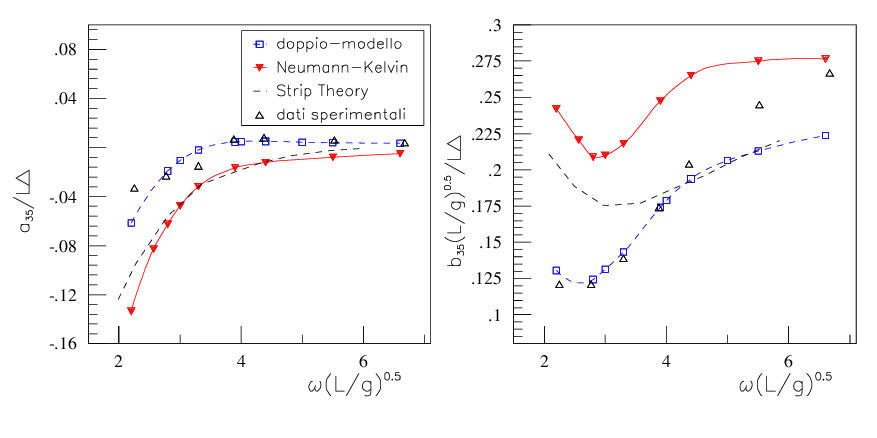}}
      \epsfxsize=\textwidth
      \makebox[\textwidth]{\epsfbox{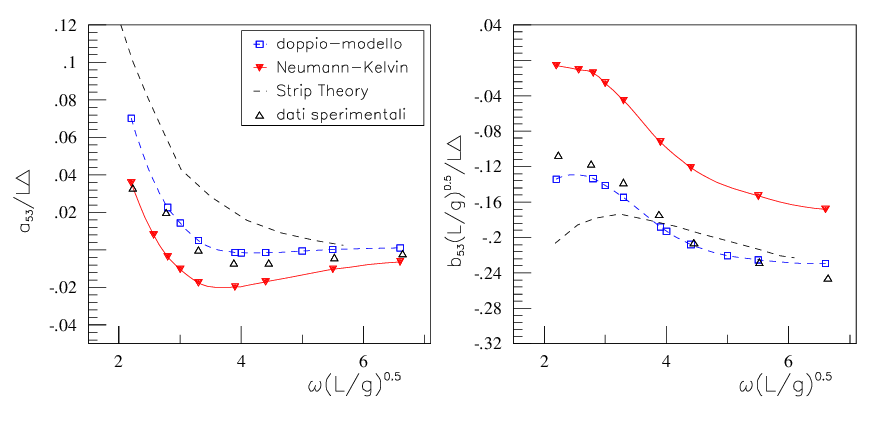}}
    \vskip -0.5cm
    \caption{Quarta carena Wigley:
             confronto numerico--sperimentale 
             per i termini di {\em cross--coupling} (Fr =0.2).
             Nel termine di adimensionalizzazione $\Delta := \rho\nabla\,$.
             \label{wjou253f2}
             }
\end{figure}

\begin{figure}[htb]
      \epsfxsize=\textwidth
      \makebox[\textwidth]{\epsfbox{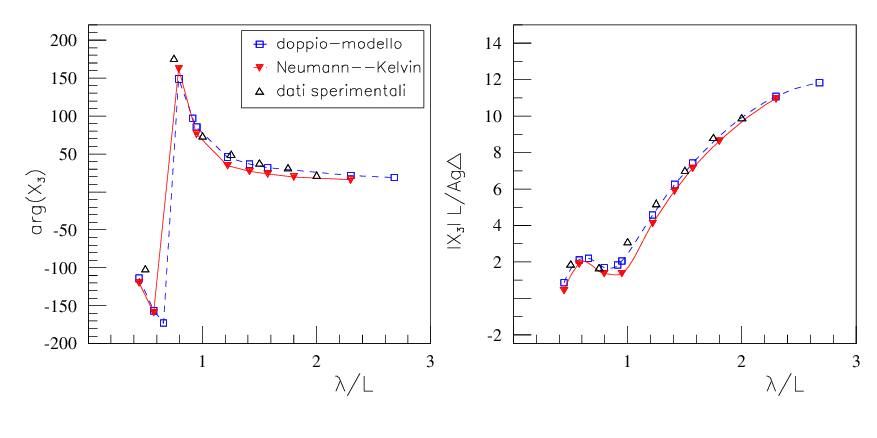}}
      \epsfxsize=\textwidth
      \makebox[\textwidth]{\epsfbox{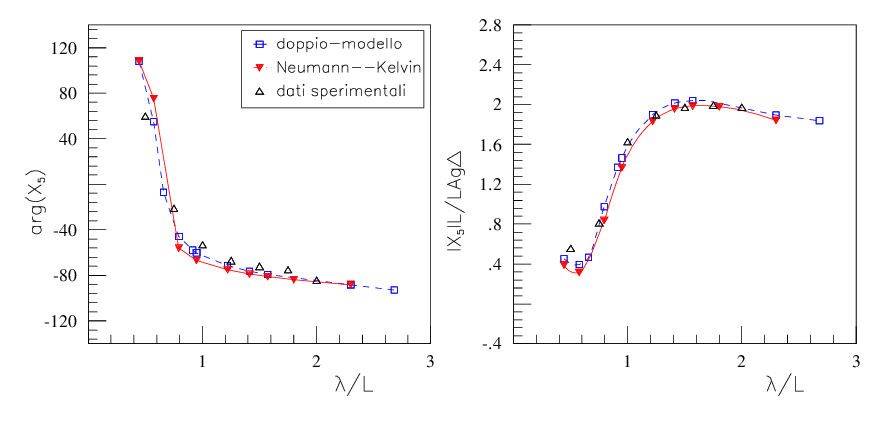}}
    \vskip -0.5cm
    \caption{Quarta carena Wigley: confronto numerico--sperimentale 
             per le forze di eccitazione in {\em heave} e {\em pitch}  (Fr =0.2).
             Nel termine di adimensionalizzazione $\Delta := \rho\nabla\,$.
             \label{jouex2f2}
             }
\end{figure}

\begin{figure}[htb]
      \epsfxsize=\textwidth
      \makebox[\textwidth]{\epsfbox{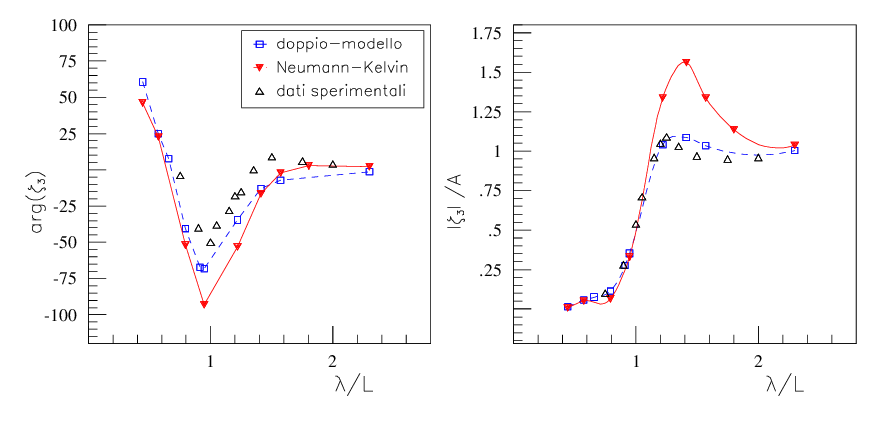}}
      \epsfxsize=\textwidth
      \makebox[\textwidth]{\epsfbox{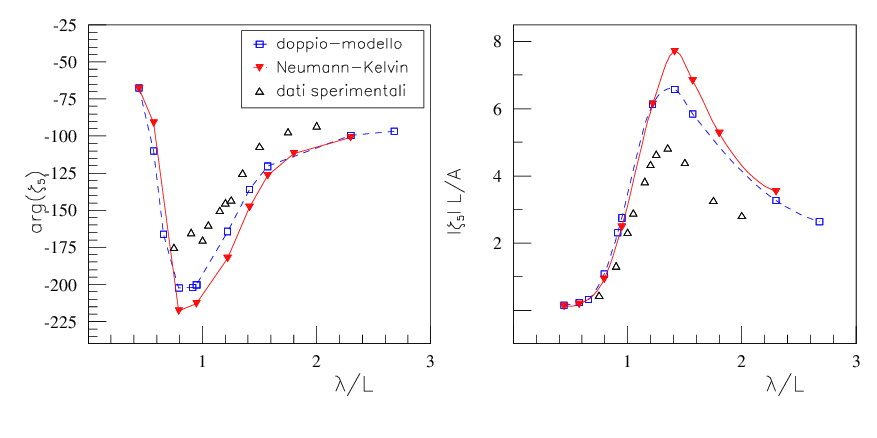}}
    \vskip -0.5cm
    \caption{Quarta carena Wigley:
             confronto numerico--sperimentale per l'ampiezza dei moti nave in
             {\em heave} e {\em pitch} (Fr =0.2).
             \label{jourao2f2}
             }
\end{figure}

 La quarta carena Wigley presenta un $Cb\simeq 0.56$, $L/B\,=\,5$
 ed inoltre $L/T\,=\,16$. Ricordiamo che questa carena appartiene alla 
 stessa famiglia della seconda carena Wigley, questo significa che le 
 sezioni trasversali sono dello stesso tipo soltanto che in questo caso sono
 allungate orizzontalmente come si pu\o vedere nella figura sottostante:   
\begin{figure}[htb]
    \vskip 0.5cm
      \epsfxsize=.5\textwidth
      \epsfxsize=.5\textwidth
      \makebox[.9\textwidth]{\epsfbox{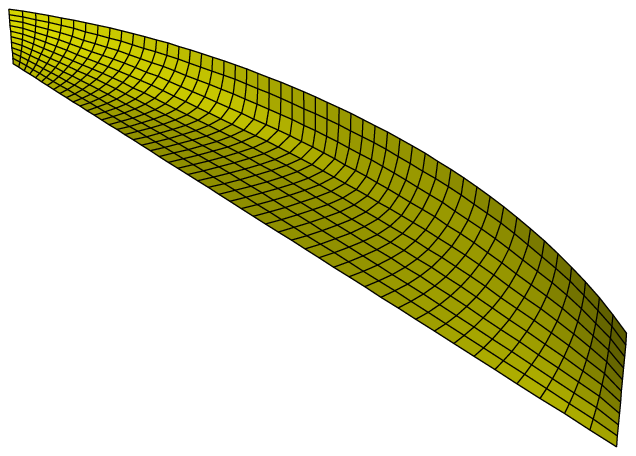}}
    \vskip -0.5cm
    \caption{Esempio di un tipico reticolo di calcolo per la quarta 
             carena Wigley 
             }
\end{figure}
\\
Per questa carena i dati sperimentali sono riportati soltanto per $Fr\,=\,0.2$ 
\footnote{ probabilmente perch\e per una geometria così fatta il confronto a $Froude$ pi\u
elevati tra i dati sperimentali e i risultati forniti dalla strip theory 
avrebbe sicuramente fornito dei pessimi risultati.}. 
Iniziamo a vedere le figure \ref{wjou235f2}, dove si pu\o notare che il valore 
sperimentale del coefficiente $a_{33}$ si discosta dalle previsioni 
date dalle linearizzazioni di Neumann Kelvin e di doppio modello,
mentre la strip-theory fornisce addirittura una previsione migliore. 
Per $a_{55}$ l'utilizzo del flusso di doppio modello non fornisce pi\u 
i risultati migliori, qui infatti \e Neumann-Kelvin a prevedere con maggior 
accuratezza i dati sperimentali, tranne che per pulsazioni molto basse dove 
il doppio modello continua a fornire risultati pi\u vicini alla realt\a fisica. 
Per il coefficiente di smorzamento $b_{33}$ si ha un'ottima previsione
per tutte e tre le tecniche esposte. Invece per $b_{55}$ si ha 
un risultato molto buono per la linearizzazione di doppio modello che per\o 
va peggiorando per pulsazioni crescenti infatti in questo caso i fenomeni dissipativi
esterni ai nostri modelli tornano ad avere effetti non trascurabili.  
Come per la carena precedente a $Fr\,=\,0.3$, si ottiene per la previsione dei 
termini di cross coupling un peggioramento dei risultati, in special modo per 
il coefficiente di smorzamento $b_{35}$  che per pulsazioni elevate 
presenta dei notevoli errori non rispettando pi\u la relazione di 
reciprocit\a con il coefficiente $b_{53}$ (figura \ref{wjou253f2}).  
Per questa carena si hanno degli ottimi risultati per quanto riguarda il 
problema della diffrazione e il relativo calcolo delle forze di eccitazione
(figure \ref{jouex2f2}) e questo fa si che anche i risultati per la previsione 
dei moti di sussulto e beccheggio siano buoni. 
Dalle figure \ref{jourao2f2} si vede che l'errore pi\u grande per il 
$R.A.O.$ \e commesso nella previsione dell'ampiezza per il moto di beccheggio,
questa infatti viene sovrastimata a causa degli errori commessi per il 
calcolo del coefficiente di smorzamento $b_{55}$. 
\newpage
clearpage
\newpage
 \clearpage
\newpage
 \clearpage
\newpage
 \clearpage
\newpage
 \clearpage
\newpage
\clearpage
\section{Casi esaminati: Quinta Carena Wigley }
\begin{figure}[htb]
      \epsfxsize=\textwidth
      \makebox[\textwidth]{\epsfbox{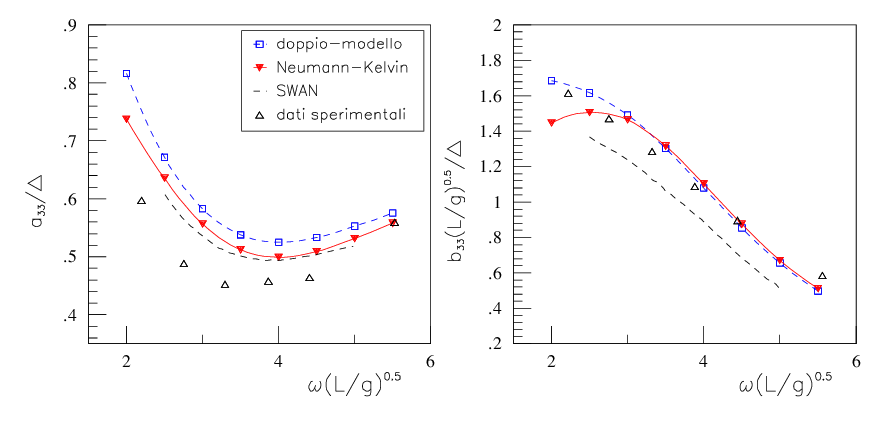}}
      \epsfxsize=\textwidth
      \makebox[\textwidth]{\epsfbox{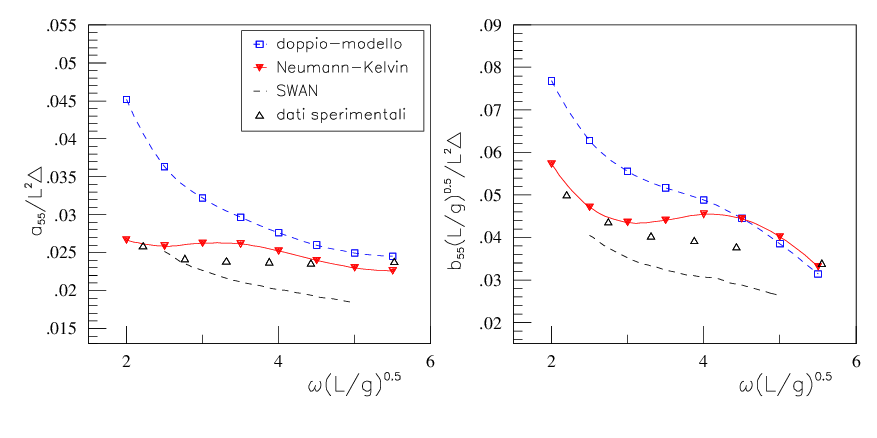}}
    \vskip -0.5cm
    \caption{Caso di una carena Wigley $Cb\simeq 0.63$: confronto numerico--sperimentale 
             per {\em massa aggiunta} e {\em smorzamento} (Fr =0.3).
             Nel termine di adimensionalizzazione $\Delta := \rho\nabla\,$,
             con $\nabla\,=\,Volume\,carena/L^3$.
             \label{wig35}
             }
\end{figure}

\begin{figure}[htb]
      \epsfxsize=\textwidth
      \makebox[\textwidth]{\epsfbox{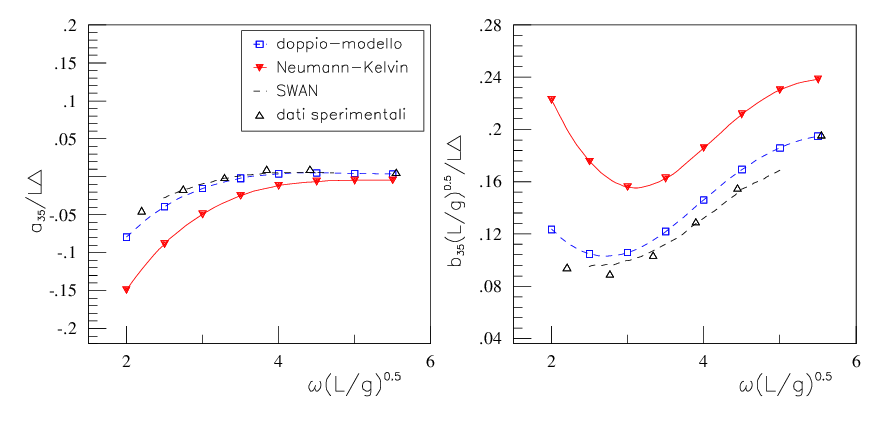}}
      \epsfxsize=\textwidth
      \makebox[\textwidth]{\epsfbox{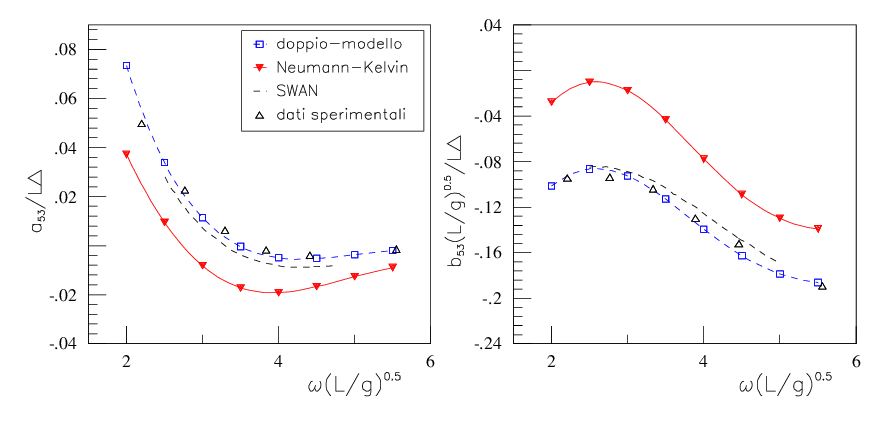}}
    \vskip -0.5cm
    \caption{Caso di una carena Wigley $Cb\simeq 0.63$: confronto numerico--sperimentale 
             per i termini di {\em cross--coupling} (Fr =0.3).
             Nel termine di adimensionalizzazione $\Delta := \rho\nabla\,$.
             \label{wig53}
             }
\end{figure}

\begin{figure}[htb]
      \epsfxsize=\textwidth
      \makebox[\textwidth]{\epsfbox{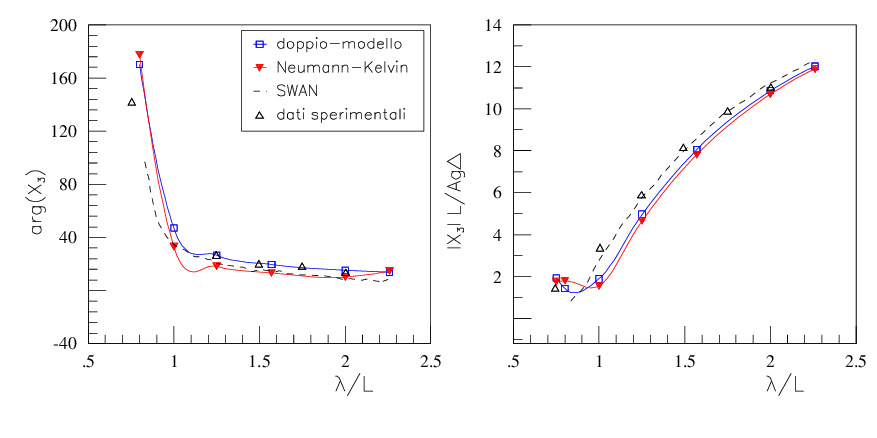}}
      \epsfxsize=\textwidth
      \makebox[\textwidth]{\epsfbox{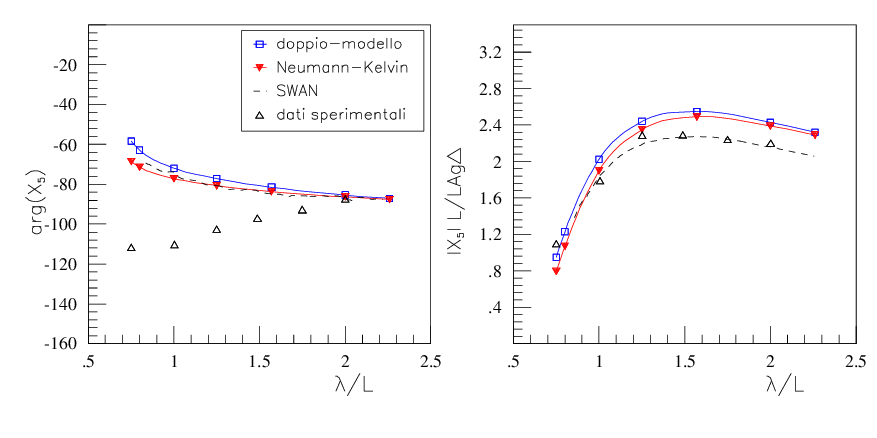}}
    \vskip -0.5cm
    \caption{Caso di una carena Wigley $Cb\simeq 0.63$: confronto numerico--sperimentale 
             per le forze di eccitazione in {\em heave} e {\em pitch}  (Fr =0.3).
             Nel termine di adimensionalizzazione $\Delta := \rho\nabla\,$.
             \label{excit}
             }
\end{figure}

\begin{figure}[htb]
      \epsfxsize=\textwidth
      \makebox[\textwidth]{\epsfbox{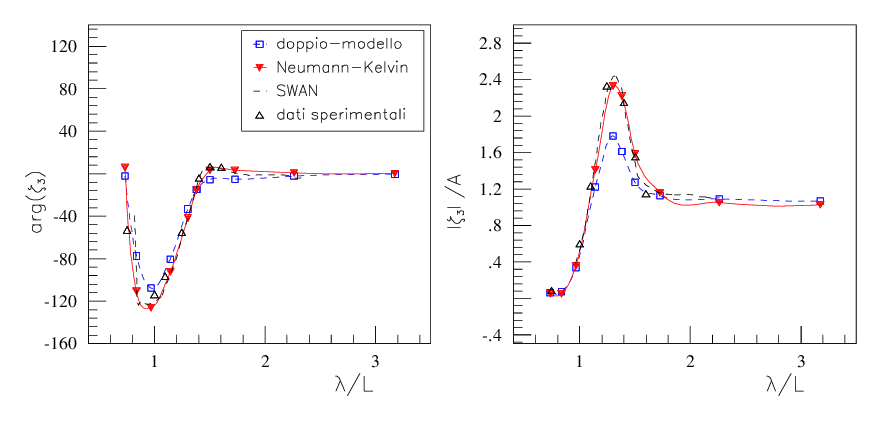}}
      \epsfxsize=\textwidth
      \makebox[\textwidth]{\epsfbox{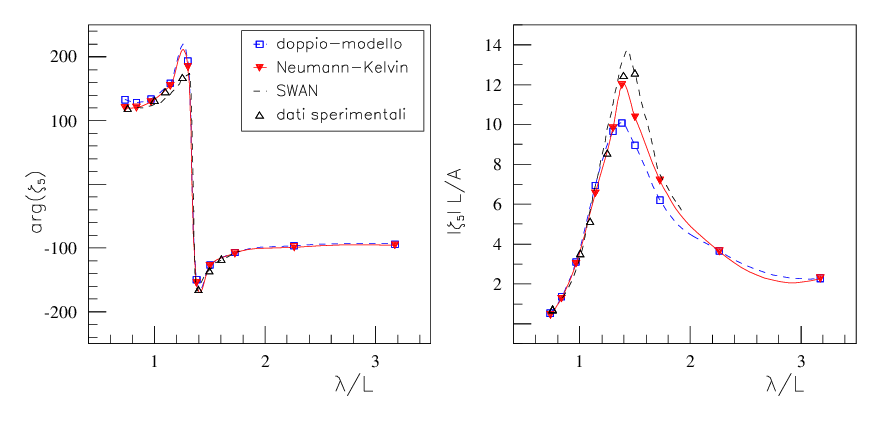}}
    \vskip -0.5cm
    \caption{Caso di una carena Wigley $Cb\simeq 0.63$: confronto numerico--sperimentale 
             per l'ampiezza dei moti nave in {\em heave} e {\em pitch} (Fr =0.3).
             \label{rao}
             }
\end{figure}

 Vengono ora illustrati i risultati numerici relativi al problema di una carena Wigley 
 avente i seguenti parametri geometrici caratteristici.
\be 
\dsty
\frac{L}{B}\,=\,10\qquad \frac{L}{T}\,=\,16\qquad Cb\,\simeq\,0.63 
\ee
L'elevato valore del coefficiente di blocco $Cb$ mostra che le sezioni 
centrali di questa carena sono prossime alla forma di un rettangolo e quindi 
presentano una "ginocchio" con un raggio di curvatura molto basso, come 
si pu\o osservare nella figura sottostante:  
\begin{figure}[htb]
      \vskip 1cm
      \epsfxsize=.5\textwidth
      \epsfxsize=.5\textwidth
      \makebox[.9\textwidth]{\epsfbox{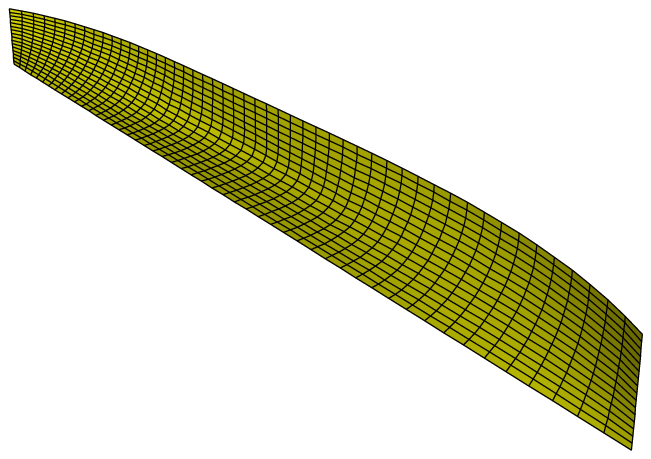}}
    \caption{Esempio di un tipico reticolo di calcolo per una Wigley con
             $Cb\simeq 0.63\,$. \label{wignak}
             }
\end{figure}
\\
L'analisi \e stata fatta soltanto per $Froude\,=\,0.3$. Per questa 
geometria si hanno dei risultati che si discostano molto da quelli visti 
per le prime quattro carene. 
Per questa carena abbiamo il confronto con la strip theory soltanto per 
i moti navi, mentre sono presentati i risultati forniti da un codice che
utilizza pannelli di ordine superiore insieme alla linearizzazione 
di doppio modello \cite{Nakos}.
Iniziamo dalle figure \ref{wig35} nelle quali 
si vede che tutte e tre le tecniche esposte sembrano sovrastimare il coefficiente  
di massa aggiunta $a_{33}$. In realt\a in questo caso i dati sperimentali per 
i coefficienti di massa aggiunta non sono stati depurati dai coefficienti 
delle forze di richiamo $c_{ij}$ quindi l'errore che si osserva \e 
in relt\a maggiore di quello reale \cite{Gerrit1}. Quanto detto vale anche 
per il coefficiente $a_{55}$, quindi l'apparente errore commesso dalla 
linearizzazione di doppio modello \e sicuramente pi\u contenuto.\\ 
\be
\dsty
a_{ij}\rightarrow\,a_{ij}-\frac{c_{ij}}{\omega^2}
\ee
Bench\h non sia stato specificato nell'articolo, dove abbiamo preso
i dati sperimentali e quelli relativi al codice con pannelli di ordine superiore,
riteniamo che anche i risultati di quest'ultimo non siano stati depurati dai 
coefficienti di richiamo.
Il coefficiente di smorzamento $b_{33}$ \e, come per tutte le carene viste, 
stimato con una buona approssimazione anzi il codice di ordine superiore 
\e quello che commette un'errore pi\u elevato.\\
Per quanto riguarda il coefficiente di smorzamento $b_{55}$ abbiamo 
dei valori e degli andamenti molto diversi da quelli ottenuti nelle 
precedenti carene. In particolare si ha che il doppio modello sovrastima 
molto questo coefficiente mentre la linearizzazione di Neumann Kelvin 
fornisce dei valori pi\u vicini a quelli sperimentali. 
Il codice "SWAN" invece sottostima questo coefficiente per tutto l'intervallo
delle pulsazioni di interesse. \\
Nelle figure \ref{wig53} abbiamo i coefficienti di cross coupling in cui 
si pu\o vedere che entrambi i codici che utilizzano il flusso base 
di doppio modello danno un'ottima stima dei valori sperimentali.\\ 
Per le forze di eccitazione abbiamo che i risutati migliori sono ottenuti dal codice SWAN
tuttavia i due codici che utilizzano pannelli all'ordine zero forniscono anche loro
dei buoni risultati. C'\e da sottolineare che tutti e tre 
danno una stima falsata della fase per il momento di eccitazione $X_5$
(figure \ref{excit}).\\
Infine per la previsione dei moti di sussulto e beccheggio otteniamo dalle 
figure \ref{rao} che il codice che utilizza pannelli di ordine superiore 
d\a una ottima stima  così come la linearizzazione di Neumann Kelvin,
mentre l'utilizzo di pannelli all'ordine zero con la linearizzazione 
di doppio modello fornisce una sottostima dei valori delle ampiezze dei moti.
Tuttavia l'accuratezza dei risultati del codice SWAN andrebbe 
verificata in un numero pi\u ampio di casi; 
infatti le stime che si ottengono per i coefficienti 
di massa aggiunta e smorzamento, non sono nettamente superiori a quelle 
fornite dai codici che utilizzano pannelli di ordine zero, anzi si \e visto
che per alcuni di questi coefficienti gli errori commessi dallo SWAN sono 
anche maggiori rispetto agli altri due codici presentati. 
\newpage
\clearpage
\newpage
 \clearpage
\newpage
 \clearpage
\newpage
 \clearpage
\newpage
 \clearpage
\section{Casi esaminati: Carena Serie $60$} 
\begin{figure}[htb]
    \vskip 0.5cm
      \epsfxsize=\textwidth
      \makebox[\textwidth]{\epsfbox{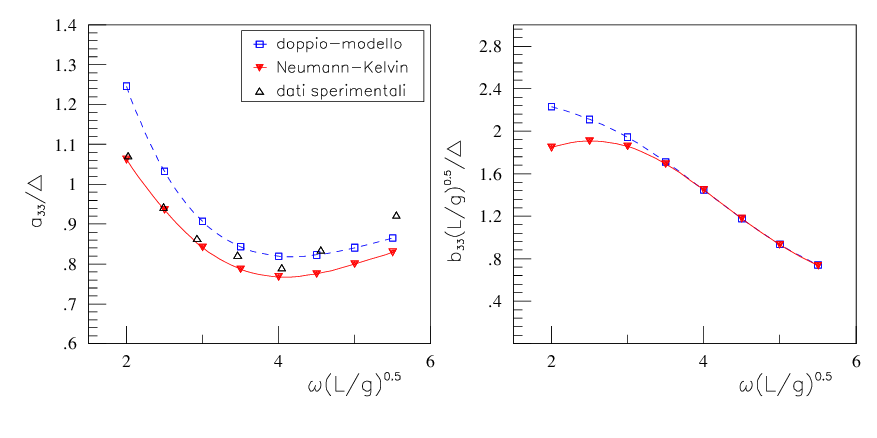}}
      \epsfxsize=\textwidth
      \makebox[\textwidth]{\epsfbox{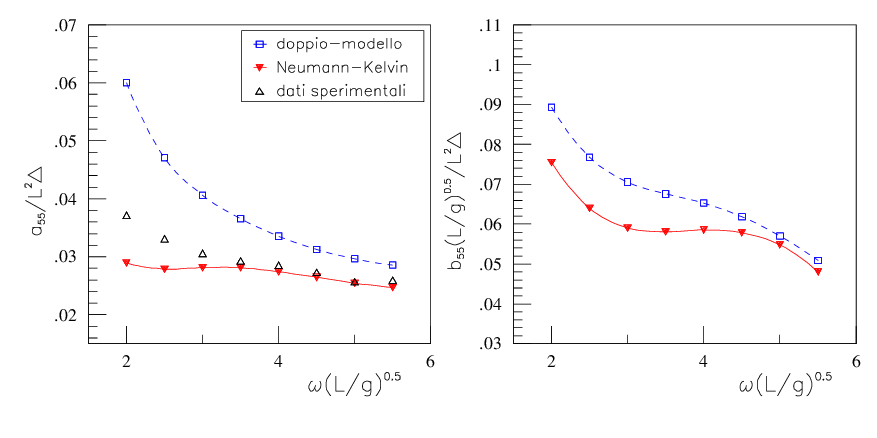}}
    \vskip -0.5cm
    \caption{Caso di una carena serie $60$ $Cb = 0.6$: Risultati
             numerici per {\em massa aggiunta} e {\em smorzamento} (Fr =0.2).
             Nel termine di adimensionalizzazione $\Delta := \rho\nabla\,$,
             con $\nabla\,=\,Volume\,carena/L^3$.
             \label{ser331}
             }
\end{figure}

\begin{figure}[htb]
    \vskip 0.5cm
      \epsfxsize=\textwidth
      \makebox[\textwidth]{\epsfbox{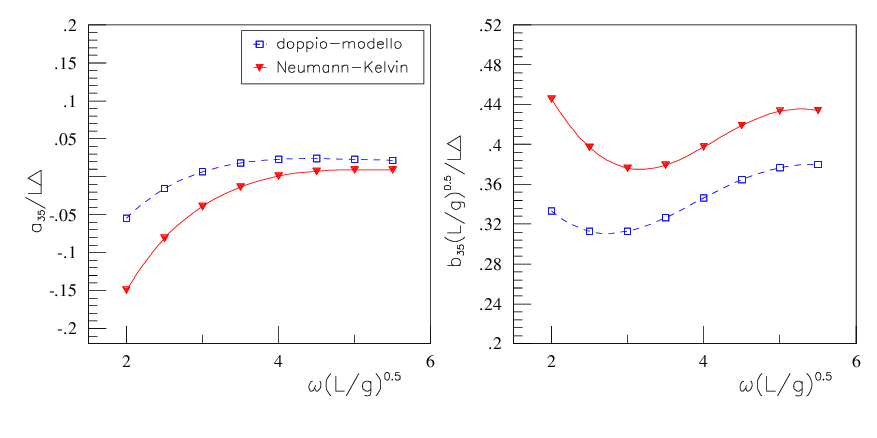}}
      \epsfxsize=\textwidth
      \makebox[\textwidth]{\epsfbox{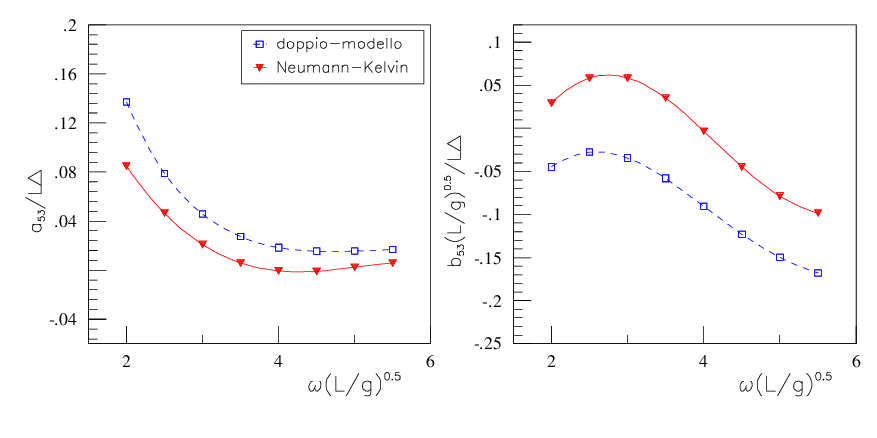}}
    \vskip -0.5cm
    \caption{Caso di una carena serie $60$ $Cb = 0.6$: risultati
             numerici per i termini di {\em cross--coupling} (Fr =0.2).
             Nel termine di adimensionalizzazione $\Delta := \rho\nabla\,$.
             \label{ser351}
             }
\end{figure}

\begin{figure}[htb]
    \vskip 0.5cm
      \epsfxsize=\textwidth
      \makebox[\textwidth]{\epsfbox{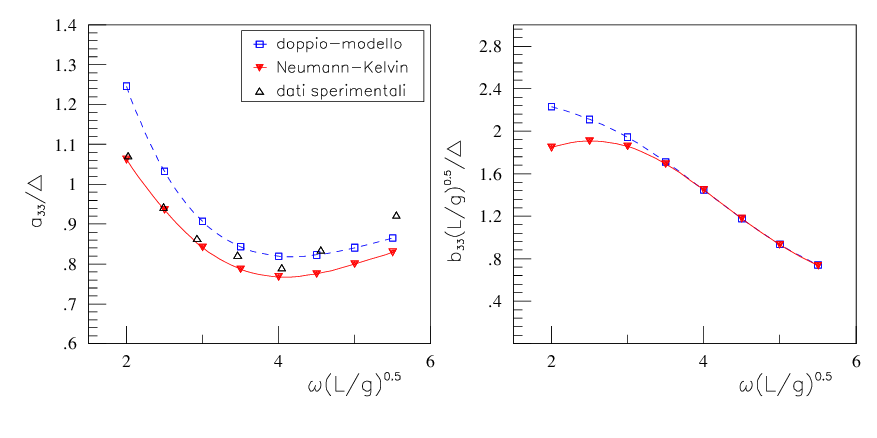}}
      \epsfxsize=\textwidth
      \makebox[\textwidth]{\epsfbox{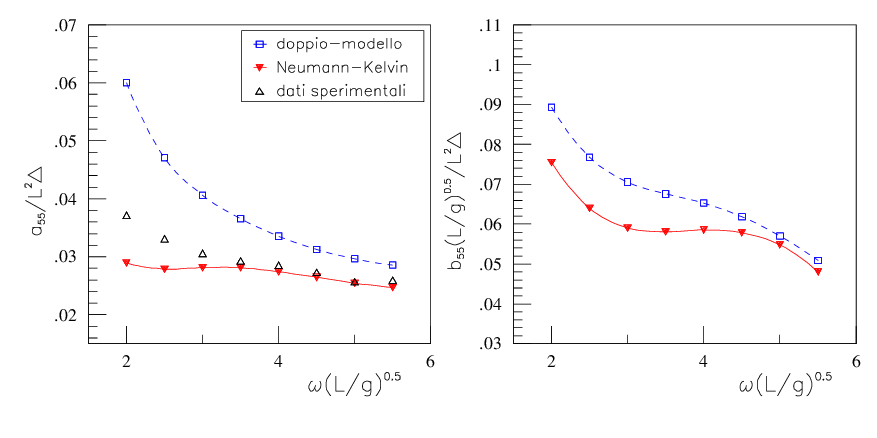}}
    \vskip -0.5cm
    \caption{Caso di una carena serie $60$ $Cb = 0.6$: 
             numerici per {\em massa aggiunta} e {\em smorzamento} (Fr =0.3).
             Nel termine di adimensionalizzazione $\Delta := \rho\nabla\,$.
             \label{ser33}
             }
\end{figure}

\begin{figure}[htb]
    \vskip 0.5cm
      \epsfxsize=\textwidth
      \makebox[\textwidth]{\epsfbox{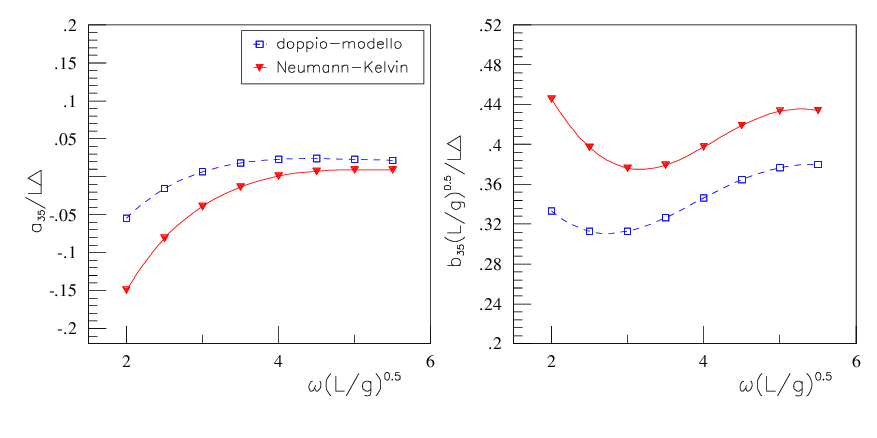}}
      \epsfxsize=\textwidth
      \makebox[\textwidth]{\epsfbox{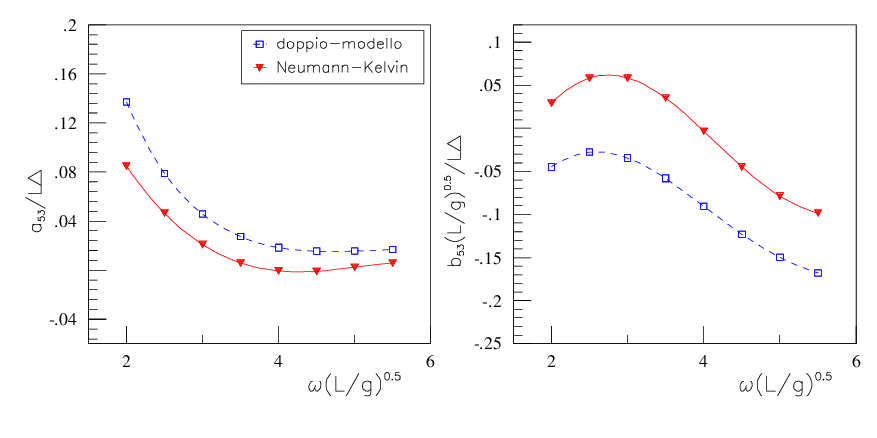}}
    \vskip -0.5cm
    \caption{Caso di una carena serie $60$ $Cb = 0.6$: risultati
             numerici per i termini di {\em cross--coupling} (Fr =0.3).
             Nel termine di adimensionalizzazione $\Delta := \rho\nabla\,$.
             \label{ser35}
             }
\end{figure}

 Viene analizzata infine una carena Serie $60$ avente: 
\be 
\dsty
\frac{L}{B}\,=\,7.5\qquad \frac{L}{T}\,\simeq\,18.8\qquad Cb\,\simeq\,0.6 
\ee
Tale geometria presenta un $Cb$ intermedio tra quello dell'ultima carena 
Wigley e quello della seconda e quarta carena Wigley. 
Rispetto alle carene Wigley analizzate in questo lavoro, si ha 
una geometria pi\u complessa soprattutto per le rilevanti differenze tra 
la zona poppiera e quella di prua, mentre in tutte le carene precedenti 
c'\e simmetria rispetto al piano verticale $yz$:  
\begin{figure}[htb]
      \vskip 0.5cm
      \epsfxsize=.5\textwidth
      \epsfxsize=.5\textwidth
      \makebox[.9\textwidth]{\epsfbox{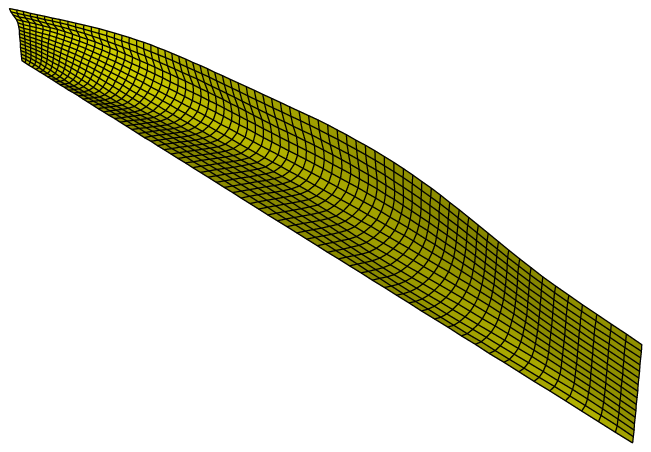}}
    \caption{Esempio di un tipico reticolo di calcolo per una Serie $60$ $Cb=0.6\,$. 
             \label{serie60}
             }
\end{figure}
\\ 
 Questo tipo di carena \e quello tra i pi\u semplici utilizzati 
 per la costruzione di navi da trasporto
 da parte dei cantieri navali e per questo risulta particolarmente interessante 
 valutare la verifica della tenuta al mare dei codici ottenuti con questa 
 geometria.  
 In particolare, nelle figure \ref{ser331}--\ref{ser351} e \ref{ser33}--\ref{ser35}
 sono riportati i coefficienti di forza idrodinamica per la radiazione in sussulto e
 in beccheggio, alle velocit\a corrispondenti a $Fr=.2$ e $Fr=.3\,$.  
 Nell'intervallo di frequenze considerato l'andamento delle diverse curve 
 \e quasi del tutto analogo a quello delle carene Wigley precedenti. 
 Soltanto per i coefficienti di massa aggiunta $a_{33}\,,a_{55}$ \e stato possibile 
 riportare i confronti con i dati sperimentali (contenuti in \cite{Gerrit2}).
 Per questi coefficienti si osserva complessivamente un discreto accordo 
 numerico--sperimentale. 
 I dati sperimentali per i due coefficienti $a_{33}$ e $a_{55}$ si mantengono 
 tra le stime fatte dalle due linearizzazioni. 
 In particolare per $a_{55}$ la linearizzazione di doppio modello mostra per 
 basse frequenze una sovrastima, al contrario delle prime quattro carene Wigley, 
 dove per basse $\omega$ si aveva una stima superiore del doppio modello rispetto 
 alla linearizzazione con Neumann-Kelvin. 
 Diversamente da quanto visto in precedenza, per questa carena non \e pi\u valida 
 la relazione di reciprocit\a dei termini di {\em cross--coupling},
 non presentando la carena una simmetria prua--poppa.
 Tuttavia vanno rispettati i limiti per $\omega\rightarrow\infty$ e infatti,
 soprattutto per la linearizzazione di doppio modello,
 si nota come i termini $a_{53}$ e $a_{35}$ convergono verso lo stesso valore, 
 mentre $\,-\,b_{53}$ tende al crescere di $\omega$ ad avvicinarsi alla curva $b_{35}$. 
\newpage
\clearpage
\newpage
 \clearpage
\newpage
 \clearpage
\newpage
 \clearpage
\newpage
 \clearpage

\section{Elevazioni d'onda} 
\begin{figure}[hp]
\vskip  2.5cm
 \vspace*{-2cm}
      \epsfxsize=0.75\textwidth
      \epsfysize=0.45\textwidth
      \makebox[\textwidth]{\epsfbox{./TESIFIG/wighvG.ps}}
      \epsfxsize=0.75\textwidth
      \epsfysize=0.45\textwidth
      \makebox[\textwidth]{\epsfbox{./TESIFIG/wigpcG.ps}}
      \epsfxsize=0.75\textwidth
      \epsfysize=0.45\textwidth
      \makebox[\textwidth]{\epsfbox{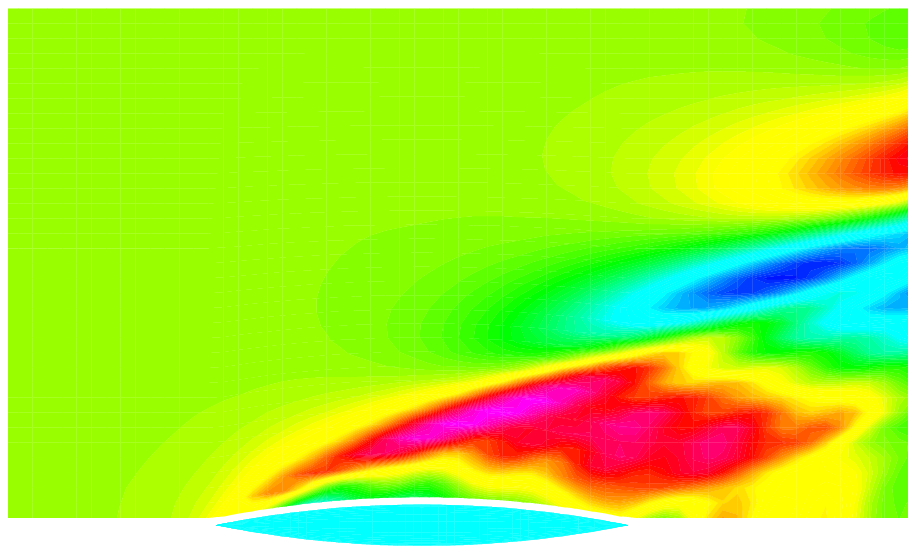}}
    \vskip -1.0cm
    \caption{Prima carena Wigley $Cb\simeq 0.46$: elevazione d'onda per i casi
             di radiazione in heave (in alto) e in pitch (al centro) e di
             {\em scattering} (in basso) per $\omega=3$ e $Fr=.3$. 
             \label{wigst_om3}
             }
\end{figure}

\begin{figure}[hp]
\vskip  2.5cm
 \vspace*{-2cm}
      \epsfxsize=0.75\textwidth
      \epsfysize=0.45\textwidth
      \makebox[\textwidth]{\epsfbox{./TESIFIG/wighvI.ps}}
      \epsfxsize=0.75\textwidth
      \epsfysize=0.45\textwidth
      \makebox[\textwidth]{\epsfbox{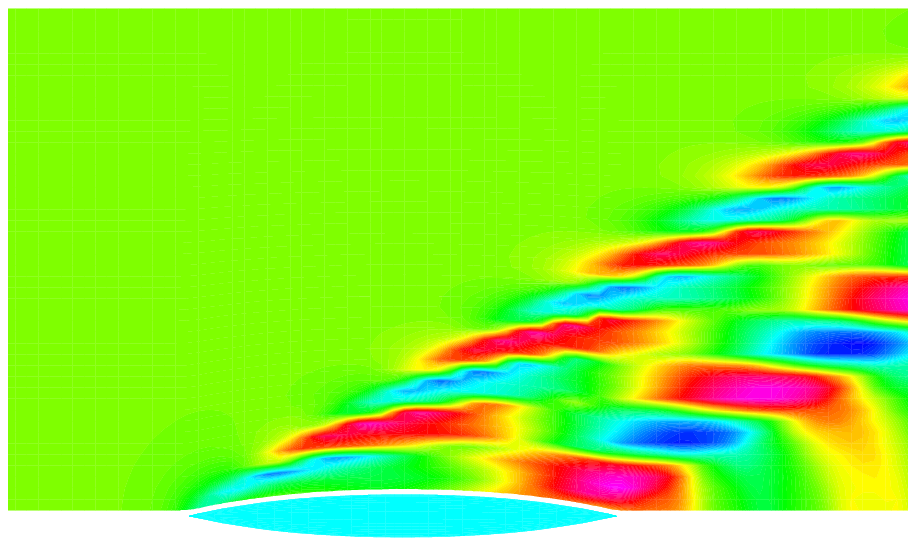}}
      \epsfxsize=0.75\textwidth
      \epsfysize=0.45\textwidth
      \makebox[\textwidth]{\epsfbox{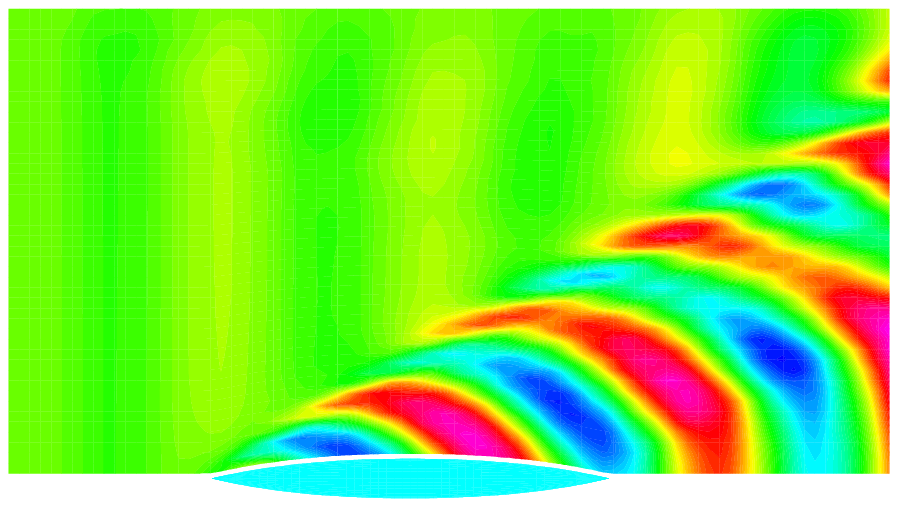}}
    \vskip -1.0cm
    \caption{Prima carena Wigley: elevazione d'onda per i casi
             di radiazione in heave (in alto) e in pitch (al centro) e di
             {\em scattering} (in basso) per $\omega=5.5$ e $Fr=.3$. 
             \label{wigst_om5}
             }
\end{figure}

\begin{figure}[hp]
\vskip  2.5cm
 \vspace*{-2cm}
      \epsfxsize=0.75\textwidth
      \epsfysize=0.45\textwidth
      \makebox[\textwidth]{\epsfbox{./TESIFIG/wighvg.ps}}
      \epsfxsize=0.75\textwidth
      \epsfysize=0.45\textwidth
      \makebox[\textwidth]{\epsfbox{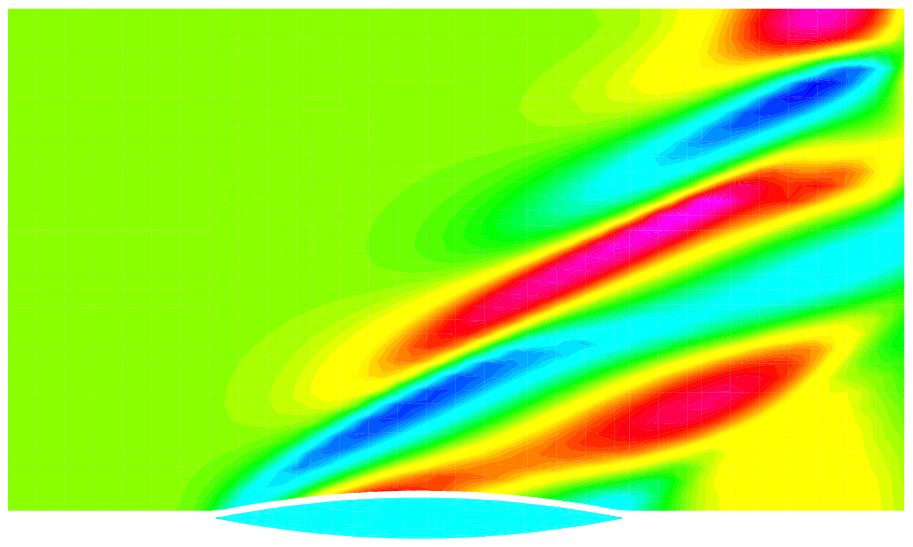}}
      \epsfxsize=0.75\textwidth
      \epsfysize=0.45\textwidth
      \makebox[\textwidth]{\epsfbox{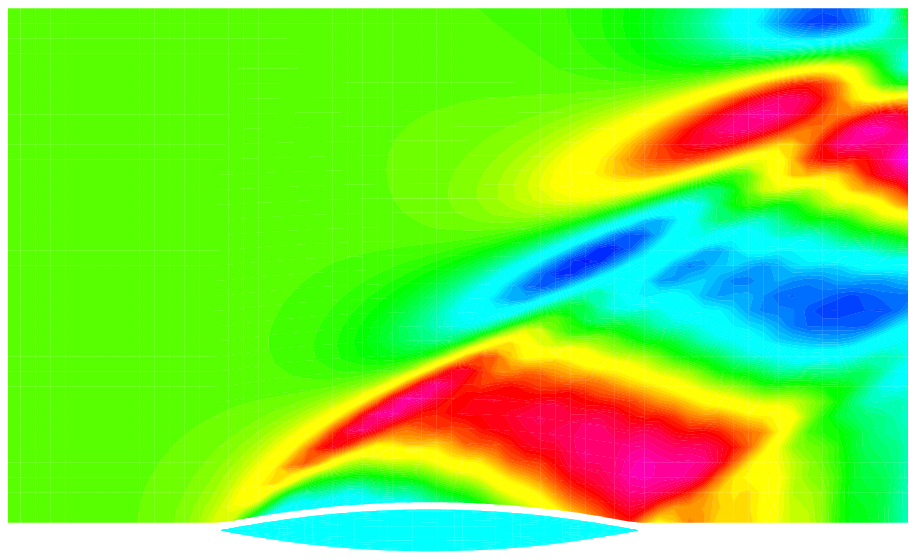}}
    \vskip -1.0cm
    \caption{Prima carena Wigley: elevazione d'onda per i casi
             di radiazione in heave (in alto) e in pitch (al centro) e di
             {\em scattering} (in basso) per $\omega=3$ e $Fr=.2$. 
             \label{wigst_om3fr2}
             }
\end{figure}

 In questo paragrafo vengono presentate le  elevazioni d'onda
 per i problemi di radiazione e di diffrazione, generate da alcune 
 delle carene analizzate.
 Tutti i casi sono ottenuti con la linearizzazione di doppio--modello.
 Vengono riportate successivamente diverse immagini in pianta del campo
 fluidodinamico non stazionario attorno alla carena.
 Tutti i casi si riferiscono a frequenze ridotte superiori al valore critico 
 ($\tau=0.25$), pertanto non si riscontrano onde che precedono la carena.
 Nelle figure \ref{wigst_om3}--\ref{wigst_om5}
 viene presentato l'elevazione d'onda associata ai problemi di
 radiazione, in sussulto e  beccheggio, e di diffrazione per differenti 
 valori della frequenza $\omega$ 
 \footnote{Ricordiamo che per i problemi di radiazione questa \e la pulsazione 
 di oscillazione della carena, mentre nel caso di diffrazione \e la pulsazione di 
 incontro carena--onda incidente}, per $Fr=.3\,$. 
 La carena \e la Wigley con $Cb\,=\,0.46$ e rapporto $L/B\,=\,10$.
 Si pu\o vedere come passando dalla pulsazione $\omega\,=\,3$ ad $\omega\,=\,5.5$ 
 si manifesta, per sussulto e beccheggio, una riduzione dell'angolo con cui 
 divergono le onde irradiate dalla carena.
 Allo stesso tempo, naturalmente, diminuisce la lunghezza d'onda trasversale 
 associata al sistema d'onde irradiato. 
 Passando invece da $Fr\,=\,0.3$ (v. fig. \ref{wigst_om3})
 a $Fr\,=\,0.2$ (v.fig. \ref{wigst_om3fr2}) si ha, in particolare, 
 un accorciamento delle onde come prescrive la relazione di dispersione, e 
 un aumento dell'angolo di divergenza del sistema ondoso irradiato. 
 L'elevazione d'onda associata al caso della diffrazione, e in particolare 
 l'elevazione d'onda relativa al potenziale di {\em scattering}, presenta un 
 analogo comportamento al variare delle grandezze precedentemente analizzate. \\
 Passiamo alla carena con $Cb\,=\,0.56$ e $L/B\,=\,5$ con una velocit\a 
 di avanzamento corrispondente a $Fr\,=\,0.2$ ed una pulsazione pari a 
 $\omega\,=\,3$. In questo caso si ha un 
 aumento dell'angolo di divergenza e la presenza, sia per il problema 
 di radiazione che per quello di diffrazione, di lunghezze d'onda pi\u piccole 
 rispetto al precedente caso a parit\a dei valori dei parametri. 
 Passando a $\omega\,=\,5.5$ si accentua la presenza di lunghezze d'onda 
 trasversali sempre pi\u piccole e quindi la necessit\a di utilizzare 
 dei reticoli sempre pi\u densi per la superficie libera.\\
 Per la carena con il massimo coefficiente di blocco $Cb\,=\,0.63$
 l'effetto dell'aumento del $Cb$ \e quello di realizzare un incremento del 
 sistema d'onda divergente a cui si aggiunge una riduzione delle onde trasversali
 come mostrato in \cite{MarHy} quindi tale effetto \e del tutto analogo a quello che
 si ottiene da una riduzione del numero di Froude. 
 Ci\o \e ben evidente se si fa il confronto con la prima Wigley per un numero 
 di $Froude$ pari a $0.2\,$ . \\ 
 Infine abbiamo le elevazioni d'onda corrispondenti alla carena appartenente alla 
 famiglia Serie 60, qui si nota che la particolare forma "aguzza" della prua permette
 la genesi di sistemi d'onda con un angolo di divergenza contenuto. Si ha inoltre 
 che a parit\a dei parametri rispetto alle altre carene il {\em pattern} ondoso 
 risulta pi\u regolare con la presenza di lunghezze d'onda trasversali pi\u ampie 
 che nei casi precedenti.
\begin{figure}[hp]
\vskip  2.5cm
 \vspace*{-2cm}
      \epsfxsize=0.75\textwidth
      \epsfysize=0.45\textwidth
      \makebox[\textwidth]{\epsfbox{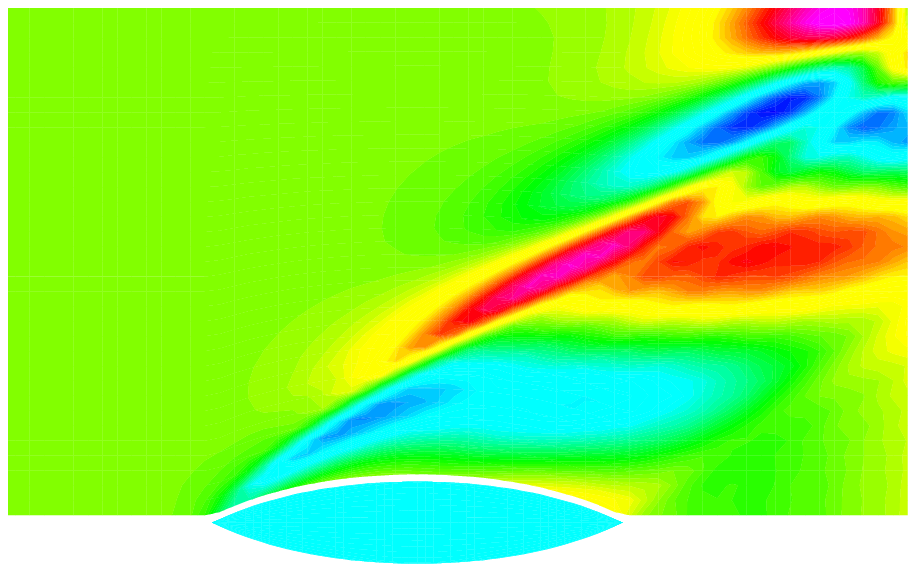}}
      \epsfxsize=0.75\textwidth
      \epsfysize=0.45\textwidth
      \makebox[\textwidth]{\epsfbox{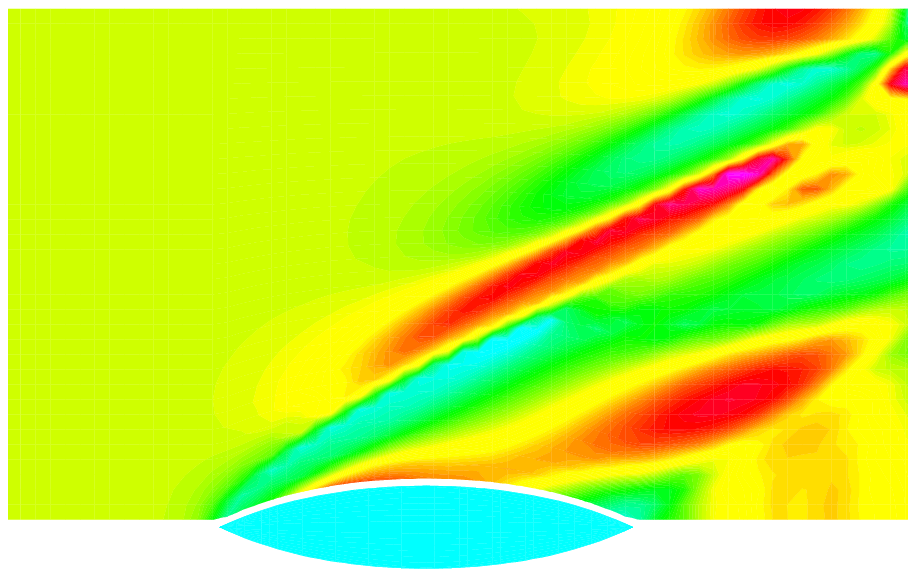}}
      \epsfxsize=0.75\textwidth
      \epsfysize=0.45\textwidth
      \makebox[\textwidth]{\epsfbox{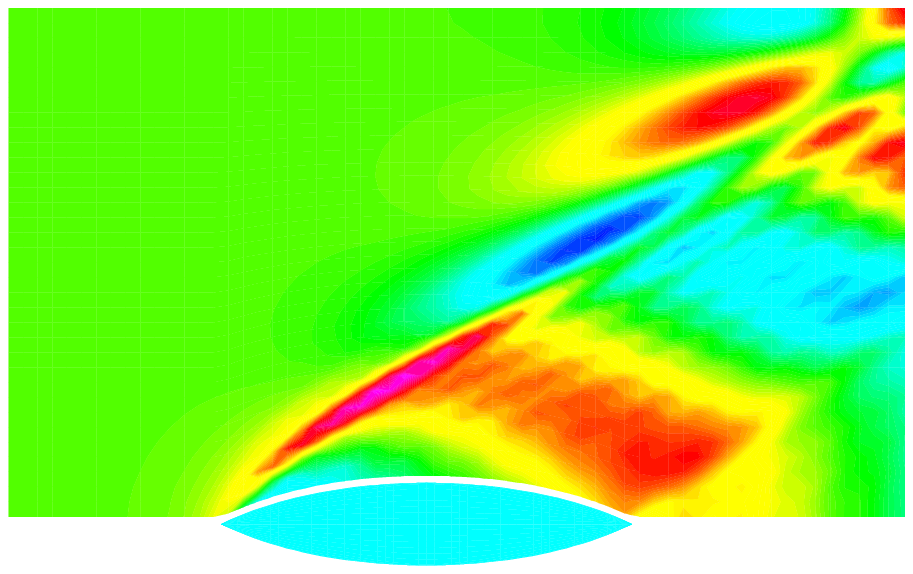}}
    \vskip -1.0cm
    \caption{Quarta carena Wigley: elevazione d'onda per i casi
             di radiazione in heave (in alto) e in pitch (al centro) e di
             {\em scattering} (in basso) per $\omega=3$ e $Fr=.2$. 
             \label{wigj2_om3}
             }
\end{figure}
\begin{figure}[hp]
\vskip  2.5cm
 \vspace*{-2cm}
      \epsfxsize=0.75\textwidth
      \epsfysize=0.45\textwidth
      \makebox[\textwidth]{\epsfbox{./TESIFIG/wigj2hvI.ps}}
      \epsfxsize=0.75\textwidth
      \epsfysize=0.45\textwidth
      \makebox[\textwidth]{\epsfbox{./TESIFIG/wigj2pcI.ps}}
      \epsfxsize=0.75\textwidth
      \epsfysize=0.45\textwidth
      \makebox[\textwidth]{\epsfbox{./TESIFIG/wigj2scI.ps}}
    \vskip -1.0cm
    \caption{Quarta carena Wigley: elevazione d'onda per i casi
             di radiazione in heave (in alto) e in pitch (al centro) e di
             {\em scattering} (in basso) per $\omega=5.5$ e $Fr=.2$. 
             \label{wigj2_om55}
             }
\end{figure}
\begin{figure}[hp]
\vskip 2.5cm 
 \vspace*{-2cm}
      \epsfxsize=0.75\textwidth
      \epsfysize=0.45\textwidth
      \makebox[\textwidth]{\epsfbox{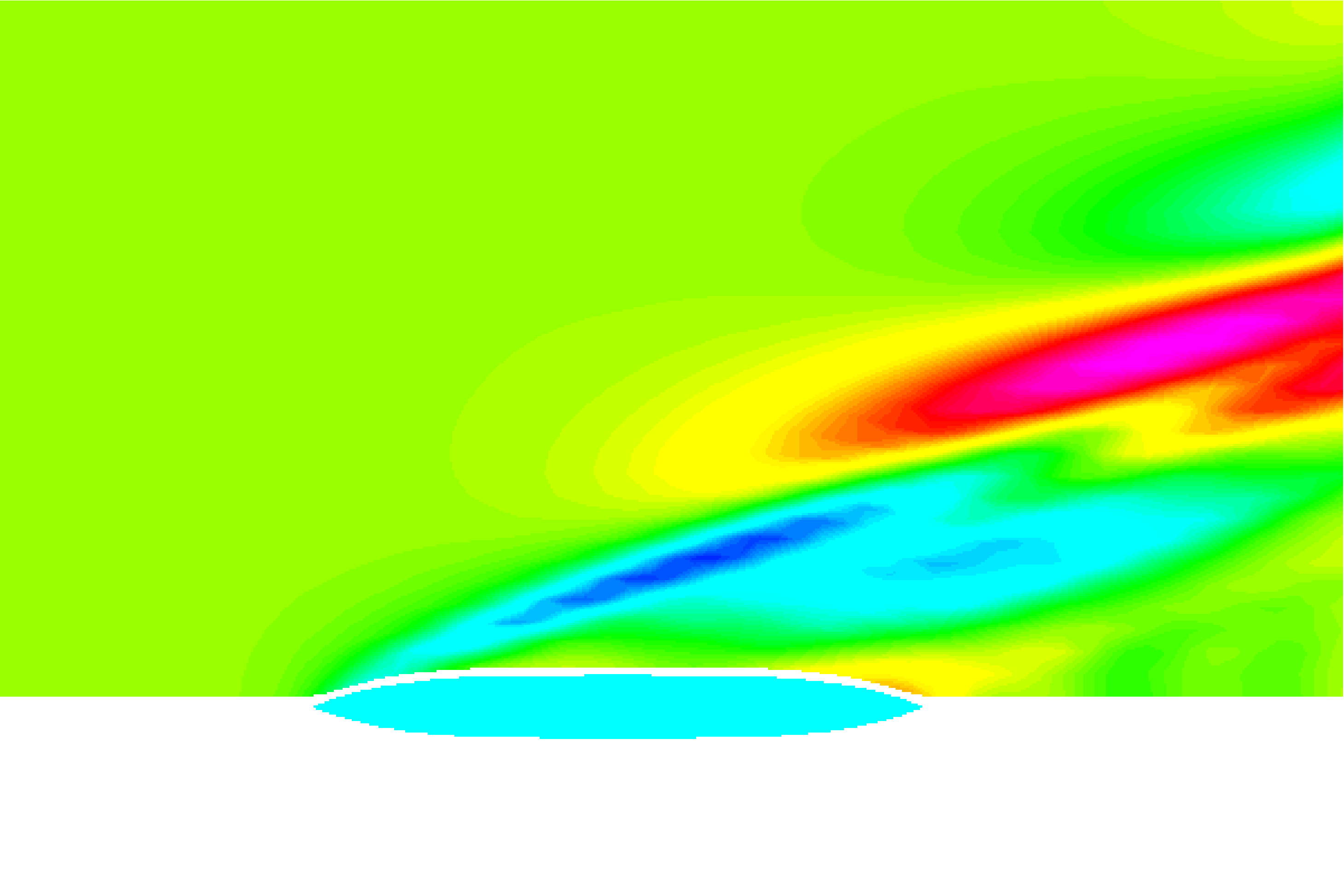}}
      \epsfxsize=0.75\textwidth
      \epsfysize=0.45\textwidth
      \makebox[\textwidth]{\epsfbox{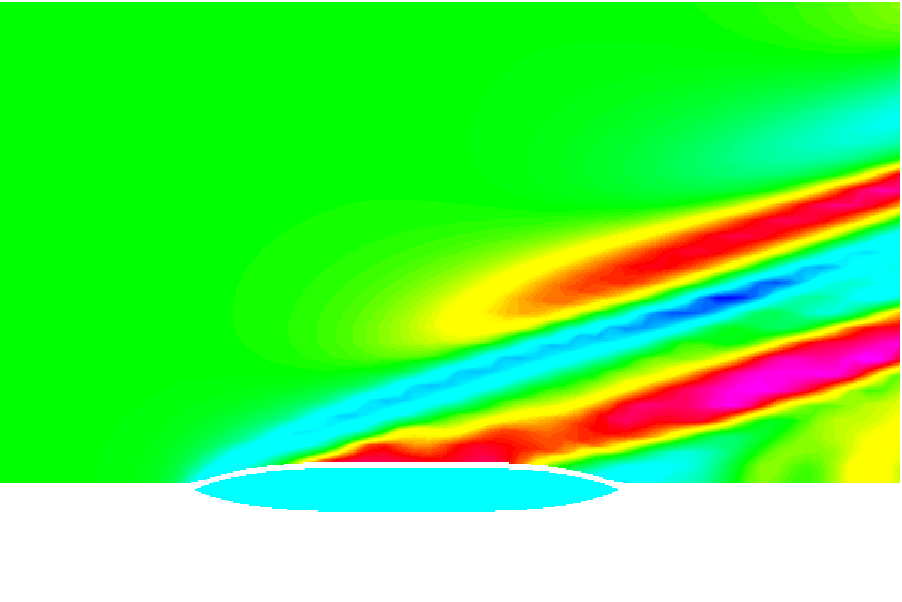}}
      \epsfxsize=0.75\textwidth
      \epsfysize=0.45\textwidth
      \makebox[\textwidth]{\epsfbox{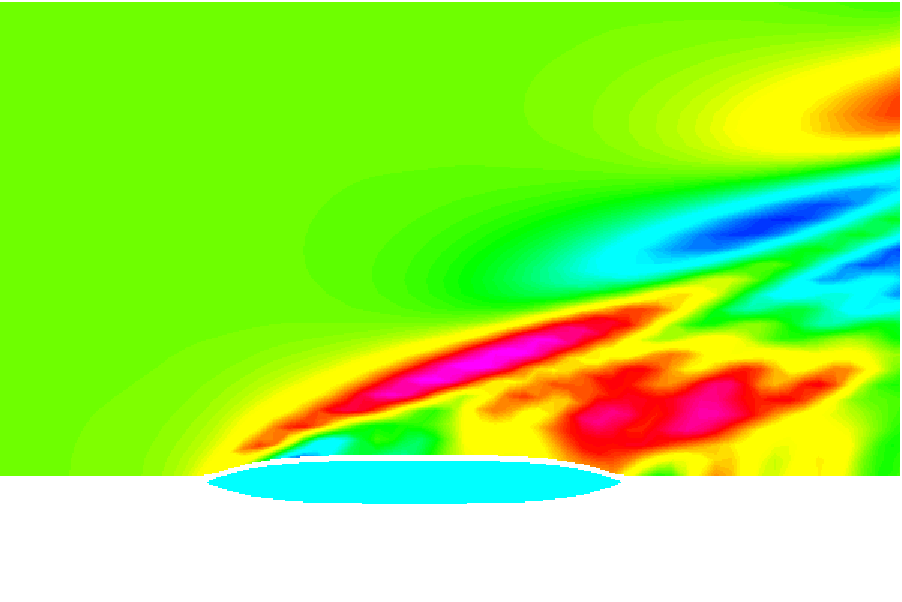}}
    \vskip -1.0cm
    \caption{Caso di una carena Wigley $Cb\simeq 0.63$: elevazione d'onda per i casi
             di radiazione in heave (in alto) e in pitch (al centro) e di
             {\em scattering} (in basso) per $\omega=3$ e $Fr=.3$. 
             \label{wn_om3}
             }
\end{figure}
\begin{figure}[hp]
\vskip  2.5cm
 \vspace*{-2cm}
      \epsfxsize=0.75\textwidth
      \epsfysize=0.45\textwidth
      \makebox[\textwidth]{\epsfbox{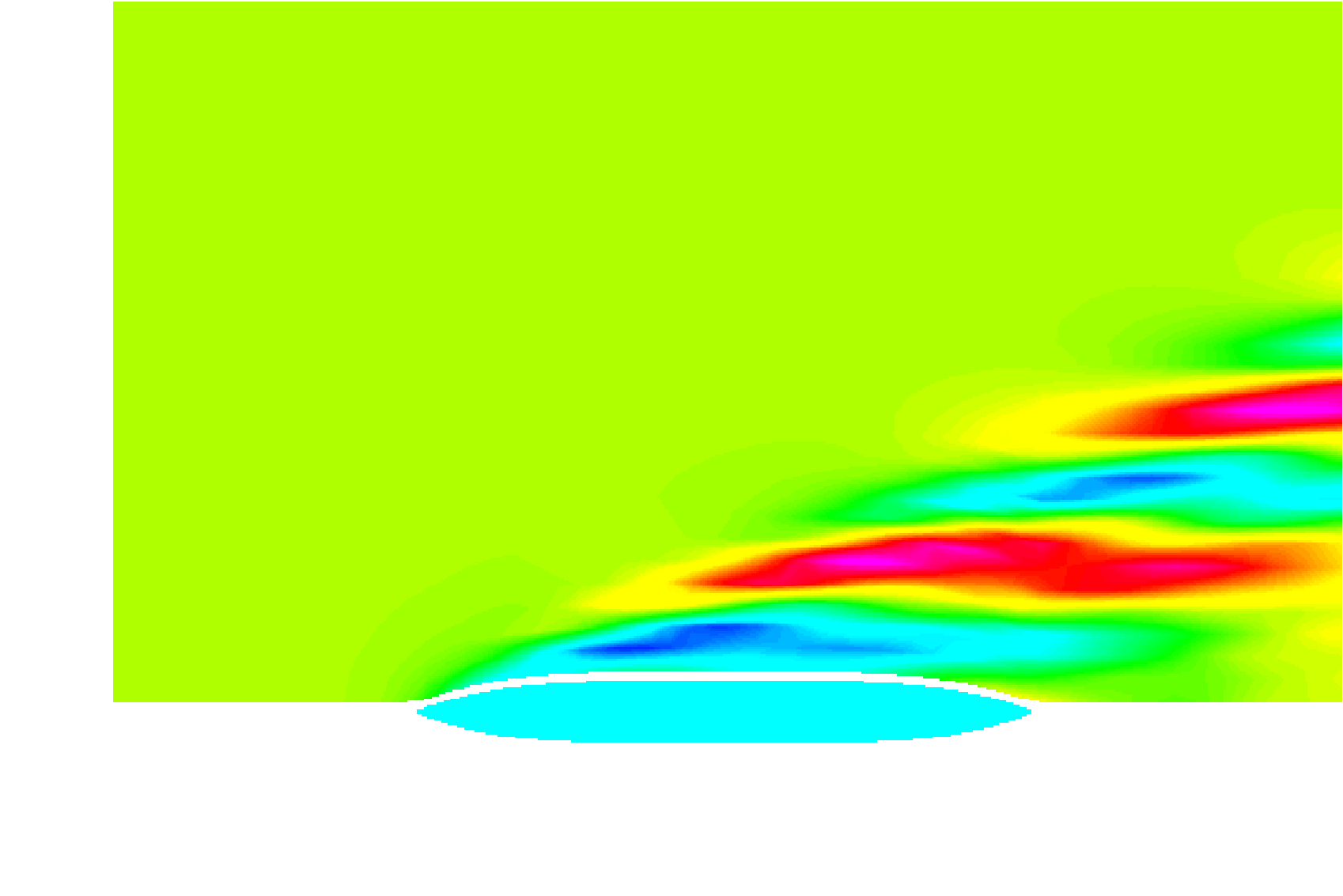}}
      \epsfxsize=0.75\textwidth
      \epsfysize=0.45\textwidth
      \makebox[\textwidth]{\epsfbox{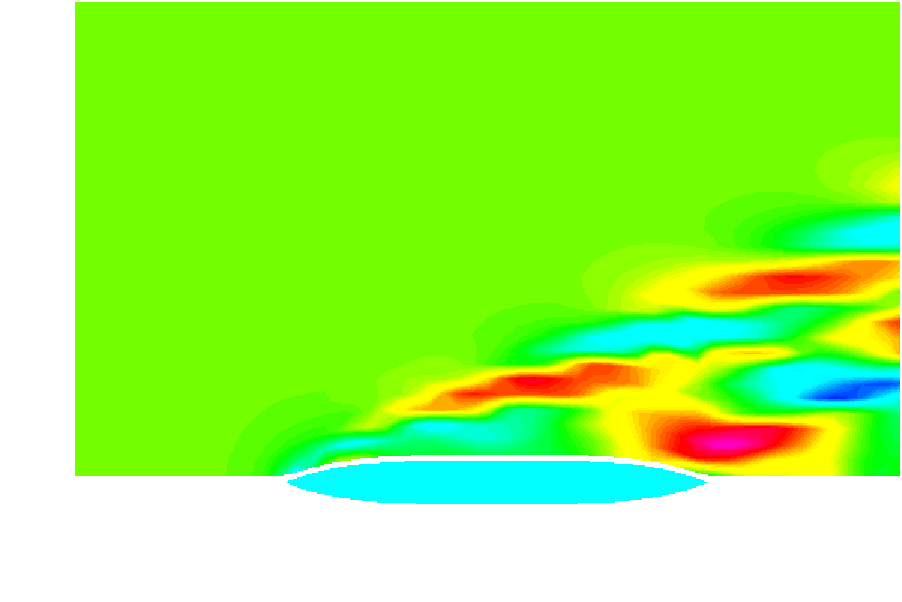}}
      \epsfxsize=0.75\textwidth
      \epsfysize=0.45\textwidth
      \makebox[\textwidth]{\epsfbox{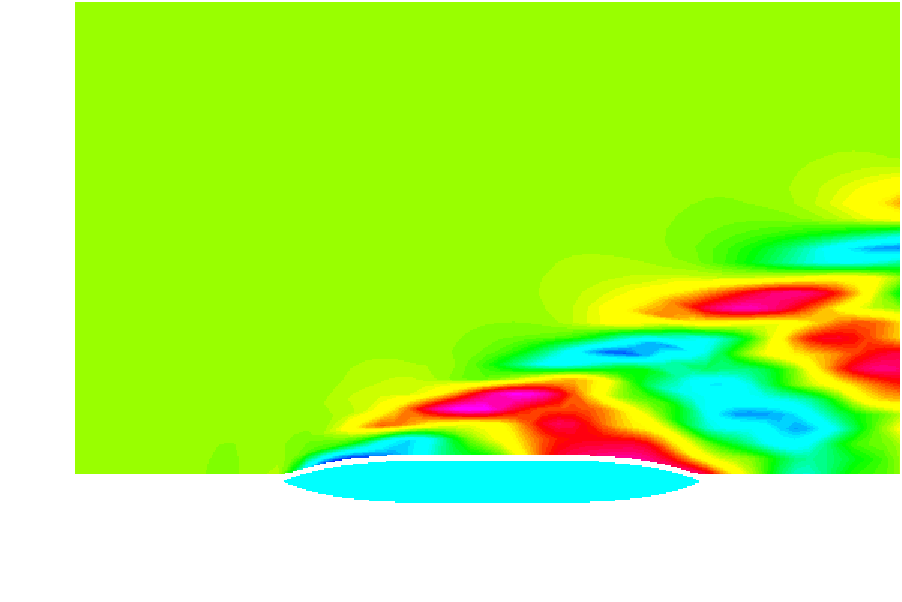}}
    \vskip -1.0cm
    \caption{Caso di una carena Wigley $Cb\simeq 0.63$: elevazione d'onda per i casi
             di radiazione in heave (in alto) e in pitch (al centro) e di
             {\em scattering} (in basso) per $\omega=5$ e $Fr=.3$. 
             \label{wn_om5}
             }
\end{figure}
\begin{figure}[hp]
\vskip  2.5cm
 \vspace*{-2cm}
      \epsfxsize=0.75\textwidth
      \epsfysize=0.45\textwidth
      \makebox[\textwidth]{\epsfbox{./TESIFIG/wnakhvg.eps}}
      \epsfxsize=0.75\textwidth
      \epsfysize=0.45\textwidth
      \makebox[\textwidth]{\epsfbox{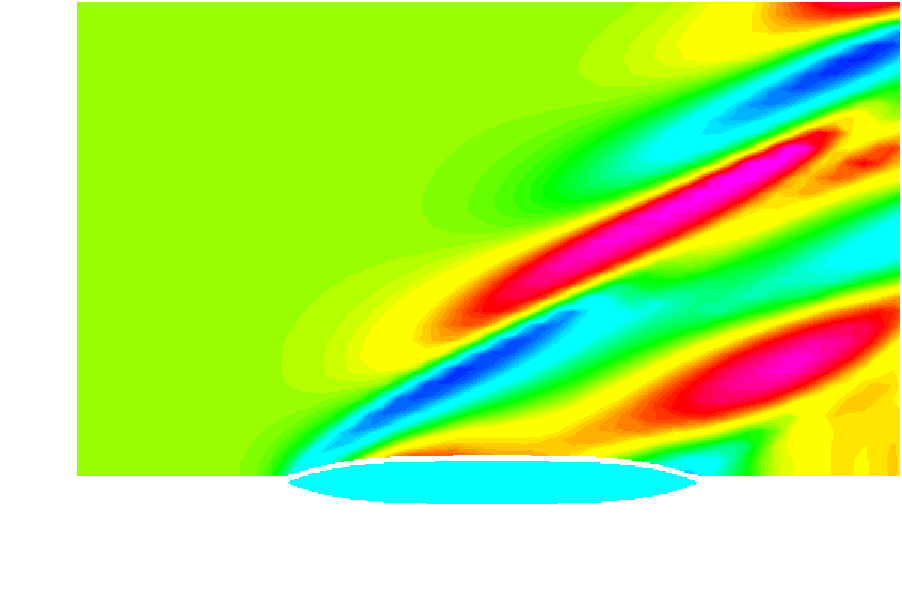}}
      \epsfxsize=0.75\textwidth
      \epsfysize=0.45\textwidth
      \makebox[\textwidth]{\epsfbox{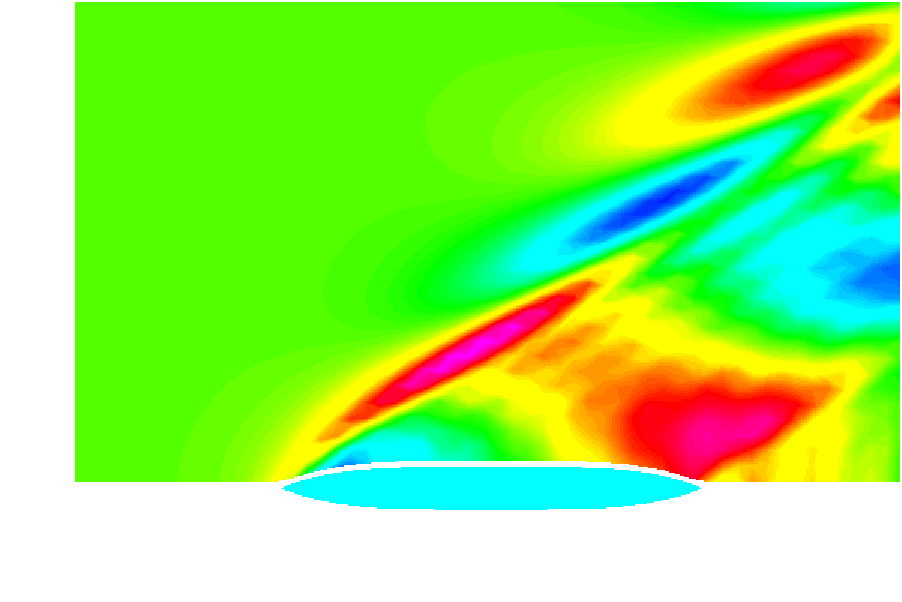}}
    \vskip -1.0cm
    \caption{Caso di una carena Wigley $Cb\simeq 0.63$: elevazione d'onda per i casi
             di radiazione in heave (in alto) e in pitch (al centro) e di
             {\em scattering} (in basso) per $\omega=3$ e $Fr=.2$. 
             \label{wn_om3fr2}
             }
\end{figure}
\begin{figure}[hp]
\vskip  2.5cm
 \vspace*{-2cm}
      \epsfxsize=0.75\textwidth
      \epsfysize=0.45\textwidth
      \makebox[\textwidth]{\epsfbox{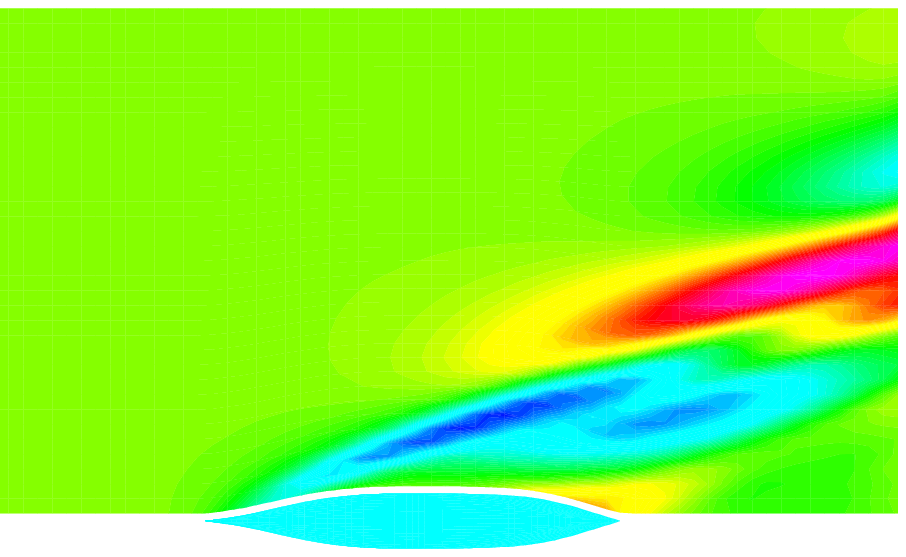}}
      \epsfxsize=0.75\textwidth
      \epsfysize=0.45\textwidth
      \makebox[\textwidth]{\epsfbox{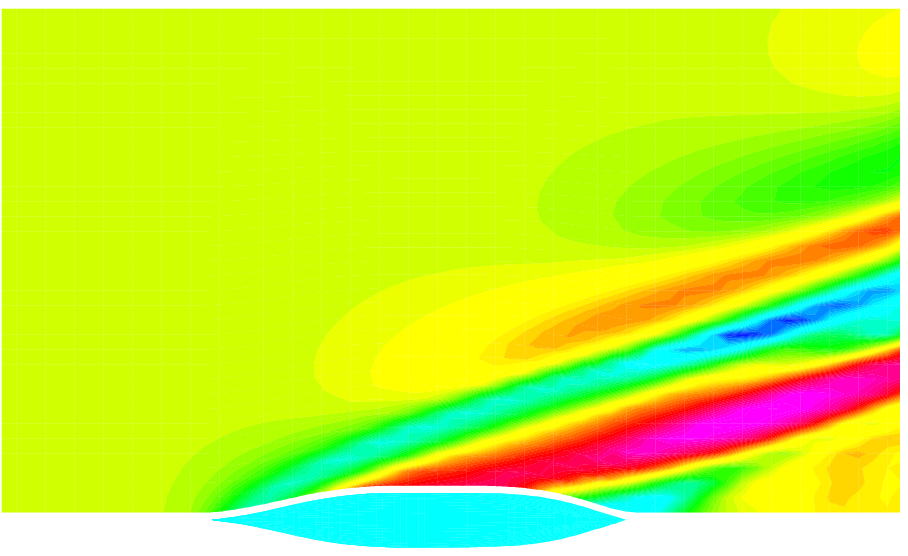}}
      \epsfxsize=0.75\textwidth
      \epsfysize=0.45\textwidth
      \makebox[\textwidth]{\epsfbox{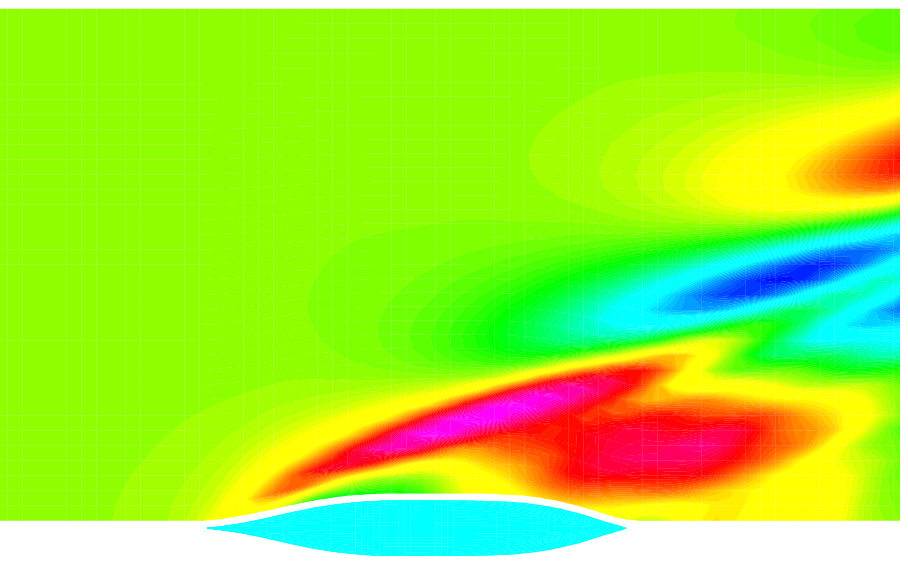}}
    \vskip -1.0cm
    \caption{Caso di una carena Serie $60$ $Cb=0.6$: elevazione d'onda per i casi
             di radiazione in heave (in alto) e in pitch (al centro) e di
             {\em scattering} (in basso) per $\omega=3$ e $Fr=.3$. 
             \label{ser_om3}
             }
\end{figure}
\begin{figure}[hp]
\vskip  2.5cm
 \vspace*{-2cm}
      \epsfxsize=0.75\textwidth
      \epsfysize=0.45\textwidth
      \makebox[\textwidth]{\epsfbox{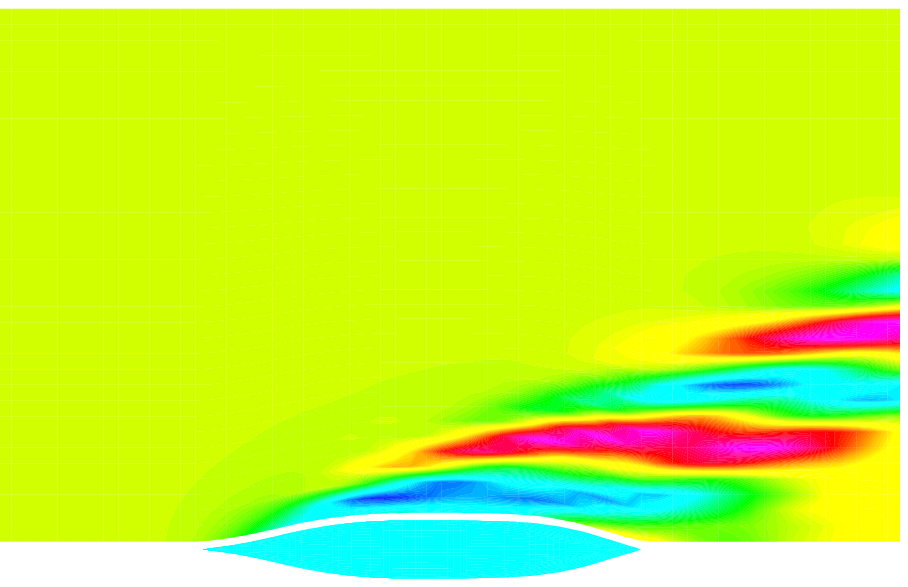}}
      \epsfxsize=0.75\textwidth
      \epsfysize=0.45\textwidth
      \makebox[\textwidth]{\epsfbox{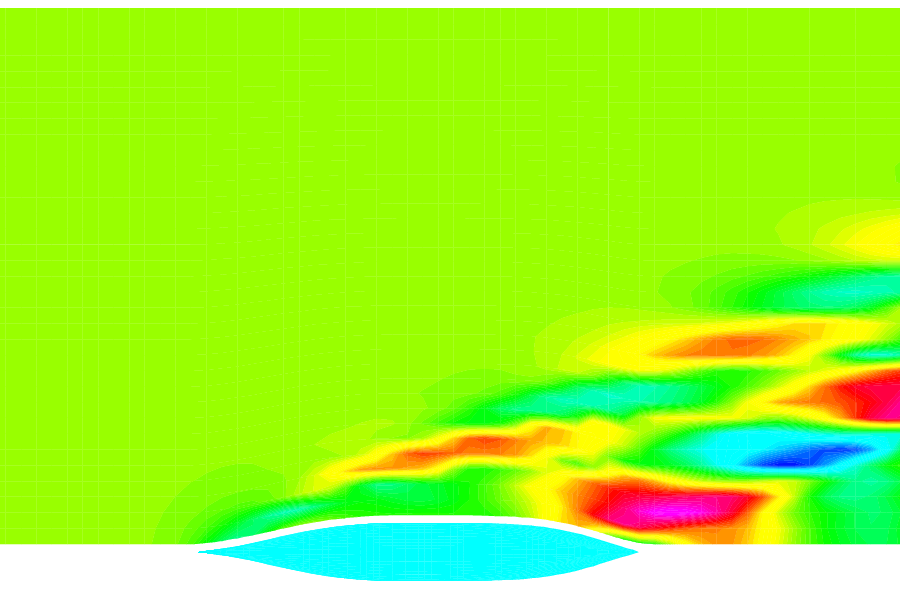}}
      \epsfxsize=0.75\textwidth
      \epsfysize=0.45\textwidth
      \makebox[\textwidth]{\epsfbox{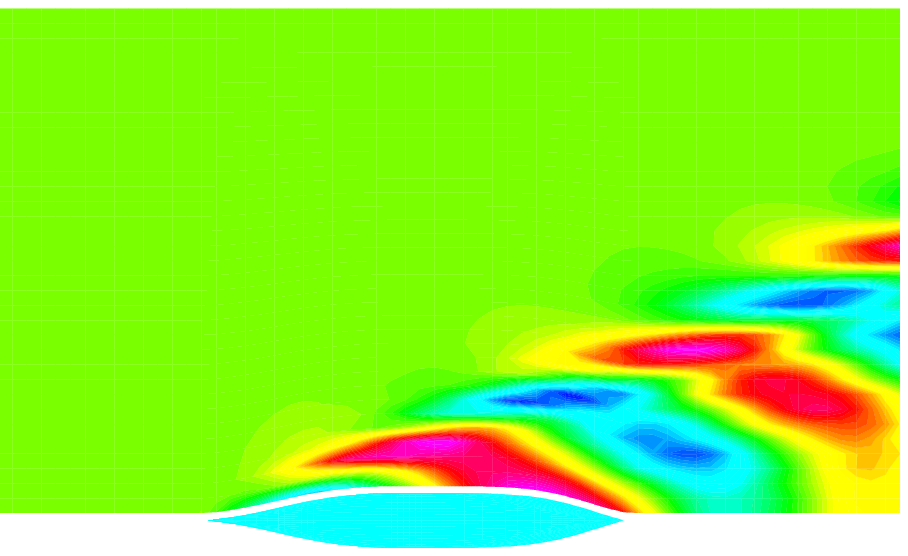}}
    \vskip -1.0cm
    \caption{Caso di una carena Serie $60$ $Cb=0.6$: elevazione d'onda per i casi
             di radiazione in heave (in alto) e in pitch (al centro) e di
             {\em scattering} (in basso) per $\omega=5$ e $Fr=.3$. 
             \label{ser_om5}
             }
\end{figure}
\begin{figure}[hp]
\vskip  2.5cm
 \vspace*{-2cm}
      \epsfxsize=0.75\textwidth
      \epsfysize=0.45\textwidth
      \makebox[\textwidth]{\epsfbox{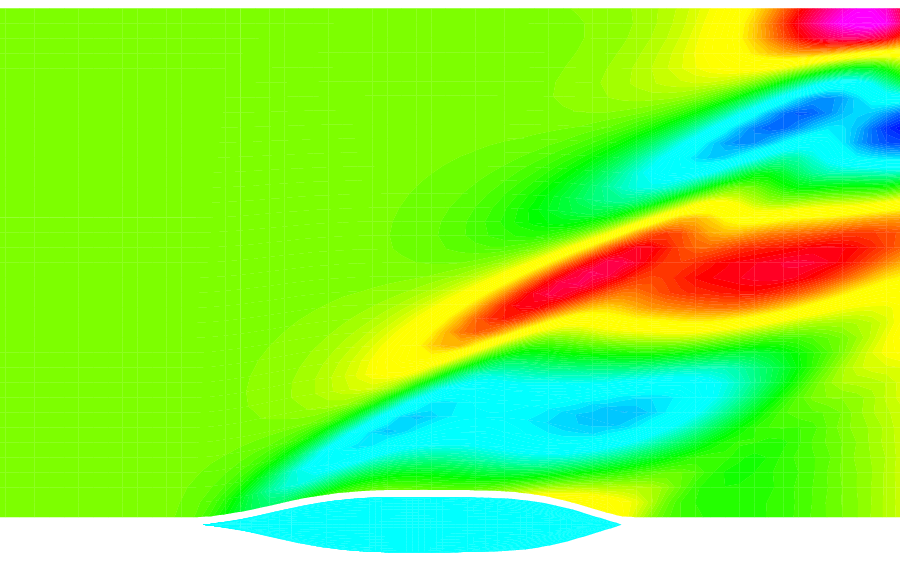}}
      \epsfxsize=0.75\textwidth
      \epsfysize=0.45\textwidth
      \makebox[\textwidth]{\epsfbox{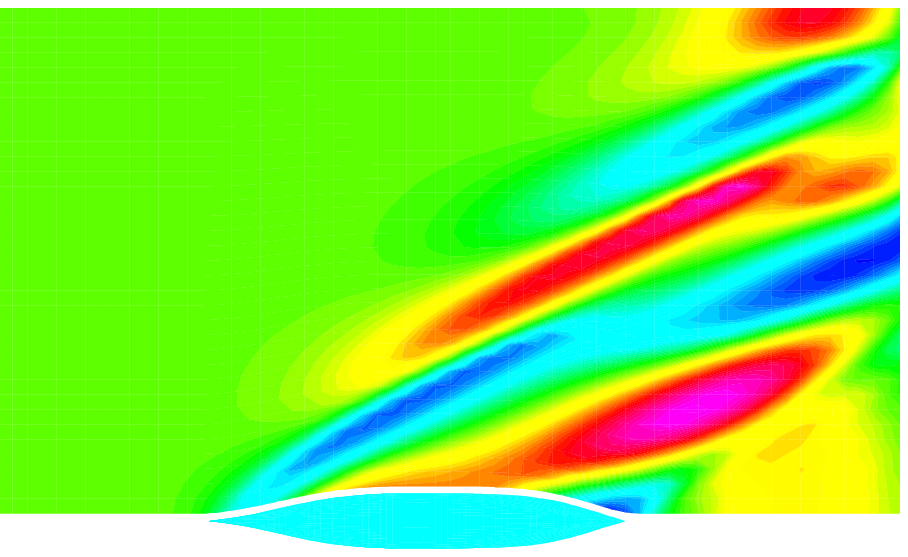}}
      \epsfxsize=0.75\textwidth
      \epsfysize=0.45\textwidth
      \makebox[\textwidth]{\epsfbox{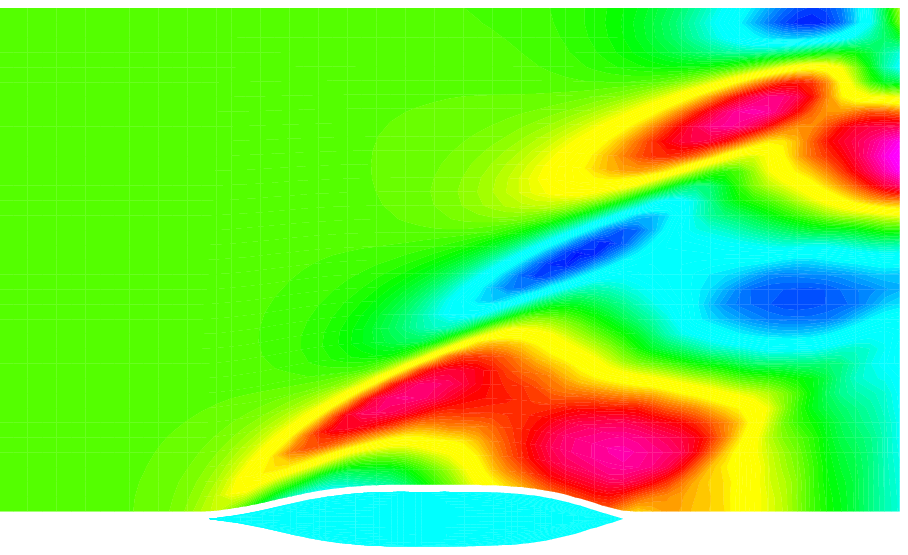}}
    \vskip -1.0cm
    \caption{Caso di una carena Serie $60$ $Cb = 0.6$: elevazione d'onda per i casi
             di radiazione in heave (in alto) e in pitch (al centro) e di
             {\em scattering} (in basso) per $\omega=3$ e $Fr=.2$. 
             \label{ser_fr2}
             }
\end{figure}

\chapter{Formulazione integrale del problema 
         nel dominio del tempo.}

Nel seguito si affronta la soluzione numerica nel dominio del
tempo del problema (linearizzato) definito nel quarto capitolo.
Prima di addentrarci in dettagli algoritmici e nella descrizione di 
alcuni risultati preliminari ottenuti, \e importante sottolineare
le finalit\a di questo differente percorso seguito.

Come si ricorder\a da quanto discusso nei precedenti capitoli,
\e teoricamente possibile, per esempio imponendo ingressi transitori
sulle variabili lagrangiane, arrivare al calcolo diretto del nucleo $\vec K(t)$
\footnote{Sottolineamo che le prove impulsive sono sulle variabili 
$\dot{\vec q}(t)$, grazie al passaggio fatto nel quarto capitolo dalla 
funzione $\vec H(s)$ alla funzione $\vec K(s)$} legato alla sollecitazione
idrodinamica in uscita tramite la 
\be
\dsty
\vec F_{idrod}(t)\,=-\left(\,\vec A\ddot{\vec q}\,+\,\vec B\dot{\vec q}\,+\,
                                              \vec C\vec q\,\right)-\,
			 \int_0^t \vec K(t\,-\,\tau)\dot{\vec q}(\tau) d\tau
\ee
Nota tale funzione che in teoria pu\o essere calcolata con una soltanto 
di queste prove, \e possibile trovare le funzioni 
$\vec A^*(\omega)$, $\vec B^*(\omega)$ attraverso le trasformate
seno e coseno  
\be 
\left \{
\begin{array}{l}
\dsty
\vec A^*(\omega):=\vec A\,-\frac{1}{\omega}\,
                  \int_0^{\infty} \vec K(t)\sin\omega t\,dt
\\ \\
\dsty
\vec B^*(\omega):=\vec B\,+\,\int_0^{\infty} \vec K(t)\cos\omega t\,dt
\end{array}
\right.
\ee 
Nel caso di soluzione nel dominio della frequenza queste funzioni vanno
calcolate per un dato insieme di valori di $\omega$ ed ognuna di queste 
pulsazioni richiede la soluzione {\em ab initio} del
problema integrale.
Da qui nasce il potenziale interesse per un modello nel dominio del
tempo coniugato con l'uso di ingressi transitori.

Per estrarre $\vec K(t)$ dalle forze idrodinamiche misurate occorre conoscere
preventivamente le matrici delle forze istantanee, $\vec A\,,\vec B\,,\vec C$. 
Una volta note le funzioni $\vec A^*(\omega)$ e $\vec B^*(\omega)$ \e possibile
scrivere la funzione di trasferimento complessiva $\vec G(s)$  
note le matrici che descrivono la dinamica libera della carena.

Ovviamente per prove {\em transitorie} o {\em impulsive}
non si devono intendere prove fondate ull'uso di impulsi di Dirac o comunque 
estremamente compatte nel tempo in quanto ci\o comporterebbe dei 
problemi numerici, così come nelle prove sperimentali reali non \e 
possibile sollecitare il modello con questi ingressi. 
Quindi sia in campo numerico che in campo sperimentale esiste
l'esigenza di trovare delle funzioni impulsive speciali che abbiano 
uno spettro sufficientemente ampio da contenere le frequenze di interesse, 
e che siano allo stesso tempo praticamente attuabili sul modello. 

Ricavata la funzione $\vec G(s)$ possiamo valutare il problema della risposta 
ad un sistema ondoso
\be
\dsty
 q_j(i\omega)\,=\,\sum_{k=1}^{6} G_{jk}(i\omega)\, A(\omega) X_{k}(i\omega)
\ee 
una volta risolto il problema della diffrazione e quindi ricavata 
la funzione $\vec X(i\omega)$.  
Per trovare queste funzioni, come abbiamo visto nei capitoli precedenti, 
occorre risolvere un certo insieme di problemi di diffrazione per diversi 
valori della  pulsazione, ad ognuno dei quali compete una certa ampiezza 
$A(\omega)$, a meno che, in analogia ai problemi radiativi,
si consideri l'interazione con sistemi ondosi compatti nello
spazio e che quindi abbiano un ampio contenuto in frequenza.
Anche in questo caso, con una sola prova nel dominio del tempo,
sarebbe possibile ricavare la funzione $\vec X(j\omega)$ \cite{Lugni} 
(oltre che per la tenuta al mare, prove analoghe si possono costruire per 
 la manovrabilit\a delle navi).

\section{Studio del problema lineare nel dominio del tempo}
Per semplicit\a ci concentreremo sulla pi\u semplice formulazione tipo
$Neumann-Kelvin$: i risultati, come visto, sono spesso soddisfacenti e
la principale attenzione potr\a essere rivolta agli
aspetti algoritmici tralasciando, per ora, le difficolt\a insite nei
termini aggiuntivi contenuti nella linearizzazione di doppio modello.

Come per la formulazione nel dominio della frequenza, anche in questo caso 
verr\a utilizzata una rappresentazione indiretta del potenziale 
in termini di distribuzioni di semplice strato sulla carena e sulla superficie 
libera.

Per quanto riguarda la discretizzazione temporale invece useremo due 
differenti approcci.
Nel primo le derivate temporali contenute nelle condizioni di
superficie libera verranno approssimate attraverso differenze 
finite. Nel secondo schema verr\a utilizzato un algoritmo tipo $Runge-Kutta$.
Una prima valutazione del 'comportamento'
degli algoritmi verr\a effettuata considerando il problema,
pi\u semplice rispetto a quello della tenuta al mare, 
della partenza da condizione di quiete fino al raggiungimento delle 
condizioni di moto rettilineo uniforme in mare calmo.
Sar\a in particolare possibile valutare la 
stabilit\a al variare della discretizzazione e dell'ampiezza del
dominio di calcolo, del passo temporale $\Delta t$, del numero di $Froude$
per citarne alcuni.
Una volta assicurata la stabilit\a
si potr\a valutare il tempo che occorre a raggiungere le condizioni
stazionarie e se esiste una particolare condizione
iniziale che consenta una riduzione del transitorio stesso,
al termine del quale si potr\a dare un giudizio di merito attraverso il 
confronto con dati sperimentali relativi alla resistenza d'onda in 
condizioni di regime.
In questo caso verr\a fatto anche un confronto con i risultati ottenuti 
mediante un codice stazionario
\footnote{Il codice per il calcolo
della resistenza d'onda WARP \cite{Campana}}. 

Rispetto all'approccio in frequenza, quello nel dominio del tempo presenta
diverse difficolt\a nella trattazione della condizione di radiazione.  
Da un punto di vista fisico ci si pu\o attendere che in un 
problema alle condizioni iniziali vengano generati sistemi d'onda che si 
propaghino in tutte le direzioni: anche a monte della carena.
Questo fenomeno, almeno per un certo transitorio
non \e in diretta relazione con la frequenza ridotta $\tau$ e
anche per valori supercritici $>0.25$ si potrebbero avere occasionali
disturbi a monte. D'altro canto, \e fisicamente intuitivo come
l'ampiezza dei sistemi d'onda generati debba attenuarsi al crescere della 
distanza. Per realizzare questa condizione 
nel caso di superficie libera di estensione finita si utilizza, 
in prossimit\a del confine del dominio discreto, una
condizione modificata di superficie libera che 'attenui' i segnali uscenti.
Pertanto, l'ampiezza della fascia in cui si applicano le condizioni di 
attenuazione ed il suo posizionamento rispetto alla carena e l'{\em intensit\a} 
dello smorzamento sono altri parametri del problema discreto che si aggiungono 
a quelli gi\a elencati in precedenza. 

La partenza da uno stato di quiete \footnote{e in generale tutti i problemi 
alle condizioni iniziali come abbiamo visto nel secondo capitolo.}
\e un problema impulsivo e questo significa che viene coinvolta 
un'ampia scala di frequenze. Per questo motivo nel dominio del tempo 
esiste un'altra difficolt\a legata alla discretizzazione della superficie 
libera. Se indichiamo con $h$ la lunghezza di un pannello di griglia di 
superficie libera, per il {\em teorema del campionamento di Nyquist} 
si ha che la massima lunghezza d'onda che pu\o essere risolta dalla 
griglia \e $2h$. Quindi l'energia delle onde che hanno una lunghezza d'onda 
fisica pi\u piccola di $2h$, che spesso \e significativa per superfici  
libere attraversate da navi, verr\a {\em distorta} dalla discretizzazione 
della superficie libera e quindi deve essere rimossa. Il rimedio a ci\o 
\e dato dall'utilizzo di un {\em filtro spaziale passa-basso}. 
Anche per questo abbiamo dei nuovi parametri per il problema numerico,
infatti il filtro deve essere tale da non smorzare troppo i disturbi
generati dalla carena. I parametri ottimi per un filtro vanno valutati
in funzione del numero di $Fr$ e dell'intervallo di frequenze di interesse.

Date le nuove complessit\`a, solo brevemente accennate, caratteristiche
di un approccio nel dominio del tempo il lavoro di questa tesi si concluder\a
 nella realizzazione di un algoritmo con i requisiti sopra elencati e
si rinvia a lavori successivi un approfondito studio delle 'prestazioni'
in termini di previsione di grandezze ingegneristicamente significative.

Pertanto, come ultima applicazione,
verranno considerate delle semplici prove di moto forzato con legge 
sinusoidale e confrontati i risultati ottenuti con 
quelli relativi al modello nel dominio della frequenza.
Al riguardo, abbiamo visto nel quarto capitolo che la trasformata 
di Fourier delle forze idrodinamiche \e 
\be
\dsty
 \vec F_{idrodyn}(j\omega)\,=\,\left[\omega^2\,\vec A^*(\omega))\,-\,
                               i\omega\,\vec B^*(\omega)\,-\,\vec C 
                               \right]\vec q(i\omega)
\ee
e
quindi trasformando le forze valutate 
nel dominio del tempo possiamo trovare i coefficienti di smorzamento
$\vec B^*$ dalla componente immaginaria, mentre da quella reale 
possiamo ricavare $\vec A^*$ (una volta note la quota
$\vec C$ delle forze idrodinamiche).
A questo punto \e possibile fare un confronto quantitativo con i 
risultati del codice in frequenza.

\section{Primo schema numerico: differenze finite}
Iniziamo a descrivere il primo schema numerico nel dominio del tempo.
A tal proposito \e utile definire i seguenti operatori lineari
\be \label {75}
\begin{array}{l}
\dsty
\frac{D\cdot}{Dt}\,=\,\pder{\cdot}{t}\,-\,U_{\infty}\pder{\cdot}{x} \qquad 
\rightarrow\qquad
\frac{D^2\cdot}{D t^2}\,=\,\frac{\partial^2\cdot}{\partial t^2}\,+\,
		   U_{\infty}^2\,\frac{\partial^2\cdot}{\partial x^2}\,-\,
                   2\,U_{\infty}\,\frac{\partial^2\cdot}{\partial x\,\partial t}
\end{array}
\ee
Nel caso della linearizzazione tipo {\em Neumann-Kelvin},
le condizioni al contorno per la superficie libera
\be
\left \{
\begin{array}{l}
\dsty
\frac{D\varphi}{D t}\,+\,g\eta\,=\,0 
\\ \\
\dsty
\frac{D\eta}{D t}\,-\varphi_z\,=\,0 
\end{array}
\right.
\ee
possono essere combinate giungendo alla condizione unificata sul potenziale
\be
\frac{D^2\varphi}{D t^2}\,+\,g\varphi_z\,=\,0 \qquad
\ee
Così come fatto nel quinto capitolo, attraverso la discretizzazione spaziale, 
possiamo scrivere il potenziale di velocit\a come:
\be 
\begin{array}{l}
\dsty
\varphi(i)\,=\,\int_F\sigma_F(j)\,G(i,j)\,dS\,+\, 
               \int_B\sigma_B(k)\,G(i,k)\,dS\,\doteq\,
           K_{0F}(i,j)\sigma_F(j)\,+\,K_{0B}(i,k)\sigma_B(k)\,
\end{array}
\ee
dove $i$ \e l'indice di un generico punto di collocazione mentre 
$j$ e $k$ indicano i pannelli sulla superficie libera o sulla carena 
rispettivamente.  Togliendo gli indici possiamo scrivere 
\be
K_{0FF}\sigma_F\,+\,K_{0FB}\sigma_B\,=\,\varphi_F    
\qquad \qquad 
K_{0BF}\sigma_F\,+\,K_{0BB}\sigma_B\,=\,\varphi_B    
\ee
e la condizione al contorno unificata per la superificie libera pu\o essere
espressa
\be \label{79}
\dsty
\frac{D^2}{D t^2}\left[\,K_{0F}\sigma_F\,+\,K_{0B}\sigma_B\,\right]\,+\, 
g\pder{}{z}\left[\,K_{0F}\sigma_F\,+\,K_{0B}\sigma_B\,\right]\,=\,0
\ee
mentre la condizione sul corpo nel caso del semplice avanzamento \e data da:
\be
\begin{array}{c}
\dsty
\pder{\varphi}{\vec n}\,=\,-\,U_{\infty}(t)\,\hat{i}\cdot\vec n
\\ \downarrow  \\ \dsty
(\vec K_{1F}\cdot\vec n)(i,j)\,\sigma_F(j)\,+\, 
(\vec K_{1B}\cdot\vec n)(i,k)\,\sigma_B(k)\,=\,
                                  -\,U_{\infty}(t)\,\hat{i}\cdot\vec n(i)
\\\\ \dsty
(\vec K_{1BF}\cdot\vec n_B)\,\sigma_F\,+\, 
(\vec K_{1BB}\cdot\vec n_B)\,\sigma_B\,=\,
                                  -\,U_{\infty}(t)\,\hat{i}\cdot\vec n_B
\end{array}
\ee
Tenendo conto della (\ref{75}), il primo membro della (\ref{79})
 pu\o essere scritto come:
\be
\dsty
\frac{D^2}{D t^2}(K_{0F}\sigma)\,=\,K_{0F}\ddot{\sigma}\,+
                                         \,U^2\,K_{0xxF}\,\sigma\,-\,
			                 \,2U\,K_{0xF}\,\dot{\sigma}\,\qquad
\ee
A questo punto approssimiamo le derivate nel tempo con 
degli operatori lineari $L_2\,,L_1$
\be
\left \{
\begin{array}{l}
\dsty
\frac{D^2}{D t^2}(K_{0F}\sigma)\,=\,K_{0F}\,L_2\sigma\,+\,U^2\,K_{0xxF}\sigma\,-\,
                                      \,2U\,K_{0xF}\,L_1\sigma\,\qquad
\\ \\
\dsty
L_2\,(\cdot)=\,\alpha_2(\cdot)_{i}\,+\,\beta_2(\cdot)_{i-1}\,+\,
               \gamma_2(\cdot)_{i-2}\,+\,........
\\ \\
\dsty
L_1\,(\cdot)=\,\alpha_1(\cdot)_{i}\,+\,\beta_1(\cdot)_{i-1}\,+\,
               \gamma_1(\cdot)_{i-2}\,+\,........
\end{array}
\right.
\ee 
dove il pedice $i-k$ ($k=0,1,...$) indica
l'istante $t_{i-k}\,=\,t_{i}\,-\,k\Delta t$.
Nella condizione al contorno per la superficie libera discretizzata portiamo
a primo membro le grandezze attuali e a secondo membro quelle {\em passate}
{\footnotesize 
\be
\begin{array}{lcl}
\left(K_{0F}\,\alpha_2\,\sigma\,+\,U^2\,K_{0xxF}\sigma\,-\,
      2U\,K_{0xF}\,\alpha_1\,\sigma\right)_F \,+\, 
\\ \\
\left(K_{0F}\,\alpha_2\,\sigma\,+\,U^2\,K_{0xxF}\sigma\,-\,
      2U\,K_{0xF}\,\alpha_1\,\sigma\right)_B \,+\, 
\\ \\
g\,K_{0zFF}\sigma_F\,+\,g\,K_{0zFB}\,\sigma_B
& = & 
\left[
           K_{0F}\,(\alpha_2-L_2)\,\sigma
      \,-\,2U\,K_{0xF}\,(\alpha_1-L_1)\,\sigma
\right]_F
\\ \\ 
& + &
\left[K_{0F}\,(\alpha_2-L_2)\,\sigma\,-\,
      2U\,K_{0xF}\,(\alpha_1-L_1)\,\sigma\right]_B
\end{array}
\ee
}
Questa relazione pu\o essere scritta in maniera compatta definendo delle 
matrici di influenza:
\be
\left \{
\begin{array}{l}
\dsty
 \vec A_{FF}\,:=\,\left(K_{0FF}\,\alpha_2\,+\,U^2\,K_{0xxFF}\,-\,
                       2U\,K_{0xFF}\,\alpha_1\,+\,g\,K_{0zFF}\,\right) 
\\ \\ \dsty
 \vec A_{FB}\,:=\,\left(K_{0FB}\,\alpha_2\,+\,U^2\,K_{0xxFB}\,-\,
                       2U\,K_{0xFB}\,\alpha_1\,+\,g\,K_{0zFB}\,\right) 
\\ \\ \dsty
 \vec B_F\,:=\,\left[K_{0F}\,(\alpha_2-L_2)\,\sigma\,-\,
                     2U\,K_{0xF}\,(\alpha_1-L_1)\,\sigma\right]_F \,+\,
               \left[K_{0F}\,(\alpha_2-L_2)\,\sigma\,-\,
                     2U\,K_{0xF}\,(\alpha_1-L_1)\,\sigma\right]_B
\\ \\ \dsty
 \vec A_{FF}\sigma_F\,+\,\vec A_{FB}\sigma_B\,=\,\vec B_F 
 \qquad \mbox{B.C. discretizzata per la superficie libera}
\\ \\ \\ \dsty
 \vec A_{BB}\,=\,\vec K_{1BB}\cdot\vec n_B 
\qquad \qquad
 \vec A_{BF}\,=\,\vec K_{1BF}\cdot\vec n_B 
\\ \\ \dsty
 \vec B_B\,=\,-\,U_{\infty}(t)\,\hat{i}\cdot\,\vec n_B
\\ \\ \dsty
 \vec A_{BF}\sigma_F\,+\,\vec A_{BB}\sigma_B\,=\,\vec B_B
\qquad \mbox{B.C. discretizzata per la superficie del corpo}
\end{array}
\right.
\ee
E' da notare che la matrice di influenza $\vec A$ dipende dalla velocit\a
di avanzamento. Questo \e svantaggioso qualora si volesse far variare la 
velocit\a di avanzamento  poich\'e occorrerebbe ricalcolare e invertire la 
matrice di influenza ad ogni passo temporale. 
Vedremo che utilizzando l'algoritmo Runge-Kutta non avremo tale 
inconveniente. 

Con lo schema ora introdotto \e possibile eseguire una prova con partenza 
impulsiva ponendo nelle condizioni iniziali,
contenute in $\vec B_F$, lo stato di quiete ossia $\sigma\,=\,0$ 
mentre in $\vec B_B$ si imporr\a la $U_{\infty}(t)\,=\,U$,
e ad ogni passo temporale si risolver\a il sistema algebrico scritto
con le $\sigma$ relative al tempo $t$ come incognite. 

Nel caso di partenza impulsiva si ottiene un transitorio molto lungo e quindi 
un notevole tempo di calcolo per raggiungere le condizioni stazionarie. 
Per ridurlo sono state utilizzate come condizioni iniziali
le $\sigma$ che competono al flusso di doppio modello con una corrente di 
intensit\a pari ad $U$.
Il risultato \e soddisfacente, ottenendo una 
notevole riduzione del transitorio per le forze idrodinamiche, 
come si pu\o vedere nei grafici riportati pi\u oltre.

Una volta note le $\sigma$ ad un dato istante \e possibile calcolare 
l'elevazione d'onda dalla condizione dinamica per la superficie libera 
e le forze idrodinamiche dalle pressioni attraverso l'equazione di
Bernoulli 
\be
\begin{array}{cc}
\mbox{{\em Elevazione d'onda:}} \qquad \qquad \qquad 
& \mbox{{\em Equazione di Bernoulli:}} 
\\\\ \dsty
\dot{\varphi}\,-\,U\varphi_x\,+\,g\eta\,=\,0 \qquad \qquad \qquad 
& p\,=\,-\rho\,\left(\,\dot{\varphi}\,-\,U_{\infty}\varphi_x\,\right)
\end{array}
\ee
ossia
\be
\left \{
\begin{array}{l}
\dsty
\eta(i)\,=\,-\,\frac{1}{g}\{\,K_{0F}(i,j)\,L_1\sigma_F(j)\,+\,
                                   K_{0B}(i,k)\,L_1\sigma_B(k)\,-\, 
\\ \\ \qquad \qquad \qquad     
                                   U\,(K_{0xF}(i,j)\,\sigma_F(j)\,+\,
                                       K_{0xB}(i,k)\,\sigma_B(k))\,\}
\\\\ \dsty
\mbox{i,j indici dei pannelli della superficie libera;  k indice dei pannelli  
della carena }
\\\\ \dsty
\eta_F\,=\,-\,\frac{1}{g}\{\,K_{0FF}\,L_1\sigma_F\,+\,
                                   K_{0FB}\,L_1\sigma_B\,-\, 
                                   U\,(K_{0xFF}\,\sigma_F\,+\,
                                       K_{0xFB}\,\sigma_B\,)\,\}
\end{array}
\right.
\ee
\be
\left \{
\begin{array}{l}
\dsty
p(i)\,=\,-\rho\,\{\,K_{0F}(i,j)\,L_1\sigma_F(j)\,+\,
                                   K_{0B}(i,k)\,L_1\sigma_B(k)\,-\,
\\ \\ \qquad \qquad \qquad     
                                   U\,(K_{0xF}(i,j)\,\sigma_F(j)\,+\,
                                       K_{0xB}(i,k)\,\sigma_B(k))\,\}
\\\\ \dsty
\mbox{i,k indice dei pannelli della carena;  j indici dei pannelli di 
superficie libera} 
\\\\ \dsty
p_B\,=\,-\rho\,\{\,K_{0BF}\,L_1\sigma_F\,+\,
                                   K_{0BB}\,L_1\sigma_B\,-\,
                                   U\,(K_{0xBF}\,\sigma_F\,+\,
                                       K_{0xBB}\,\sigma_B)\,\}
\end{array}
\right.
\ee  \\
Notiamo che sia per il calcolo del termine noto che per quello dell'elevazione
d'onda occorre memorizzare le matrici di influenza $K_{0FF}\,e\,K_{0FB}$ 
mentre per il calcolo della pressione occorre memorizzare $K_{0BF}\,e\,K_{0BB}$
e quindi insieme alla matrice di influenza $\vec A$ occorre memorizzare 
l'intera matrice $K_0(i,j)$ con $i,j$ indici che variano sia sul corpo che 
sulla superficie libera. 

Fin qui l'algoritmo illustrato non tiene conto della 
condizione di radiazione che occorre imporre per evitare riflessioni
non fisiche al contorno del dominio.
Per far ci\o una certa porzione 
della griglia di superficie libera viene 'dedicata'
allo smorzamento delle onde uscenti. 
In questa zona le condizioni di superficie libera vengono alterate
introducendo una certa viscosit\a artificiale per l'attenuazione 
delle onde. 
La spiaggia numerica deve essere fatta in modo da non generare 
onde riflesse sulla superficie libera che altrimenti altererebbero tutto 
il campo e quindi anche le forze sulla carena.
Questo smorzamento pu\o essere dato in diversi modi, sulla condizione al
contorno cinematica, su quella dinamica o su entrambe.
In una prima modalit\a
si introduce uno smorzamento nella condizione al contorno dinamica
ottenendo
\be
\left \{
\begin{array}{l}
\dsty
\dot{\varphi}\,-\,U\,\varphi_x\,+\,g\eta\,+\nu(\vec x)\,\varphi_z\,=\,0
\\ \qquad \qquad  + \\ \dsty
\dot{\eta}\,-\,U\,\eta_x\,-\,\varphi_z\,=\,0
\\ \qquad \qquad  = \\ \dsty
\frac{D^2\varphi}{D t^2}\,+\,g\,\varphi_z\,=\,-\nu\,\dot{\varphi_z}\,+\,
                             U\,\nu\,\varphi_{xz}\,+\,U\,\nu_x\,\varphi_z\,=\,
			     \,-\,\frac{D (\nu\varphi_z)}{D t}
\end{array}
\right.
\ee 
Dove il coefficiente di smorzamento $\nu(\vec x)$ \e zero sulla superficie 
libera e diventa diverso da zero quando si entra nella zona di griglia 
dedicata alla spiaggia.
In particolare il suo valore crescer\a monotonicamente man mano che si 
considerano punti sempre pi\u lontani dalla carena.
In una seconda modalit\a si considera uno smorzamento sulla condizione 
cinematica
\be
\left \{
\begin{array}{l}
\dsty
\dot{\varphi}\,-\,U\,\varphi_x\,+\,g\eta\,=\,0
\\ \qquad \qquad  + \\ \dsty
\dot{\eta}\,-\,U\,\eta_x\,-\,\varphi_z\,=\,-2\,\nu\,\eta\,+\,\frac{\nu^2}{g}\varphi
\\ \qquad \qquad  = \\ \dsty
\frac{D^2\varphi}{D t^2}\,+\,g\,\varphi_z\,=\,-2\,\nu\,\frac{D\varphi}{D t}\,
					      -\,\nu^2\varphi
\end{array}
\right.
\ee 
Entrambe queste condizioni modificate cambiano la matrice di influenza e 
i termini noti, ad esempio nel secondo caso si ha
\be
\left \{
\begin{array}{l}
\dsty
 \vec A_{FF}\,:=\,K_{0FF}\,\alpha_2\,+\,U^2\,K_{0xxFF}\,-\,
                       2U\,K_{0xFF}\,\alpha_1\,+\,g\,K_{0zFF}\,+\, 
\\[0.5cm]
\qquad \qquad 
                       2\nu\,K_{0FF}\,\alpha_1\,-\,2\nu\,U\,K_{0xFF}\,+
                       \nu^2\,K_{0FF}
\\\\ \dsty
 \vec A_{FB}\,:=\,K_{0FB}\,\alpha_2\,+\,U^2\,K_{0xxFB}\,-\,
                       2U\,K_{0xFB}\,\alpha_1\,+\,g\,K_{0zFB}\,+\, 
\\[0.5cm]
\qquad \qquad 
                       2\nu\,K_{0FB}\,\alpha_1\,-\,2\nu\,U\,K_{0xFB}\,+
                       \nu^2\,K_{0FB} 
\\ \\ \dsty
 \vec B_F\,:=\,(-\,2U\,K_{0xFF}\,+\,2\nu\,K_{0FF})(\alpha_1-L_1)\,\sigma_F \,+\,
                K_{0FF}(\alpha_2\,-\,L_2)\,\sigma_F\,+\,  \\\\       
\qquad \qquad  (-\,2U\,K_{0xFB}\,+\,2\nu\,K_{0FB})(\alpha_1-L_1)\,\sigma_B \,+\,
                K_{0FB}(\alpha_2\,-\,L_2)\,\sigma_B         
\end{array}
\right.
\ee
\section{Secondo schema numerico: Runge Kutta} 
La prima differenza di questo secondo schema \e che la matrice 
di influenza per il calcolo delle $\sigma$ non viene costruita con la 
condizione di superficie libera unificata. In questo caso il calcolo delle 
distribuzioni di semplice strato, nell'istante attuale, sulla carena e sulla 
superficie libera viene fatto imponendo la sola impermeabilit\a sul corpo,
e la conoscenza del potenziale di velocit\a sulla superficie libera.
Riprendiamo le formule per la discretizzazione del potenziale di velocit\a
sulla superficie libera e sul corpo.
\be 
K_{0FF}\sigma_F\,+\,K_{0FB}\sigma_B\,=\,\varphi_F    
\qquad \qquad 
K_{0BF}\sigma_F\,+\,K_{0BB}\sigma_B\,=\,\varphi_B    
\ee
E quindi possiamo scrivere il sistema algebrico:
\be \label{sistema}
 \left \{
 \begin{array}{l}
 \dsty
 \vec A_{FF}\,:=\,K_{0FF} 
 \qquad \qquad
 \vec A_{FB}\,:=\,K_{0FB}
 \\[0.5cm] \dsty
 \vec A_{BB}\,:=\,\vec K_{1BF}\cdot\vec n_B  
 \qquad 
 \vec A_{BF}\,:=\,\vec K_{1BB}\cdot\vec n_B 
 \\\\ \dsty
 \vec B_F\,:=\,\varphi_F
 \qquad \qquad
 \vec B_B\,=\,-\,U(t)\,\hat{i}\cdot\,\vec n_B 
\end{array}
\right.
\left \{
\begin{array}{l}
 \vec A_{FF}\sigma_F\,+\,\vec A_{FB}\sigma_B\,=\,\vec B_F 
 \\ \\ \dsty
 \vec A_{BF}\sigma_F\,+\,\vec A_{BB}\sigma_B\,=\,\vec B_B
\end{array}
\right.
\ee
Come si pu\o vedere quasta matrice di influenza non dipende dalla 
velocit\a di avanzamento $\vec U(t)\,=\,U(t)\,\hat{i}$.\\
Per questo schema occorre inoltre {\em memorizzare} altre tre matrici, e questo 
\e uno svantaggio rispetto allo schema precedente il quale  richiede 
una quantit\a di memoria pi\u bassa:
\be  \label{phix}
\left \{
 \begin{array}{l}
 \dsty
 \varphi_x(i)\,=\,\int_F\sigma_F(j)\,\pder{G}{x}(i,j)\,dS\,+\, 
                        \int_B\sigma_B(k)\,\pder{G}{x}(i,k)\,dS\,\doteq\,
                         K_{0xF}(i,j)\sigma_F(j)\,+\,K_{0xB}(i,k)\sigma_B(k)
\\\\ \dsty
K_{0xFF}\sigma_F\,+\,K_{0xFB}\sigma_B\,=\,\varphi_{xF}
\\\\ \dsty
K_{0xBF}\sigma_F\,+\,K_{0xBB}\sigma_B\,=\,\varphi_{xB}
\end{array}
\right.
\ee
\be \label{phiz}
\left \{
 \begin{array}{l}
 \dsty
\varphi_z(i)\,=\,\int_F\sigma_F(j)\,\pder{G}{z}(i,j)\,dS\,+\, 
                        \int_B\sigma_B(k)\,\pder{G}{z}(i,k)\,dS\,\doteq\,
                         K_{0zF}(i,j)\sigma_F(j)\,+\,K_{0zB}(i,k)\sigma_B(k)
\\\\ \dsty
K_{0zFF}\sigma_F\,+\,K_{0zFB}\sigma_B\,=\,\varphi_{zF}
\\\\ \dsty
K_{0zBF}\sigma_F\,+\,K_{0zBB}\sigma_B\,=\,\varphi_{zB}
\end{array}
\right.
\ee
e per ultima la matrice $[\,K_{0BF}\,\,K_{0BB}\,]$ che servir\a per il calcolo delle pressioni
sul corpo come per il precedente schema:
\be 
\left \{
 \begin{array}{l}
 \dsty
\varphi(i)\,=\,\int_F\sigma_F(j)\,G(i,j)\,dS\,+\, 
               \int_B\sigma_B(k)\,G(i,k)\,dS\,\doteq\,
              K_{0F}(i,j)\sigma_F(j)\,+\,K_{0B}(i,k)\sigma_B(k)
\\\\ \dsty
K_{0BF}\sigma_F\,+\,K_{0BB}\sigma_B\,=\,\varphi_B
\end{array}
\right.
\ee
L'utilizzo del metodo di Runge-Kutta porta a suddividere ogni passo temporale 
in pi\u sotto-passi; maggiore \e il numero di questi e maggiore sar\a 
l'ordine di accuratezza nel tempo a scapito per\o dell'onere
computazionale. 
Nel metodo precedente invece l'accuratezza \e nella scelta degli operatori 
$L1$ e $L2$ e quindi una maggiore accuratezza modifica la matrice di influenza 
ma il tempo di calcolo non ne risente.
Per la condizione al contorno sulla superficie libera non utilizzeremo quella 
unificata, in quanto per l'algoritmo di Runge-Kutta, occorre avere un 
sistema di equazioni differenziali al primo ordine, per questo 
considereremo le due condizioni al contorno cinematica e dinamica 
nelle variabili $\varphi$ e $\eta$:
\be
 \left \{
 \begin{array}{l}
 \dsty
 \dot{\eta}\,=\,f(t\,,\eta\,,\varphi)\,=\,\varphi_z\,+\,U\,\eta_x\,
                         				 -2\nu\,\eta\,+\,\frac{\nu^2}{g}\varphi 
 \\\\ \dsty
 \dot{\varphi}\,=\,g(t\,,\eta\,,\varphi)\,=\,-\,g\eta\,+\,U\,\varphi_x
\end{array}
 \right.
\ee
Dove la condizione cinematica \e stata modificata per introdurre la spiaggia numerica,
così come \e stato fatto nel precedente schema numerico. \\
Per la descrizione utilizziamo un metodo di Runge-Kutta al secondo ordine:
\be
 \left \{
 \begin{array}{l}
 \dsty
 \dot{y}\,=\,F(t,y) \\\\ \dsty
 y_{i+1}\,=\,y_i\,+\,\Delta t\,F\,\left\{\,t_i\,+\oneh\,\Delta t\,,y_i\,+\,
             			               \oneh\Delta t\,F(t_i\,,y_i)\,\right\}
\end{array}
\right.
\ee
Nel sottointervallo $t_k\,=\,(t_i\,+rk(l)\,\Delta t)$ si calcolano le grandezze:
\be 
 \left \{
 \begin{array}{l}
 \dsty
 \eta^k\,=\,\eta^i\,+\,rk(k)\,\Delta t\,\eta_t^i \\\\ \dsty
 \varphi^k\,=\,\varphi^i\,+\,rk(k)\,\Delta t\,\varphi_t^i 
 \end{array}
\right.
\ee
Nel nostro caso con uno schema al secondo ordine abbiamo $rk(1)\,=\,0\,\,rk(2)\,=\,1$.
Sempre in questo sotto-intervallo si calcola la velocit\a $U(t^k)$ e l'accelerazione
$U_t(t^k)$ si costruisce quindi il termine noto:
\be
\left [\,B_F\,,\, B_B\,\right]\,=\,
\left[\,\varphi_F\,,\,  -\,U(t)\,\hat{i}\cdot\vec n_B\,\right]^k
\ee
e si risolve il sistema \ref{sistema} nelle incognite $\sigma(t_k)$ a questo punto 
si utilizzano le due condizioni al contorno per la superficie libera:
\be \label{bcslib}
\left \{
\begin{array}{l}
\dsty
\dot{\varphi}^k\,=\,\left(\,U\,\varphi_x\,-\,g\,\eta\,\right)^k
\\\\ \dsty
 \dot{\eta}^k\,=\,\left(\,U\eta_x\,+\,\varphi_z\,
                                 -2\nu\,\eta\,+\,\frac{\nu^2}{g}\varphi \,\right)^k
 \end{array}
\right.
\ee
Qui le grandezze $\varphi_z^k\,,\varphi_x^k$ possono essere calcolate 
moltiplicando le matrici di influenza \ref{phix}, \ref{phiz} con le $\sigma_F^k$.
Ancora una volta va sottolineato il vantaggio della formulazione indiretta e il 
conseguente utilizzo delle formule di {\em Hess-Smith} per il calcolo analitico
delle derivate spaziali del potenziale di velocit\a. \\
Il calcolo della $\dot{\varphi}$ pu\o essere quindi svolto, mentre per 
la condizione cinematica rimane da valutare la derivata $\eta_x$,
che pu\o essere approssimata attraverso una derivata numerica fatta sui 
nodi della griglia di superficie libera. Questa griglia risulta molto deformata
lungo la direzione trasversale in prossimit\a della carena, e poco 
deformata lungo la direzione longitudinale, e ci\o comporta l'introduzione 
di errori di troncamento, attraverso il giacobiano, nel calcolo del $\nabla\eta$  
che rendono instabile lo schema numerico. Per questo abbiamo dovuto
utilizzare un'altra strada ottenuta calcolando la  $\eta_x$ attraverso la 
derivata rispetto a $x$ della condizione dinamica: 
\be
\begin{array}{lc}
\dsty
\eta_x\,=\,-\,{\frac{D\varphi_x}{Dt}}\,/\,{g}
\end{array}
\ee
Possiamo interpretare questa nuova relazione osservando come la curvatura 
$\eta_x$ della superficie libera sia legata al rapporto di due accelerazioni. \\
La condizione cinematica diventa pertanto:
\be
\dsty
\dot{\eta}\,=\,\frac{U}{g}\left(\,\varphi_{tx}\,-\,U\,\varphi_{xx}\right)\,+\,\varphi_z
		  \,-\,2\nu\,\eta\,+\,\frac{\nu^2}{g}\varphi 
\ee
Il problema si complica per il calcolo delle grandezze $\varphi_{tx}\,,\varphi_{xx}$.
Iniziamo dalla prima, che possiamo scrivere come:
\be
\varphi_{txF}\,\doteq\,\pder{}{t} \left[\,K_{0xFF}\,\sigma_F\,+\,K_{0xFB}\,\sigma_B\,\right]
                  \,=\,\left[\,K_{0xFF}\,\dot{\sigma}_F\,+\,K_{0xFB}\,\dot{\sigma}_B\,\right]
\ee
Il calcolo delle $\dot{\sigma}$ pu\o essere fatto derivando il sistema \ref{sistema}
dove in particolare la conoscenza delle $\dot{\varphi}$ nel termine noto 
\e data dalla condizione dinamica \ref{bcslib} che \e gi\a stata risolta:
\be \label{dotsistema}
\dsty
 \vec A\, \dot{\sigma}\,=\,\left[\, \dot{\varphi}_F\,,\,-\,U_t\,\hat{i}\cdot\vec n_B\,\right]
\ee
Per la $\varphi_{xxF}$ occorre invece calcolare un'altra matrice di influenza e
ci\o comporta un ulteriore aggravio sulla memoria necessaria:
\be  \label{phixx}
\left \{
 \begin{array}{l}
 \dsty
 \varphi_{xx}(i)\,=\,\int_F\sigma_F(j)\,\pder{^2G}{x^2}(i,j)\,dS\,+\, 
                        \int_B\sigma_B(k)\,\pder{^2G}{x^2}(i,k)\,dS\,
\\\\ \dsty
 \qquad \qquad  \doteq\,K_{0xxF}(i,j)\sigma_F(j)\,+\,K_{0xxB}(i,k)\sigma_B(k)
\\\\ \dsty
K_{0xxFF}\sigma_F\,+\,K_{0xxFB}\sigma_B\,=\,\varphi_{xxF}
\end{array}
\right.
\ee
Il calcolo di questa componente del gradiente di velocit\a non presenta i 
problemi visti nel caso dei {\em Termini $m_j$} perch\e siamo sulla superficie libera 
indeformata e quindi anche con pannelli di ordine 0 le formule di {\em Hess-Smith}
forniscono questa grandezza con una accuratezza sufficientemente buona. \\
Riassumiamo il calcolo delle derivate $\dot{\varphi}\,,\dot{\eta}$:
\be
 \left \{
 \begin{array}{l}
 \dot{\varphi_F}\,=\,-\,g\,\eta\,+
                             \,U\,\left(\,K_{0xFF}\sigma_F\,+\,K_{0xFB}\sigma_B\,\right)
 \\\\ \dsty
 \dot{\eta_F}\,=\,\left(\,K_{0zFF}\sigma_F\,+\,K_{0zFB}\sigma_B\,\right)\,+\,
 \\\\ \dsty
 \qquad \,                      \frac{U}{g}\,\left\{
                   \left[\,K_{0xFF}\dot{\sigma}_F\,+\,K_{0xFB}\dot{\sigma}_B\,\right]
                   \,-\,U\left[\,K_{0xxFF}\sigma_F\,+\,K_{0xxFB}\sigma_B\,\right]
                        \,\right\}\,
\\\\ \dsty
\qquad \,        -\,2\nu\eta\,+\,\frac{\nu^2}{g}
                       \left[\,K_{0FF}\sigma_F\,+\,K_{0FB}\sigma_B\,\right]
\end{array}
\right.
\ee
\\
 Calcolate le grandezze $\dot{\varphi}\,,\dot{\eta}$ nel sottoistante $t^k$ possiamo 
 calcolare l'incremento delle grandezze $\varphi\,,\eta$:
 \be 
 \left \{
 \begin{array}{l}
 \dsty
 d\eta^{k+1}\,=\,d\eta^k\,+\,rk_2(k)\,\Delta t\,\dot{\eta}^k \\\\ \dsty
 d\varphi^{k+1}\,=\,d\varphi^k\,+\,rk_2(k)\,\Delta t\,\dot{\varphi}^k 
 \end{array}
\right.
\ee
Nel nostro caso per uno schema al secondo ordine abbiamo 
$rk_2(1)\,=\,0.5\,\,rk_2(2)\,=\,0.5$.
Terminati i sottopassi temporali possiamo sommare alle $\eta^i\,,\,\varphi^i$ 
gli incrementi $d\eta\,e\,d\varphi$ per conoscere i valori del'elevazione  
d'onda e del potenziale nel nuovo istante temporale $i+1$. \\ \\  
\indent
Come ultimo rimane il calcolo delle forze idrodinamiche nell'istante $t^{i+1}$. \\
\be
\left \{
\begin{array}{l}
\mbox{{\em Equazione di Bernoulli:}} 
\\\\ \dsty
p\,=\,-\rho\,\left(\,\dot{\varphi}\,-\,U_{\infty}\varphi_x\,\right)
\\\\ \dsty
p\,=\,-\rho\,\left\{
                   \left[\,K_{0BF}\dot{\sigma}_F\,+\,K_{0BB}\dot{\sigma}_B\,\right]
           \,-\,U\left[\,K_{0xBF}\sigma_F\,+\,K_{0xBB}\sigma_B\,\right]\,
                   \right\}
\end{array}
\right.
\ee
Come si vede  anche qui occorre calcolare le $\dot{\sigma}$ attraverso il sistema
\ref{dotsistema}, questo calcolo quindi prescinde dalla tecnica utilizzata per 
il calcolo della grandezza $\eta_x$, rimane per\o da sottolineare che per 
quest'ultima occorre risolvere il sistema per ogni sottopasso temporale, mentre
le forze possono essere calcolate in ogni passo temporale.\\
Possiamo riassumere dicendo che, rispetto al precedente schema, 
si hanno gli svantaggi di una quantit\a di memoria pi\u alta 
e di avere, a parit\a di accuratezza, un tempo di calcolo pi\u grande.
A quest'ultimo punto si deve aggiungere che per questo schema al crescere del 
numero di $Froude$ occorre utilizzare dei $\Delta t$ pi\u piccoli rispetto
all'altro schema per non avere problemi di instabilit\a. Pertanto il precedente 
schema presenta una {\em migliore stabilit\a}. Ci\o non toglie che lo
schema che utilizza {\em Runge-Kutta} porta a dei risultati molto buoni come
vedremo nel prossimo paragrafo. 
\newpage
\section{Alcuni risultati nel dominio del Tempo} 
\begin{figure}[hp]
\vskip  2.5cm
 \vspace*{-2cm}
      \epsfxsize=0.75\textwidth
      \epsfysize=0.45\textwidth
      \makebox[\textwidth]{\epsfbox{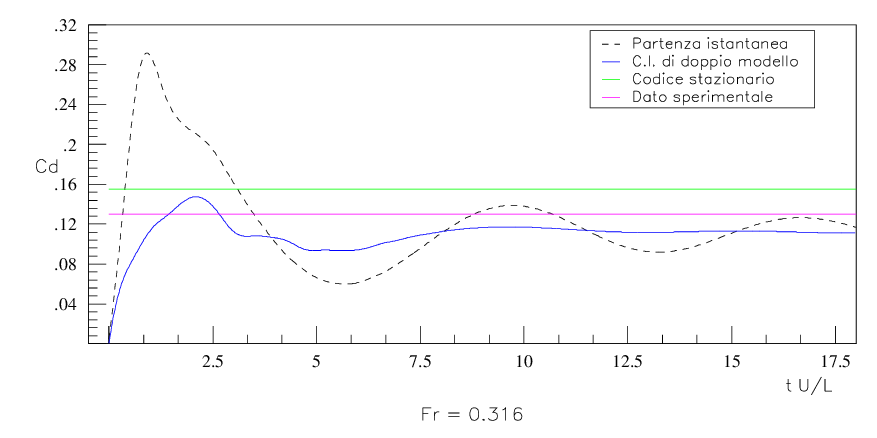}}
      \epsfxsize=0.75\textwidth
      \epsfysize=0.45\textwidth
      \makebox[\textwidth]{\epsfbox{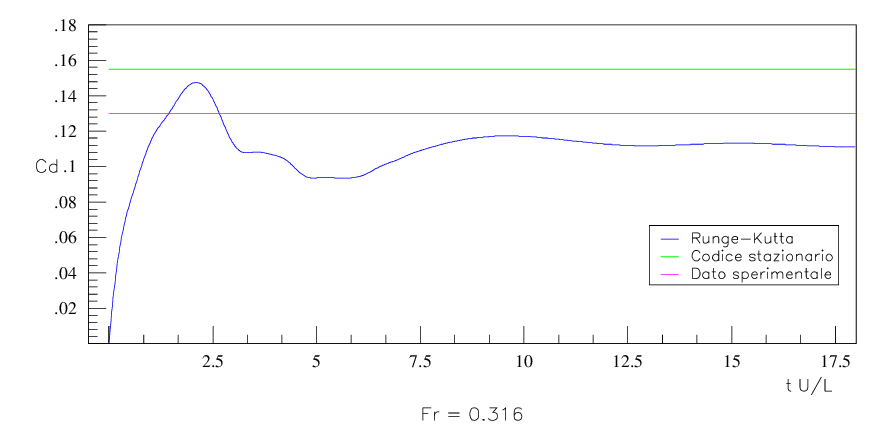}}
      \epsfxsize=0.75\textwidth
      \epsfysize=0.45\textwidth
      \makebox[\textwidth]{\epsfbox{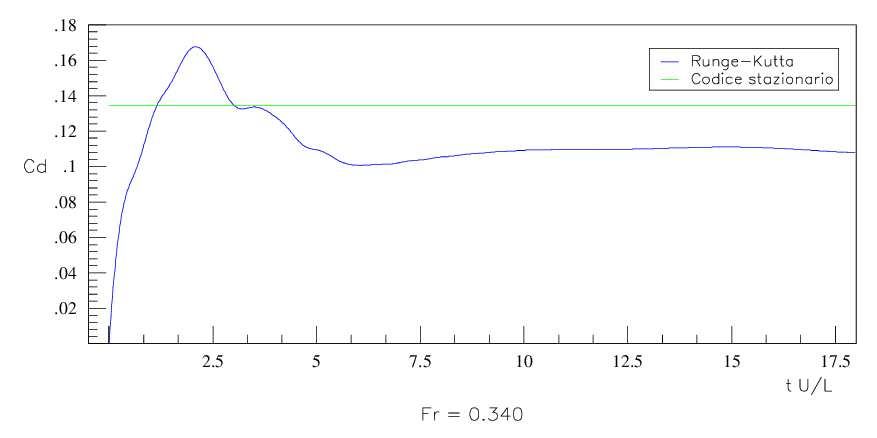}}
    \vskip -0.5cm
    \caption{ Risultati nel dominio del tempo per il semplice avanzamento
             } \label{staz1}
\end{figure}

\begin{figure}[hp]
\vskip  2.5cm
 \vspace*{-2cm}
      \epsfxsize=0.75\textwidth
      \epsfysize=0.45\textwidth
      \makebox[\textwidth]{\epsfbox{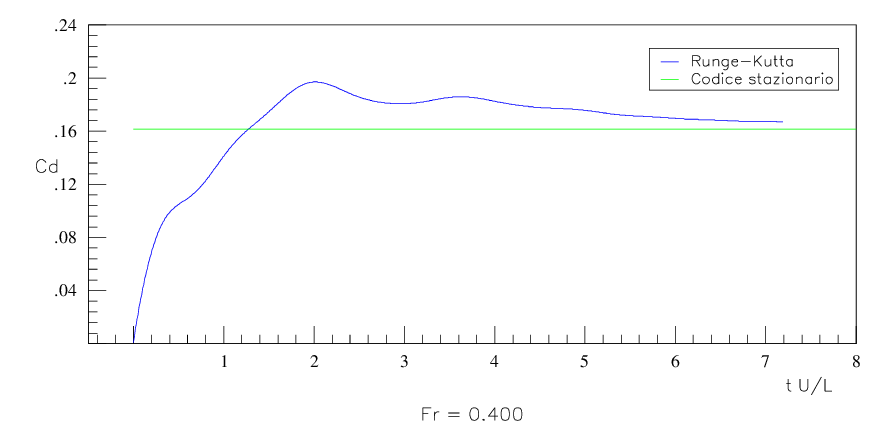}}
      \epsfxsize=0.75\textwidth
      \epsfysize=0.45\textwidth
      \makebox[\textwidth]{\epsfbox{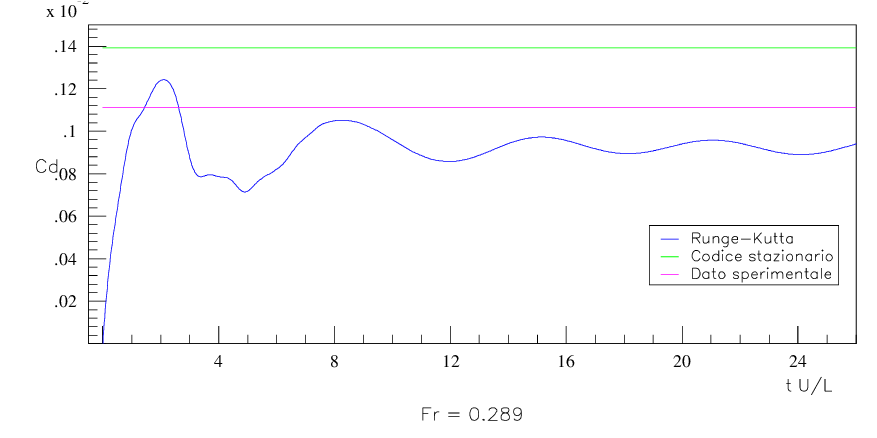}}
      \epsfxsize=0.75\textwidth
      \epsfysize=0.45\textwidth
      \makebox[\textwidth]{\epsfbox{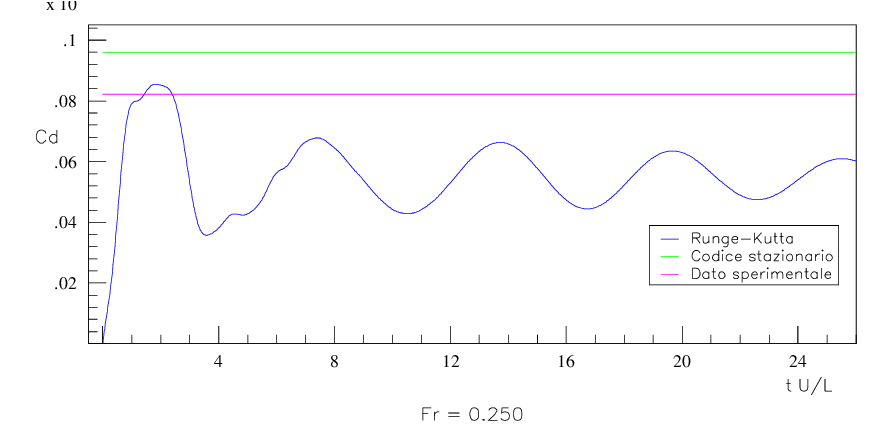}}
    \vskip -0.5cm
    \caption{ Risultati nel dominio del tempo per il semplice avanzamento
             } \label{staz2}
\end{figure}

Le prime prove per il codice nel dominio del tempo sono state 
quelle della partenza da uno stato di quiete fino al raggiungimento 
di un semplice moto di avanzamento ad un dato numero di $Froude$. 
Abbiamo scelto per tali prove la prima carena {\em Wigley} descritta
nel sesto capitolo. \\ 
I risultati riportati in questo capitolo sono relativi allo schema 
numerico che utilizza Runge-Kutta avendo ottenuto con questo 
i risultati migliori. \\ 
Nella prima figura delle \ref{staz1} viene mostrato l'andamento del coefficiente 
di resistenza relativo a due casi con diverse condizioni iniziali sul
potenziale di velocit\a e sull'elevazione d'onda $\eta$. 
In questa figura sono riportati sia il valore sperimentale della resistenza
d'onda, sia il valore fornito da un codice stazionario che utilizza 
una linearizzazione di doppio modello e una discretizzazione spaziale simile
a quella utilizzata per il codice nel dominio del tempo. 
Come si pu\o osservare il valore raggiunto a regime dal codice nel dominio del
tempo \e sufficientemente vicino al dato sperimentale.
Se come condizioni iniziali diamo quelle che competono al flusso base, si ha
una notevole riduzione del transitorio, e questo \e un notevole vantaggio 
se si sceglie di fare una prova armonica o una prova impulsiva 
utilizzando come flusso base proprio lo stato stazionario ricavato con
queste prove. \\
\indent
Osservando la figura \ref{staz1} si nota come il livello stazionario 
sia raggiunto attraverso delle oscillazioni smorazate, \e possibile 
dimostrare che per $t\rightarrow\infty$ tutte le forze idrodinamiche 
hanno un andamento asintotico del tipo:
\be
\dsty
F_{idro}(t)\simeq\ A_0\,+\,\frac{A_1}{t}\,\sin(\,\omega_c\,t\,+\,A_2\,)
\ee
Dove $\omega_c$ \e proprio la pulsazione critica per la quale si ha 
$\tau\,=\,1/4$. \\
\indent
Sempre per il semplice avanzamento abbiamo fatto diverse prove al variare 
del numero di $Froude$ (figure \ref{staz2}). 
All'aumentare di quest'ultimo si ottiene un 
transitorio sempre pi\u piccolo ma, per la stabilit\a dello schema\, occorre
diminuire sensibilmente l'incremento temporale. Per bassi numeri di $Froude$
si hanno invece dei transitori pi\u lunghi. Inoltre, dati gli ampi angoli di 
divergenza del sistema d'onda, occorre utilizzare delle griglie sufficientemente
ampie nella direzione trasversale onde evitare delle riflessioni indesiderate
all'interno del dominio di interesse. 

\begin{figure}[hp]
\vskip  2.5cm
 \vspace*{-2cm}
      \epsfxsize=0.85\textwidth
      \epsfysize=0.55\textwidth
      \makebox[\textwidth]{\epsfbox{./TESIFIG/steady.ps}}
      \epsfxsize=0.85\textwidth
      \epsfysize=0.55\textwidth
      \makebox[\textwidth]{\epsfbox{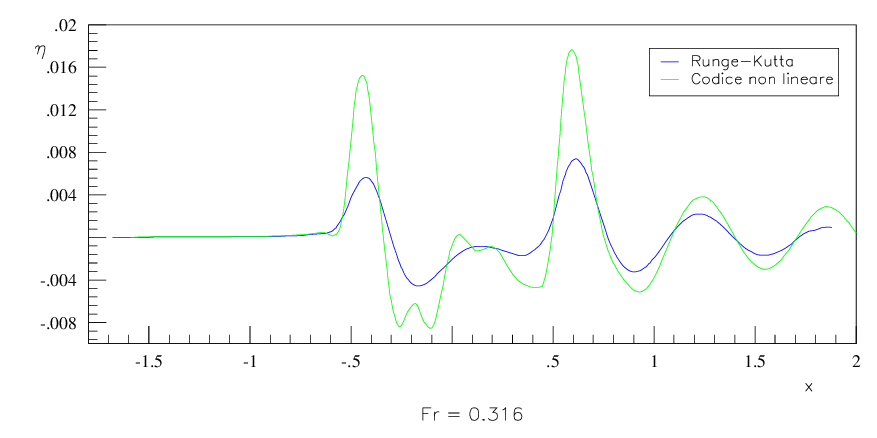}}
    \vskip -0.5cm
    \caption{ Confronti tra un codice lineare e un codice non lineare sulle 
              elevazioni d'onda. Nella figura in alto si ha una vista in pianta
              del sistema ondoso stazionario per un $Fr\,=\,0.316$ ottenuto 
              con il codice lineare (met\a superiore) e con un codice non lineare    
              (met\a inferiore). Nella figura in basso si ha invece un confronto 
              sui tagli d'onda appartenti al piano $xz$.  
             } \label{staz3}
\end{figure}

Nelle figure \ref{staz3} vengono mostrate le linee di livello 
dell'elevazione d'onda ed un taglio d'onda longitudinale, ossia
eseguito mediante intersezione della superficie libera con il piano 
$y\,=\,0$.
L'elevazione d'onda riportata \e quella relativa alla 
fine del transitorio ossia l'elevazione d'onda che compete allo stato 
stazionario di semplice avanzamento ad un dato $Froude$.
La figura mostra un confronto con i risultati ottenuti con un codice non 
lineare (cfr. \cite{mari},\cite{aimeta}), quest'ultimo fornisce una 
stima molto accurata dei dati sperimentali relativi alle forze e ai tagli 
d'onda.
Dal confronto fra i  due risultati \e evidente come i sistemi ondosi 
siano qualitativamente molto simili ma le elevazioni d'onda ottenute con il 
codice lineare siano sensibilmente pi\u piccole rispetto a quelle fornite dal 
codice non lineare.
Oltre che intuitivo, questo \e ben noto in letteratura. l'uso di una
linearizzazione di doppio modello consentirebbe una migliore previsione
dei valori locali dell'altezza d'onda.
La minor {\em intensit\a} del sistema ondoso prodotto dal codice lineare 
motiva la sotto stima della resistenza d'onda rispetto a quella misurata
sperimentalmente.

I risultati appaiono soddisfacenti ed il recupero dei termini
di doppio modello (pur se oneroso) consentirebbe una migliore accuratezza. \\
\indent
Tuttavia, in questa tesi, l'enfasi \e posta sul problema della tenuta al
mare e nel seguito si \e piuttosto preferito
considerare il caso di moti armonici di sussulto e beccheggio imposti alla carena 
in avanzamento.
Per questi due problemi di radiazione 
la condizione al contorno di impermeabilit\a per la carena diventa: 
\be
\left \{
\begin{array}{l}
\dsty
\pder{\Phi_j}{\vec n}\,=\,n_j\dot{q}_j\,+\,m_j\,q_j
\\ \\
\dsty
q_j(t)\,=\,q_j\,\cos(\omega t) \qquad con j\,=\,3\,,5
\end{array}
\right.
\ee
e quindi va modificato il termine noto per i sistemi algebrici,
visti nel precedente paragrafo \ref{sistema}--\ref{dotsistema}, che 
calcolano rispettivamente le $\sigma$ e le $\dot{\sigma}$. 
Per contenere il transitorio, l'ampiezza del moto imposto sar\a inizialmente 
alterata mediante una funzione di crescita graduale.

\begin{figure}[hp]
\vskip  2.5cm
 \vspace*{-2cm}
      \epsfxsize=0.75\textwidth
      \epsfysize=0.45\textwidth
      \makebox[\textwidth]{\epsfbox{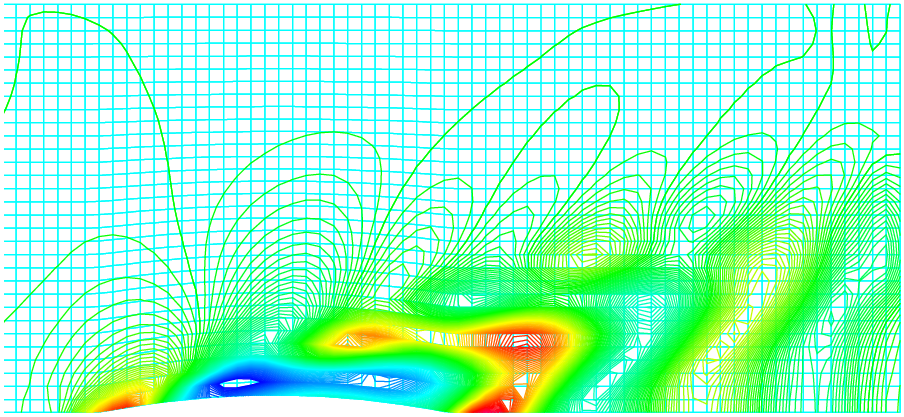}}
      \epsfxsize=0.75\textwidth
      \epsfysize=0.45\textwidth
      \makebox[\textwidth]{\epsfbox{./TESIFIG/heave0500.ps}}
      \epsfxsize=0.75\textwidth
      \epsfysize=0.45\textwidth
      \makebox[\textwidth]{\epsfbox{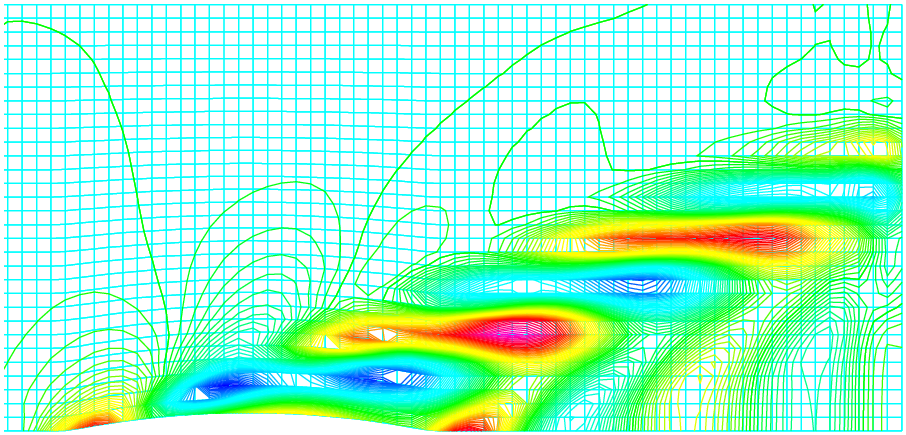}}
    \vskip -0.5cm
    \caption{ Risultati nel dominio del tempo: moto in avanzamento a 
              $Fr\,=\,0.316$
              con moto di sussulto armonico imposto 
              con una pulsazione $\omega\,=\,4.375$. 
              Sono riportati tre diversi istanti del transitorio. 
             } \label{heave}
\end{figure}

\begin{figure}[hp]
\vskip  2.5cm
      \epsfxsize=0.84\textwidth
      \epsfysize=0.54\textwidth
      \makebox[\textwidth]{\epsfbox{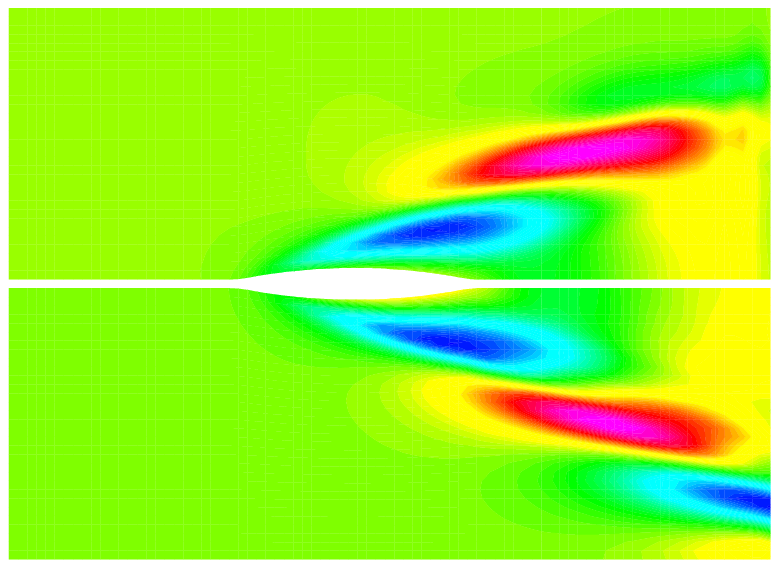}}
      \epsfxsize=0.85\textwidth
      \epsfysize=0.55\textwidth
      \makebox[\textwidth]{\epsfbox{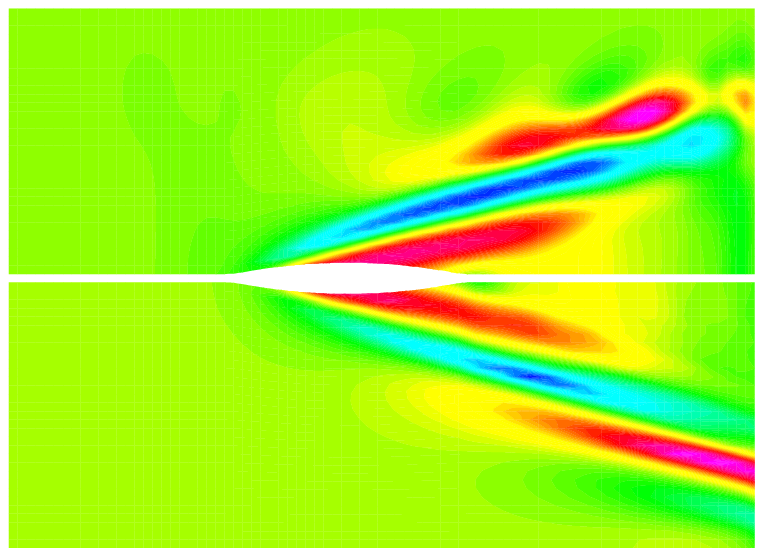}}
    \vskip -0.5cm
    \caption{ Confronto delle elevazioni d'onda prodotte dal codice in frequenza
              con quelle prodotte dal codice nel dominio del tempo per una
              prova armonica con pulsazione $\omega\,=\,3$ e con 
              una velocit\`a di avanzamento data da $Fr\,=\,0.316$.
              Nella prima figura in alto \`e riportata la prova armonica in 
              sussulto, mentre nella seconda figura \`e riportata la prova 
              armonica in beccheggio.
              La parte superiore delle due figure \`e data 
              dal codice nel tempo, quella inferiore dal codice in frequenza.
             \label{pora}}
\end{figure}

\begin{figure}[hp]
\vskip  2.5cm
 \vspace*{-2cm}
      \epsfxsize=0.75\textwidth
      \epsfysize=0.45\textwidth
      \makebox[\textwidth]{\epsfbox{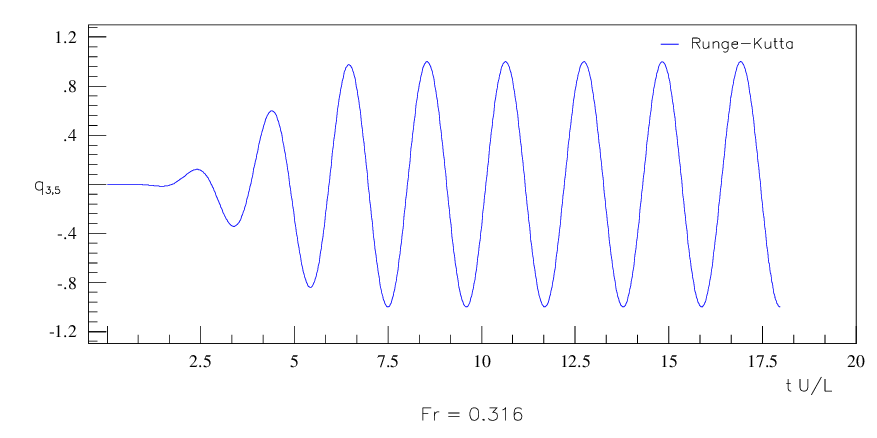}}
      \epsfxsize=0.75\textwidth
      \epsfysize=0.45\textwidth
      \makebox[\textwidth]{\epsfbox{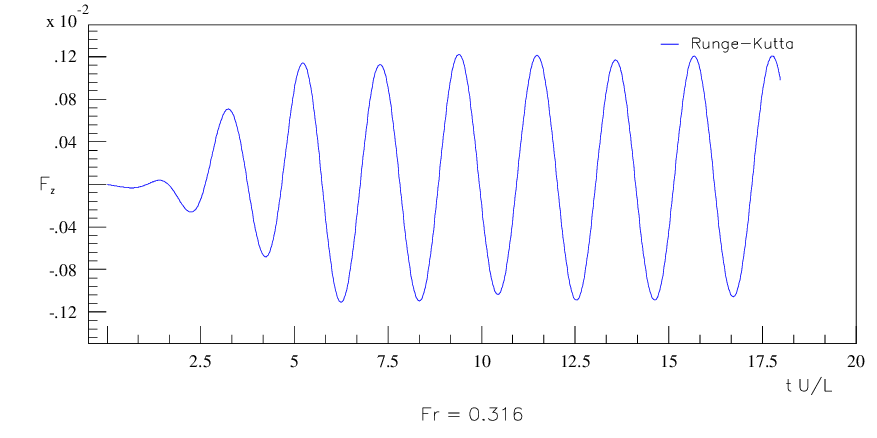}}
      \epsfxsize=0.75\textwidth
      \epsfysize=0.45\textwidth
      \makebox[\textwidth]{\epsfbox{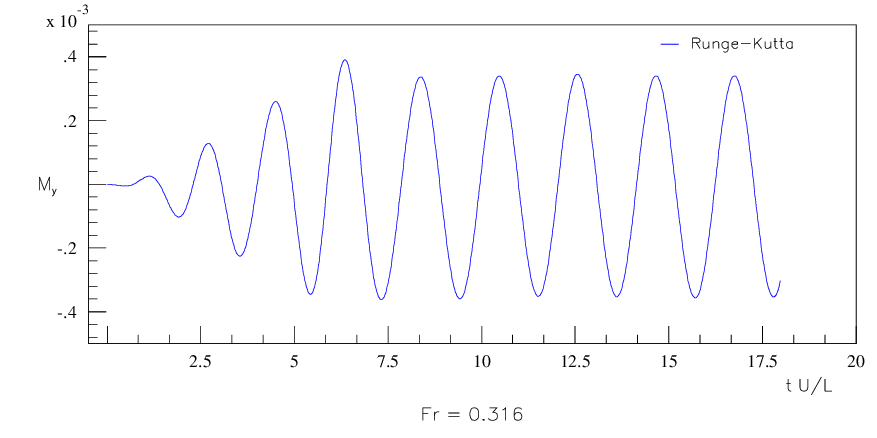}}
    \vskip -0.5cm
    \caption{ Risultati nel dominio del tempo per una prova armonica in sussulto
              e in beccheggio: la pulsazione \`e pari a 
              $\omega\,=\,3.00$. Nella prima figura \`e riportato l'andamento 
              della $q_3$ e della $q_5$ imposte alle carena. Nella seconda 
              figura \`e si ha invece la forza verticale nel moto di sussulto e  
              nella terza il momento di beccheggio nel rispettivo moto. 
             } \label{nostaz}
\end{figure}

\begin{figure}[hp]
\begin{center} 
      \epsfxsize=\textwidth
      \makebox[\textwidth]{\epsfbox{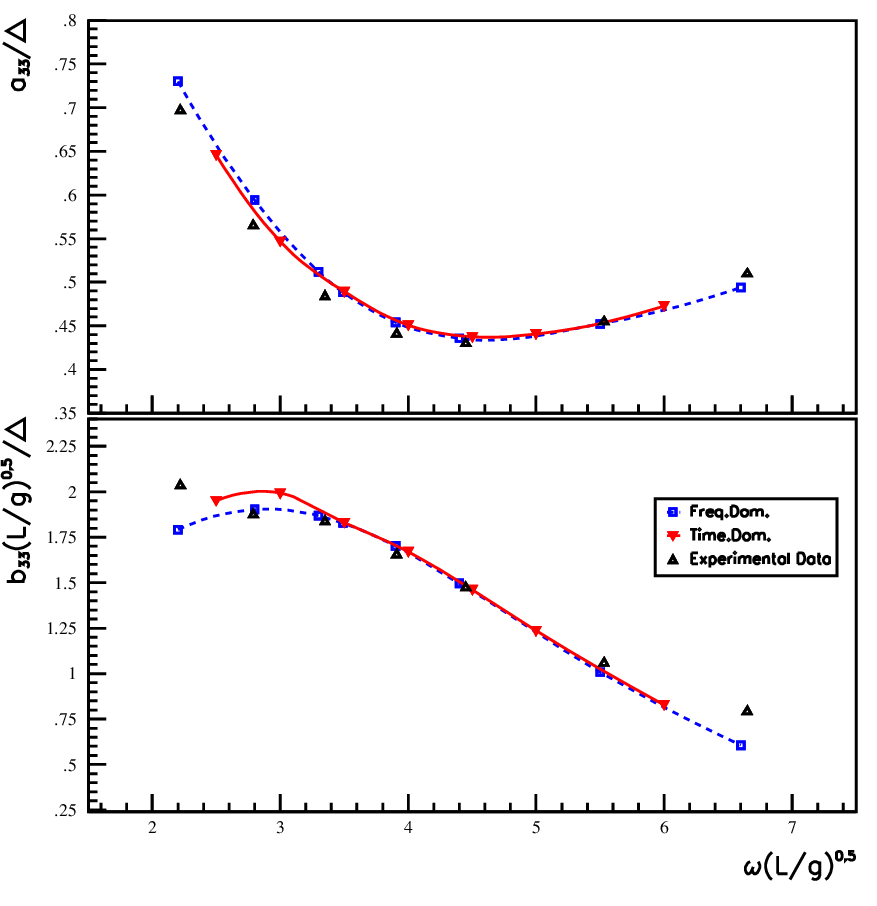}}
    \caption{ Coefficienti di massa aggiunta $a_{33}$ e smorzamento $b_{33}$ 
              per il moto di sussulto. Viene qui fatto un confronto fra i 
              dati sperimentali e i risultati ottenuti con il codice in 
              Frequenza e con il codice nel dominio del tempo.
              Per quest'ultimo gli ingressi sono monocromatici. 
              La carena utilizzata \e una Wigley con coefficiente di 
              blocco pari a 0.46 e un volume pari a $\nabla$, 
              $\Delta\,:=\,\rho\,\nabla$. Il numero di Froude \e pari a 0.3.
             }\label{heavemono}  
\end{center}
\end{figure}

\begin{figure}[hp]
\begin{center} 
      \epsfxsize=\textwidth
      \makebox[\textwidth]{\epsfbox{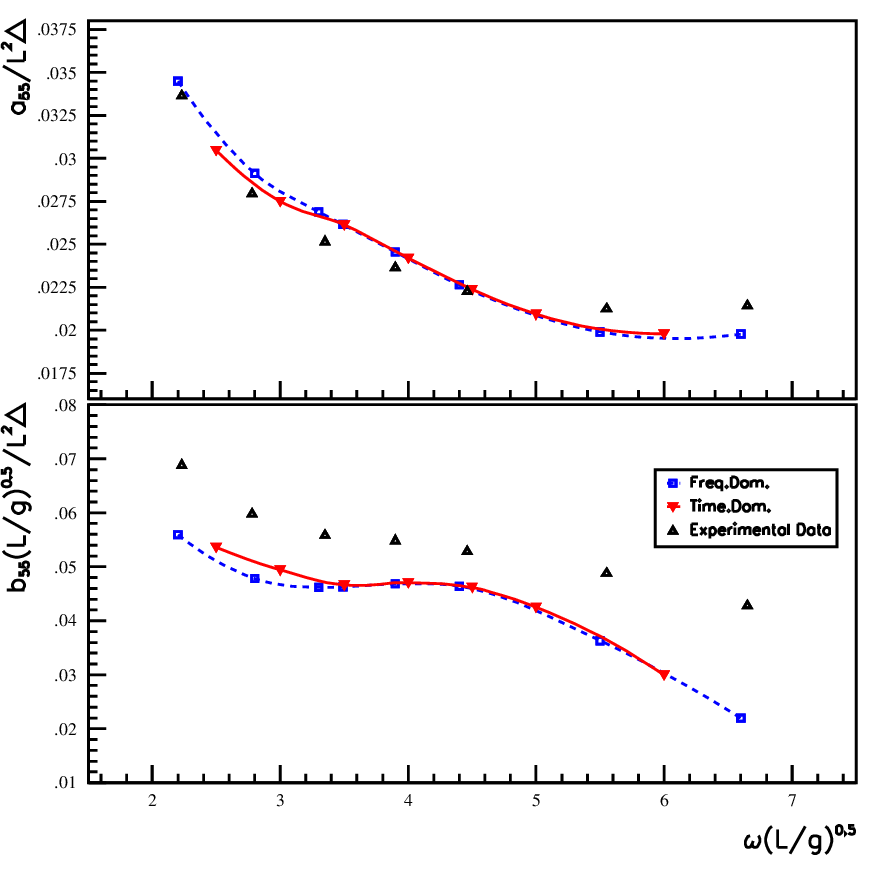}}
    \caption{ Coefficienti di massa aggiunta $a_{55}$ e smorzamento $b_{55}$ 
              per il moto di beccheggio. Viene qui fatto un confronto fra i 
              dati sperimentali e i risultati ottenuti con il codice in 
              Frequenza e con il codice nel dominio del tempo.
              Per quest'ultimo gli ingressi sono monocromatici. 
              La carena utilizzata \e una Wigley con coefficiente di 
              blocco pari a 0.46 e un volume pari a $\nabla$, 
              $\Delta\,:=\,\rho\,\nabla$. Il numero di Froude \e pari a 0.3.
             }\label{pitchmono}  
\end{center}
\end{figure}

Nella figura \ref{heave} viene mostrata la nascita e lo sviluppo 
di un sistema ondoso quando viene imposto un moto di sussulto alla
carena.
In questo caso nella condizione al contorno di impermeabilit\a 
della carena \e stato tenuto anche il termine relativo alla corrente, 
quindi il potenziale ottenuto \e la somma di quello stazionario con 
quello non stazionario (e così per le elevazioni d'onda). \\
Il risultato qualitativo ottenuto \e molto soddisfacente.\\ \\
Una seconda prova \e stata quella di confrontare le elevazioni  
d'onda ottenute con il codice nel tempo con quelle ottenute con 
il codice nel dominio della frequenza.
In particolare questo confronto \e stato fatto sia nel caso
di sussulto che in quello di beccheggio e nelle figure \ref{pora} \e 
riportato il soddisfacente confronto. \\ \\
Analizziamo ora le forze idrodinamiche che ci daranno
un risultato quantitativo per il codice nel dominio del tempo. 
Nella figura \ref{nostaz} \e mostrata la storia temporale per 
le forze idrodinamiche: nel primo diagramma si ha l'andamento temporale
della variabile lagrangiana, nel secondo l'andamento della forza verticale 
(moto di sussulto) e nell'ultimo in basso l'andamento temporale del momento 
di beccheggio (moto di beccheggio).
Da queste storie temporali possiamo 
ricavare i coefficienti di massa aggiunta e smorzamento  per fare un
confronto con i risultati forniti dal codice in frequenza.
Per ricavare questi coefficienti abbiamo sviluppato in serie di Fourier 
le storie temporali delle forze idrodinamiche, escludendo la parte relativa 
al transitorio. 
Per la pulsazione uguale a quella della forzante ricaviamo 
la parte reale e quella immaginaria, queste ci danno le grandezze:
\be 
\left \{
\begin{array}{l}
Re{F_i}\,=\,\omega^2\,a_{ij}\,-\,c_{ij}  \\\\  
Im{F_i}\,=\,\omega\,b_{ij} \qquad 
\mbox{ Dove i=1,3,5 mentre  j=3,5 indica il tipo di moto imposto}  
\end{array}
\right.
\ee 
I coefficienti di smorzamento $b_{33}\,,b_{55}$ possono essere ricavati
anche attraverso il valor medio della potenza erogata dalle di forze 
idrodinamiche. Infatti, come abbiamo visto nell'ultimo paragrafo del 
quarto capitolo, vale la relazione: 
\be
\dsty
b_{jj}\,=\,\frac{2\,\bar{W}}{q_j^2\omega^2}
\qquad dove \qquad \bar{W} \,=\,F_j\,\dot{q_j}
\ee
Nelle figure \ref{heavemono} \ref{pitchmono} sono riportati i grafici
per i coefficienti di massa aggiunta e smorzamento relativi a problemi
di radiazione in sussulto e beccheggio ottenuti con il codice nel dominio 
del tempo. Si pu\o qui osservare l'ottimo 
accordo tra i risultati forniti da quest'ultimo con quelli ottenuti 
con il codice in frequenza
\footnote{ Per coerenza abbiamo utilizzato anche 
per questo la linearizzazione di {\em Neumann - Kelvin} }.// 
\indent
In conclusione, i buoni risultati ottenuti per le prove armoniche
suggeriscono l'applicazione del codice nel dominio del tempo all'analisi 
delle risposte ad "opportuni" ingressi transitori.  
Sar\a quindi possibile stimare il nucleo $\vec K(t)$ delle
forze idrodinamiche attraverso un'unica prova e si potr\a ottenere
l'intera curva dei coefficienti di massa aggiunta e smorzamento al variare 
della pulsazione $\omega$ (ovviamente a parit\a di altri parametri quali la 
velocit\a di avanzamento e la geometria di carena analizzata).
Il calcolo dei suddetti coefficienti sar\a fatto attraverso la parte reale e immaginaria della trasformata di Fourier della matrice $\vec K(t)$ 
alle quali per\o devono essere sommate le matrici $\vec A$ e $\vec B$ 
delle masse aggiunte e degli smorzamenti valutate per 
$\omega\rightarrow\infty$, queste potranno essere calcolate attraverso 
l'analisi di flussi potenziali intorno al doppio modello 
generato dalla carena e dalla sua immagine riflessa
\footnote{Di ci\o abbiamo gi\a discusso nell'ultimo 
paragrafo del quarto capitolo} 
rispetto al piano $xy$.

Studiati e superati i problemi numerici connessi con la simulazione di
ingressi transitori, si potr\a valutare se l'approccio nel dominio del tempo 
sia in grado di fornire delle stime dei dati sperimentali in 
un tempo di calcolo ridotto rispetto ad analoghi codici sviluppati nel 
dominio della frequenza.

\chapter{Conclusioni e Sviluppi per il futuro.}

In questa tesi \e stato studiato numericamente il campo idrodinamico attorno 
ad una nave in avanzamento in mare formato.
L'elemento che ha guidato l'impostazione generale del lavoro \e 
essenzialmente la volont\a di sviluppare codici di calcolo in grado di 
analizzare problemi geometricamente complessi e, quindi, fornire strumenti 
di progettazione e verifica di una nave reale.
Pertanto, come necessario compromesso fra aderenza al fenomeno fisico e
accettabili costi computazionali, il problema \e stato formulato nell'ambito di 
un modello di fluido non viscoso in moto irrotazionale e le non linearit\a 
insite nel fenomeno sono state eliminate seguendo una procedura di 
linearizzazione basata sul concetto di flusso base arbitrario.

In questo contesto, dopo una rilettura degli aspetti teorici fondamentali,
sono stati sviluppati due differenti approcci per la soluzione del problema 
matematico così definito.

In un primo caso il problema non stazionario al contorno \e stato ricondotto 
ad un problema nel dominio della frequenza.
In tal modo \e stato possibile riprodurre con un ottimo accordo dati 
sperimentali relativi a carene di complessit\a geomtrica crescente.
Sulla base di questi incoraggianti risultati sono stati individuati
alcuni futuri sviluppi applicativi del modello di calcolo, sostanzialmente
mirati a renderlo un strumento idoneo allo studio di moderne carene
commerciali.
Infatti, come risultato della tendenza verso lo sviluppo di naviglio
veloce, le carene sono sempre pi\u caratterizzate da prue snelle a
'V' pronunciata (bow flare) con bulbo prodiero per 
ridurre la resistenza all'avanzamento e poppa troncata (transom) per 
una sua ulteriore diminuzione.
Tali geometrie andranno sistematicamente studiate, verificando così
il grado di attendibilit\a del modello mediante il confronto con dati 
sperimentali.

Il secondo approccio, potenzialmente di estremo interesse, consiste 
nella soluzione del problema non stazionario e nell'interpretazione dei
risultati mediante analisi di Fourier per determinare l'operatore di
risposta. 
Infatti, almeno per sistemi  lineari, \e teoricamente possibile
determinare l'operatore di risposta di un veicolo in mare con un limitato
numero di prove transitorie.

Su questa base \e stato sviluppato un algoritmo di
calcolo per la soluzione del problema non stazionario nel dominio del
tempo.
Simulando l'evoluzione del fenomeno sin dalla quiete, 
sono stati riprodotti con successo sia risultati relativi al problema della
resistenza d'onda in condizioni di moto rettilineo a velocit\a costante,
sia prove di moto armonico forzato mediante le quali sono stati riprodotti
i risultati ottenuti con il codice in frequenza.
Su questa base verr\a nel prossimo futuro sviluppato lo studio 
numerico della tenuta al mare mediante prove transitorie: se coronato da
successo il metodo dovrebbe consentire una drastica riduzione del 
tempo totale per la valutazione teorica dell'operatore di risposta e
lo strumento di calcolo sviluppato potrebbe divenire definitivamente
idoneo alla progettazione navale.

Nello sviluppo dei metodi di calcolo sono state individuate alcune 
problematiche di natura numerica degne di attenzione per il futuro. 

In particolare, malgrado lo sforzo positivo mirato ad una corretta 
valutazione dei termini $m_j$ del problema di doppio modello, 
la necessit\a di valutare il gradiente del campo di velocit\a su
superfici geometricamente complesse richiede lo sviluppo di tecniche di 
soluzione per le equazioni agli integrali di contorno che siano pi\u accurate 
di quelle usate in questo lavoro.

Una ulteriore riduzione dei tempi di calcolo potrebbe essere ottenuta
mediante tecniche di decomposizione in sotto--domini del campo fluidodinamico.
In alcune esperienze \e stato mostrato come sia possibile avere un
costo computazionale dell'ordine del numero totale delle incognite e non
del suo quadrato come nei convenzionali solutori BEM.

Sembra inoltre essenziale sottolineare l'importanza di una accurata analisi 
di sensibilit\a della soluzione al variare dei parametri discreti.
Infatti, con riferimento alla simulazione nel dominio del tempo,
ci si pu\o aspettare che i risultati relativi ad ingressi compatti nel tempo
possano essere falsati da una non corretta relazione di dispersione.

A conclusione si vuole evidenziare il valore propedeutico, se non
sinergico, del presente studio al fine dello sviluppo di modelli completamente
non lineari.
\newpage
\clearpage
\appendix
\chapter{Calcolo numerico dei termini $m_j$}

\section{Introduzione}
 In questa appendice vengono presentati alcuni risultati numerici ottenuti con 
 la formulazione linearizzata sviluppata in \cite{ASeak} per l'analisi in frequenza
 della {\em tenuta al mare} di carene convenzionali.
 Sono esaminate entrambe le linearizzazioni che essa prevede, in particolare, 
 quella alla Neumann--Kelvin e quella di doppio--modello. 

 Lo studio della risposta a sistemi ondosi
 preesistenti, e pertanto dell'accoppiamento campo fluidodinamico--moto della nave, 
 \`e condizionato in modo sostanziale dal calcolo dell'azione idrodinamica 
 agente sulla carena che dipende, a sua volta, dal gradiente del potenziale $\Phi$ e
 dal gradiente di velocit\`a del flusso 'base' sul corpo.
 Tale legame si ritrova nei termini $m_j$ introdotti mediante la (2.30) e
 nei coefficienti di 'richiamo' $c_{ij}$ definiti dalla (2.40), entrambe contenute
 nel rapporto tecnico citato, e pu\`o essere fonte di errori numerici notevoli. 
 In particolare, una certa difficolt\`a \`e associata al calcolo di
 $\nabla\nabla\Phi$.
 Naturalmente quanto detto vale per la linearizzazione di doppio--modello, 
 essendo il gradiente della velocit\`a identicamente nullo per la linearizzazione
 alla Neumann--Kelvin. 
 Se si considera, pertanto, la prima delle due semplificazioni si ha la necessit\`a
 di superare prevedibili inconvenienti numerici.
 A tal proposito,  ma solo sotto certe ipotesi, per i termini
 $m_j$ il calcolo di $\nabla\nabla\Phi$ pu\`o essere
 evitato, potendo determinare tali termini mediante una relazione integrale che
 elimina la dipendenza da questa variabile.
 Il problema rimane tuttavia per i coefficienti di 'richiamo'.

 Per calcolare in modo
 accurato $\nabla\nabla\Phi$ sulla superficie del corpo, risulta conveniente
 ricorrere ad una tecnica di estrapolazione.
 Per tale ragione, nella prima sezione di questa appendice viene presentata una 
 analisi approfondita riguardante l'influenza della geometria, della
 discretizzazione e di altri parametri numerici nella valutazione delle derivate
 di $\Phi$ e dei conseguenti termini $m_j$. In particolare, sono considerati i casi
 di avanzamento di una sfera e di un ovoide di Rankine, per i quali
 esistono soluzioni analitiche di confronto.
 Questa indagine di tipo euristico permette di giungere ad 
 importanti conclusioni, alcune delle quali sono da ritenersi di carattere
 generale. Fornisce, inoltre, i criteri per una valutazione opportuna delle
 variabili d'interesse, limitata tuttavia dall'uso di pannelli di ordine zero 
 per la discretizzazione di entrambe le porzioni della frontiera di $\Omega$.      

 Nella seconda parte, la metodologia viene applicata all'analisi del
 {\em seakeeping} di alcune carene a geometrie semplici. 
 In particolare, sono considerate
 tre carene Wigley modificate ed una Serie 60 (Cb = 0.6). Nei casi trattati
 le carene sono sottoposte a sistemi d'onda incidente di varia lunghezza e 
 vengono inoltre analizzate diverse velocit\`a di avanzamento.
 Sono valutati i coefficienti di 'massa aggiunta' ($a_{ij}$), di 
 'smorzamento' ($b_{ij}$) e di 'richiamo' ($c_{ij}$), inoltre, la forza di 
 'eccitazione' ($X_i$) e l'ampiezza dei moti nave che ne derivano ($\zeta_i$). 
 Per alcuni di questi risultati sono infine mostrati i
 confronti con dati sperimentali disponibili in letteratura.   

\section{Calcolo accurato del gradiente di velocit\`a sul corpo} \label{Grad}
 I termini $m_j$\footnote{$j = 1,..,6$ come i gradi di libert\`a associati alla 
 carena per la quale si assume un comportamento rigido.} risultano definiti dalla 
 relazione
 \be \label{Def_mj}
  \left\{
  \begin{array}{lcl}  
  (m_1,m_2,m_3)\,:=\,-\,(\vec n\,\cdot\,\nabla)\nabla \Phi \\[.5cm]   
  \dsty
  (m_4,m_5,m_6)\,:=\,-\,(\vec n\,\cdot\,\nabla)(\vec x\,\times\,
   \nabla \Phi)    
  \end{array}  
   \forall \vec x \in \overline{\cal B}
   \right.
 \ee
 per ogni $\vec x$ della configurazione media $\overline{\cal B}$ occupata
 dalla nave\footnote{Nella (\ref{Int_mj}) $\vec n$ \`e la normale alla carena mentre
 $\Phi$ indica il potenziale del flusso 'base' (cfr. \cite{ASeak}) associato al
 problema.} e dipendono, pertanto, in generale dalla variabile gradiente di
 velocit\`a.
 In alcuni casi tuttavia, \`e possibile procedere per altra via.
 Infatti, sotto alcune  ipotesi, tra le quali: superficie del corpo continua e
 {\em wall--sided}, si pu\`o dimostrare (\cite{AOgTu}) la validit\`a della
 relazione integrale  
 \be \label{Int_mj}
  \begin{array}{lcl}  
   \dsty
   \int_{\overline{\cal B}} m_j (\vec x^*)\, G (\vec x, \vec x^*)\,
   \dsty
    d S_{\overline{\cal B}}(\vec x^*)\,=\,-\,    
   \dsty
   \int_{\overline{\cal B}} \nabla \Phi (\vec x^*)\,\cdot\,\nabla_{\vec x^*} G
   \dsty
    (\vec x, \vec x^*)\,n_j d S_{\overline{\cal B}}(\vec x^*)\,,
   \dsty
  \end{array}  
 \ee
  in cui $G$ indica la funzione di Green di spazio 
  libero (cfr. \cite{ASeak}). Il vantaggio appare evidente non comparendo in essa 
  il termine $\,\nabla \nabla \Phi$.
  Vengono così evitati i problemi numerici associati al calcolo di questo, 
  tuttavia, se ci si allontana dalle ipotesi di validit\`a della
  (\ref{Int_mj}) non resta che applicare la definizione.

  Inoltre la dipendenza da tale grandezza \e presente   
  anche per i coefficienti di richiamo $c_{ij}$, pertanto, \`e opportuno 
  valutare il tipo di problemi numerici che il suo calcolo comporta e procedere verso
  una metodologia capace di garantire un'opportuna accuratezza.

\section{Applicazione di un processo di estrapolazione} 
  La formulazione in oggetto \`e caratterizzata dall'introduzione di pannelli di
  ordine zero in corrispondenza della porzione {\em bagnata} ${\overline{\cal B}}$
  della carena e sulla {\em superficie libera} indisturbata.
  Questo rende particolarmente semplice la formulazione, riducendo il numero di 
  variabili da calcolare rispetto ad un metodo di ordine superiore. Tuttavia, 
  risultano diminuite anche l'accuratezza e la variet\`a di informazioni rese
  disponibili. In particolare, se ci si avvicina progressivamente alla frontiera
  $\partial\Omega$, viene persa l'informazione associata alla sua curvatura.
  Naturalmente ci\`o non \`e rilevante per $\SL$, almeno per i problemi discussi in 
  questa sede, per i quali, la condizione di {\em superficie libera} viene imposta
  sul piano $y=0$. La cosa risulta diversa per la carena. In questo caso il
  calcolo dei gradienti comporta errori numerici rilevanti,
  in quanto, per $\vec x\rightarrow \overline{\cal B}$, diventa decisivo
  l'ordine dei pannelli utilizzati. In particolare per quanto riguarda le 
  derivate di ordine superiore al primo. Per ovviare a questo
  problema, si decide di introdurre una tecnica di estrapolazione che consente
  di ricavare le derivate sul corpo mediante il valore da queste assunto nelle
  sue vicinanze, dove risultano contenuti i relativi errori numerici.

\noindent
  L'applicazione di un tale processo richiede la scelta di alcuni
  parametri, in particolare, il grado del polinomio estrapolante, il primo
  nodo per  l'estrapolazione e la distanza tra i nodi successivi. Tali elementi,
  come si vedr\`a, sono decisivi per ottenere una soluzione sufficientemente
  accurata. 
\section{Casi esaminati: Sfera}
\begin{figure}[htb]
      \vskip 1cm
      \epsfxsize=.5\textwidth
      \epsfxsize=.5\textwidth
      \makebox[.9\textwidth]{\epsfbox{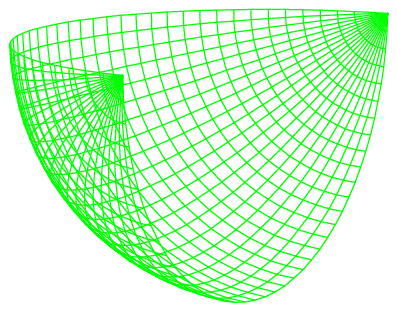}}
    \vskip -0.5cm
    \caption{Esempio di un tipico reticolo di calcolo per una sfera. 
             \label{sfe}
             }
\end{figure}
\begin{figure}[htb]
      \epsfxsize=\textwidth
      \makebox[\textwidth]{\epsfbox{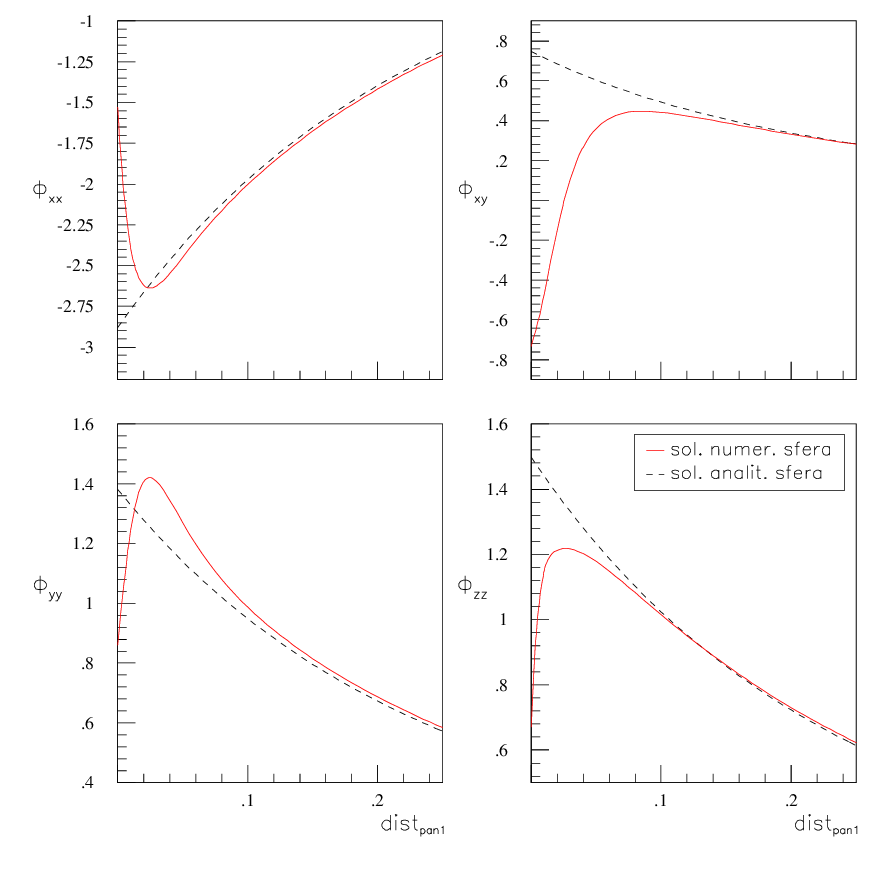}}
    \vskip -0.5cm
    \caption{Sfera di raggio $R = 1$: confronto analitico--numerico di alcune
             componenti di $\nabla \nabla\Phi$ al variare della distanza normale dal
             pannello in prossimit\a del punto di ristagno ($pan1$).
             \label{con1}
             }
\end{figure}
\begin{figure}[htb]
      \epsfxsize=\textwidth
      \makebox[\textwidth]{\epsfbox{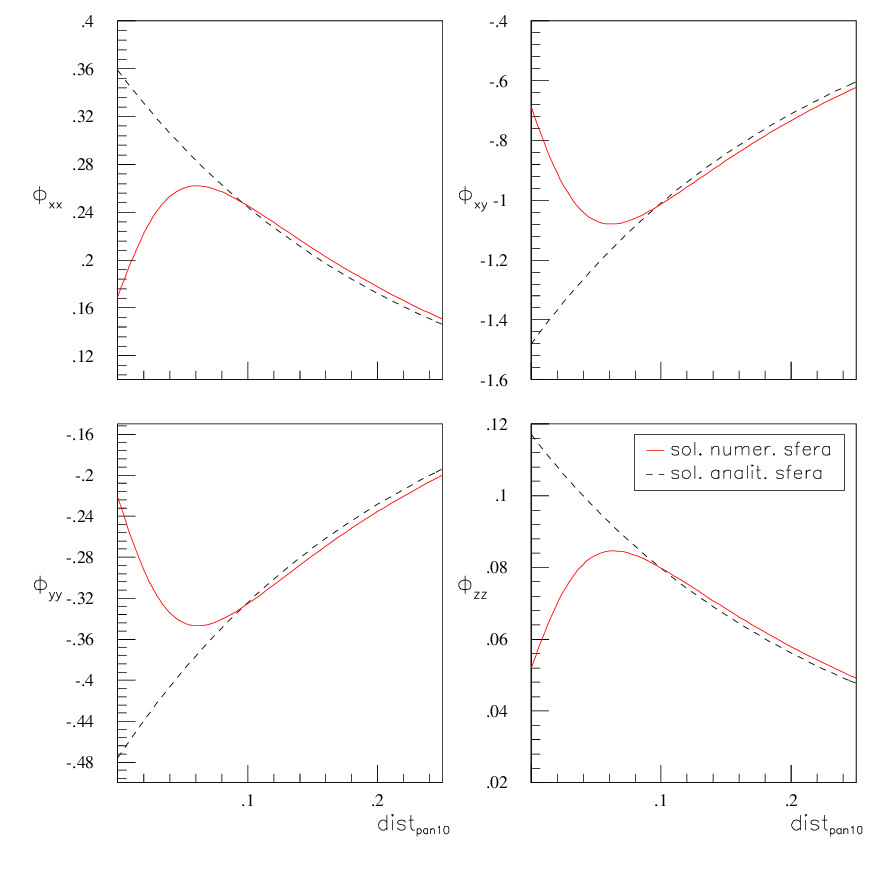}}
    \vskip -0.5cm
    \caption{Sfera di raggio $R = 1$: confronto analitico--numerico di alcune
             componenti di $\nabla \nabla\Phi$ al variare della distanza normale dal
             pannello in corrispondenza del massimo valore della velocit\`a
             ($pan10$).
             \label{con2}
             }
\end{figure}
\begin{figure}[htb]
      \epsfxsize=\textwidth
      \makebox[\textwidth]{\epsfbox{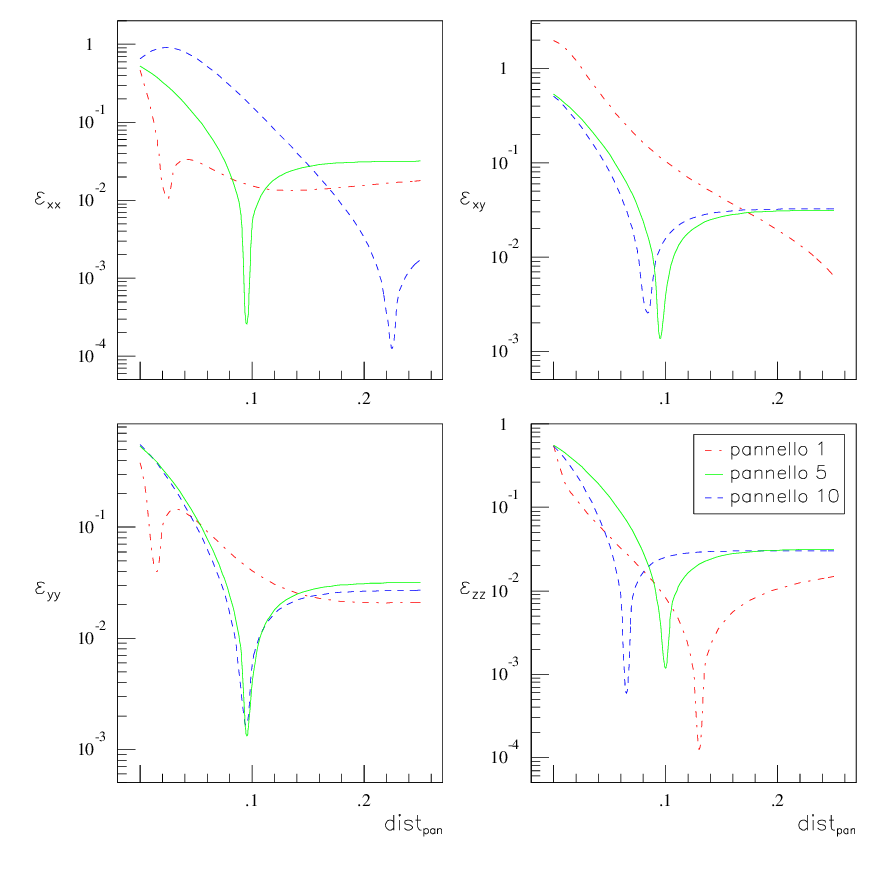}}
    \vskip -0.5cm
    \caption{Caso di una sfera di raggio $R=1$: andamento dell'errore relativo 
             $\epsilon$ al variare della distanza normale dai pannelli 1, 5 e 10. 
             \label{err}
             }
\end{figure}

  Il nostro scopo \`e quello di studiare i problemi numerici 
  associati al calcolo delle componenti di $\nabla \nabla \Phi$ sulla superficie 
  del corpo, con l'intenzione di ricavarne strumenti adatti alla relativa riduzione.
  Si considera come primo esempio il caso di una sfera di raggio $R = 1$ che avanza
  in un fluido infinito (v. fig. \ref{sfe}). Nelle figure \ref{ana1}--\ref{ana6},
  vengono riportati, a tal proposito,  gli
  andamenti analitici delle derivate seconde del potenziale 
  per i punti appartenenti al piano $xy$ (cfr. \cite{Batch}). 
  Con $N_{pan}$ \e indicato il numero del pannello dove vengono calcolate 
  le grandezze in considerazione, $N_{pan}=1$ \e il pannello pi\u vicino
  al punto di ristagno, mentre $N_{pan}=20$ corrisponde al punto di massima
  velocit\a. 
  In figura \ref{con1} \`e mostrato invece il comportamento delle derivate 
  seconde del potenziale all'interno
  del dominio fluido, al diminuire della distanza dal pannello considerato. 
  I valori numerici si riferiscono ad una discretizzazione di $40\mbox{x}40$ 
  pannelli. Come si pu\`o osservare, le curve rimangono quasi sovrapposte a quelle
  analitiche per poi separarsene definitivamente.
  La derivazione numerica sulla sfera introduce pertanto errori
  rilevanti e rende necessaria l'applicazione di un processo di estrapolazione.
  Affinch\'e quest'ultimo sia utile risulta, tuttavia, fondamentale restare al
  di fuori della zona di allontanamento critico.
  Cosa non banale in quanto tale zona non ha la medesima estensione per ogni
  pannello. Ci\`o pu\`o essere rilevato per esempio dalle figure \ref{con1} e
  \ref{con2} che si riferiscono a pannelli posti sulla medesima linea d'acqua, a
  $0$ e $90$ gradi, rispettivamente, dalla direzione della corrente. 
  Nella successiva figura \ref{err} viene invece riportato l'errore
  relativo 
\[
   \epsilon\,=\,\frac{| Sol_{anal}\,-\,Sol_{num}|}{|Sol_{anal}|}\,
\]
  associato al calcolo di alcune componenti del gradiente di velocit\`a, sempre al
  decrescere della distanza dal pannello considerato.
  I grafici si riferiscono a pannelli posti a $0$, $45$ e $90$ gradi rispetto alla
  corrente.
  In essi si nota l'esistenza di un punto di minimo a causa del cambiamento di segno
  della differenza tra soluzione analitica e soluzione numerica, la cui posizione
  varia cambiando componente nonch\'e pannello.
  Infine, esso dipende dalla dimensione di quest'ultimo e dalla geometria ad esso
  associata (triangolare o quadrangolare).
  In questa sede viene considerata, in particolare, una legge funzionale del tipo
\be \label{Pos}
     d^*_{pan}\,=\,c(\Phi_{,ij}\,,\, \alpha\,,\beta\,,\mbox{\em forma})\,
                {\cal A}^{\frac{1}{2}}\,,
\ee
 in cui, $\alpha$ e $\beta$ sono gli angoli che individuano
 la posizione del pannello fissandone, rispettivamente, la sezione e la linea
 d'acqua. A tal proposito, le figure sopra indicate si riferiscono alla prima linea
 d'acqua ($z=0$).
 E' necessario mantenersi il pi\`u possibile nell'intorno del punto di minimo per 
 la scelta del primo nodo dell'estrapolazione. 

\noindent
  Nelle figure \ref{con1} e \ref{con2} si pu\`o inoltre osservare, per
  tutte le derivate numeriche, la presenza di un flesso in prossimit\`a della sfera;
  punto non esistente invece nelle curve analitiche, prive pertanto di variazione di
  concavit\`a. 
  Se si sceglie il flesso come primo nodo di un'estrapolazione si 
  riporta la soluzione numerica ad un andamento analogo a quello analitico e questo,
  come \`e facile intuire,  
  realizza nel complesso una differenza minore tra i valori numerico--analitici di
  $\nabla\nabla\Phi$ sul corpo. Tale punto, inoltre, si trova relativamente vicino
  a quello di minimo, infatti, utilizzando la legge (\ref{Pos}) per individuare la
  sua posizione, si trova come per il minimo un parametro $c$ dell'ordine 
  dell'unit\`a. 

\noindent
  A seguito di quanto finora descritto, e' interessante confrontare le
  soluzioni numeriche relative ai due differenti processi di estrapolazione che 
  assumono il punto di minimo dell'errore e, rispettivamente, quello di flesso come
  primo nodo dell'estrapolazione.

\noindent
  Vengono utilizzati per entrambi $9$ nodi con distanza reciproca pari a  
   $\sqrt{2}\,{\cal A}^{\frac{1}{2}}_{pan}\,$,
  essendo tali valori il risultato di un'attenta analisi dei parametri numerici.
  Entrambe le scelte comportano errori relativi contenuti, come \`e possibile
  osservare in tabella \ref{Tab1}.  Pi\`u in dettaglio essi 
  ($\epsilon_{ij estr}$) sono decisamente inferiori a quelli ottenuti senza
  estrapolazione ($\epsilon_{ij}$) e risultano inoltre del medesimo
  ordine\footnote{Si noti che nell'ultimo caso (v. figura \ref{con1})
  la curva numerica non interseca  
  quella analitica (non esiste un punto di minimo) ed il flesso \`e molto distante 
  dal corpo. 
  Tuttavia, questo non \`e penalizzante poich\'e il valore di $\Phi_{xy}$ nella 
  zona di ristagno \`e piccolo rispetto alle altre componenti di
  $\nabla\nabla\Phi$.}.
  Nella stessa tabella sono infine riportati i valori analitici dei
  gradienti di velocit\`a ed i coefficienti $c$ della (\ref{Pos}) per le due 
  estrapolazioni, indicati rispettivamente come $c_{min}$ e $c_{flesso}$. 

 Poich\'e in generale la soluzione analitica di un problema non \`e
 nota, \`e impossibile valutare dove venga commesso numericamente il minimo errore,
 si pu\`o soltanto determinarne un'approssimazione.
 Al contrario, la strada basata sul flesso \`e sempre applicabile, perch\'e se questo
 punto esiste pu\`o essere individuato con precisione.
 Se invece la curva numerica non presenta flesso, l'informazione
 relativa alla curvatura diventa poco rilevante, in altri termini risulta modesto il
 gradiente da valutare. In tale circostanza si pu\`o arrivare molto vicini alla 
 superficie del corpo per il primo nodo dell'estrapolazione, in particolare, un
 valore di $c$ pari a $\frac{1}{\sqrt{2}}$ corrisponde ad una distanza soddisfacente.
 Non si commettono, tuttavia, errori notevoli valutando la grandezza
 direttamente sul corpo.   

 Pertanto, avendo verificato, mediante le precedenti figure, che le due 
 tecniche descritte forniscono risultati confrontabili, si decide di utilizzare nel
 seguito l'estrapolazione che adopera il flesso.        

\begin{figure}[htb]
    \vskip -0.0cm
      \epsfxsize=.9\textwidth
      \epsfysize=.7\textwidth
      \makebox[.8\textwidth]{\epsfbox{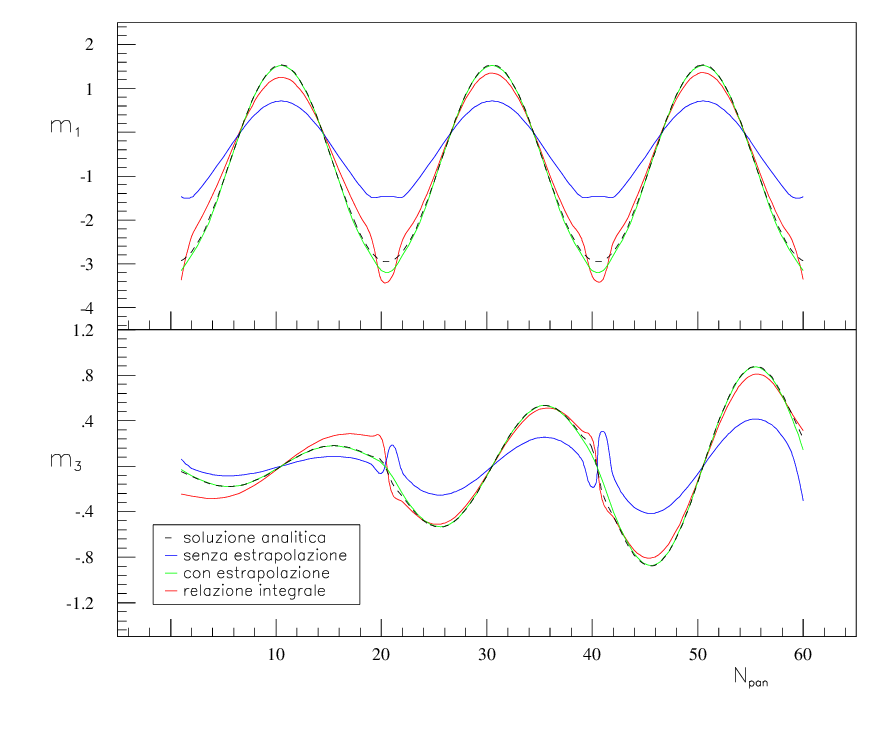}}
    \vskip -0.5cm
    \caption{Caso di una sfera di raggio $R=1$:
             confronto analitico--numerico dell'andamento di
             $m_1$ e $m_3$. Le curve numeriche si riferiscono all'uso della
             definizione (\ref{Def_mj}), senza estrapolazione e con una tecnica
             a $9$ nodi, e all'uso della relazione integrale (\ref{Int_mj}).
             \label{sfmj}
             }
     \vskip -4.cm
     \vspace*{-4.cm}
         
\end{figure}

 Come ultimo risultato, in figura \ref{sfmj} \`e mostrato
 il confronto relativo all'andamento di alcuni termini $m_j$, in particolare $m_{1}$
 e $m_{3}$.
 La curva tratteggiata rappresenta la soluzione analitica, 
 mentre le altre sono relative all'applicazione della definizione \ref{Def_mj},
 con e senza estrapolazione, e all'utilizzo della relazione integrale \ref{Int_mj}.
 Dal grafico appare evidente che il calcolo diretto basato sulla  
 definizione sia affetto da errori numerici di una certa entit\`a. 
 L'introduzione, invece, di una opportuna estrapolazione
 fornisce risultati che si sovrappongono a quelli analitici.
 Dal canto suo, la relazione (\ref{Int_mj}) presenta un accordo complessivamente
 buono con il risultato analitico. 
 Sbaglia tuttavia gli $m_j$ relativi alla prima 
 linea d'acqua; si pu\o dare una giustificazione di ci\o considerando che stiamo
 discretizzando un problema integrale su una superficie aperta, e ci\o   
 comporta dei problemi sul contorno di questa (\cite{AMuskhelishvili}). 
 A tal riguardo, in tabella \ref{Tab3} viene 
 riportato lo scarto quadratico medio dei termini $m_1$ $m_2$ e $m_3$ 
\[ 
      \sqrt{\sum_{i=1}^{N_{pan}} \frac{(m_{j anal}\,-\,m_{j num})_i^2}{N_{pan}}}\,,
\]
 Come si pu\`o osservare, all'aumentare del numero dei pannelli sulla superficie del
 corpo, i valori numerici approssimano sempre meglio quelli
 analitici\footnote{Il numero 
 di pannelli indicato in tabella si riferisce a $\frac{1}{4}$ di sfera (20x10).
 Essi sono ordinati seguendo le linee d'acqua a partire da quella a $z=0$.}.
 La tabella \ref{Tab4} mostra il medesimo scarto quadratico medio per le diverse
 soluzioni numeriche.
 In particolare, le due tecniche di estrapolazione, entrambe basate sul punto
 di flesso come primo nodo, sono caratterizzate rispettivamente da $9$ e $4$ 
 nodi. 
 L'estrapolazione con un polinomio dell'ottavo ordine fornisce, tra tutte
 le soluzioni presentate, i valori pi\`u piccoli degli errori.

\section{Casi esaminati: Ovoide di Rankine}
 La curva delle derivate numeriche del potenziale intorno ad una sfera, al variare
 della distanza dal pannello, rimane qualitativamente invariata passando al caso di
 una carena. 
 A tal proposito, in figura \ref{wigconfr} sono mostrati gli andamenti di alcune 
 componenti del gradiente di velocit\`a per una carena Wigley, i quali, 
 analogamente al caso della sfera, presentano un punto di flesso.
 Pertanto, la logica della tecnica di estrapolazione discussa nel precedente
 paragrafo pu\`o essere applicata anche al caso di carene realistiche.
 Tuttavia, prima di considerare problemi relativi a carene di navi, \`e utile
 analizzare una geometria pi\`u simile a questa ma per la quale esistano soluzioni
 analitiche. In tal modo sar\`a pi\`u facile selezionare i valori dei parametri
 per la procedura di estrapolazione.   
 \vskip .25cm
  
\begin{figure}[htb]
      \epsfxsize=.5\textwidth
      \epsfxsize=.5\textwidth
      \makebox[.9\textwidth]{\epsfbox{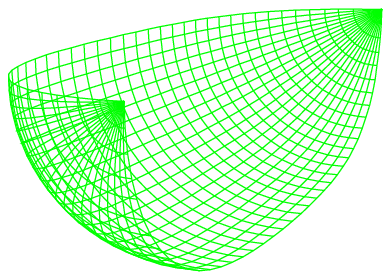}}
    \vskip -0.5cm
    \caption{Esempio di un tipico reticolo di calcolo per un ovoide di Rankine. 
             \label{ovoide}
             }
\end{figure}
\begin{figure}[htb]
      \epsfxsize=\textwidth
      \makebox[\textwidth]{\epsfbox{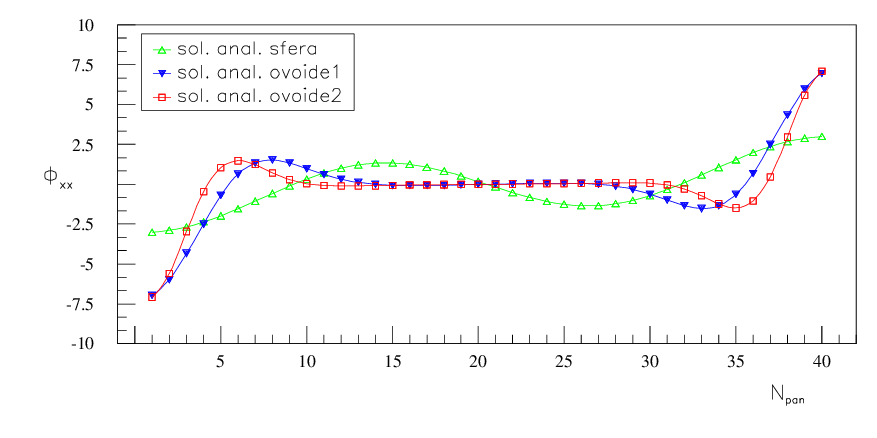}}
    \vskip -0.5cm
    \caption{ Andamenti analitici di $\Phi_{xx}$ per una sfera di raggio
              $R = 1$ e per due  geometrie di ovoide aventi entrambe $m$ pari
              a $1$; $d$ vale invece $1$ per la prima (ovoide1) e $2$ per 
              la seconda (ovoide2).  
             \label{ana1}
             }
\end{figure}
\begin{figure}[htb]
    \vskip 1.0cm
      \epsfxsize=\textwidth
      \makebox[\textwidth]{\epsfbox{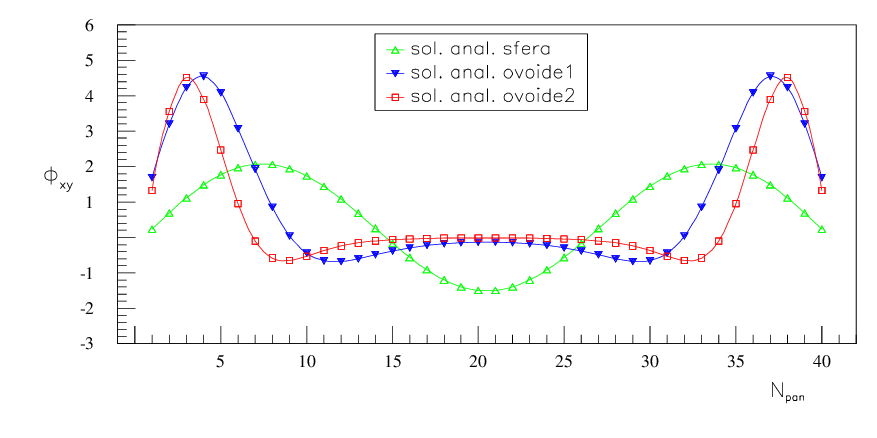}}
    \vskip -0.5cm
    \caption{ Andamenti analitici di $\Phi_{xy}$ per una sfera di raggio
              $R = 1$ e per due  geometrie di ovoide aventi entrambe $m$ pari
              a $1$; $d$ vale invece $1$ per la prima (ovoide1) e $2$ per 
              la seconda (ovoide2).  
             \label{ana2}
             }
    \vskip -4.0cm
\end{figure}
\begin{figure}[htb]
      \epsfxsize=\textwidth
      \makebox[\textwidth]{\epsfbox{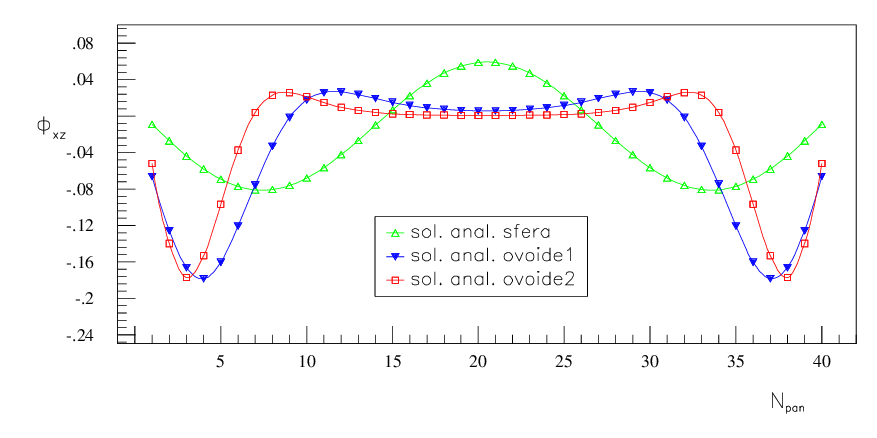}}
    \vskip -0.5cm
    \caption{ Andamenti analitici di $\Phi_{xz}$ per una sfera di raggio
              $R = 1$ e per due  geometrie di ovoide aventi entrambe $m$ pari
              a $1$; $d$ vale invece $1$ per la prima (ovoide1) e $2$ per 
              la seconda (ovoide2).  
             \label{ana3}
             }
\end{figure}
\begin{figure}[htb]
    \vskip 1.0cm
      \epsfxsize=\textwidth
      \makebox[\textwidth]{\epsfbox{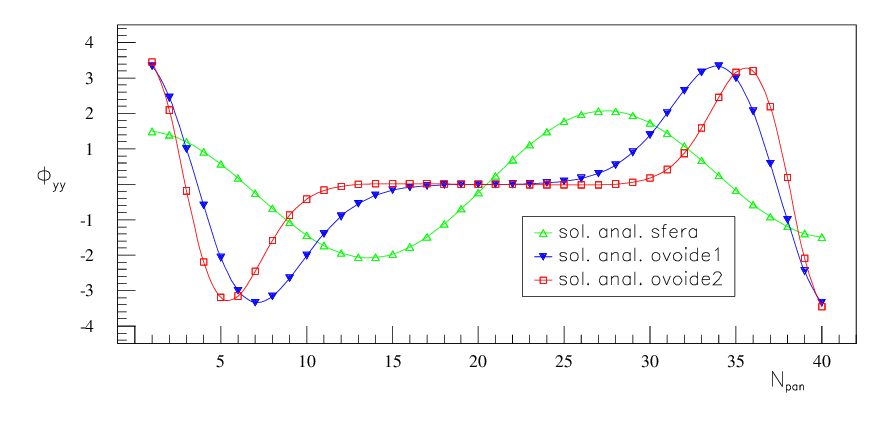}}
    \vskip -0.5cm
    \caption{ Andamenti analitici di $\Phi_{yy}$ per una sfera di raggio
              $R = 1$ e per due  geometrie di ovoide aventi entrambe $m$ pari
              a $1$; $d$ vale invece $1$ per la prima (ovoide1) e $2$ per 
              la seconda (ovoide2).  
             \label{ana4}
             }
    \vskip -4.0cm
\end{figure}
\begin{figure}[htb]
      \epsfxsize=\textwidth
      \makebox[\textwidth]{\epsfbox{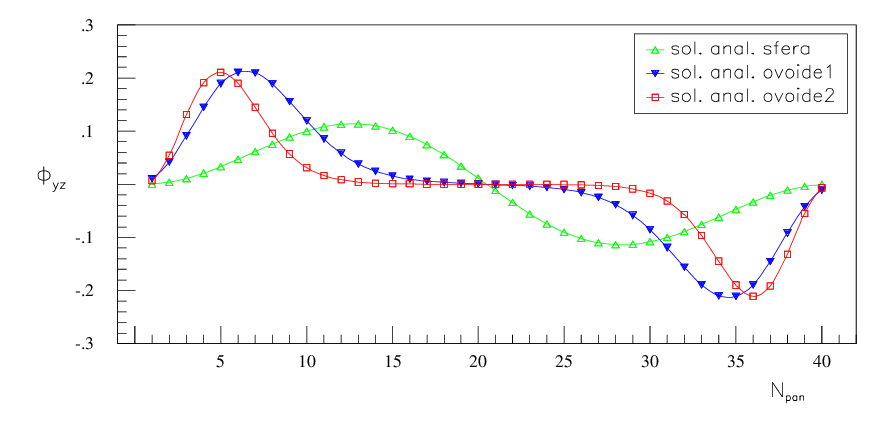}}
    \vskip -0.5cm
    \caption{ Andamenti analitici di $\Phi_{yz}$ per una sfera di raggio
              $R = 1$ e per due  geometrie di ovoide aventi entrambe $m$ pari
              a $1$; $d$ vale invece $1$ per la prima (ovoide1) e $2$ per 
              la seconda (ovoide2).  
             \label{ana5}
             }
\end{figure}
\begin{figure}[htb]
    \vskip 1.0cm
      \epsfxsize=\textwidth
      \makebox[\textwidth]{\epsfbox{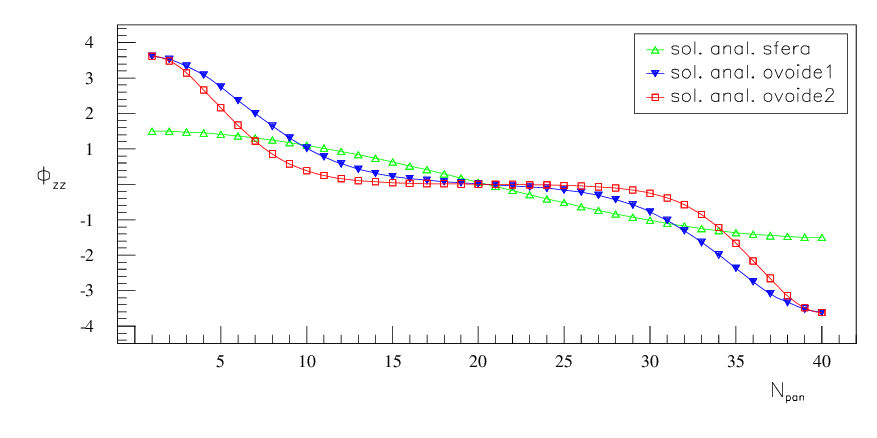}}
    \vskip -0.5cm
    \caption{ Andamenti analitici di $\Phi_{zz}$ per una sfera di raggio
              $R = 1$ e per due  geometrie di ovoide aventi entrambe $m$ pari
              a $1$; $d$ vale invece $1$ per la prima (ovoide1) e $2$ per 
              la seconda (ovoide2).  
             \label{ana6}
             }
    \vskip -4.0cm
\end{figure}

 Si esamina, in particolare, il caso di un ovoide (v. fig. \ref{ovoide}) il
 quale si configura come corpo di rivoluzione generato da una coppia
 pozzo--sorgente investita da una corrente uniforme \cite{ABatch}.
 Le caratteristiche di tale corpo dipendono dall'intensit\`a delle
 singolarit\`a ($m$) e dalla loro distanza relativa ($d$), che possono renderlo 
 pi\`u o meno affusolato. A tal proposito, nelle figure \ref{ana1}--\ref{ana6},
 gi\`a commentate in precedenza, vengono confrontate le derivate seconde di $\Phi$
 per due ovoidi con diverso rapporto dei semiassi.
 Si considerano naturalmente le medesime condizioni di
 avanzamento in un fluido infinitamente esteso. Come si pu\`o osservare,
 all'allungarsi della geometria la zona associata a valori elevati dei
 gradienti si localizza agli estremi del corpo, in corrispondenza dei punti di
 ristagno.  Le curve relative diventano pertanto meno regolari.

\begin{figure}[htb]
      \epsfxsize=\textwidth
      \makebox[\textwidth]{\epsfbox{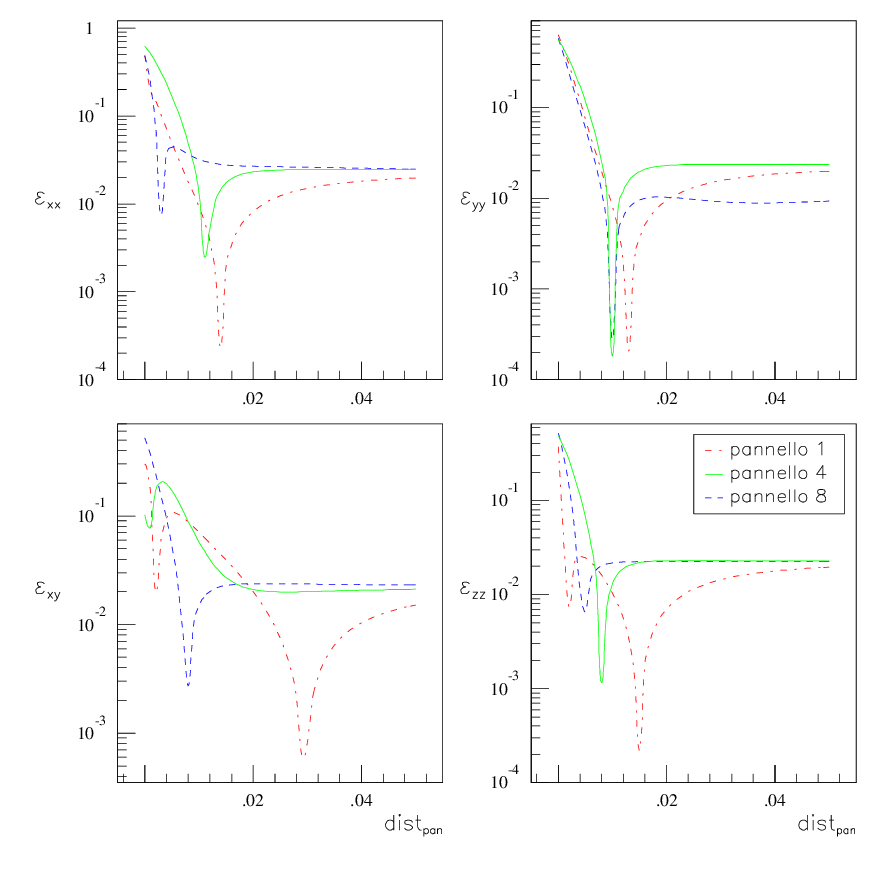}}
    \vskip -0.5cm
    \caption{Caso di un ovoide con $m = 0.04$ e $d = 0.45$: andamento dell'errore
             relativo $\epsilon$ al variare della distanza normale dai pannelli 1,
             4 e 8. 
             \label{errov}
             }
\end{figure}

 In particolare,
 per $m=0.04$ e $d=0.45$, valori che corrispondono ad un ovoide
 abbastanza affusolato, si realizza una forma che ha un $L/B$ 
 uguale a quello di una carena Wigley.
 Considerando pertanto tale geometria, nella figura
 \ref{errov} vengono riportati gli errori relativi delle componenti di
 $\nabla \nabla \Phi$ ad variare della distanza dal pannello. I risultati si 
 riferiscono ad un reticolo con $100$ pannelli longitudinali e $80$ trasversali, 
 inoltre, riguardano pannelli del corpo situati sulla prima linea d'acqua
 ($\beta=0$) con un angolo $\alpha$ all'incirca pari, in sequenza, a  $0$, $15$ e
 $30$ gradi.
 Come si pu\`o osservare gli andamenti sono qualitativamente analoghi a quelli 
 visti per la sfera.

\begin{figure}[htb]
 \vskip -1.cm
      \epsfxsize=\textwidth
      \makebox[.9\textwidth]{\epsfbox{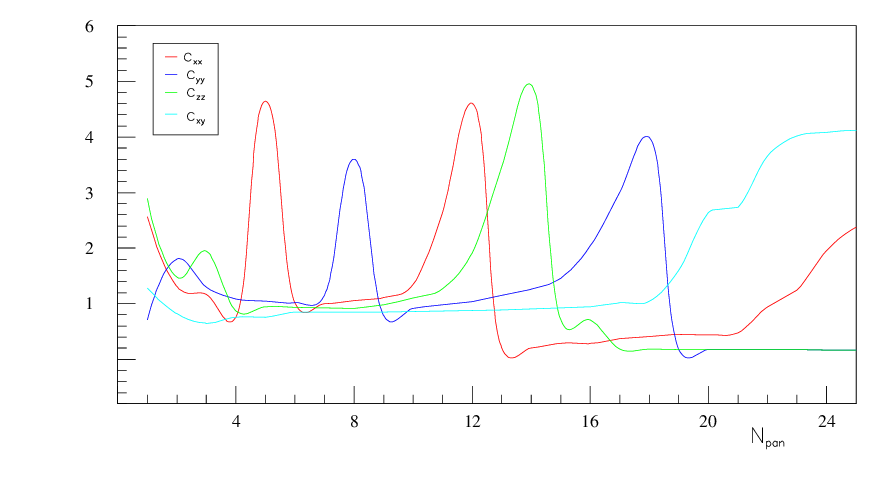}}
    \vskip -1.0cm
    \caption{Caso di un ovoide con $m = 0.04$ e $d = 0.45$: andamento del parametro
             $c$ della (\ref{Pos}) per le componenti di $\nabla\nabla \Phi$ della 
             tabella \ref{Tab2}, al variare del pannello. 
             \label{coef_dist}
             }
    \vskip -4.0cm
    \vspace*{-4.cm} 
\end{figure}

 Per lo stesso ovoide viene inoltre riportata la tabella \ref{Tab2} analoga alla 
 \ref{Tab1}, mentre nella figura \ref{coef_dist} \`e riportata la curva del 
 coefficiente $c$ del flesso per alcune componenti di $\nabla\nabla\Phi$, lungo
 la prima linea d'acqua e con $\alpha$ che va da $0$ a $90$ gradi. 

\begin{figure}[htb]
    \vskip -1.cm
      \epsfxsize=\textwidth
      \makebox[.9\textwidth]{\epsfbox{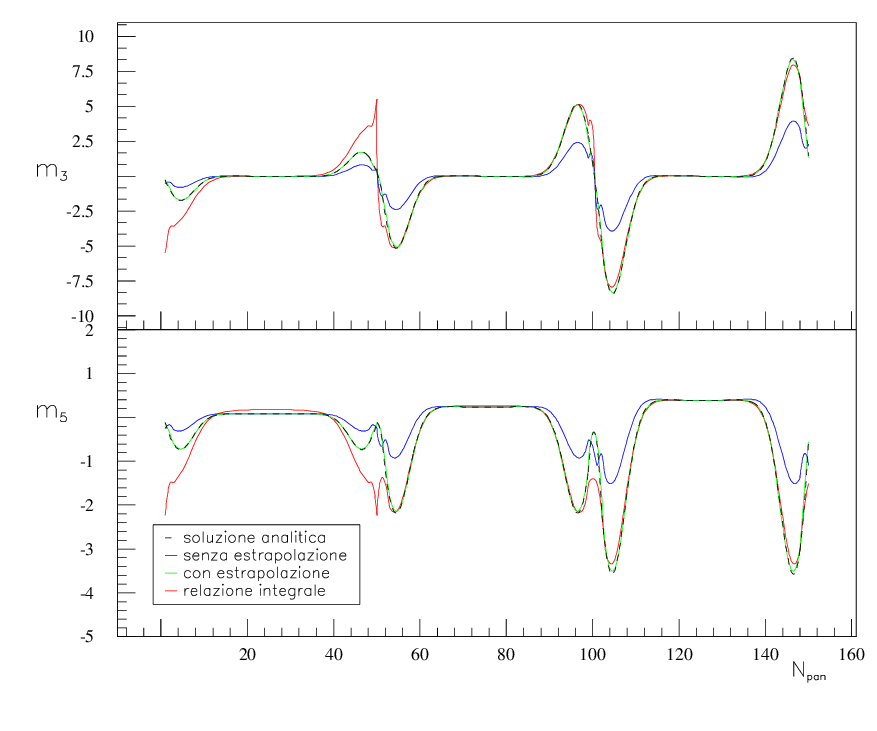}}
    \vskip -1.0cm
    \caption{Caso di un ovoide con $m = 0.04$ e $d = 0.45$: confronto 
             analitico--numerico dell'andamento di $m_3$ e $m_{5}$. Le curve
             numeriche si riferiscono all'uso della definizione (\ref{Def_mj}), senza
             estrapolazione e con una tecnica a $9$ nodi, e della relazione
             integrale (\ref{Int_mj}).
             \label{ovmj}
             }
\end{figure}

 Da ultimo, in figura \ref{ovmj} \`e mostrato il confronto dei termini $m_{3}$ e
 $m_{5}$, al variare del pannello. Come per la sfera, la curva tratteggiata si 
 riferisce alla soluzione analitica, le altre invece all'applicazione della 
 definizione \ref{Def_mj}, con e senza estrapolazione, e all'uso della relazione 
 (\ref{Int_mj}).
 Anche in questo caso l'applicazione della  
 \ref{Def_mj} comporta  errori numerici rilevanti e similmente al caso della sfera  
 la relazione integrale presenta errori non trascurabili in corrispondenza della
 prima linea d'acqua.  
 Infine, in tabella \ref{Tab5} viene riportato lo scarto quadratico medio dei 
 termini $m_j$,  per le diverse soluzioni numeriche\footnote{Non \`e presente il
 valore relativo a $m_4$ essendo questo nullo per un solido di rivoluzione.}. 
 Come si pu\`o osservare, gli scarti quadratici per $m_2$ e $m_3$ e, rispettivamente,
 $m_5$ e $m_6$ sono identici per quasi tutte le metodologie.
 Essendo, infatti, l'ovoide un solido di rivoluzione,
 {\em sway} e {\em heave}, da una parte, e {\em pitch} e
 {\em yaw}, dall'altra, risultano indistinguibili per il flusso di doppio--modello. 
 Diversamente, nel caso della relazione integrale (\ref{Int_mj}) questa simmetria
 pu\`o essere realizzata solo per reticoli molto fitti, tali da soddisfare l'ipotesi
 di superficie {\em wall--sided} sul piano $z=0\,$.  

 Ancora una volta l'estrapolazione con un polinomio dell'ottavo grado permette di
 ottenere i valori numerici pi\`u vicini alla soluzione analitica.

\newpage
 \clearpage
    \vskip -4.0cm
\newpage
 \clearpage
    \vskip -4.0cm
\newpage
 \clearpage
    \vskip -4.0cm
\newpage
 \clearpage
\newpage
 \clearpage
\newpage
  \clearpage
\newpage
  \clearpage
{\large
\begin{table}[htb]
 \hskip 15mm \begin{tabular}{||l||l|l|l||} \hline \hline
                     & {\sc pan 1}  & {\sc pan 5} & {\sc pan 10} \\ \hline \hline\hline 
  {\sc $\Phi_{xx}$}             & $-0.2880\cdot$E$+01$ & $+0.1361$            & $+0.3587$          \\ \hline
  {\sc $\epsilon_{xx}$}         & $+0.4686$            & $+0.6557$            & $+0.5275$           \\ \hline
  {\sc $c_{xx\, min} $}    & $+0.4564$             & $+0.1744\cdot$E$+01$ & $+0.6393$           \\ \hline
  {\sc $\epsilon_{xx\, estr} $} & $+0.5365\cdot$E$-01$ & $+0.3835\cdot$E$-01$ & $+0.3299\cdot$E$-01$ \\ \hline
  {\sc $c_{xx\, flesso} $} & $+0.1141\cdot$E$+01$ & $+0.1467\cdot$E$+01$ & $+0.8311$           \\ \hline
  {\sc $\epsilon_{xx\, estr} $} & $-0.8399\cdot$E$-01$ & $-0.8389\cdot$E$-01$ & $+0.9964\cdot$E$-02$ \\ \hline\hline
  {\sc $\Phi_{yy}$}             & $+0.1381\cdot$E$+01$ & $-0.1280\cdot$E$+01$ & $-0.4758$           \\ \hline
  {\sc $\epsilon_{yy}$}         & $+0.3773$            & $+0.5554$            & $+0.5347$            \\ \hline
  {\sc $c_{yy\, min} $}    & $+0.2282$            & $+0.7134$            & $+0.6393$ \\ \hline
  {\sc $\epsilon_{yy\, estr} $} & $-0.1920$            & $+0.4457\cdot$E$-01$ & $+0.3460\cdot$E$-01$ \\ \hline
  {\sc $c_{yy\, flesso} $} & $+0.8269\cdot$E$+00$ & $+0.9513$            & $+0.8311$            \\ \hline
  {\sc $\epsilon_{yy\, estr} $} & $+0.1346\cdot$E$-01$  & $+0.1082\cdot$E$-01$ & $+0.1072\cdot$E$-01$ \\ \hline\hline
  {\sc $\Phi_{zz}$}             & $+0.1499\cdot$E$+01$ & $+0.1144\cdot$E$+01$ & $+0.1171$         \\ \hline
  {\sc $\epsilon_{zz}$}         & $+0.5527$            & $+0.5434$            & $+0.5567$         \\ \hline
  {\sc $c_{zz\, min} $}    &                      & $+0.5549$            & $+0.6393$         \\ \hline
  {\sc $\epsilon_{zz\, estr} $} &                      & $+0.1641\cdot$E$-01$ & $+0.3954\cdot$E$-01$ \\ \hline
  {\sc $c_{zz\,flesso} $}  & $+0.1940$             & $+0.7927$            & $+0.8311$           \\ \hline
  {\sc $\epsilon_{zz\, estr} $} & $+0.8741\cdot$E$-01$ & $-0.1077\cdot$E$-01$ & $+0.1304\cdot$E$-01$ \\ \hline\hline
  {\sc $\Phi_{xy}$}             & $+0.7462$            & $+0.1864\cdot$E$+01$ & $-0.1481\cdot$E$+01$ \\ \hline
  {\sc $\epsilon_{xy}$}         & $+0.1981\cdot$E$+01$ & $+0.5112$            & $+0.5359$         \\ \hline
  {\sc $c_{xy\, min} $}    &                     & $+0.6342$            & $+0.6393$         \\ \hline 
  {\sc $\epsilon_{xy\, estr} $} &                     & $+0.4508\cdot$E$-01$ & $+0.3445\cdot$E$-01$ \\ \hline 
  {\sc $c_{xy\, flesso} $} &  $+0.2233\cdot$E$+01$  & $+0.9116$            & $+0.8311$            \\ \hline 
  {\sc $\epsilon_{xy\, estr} $} & $-0.1569\cdot$E$-01$  & $+0.2084\cdot$E$-02$ & $+0.1056\cdot$E$-01$ \\ \hline 
  \end{tabular}  
    \vskip 0.5cm
       \caption{Caso di una sfera di raggio $R=1$:
         per alcune componenti di $\nabla\nabla\Phi$
         sono mostrati in sequenza il valore; l'errore relativo associato al 
         calcolo senza estrapolazione; il parametro $c$ e l'errore commesso 
         usando il punto di minimo come primo nodo dell'estrapolazione; 
         il parametro $c$ e l'errore commesso usando il punto
         di flesso come primo nodo dell'estrapolazione. \label{Tab1}}
\end{table}
}
\newpage
 \clearpage
{\Large
\begin{table}[htb]
 \vskip -0cm
\hskip 15mm \begin{tabular}{||l||l|l|l||} \hline 
                  & 20x10  & 40x20 & 60x30 \\ \hline\hline
 {\sc $m_1$}  & $ 0.2962$ & $ 0.1112$  & $ 0.9526\cdot$E$-01$  \\ \hline
 {\sc $m_2$}  & $ 0.1535$ & $ 0.8187\cdot$E$-01$  & $ 0.5644\cdot$E$-01$  \\ \hline
 {\sc $m_3$}  & $ 0.1192$ & $ 0.6189\cdot$E$-01$  & $ 0.4169\cdot$E$-01$  \\ \hline
  \end{tabular}  
    \vskip 0.5cm
   \caption{Caso di una sfera di raggio $R=1$:
            scarto quadratico medio analitico--numerico
            relativo ai primi $3$ termini $m_j$ valutati con la relazione integrale
            (\ref{Int_mj}) per reticoli ad infittimento crescente.
            \label{Tab3}}
\end{table}
}
{\large
\begin{table}[htb]
 \hskip 25mm \begin{tabular}{||l||l|l|l|l||} \hline 
                  & {\sc senza estrap.}  & {\sc rel. int.} & {\sc estrap. 1} & {\sc estrap. 2}\\ \hline \hline
 {\sc $m_1$}  & $ 0.8582$ & $ 0.2030$  & $ 0.7942\cdot$E$-01$   & $ 0.7319\cdot$E$-01$ \\ \hline
 {\sc $m_2$}  & $ 0.7134$ & $ 0.1455$  & $ 0.6248\cdot$E$-01$  & $ 0.8665\cdot$E$-01$ \\ \hline
 {\sc $m_3$}  & $ 0.7134$ & $ 0.1157$  & $ 0.6297\cdot$E$-01$  & $ 0.8671\cdot$E$-01$ \\ \hline
  \end{tabular}  
    \vskip 0.5cm
   \caption{Caso di una sfera di raggio $R=1$:
            scarti quadratici medi relativi ai primi $3$ termini $m_j$ valutati
            senza estrapolazione, con
            la relazione integrale, con un'estrapolazione a $9$ nodi ed infine 
            con un'estrapolazione a $4$ nodi.  
            \label{Tab4}}
\end{table}
}
    \vskip -4.cm
     \vspace*{-4.cm}
\newpage
 \clearpage
\newpage
 \clearpage
{\large
\begin{table}[htb]
 \vskip -0.0cm
\hskip 15mm \begin{tabular}{||l||l|l|l||} \hline \hline
                     & {\sc pan 1}  & {\sc pan 5} & {\sc pan 10} \\ \hline \hline\hline
  {\sc $\Phi_{xx}$}             & $-0.3553\cdot$E$+02$ & $+0.3653$            & $+0.2427\cdot$E$+01$  \\ \hline
  {\sc $\epsilon_{xx}$}         & $+0.4905$            & $+0.1936\cdot$E$+01$ & $+0.7687$           \\ \hline
  {\sc $c_{xx\, flesso} $} & $+0.2568\cdot$E$+01$ & $+0.4644\cdot$E$+01$ & $+0.1340\cdot$E$+01$ \\ \hline
  {\sc $\epsilon_{xx\, estr} $} & $+0.8080\cdot$E$-01$ & $+0.3031\cdot$E$+01$ & $-0.6810\cdot$E$-01$ \\ \hline\hline
  {\sc $\Phi_{yy}$}             & $+0.1750\cdot$E$+02$ & $-0.1280\cdot$E$+02$ & $-0.5914\cdot$E$+01$  \\ \hline
  {\sc $\epsilon_{yy}$}         & $+0.6313$            & $+0.5593$            & $+0.5792$            \\ \hline
  {\sc $c_{yy\, flesso} $} & $+0.2889\cdot$E$+01$ & $+0.9477$            & $+0.1102\cdot$E$+01$ \\ \hline
  {\sc $\epsilon_{yy\, estr} $} & $+0.1086$            & $+0.3153\cdot$E$-01$ & $-0.3769\cdot$E$-01$ \\ \hline\hline
  {\sc $\Phi_{zz}$}             & $+0.1803\cdot$E$+02$ & $+0.1244\cdot$E$+02$ & $+0.3487\cdot$E$+01$ \\ \hline
  {\sc $\epsilon_{zz}$}         & $+0.3731$            & $+0.5189$            & $+0.4473$         \\ \hline
  {\sc $c_{zz\,flesso} $}  & $+1.2842\cdot$E$+01$ & $+0.7581$            & $+0.8569$           \\ \hline
  {\sc $\epsilon_{zz\, estr} $} & $+0.5271\cdot$E$-01$ & $-0.8483\cdot$E$-02$ & $-0.1379\cdot$E$-01$ \\ \hline\hline
  {\sc $\Phi_{xy}$}             & $+0.5300\cdot$E$+01$ & $+0.1763\cdot$E$+02$ & $-0.3274\cdot$E$+01$ \\ \hline
  {\sc $\epsilon_{xy}$}         & $-0.3001$            & $+0.5112$            & $+0.5100$         \\ \hline
  {\sc $c_{xy\, flesso} $} &                      & $+0.1042\cdot$E$+01$ & $+0.9181$ \\ \hline 
  {\sc $\epsilon_{xy\, estr} $} &                  & $-0.8316\cdot$E$-02$ & $+0.1751\cdot$E$+00$  \\ \hline 
  \end{tabular}  
  \vskip 0.5cm
       \caption{Caso di un ovoide con $m = 0.04$ e $d = 0.45$:
         per alcune componenti di $\nabla\nabla\Phi$
         sono mostrati in sequenza il valore, l'errore relativo associato al 
         calcolo senza estrapolazione, il parametro $c$ e l'errore commesso 
         usando il flesso come primo nodo dell'estrapolazione. 
         \label{Tab2}}
\end{table}
}
\newpage 
 \clearpage
{\large
\begin{table}[htb]
 \hskip 15mm \begin{tabular}{||l||l|l|l|l||} \hline 
            & {\sc senza estrap.}  & {\sc rel. int.} & {\sc estrap. 1} & {\sc estrap. 2} \\ \hline \hline
 {\sc $m_1$}  & $ 6.355$ & $ 1.336$  & $ 0.6644$   & $ 1.107 $ \\ \hline
 {\sc $m_2$}  & $ 3.672$ & $ 0.9862$  & $ 0.2836$  & $ 0.8119$ \\ \hline
 {\sc $m_3$}  & $ 3.672$ & $ 0.8035$  & $ 0.2845$  & $ 0.8119$ \\ \hline
 {\sc $m_5$}  & $ 1.701$ & $ 0.3440$  & $ 0.1379$  & $ 0.3870$ \\ \hline
 {\sc $m_6$}  & $ 1.701$ & $ 0.4237$  & $ 0.1374$  & $ 0.3870$ \\ \hline
  
  \end{tabular}  
  \vskip 0.5cm
   \caption{Caso di un ovoide con $m = 0.04$ e $d = 0.45$: 
            scarti quadratici medi relativi ai termini $m_j$ valutati senza
            estrapolazione, con la relazione integrale, con un'estrapolazione a $9$
            nodi ed infine con un'estrapolazione a $3$ nodi.  
            \label{Tab5}}
    \vskip -4.0cm
    \vspace*{-4.cm} 
\end{table}
}
\newpage
 \clearpage
\section{Casi esaminati: Carene di Superficie}
\begin{figure}[htb]
 \vskip  1.5cm
      \epsfxsize=.9\textwidth
      \makebox[.9\textwidth]{\epsfbox{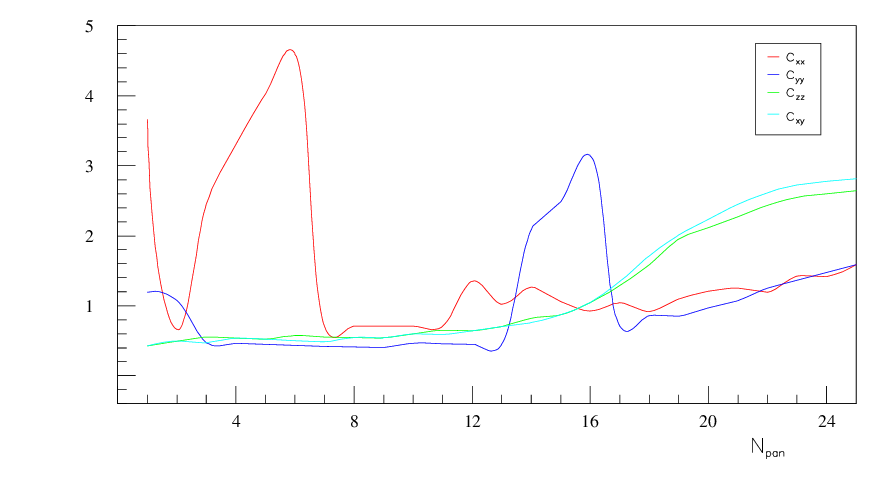}}
    \vskip -1.0cm
    \caption{Caso di una carena Wigley: andamento del parametro $c$ della
             (\ref{Pos}) relativo ad alcune componenti di $\nabla\nabla \Phi$, al
             variare del pannello. 
             \label{coef_distw}
             }
    \vskip -4.0cm
    \vspace*{-4.cm} 
\end{figure}

\begin{figure}[htb]
 \vskip  1.0cm
      \epsfxsize=.9\textwidth
      \epsfysize=.9\textwidth
      \makebox[.95\textwidth]{\epsfbox{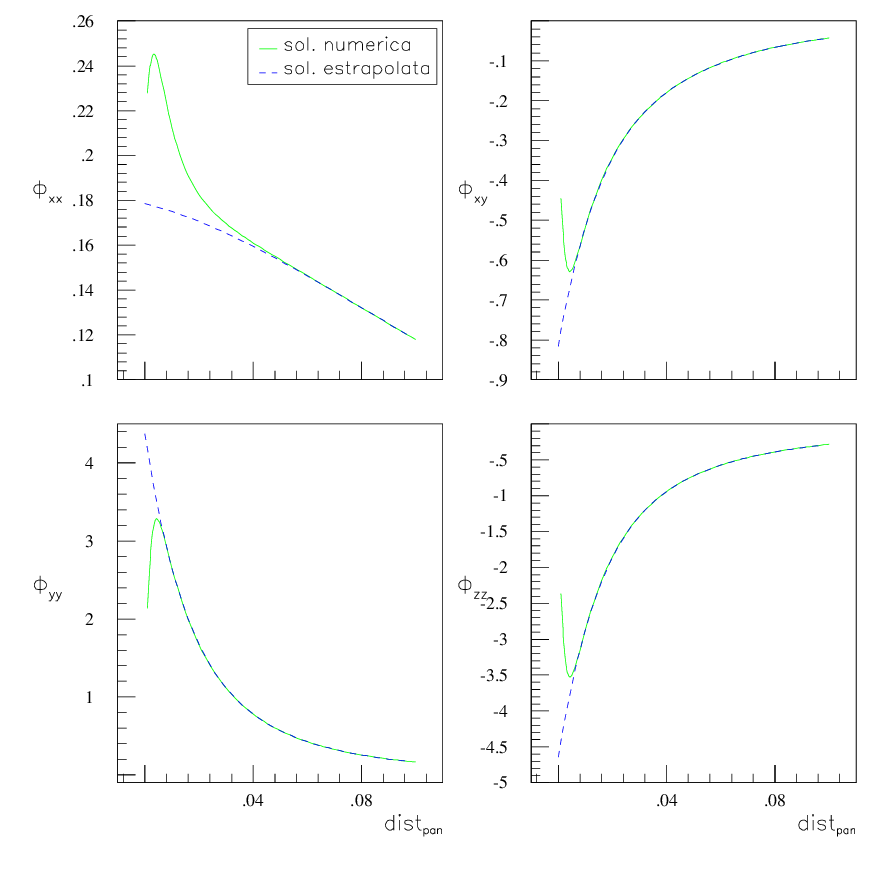}}
    \vskip -1.0cm
    \caption{ Caso di una carena Wigley: andamento numerico di alcune componenti di
              $\nabla \nabla\Phi$ al variare della distanza normale da un  pannello,
              con (tratto discontinuo) e senza estrapolazione (tratto continuo). 
             \label{wigconfr}
             }
\end{figure}

 La tecnica di estrapolazione utilizzata per il calcolo di $\nabla\nabla\Phi$, 
 applica un polinomio di ottavo grado con il primo nodo sul punto di flesso.
 La posizione di quest'ultimo viene cercata entro una
 distanza dal generico pannello caratterizzata da un parametro $c$
 dell' ${\cal O}(1)$ (v. rel. (\ref{Pos})).
 Infine, la distanza tra i nodi successivi \`e assunta pari a
   $\sqrt{2}\,{\cal A}^{\frac{1}{2}}_{pan}\,$.
 La figura \ref{wigconfr} mostra, in particolare, l'andamento numerico di alcune 
 componenti del gradiente di velocit\`a per una carena Wigley in funzione 
 della distanza da un generico pannello. Come si pu\`o osservare, le curve 
 presentano un flesso. Le linee tratteggiate dei grafici si riferiscono inoltre
 all'estrapolazione basata su tale punto, la cui posizione varia
 al variare del pannello secondo la legge rappresentata in figura \ref{coef_distw}.
\newpage
 \clearpage
\newpage
 \clearpage


\begin{thebibliography}{99}

\bibitem{LoSchiavo} 
 M. Lo Schiavo,  {\em Sistemi Dinamici}  Appunti per il corso. Universit\a "La Sapienza", Roma 
A.A. 1995/96

\bibitem{Mastroddi}
F. Mastroddi,  {\em Aeroelasticit\a applicata}  Appunti per il corso.  Universit\a "La Sapienza", Roma A.A. 1996/97

\bibitem{Nelson}
Nelson, {\em Flight Stability and Automatic Control} McGraw-Hill Series in Aeronautical and 
Aerospace Engineering.

\bibitem{Newman2}
 J.N. Newman, {\em Marine Hydrodynamics}, The Mit Press Cambridge,  
 Massachusetts, and London, England.
\end{thebibliography}

\begin{thebibliography}{99}

\bibitem{Newman3}
 J.N. Newman, {\em Marine Hydrodynamics}, The Mit Press Cambridge,  
 Massachusetts, and London, England.

\bibitem{LAMB3}
 H. Lamb, {\em Hydrodynamics}, Dover, 1932.

\bibitem{Anderson}
J.D. Anderson Jr., {\em Fundamentals of Aerodynamics}, McGraw Hill, Inc.

\bibitem{Touvia}
{\em Mathematical Approaches in Hydrodynamics}, T. Miloh ed.
Society for Industrial and Applied Mathematics. SIAM Philadelphia.
\end{thebibliography}

\begin{thebibliography}{99}

\bibitem{MarHy14}
 J. N. Newman, {\em The Theory of Ship Motions}, Advances In Applied Mechanics, 
 18.

\bibitem{Ogilvie4}
T. Francis Ogilvie,
 {\em Recent Progress Toward The Understanding and Prediction of Ship Motions.}
 David Taylor Model Basin Washington, D.C.


\bibitem{Seak4}
  M. Landrini, M. Greco, G. Graziani, 1997
  {\em Modelli in Frequenza per la Tenuta al Mare di Carene Convenzionali},
  Parte I: {\em Impostazione del problema e Metodo di Soluzione.} 
  Rapporto INSEAN 1996 - 33.

\end{thebibliography}

\begin{thebibliography}{99}

\bibitem{LAMB5}
 H. Lamb, {\em Hydrodynamics}, Dover, 1932.

\bibitem{Hess}
  J. L. Hess \& A. M. O. Smith, 1966
  {\em Calculation of non--lifting potential flow about arbitrary bodies},
   Prog. Aero. Sci. {\bf 8},1--138.

\bibitem{Chen}
 G. Chen and J. Zhou, {\em Boundary Element Methods}  
\end{thebibliography}

\begin{thebibliography}{99}
  
\bibitem{MarHy1}
 J. N. Newman, {\em The Theory of Ship Motions}, Advances In Applied Mechanics, 
 18.

\bibitem{Sclav}
P.D. Sclavounos, 
\newblock {\em Computation of wave ship interaction}, Departemenent of Ocean
       Engineering,M.I.T.,Cambridge,MA 02139,USA.

\bibitem{LAMB}
 H. Lamb, {\em Hydrodynamics}, Dover, 1932.

\bibitem{Hess_Smith}
  J. L. Hess \& A. M. O. Smith, 1966
  {\em Calculation of non--lifting potential flow about arbitrary bodies},
   Prog. Aero. Sci. {\bf 8},1--138.

\bibitem{Seak}
  M. Landrini, M. Greco, G. Graziani, 1997
  {\em Modelli in Frequenza per la Tenuta al Mare di Carene Convenzionali},
  Parte I: {\em Impostazione del problema e Metodo di Soluzione.} 
  Rapporto INSEAN 1996 - 33.

\bibitem{TimNew}
  R. Timman \& J. N. Newman, {\em The coupled damping coefficients of symmetric 
  ships}, Journal of Ship Research, Vol. 5, No. 4, pp. 34--55, 1962.  

\bibitem{MarHy}
 J. N. Newman, {\em Marine Hydrodynamics}, Cambridge University Press, 1977.

\bibitem{Journ}
 J. M. J. Journ\'ee, {\em Experiments and Calculations on Four Wigley Hull Forms},
 Report No. 909, Ship Hydromechanics Laboratory, Delft University of Technology,
 Delft, The Netherlands, 1992.

\bibitem{Gerrit1}
  J. Gerritsma, 
  {\em Measurements of Hydrodynamic Forces and Motions for a Modified Wigley Model},
  unpublished, 1986.

\bibitem{Nakos}
 D. E. Nakos, {\em Ship Wave Patterns and Motions by a Three Dimensional Rankine
 Panel Method}, MIT.   

\bibitem{Gerrit2}
  J. Gerritsma, 
  {\em Ship Motions in Longitudinal Waves}, International Shipbuilding Progress,
  Vol. 7, No. 66, 1960.

\bibitem{OgTu}
  T. F. Ogilvie, $\&$ E. O. Tuck,
  {\em  A rational Strip Theory for Ship Motions } -- Part 1,
  Report No. 013, Dept of Naval Architecture and Marine Ingineering, Univ. of 
  Michigan, USA, 1969.

\bibitem{Batch}
  G. K. Batchelor, 
  {\em An Introduction to Fluid Dynamics}, Cambridge University Press, 1967.

\bibitem{Muskhelishvili}
  N.I. Muskhelishvili,
  {\em Singular Integral Equations} 
  Dover Publication,Inc. Second Edition 1992.

\bibitem{Ogilvie}
T. Francis Ogilvie,
 {\em Recent Progress Toward The Understanding and Prediction of Ship Motions.}
 David Taylor Model Basin Washington, D.C.

\end{thebibliography}

\newpage
\clearpage
\begin{thebibliography}{99}
\bibitem{Lugni}
  G.Calcagno, C.Lugni, M. Landrini.
\newblock  {\em Identificazione delle caratteristiche 
di manovrabilit\a e tenuta al mare di veicoli marini 
mediante tecniche di prova impulsive.}
  Rapporto INSEAN 1993 - 54, Febbraio 1997.

\bibitem{Sclav7}
P.D. Sclavounos, 
\newblock {\em Computation of wave ship interaction.}, Departemenent of Ocean
       Engineering,M.I.T.,Cambridge,MA 02139,USA.

\bibitem{Ogilvie7}
T. Francis Ogilvie,
 {\em Recent Progress Toward The Understanding and Prediction of Ship Motions.}
 David Taylor Model Basin Washington, D.C.

\bibitem{Price}
H.J. Prins and A.J. Hermans,
{\em Time Domain Calculation of Drift Forces on Floating
Two Dimensional Object in Current and Waves.}
Delft Unifersity of Technology, The Netherlands.

\bibitem{Nakos time}
D.  E. Nakos,  D. Kring and P.D. Sclavounos,
{\em Rankine Panel Methods for Free-Surface
Flows.} Massachusetts Institute of Technology , USA

\bibitem{Campana}
E.F. Campana, D. Peri, 1997 
{\em "Simulazione Numerica delle Prove di Rimorchio in Acqua Calma. 
Parte I: Prove con Assetto Bloccato"} 
Rapporto INSEAN 1993 - 30.

\bibitem{aimeta}
M. Landrini, M. Greco, G. Graziani,
{\em Aspetti numerici della simulazione del flusso attorno ad una carena investita
     da onde}, Atti AIMETA, vol. I, 1997, Rapporto INSEAN 1996-29.

\bibitem{mari}
M.Greco, {\em Forze idrodinamiche e formazione ondosa nel moto di veicoli marini}
\\
Dipartimento di Meccanica e Aeronautica. Universit\a di Roma {\em La Sapienza},\\ 
Rapporto INSEAN 1993 - 29
\end{thebibliography}

\begin{thebibliography}{99}

\bibitem{ALAMB}
 H. Lamb, {\em Hydrodynamics}, Dover, 1932.

\bibitem{AHess_Smith}
  J. L. Hess \& A. M. O. Smith, 1966
  {\em Calculation of non--lifting potential flow about arbitrary bodies},
   Prog. Aero. Sci. {\bf 8},1--138.

\bibitem{ASeak}
  M. Landrini, M. Greco, G. Graziani, 1997
  {\em Modelli in Frequenza per la Tenuta al Mare di Carene Convenzionali},
  Parte I: {\em Impostazione del problema e Metodo di Soluzione.} 
  Rapporto INSEAN 1996 - 33.

\bibitem{AOgTu}
  T. F. Ogilvie, $\&$ E. O. Tuck,
  {\em  A Rational Strip Theory for Ship Motions } -- Part 1,
  Report No. 013, Dept of Naval Architecture and Marine Ingineering, Univ. of 
  Michigan, USA, 1969.

\bibitem{ABatch}
  G. K. Batchelor, 
  {\em An Introduction to Fluid Dynamics}, Cambridge University Press, 1967.

\bibitem{AMuskhelishvili}
  N.I. Muskhelishvili,
  {\em Singular Integral Equations} 
  Dover Publication,Inc. Second Edition 1992.

\end{thebibliography}
\end{document}